Skolkovo Institute of Science and Technology

# COMPUTATIONAL DESIGN OF NEW SUPERCONDUCTING MATERIALS AND THEIR TARGETED EXPERIMENTAL SYNTHESIS

*Doctoral Thesis*

by

DMITRII V. SEMENOK

DOCTORAL PROGRAM IN MATERIALS SCIENCE AND ENGINEERING

Supervisors
Professor Artem R. Oganov
Assistant Professor Alexander G. Kvashnin

Moscow - 2022

I hereby declare that the work presented in this thesis was carried out by myself at Skolkovo Institute of Science and Technology, Moscow, except where due acknowledgement is made, and has not been submitted for any other degree.

Candidate (Dmitrii Semenok)
Supervisors (Professor Artem R. Oganov, Assistant Professor Alexander G. Kvashnin)

## Abstract

In the last six years (2015–2021), many superconducting hydrides with critical temperatures $T_C$ of up to 253 K, a record for today, have been discovered. Now, a special field of hydride superconductivity at ultrahigh pressures has developed. For the most part, the properties of superhydrides are well described by the Migdal–Eliashberg theory of strong electron–phonon interaction, especially when anharmonicity of phonons is taken into account. The isotope effect, the effect of the magnetic field (up to 60–70 T) on the critical temperature and critical current in the hydride samples, the dependence of $T_C$ on the pressure and degree of doping — all data indicate that polyhydrides are conventional superconductors, the theory of which was created by Bardeen, Cooper, and Schrieffer in 1957.

This thesis presents a retrospective analysis of data for 2015–2021 and describes the main directions for future research in the field of hydride superconductivity. The thesis consists of six chapters devoted to the study of the structure and superconductivity of binary and ternary superhydrides of thorium ($ThH_9$ and $ThH_{10}$), yttrium ($YH_6$ and $YH_9$), europium and other lanthanides (Ce, Pr, Nd), and lanthanum-yttrium (La-Y). This work describes the physical properties of cubic decahydrides, hexahydrides, and hexagonal metal nonahydrides, demonstrates high efficiency of evolutionary algorithms and density functional methods in predicting the formation of polyhydrides under high-pressure and high-temperature conditions. We proposed a theoretical-experimental algorithm for analyzing the superconducting properties of hydrides, which makes it possible to systematize the accumulated experimental data. In general, this research is a vivid example of the effectiveness and synergy of modern theoretical and experimental methods for studying the condensed state of matter under high pressures.

## Publications

## Other publications related to the superconductivity in polyhydrides

1. D. V. Semenok et al. (2022), Effect of magnetic impurities on superconductivity in $LaH_{10}$. *Adv. Mater.* 2204038, DOI: 10.1002/adma.202204038.

2. W. Chen, X. Huang, D. V. Semenok et al. Enhancement of the superconducting critical temperature realized in the La-Ce-H system at moderate pressures, *ArXiv* 2203.14353, 2022.

3. I.A. Trojan, D.V. Semenok et al. "High-temperature superconductivity in hydrides", *Phys. Usp.*, 2022, DOI: 10.3367/UFNe.2021.05.039187.

4. D. V. Semenok et al. Sr-Doped Superionic Hydrogen Glass: Synthesis and Properties of $SrH_{22}$. *Adv. Mater.* 2022, 34, 2200924, DOI: 10.1002/adma.202200924.

5. W. Chen, D. V. Semenok et al. "High-Temperature Superconducting Phases in Cerium Superhydride with a Tc up to 115 K below a Pressure of 1 Megabar", *Physical Review Letters*, 127, 117001, 2021, DOI: 10.1103/PhysRevLett.127.117001.

6. W. Chen, D.V. Semenok et al., "High-Pressure Synthesis of Barium Superhydrides: Pseudocubic $BaH_{12}$", *Nature Communications* 12, 273 (2021), DOI: 10.1038/s41467-020-20103-5.

7. Di Zhou, D. Semenok et al. "Superconducting Praseodymium Superhydrides", *Science Advances*, 2020, vol. 6, no. 9, eaax6849. DOI: 10.1126/sciadv.aax6849.

8. Di Zhou, D. Semenok et al. "High-Pressure Synthesis of Magnetic Neodymium Polyhydrides". *Journal of the American Chemical Society*, 2020, 142 (6), 2803-2811. DOI: 10.1021/jacs.9b10439.

9. W. Chen, D. Semenok et al. "Superconductivity and Equation of State of Distorted fcc-Lanthanum above Megabar Pressures", *Phys. Rev. B* 102, 134510 (2020), DOI: 10.1103/PhysRevB.102.134510.

10. I. Kruglov, D. Semenok et al. "Superconductivity of $LaH_{10}$ and $LaH_{16}$ polyhydrides", *Phys. Rev. B* 101, 024508 (2020), DOI: 10.1103/PhysRevB.00.004500.

11. D.V. Semenok et al. "Actinium Hydrides $AcH_{10}$, $AcH_{12}$, and $AcH_{16}$ as High-Temperature Conventional Superconductors", *The Journal of Physical Chemistry Letters*, 9 (8), pp. 1920-1926 (2018). DOI: 10.1021/acs.jpclett.8b00615.

12. A.G. Kvashnin, I. A. Kruglov, D. V. Semenok et al. "Iron Superhydrides $FeH_5$ and $FeH_6$: Stability, Electronic Properties, and Superconductivity", *The Journal of Physical Chemistry C*, 122 (8), pp. 4731-4736 (2018). DOI: 10.1021/acs.jpcc.8b01270.

13. A. G. Kvashnin, D. V. Semenok et al. "High-Temperature Superconductivity in a Th–H System under Pressure Conditions", *ACS Appl. Mater. Interfaces*, 2018, 10 (50), pp. 43809-43816. DOI: 10.1021/acsami.8b17100.

14. High-temperature superconducting hydride and method for production thereof, patent RU2757450C1 (2020).

## Acknowledgments

This work was supported by the Russian Foundation for Basic Research (RFBR) Grant No. 20-32-90099 and the Russian Science Foundation (RSF) Grants No. 22-22-00570 and 19-72-30043. The author expresses his deepest gratitude to I. A. Troyan, I. S. Lyubutin, and A. G. Ivanova of the Federal Research Center “Crystallography and Photonics” for their invaluable help in preparing samples of metal superhydrides; researchers of the V. L. Ginzburg Center for High-Temperature Superconductivity and Quantum Materials (part of LPI, Moscow), A. V. Sadakov, V. M. Pudalov, O. A. Sobolevsky, and K. S. Pervakov for their help in carrying out the transport studies of superhydrides; College of Physics, Jilin University, and a group of Professors Tian Cui and Xiaoli Huang, Dr. Wuhao Chen, and, especially, Dr. Di Zhou for the 2018–2020 internship opportunity in China.

The author thanks Prof. Niu Haiyang (Northwestern Polytechnical University, China) for help in studying the C–S–H system, Dr. Mikhail Kuzovnikov (University of Edinburgh) for providing as yet unpublished data on the low actinide hydrides, Dr. Di Zhou for her help with the reference list, Dr. Jorge Hirsch (University of California) for many critical remarks on superconductivity in polyhydrides, Prof. P. Zinin (STC UI RAS, Moscow) for the measurements of the temperature in DACs during laser heating, and Igor Grishin (Skoltech) for proofreading and editing the text of the thesis.

The author acknowledges the teams at the European Synchrotron Radiation Facility (ESRF) high-pressure stations ID27 and ID15b, and at the Japanese synchrotron source (Spring-8) station BL10XU, namely Dr. M. Hanfland, Dr. G. Garbarino, Dr. M. Mezouar, Prof. K. Shimizu, Dr. Y. Nakamoto, and Dr. A. Yu. Seregin for their help in the diffraction experiments; and the team of the Dresden High Magnetic Field Laboratory (HZDR HLD), especially Dr. Toni Helm and Dr. Stanley Tozer (National High Magnetic Field Laboratory, US), for help with the pulsed magnetic measurements.

## Personal contribution

The author’s personal contribution consists of the experimental X-ray diffraction measurements on the synchrotron sources (ESRF, SPring-8, SSRF, Elettra, and KISI) and pulse magnets (HZDR); routine measurements of Raman spectra; preparation of high-pressure diamond anvil cells, their loading and laser heating; synthesis and purification of precursors (ammonia borane, alloys), analysis of their elemental composition (EDS) and structure; and preparation of diamond anvils and electrodes using a Xe focused ion beam.

Investigating the Th–H system, the author personally performed the experimental X-ray diffraction studies at the ESRF (station ID27), laser heating of high-pressure cells, pressure changes, and Raman spectra measurements, interpreted all obtained results, and wrote experimental reports and the research paper. When studying the Eu–H system, the author personally performed the preparation and laser heating of high-pressure diamond cells, and did their X-ray phase study on the synchrotron sources SSRF (Shanghai, China) and Spring-8 (Osaka, Japan).

In this thesis, almost all calculations of the electron–phonon interaction and superconducting parameters of the obtained polyhydrides were performed by the author. In many cases, the calculations of the equations of state and the analysis of the dynamic and mechanical stability of compounds were also done by the author. All the articles on which this thesis is based, the entire analysis of theoretical and experimental data in them, and the idea of conducting the abovementioned studies of polyhydrides belong to the author.

# Table of Contents

## List of Symbols and Abbreviations

DAC – diamond anvil cell
GLAG – Ginzburg-Landau-Abrikosov-Gor'kov theory
BCS – Bardeen-Cooper-Schrieffer theory of superconductivity
HTSC – high-temperature superconductivity
ZPE – zero-point energy
Megabar – 100 GPa, GPa – gigapascals.
T – Tesla
K - Kelvin
EPC – electron-phonon coupling
SOC – spin-orbit interaction
$T_C$ – critical temperature of superconductivity
$\mu_0 H_{C2}$ – upper critical magnetic field
SCDFT – superconducting density functional theory
DFT – Density Functional Theory
XRD – X-ray diffraction
SSCHA – stochastic self-consistent harmonic approximation
AB – ammonia borane ($NH_3BH_3$)
EPW – open-source community code for ab initio calculations of electron-phonon interactions using the density-functional perturbation theory and maximally localized Wannier functions.
A-D – Allen-Dynes formula

# Chapter 1. Introduction

## 1.1 Discovery of polyhydrides and superconductivity in them

Superconductivity (SC) is the property of some materials to possess strictly zero electrical resistance at temperatures below a certain $T_C$ value (critical temperature). Despite more than 100 years of research of this phenomenon, the engineering potential of superconductivity is not yet fully realized [1, 2]. Notwithstanding the prospects of reducing the electric energy losses and leveling the heat generation in the windings of powerful magnets, wide application of superconductors in motors and power lines is still difficult due to the significant cost of cooling them well below the critical temperature. The search for new, higher-temperature superconductors at normal pressure so far has yielded no new results since the discovery of $HgBa_2CaCuO_{6+x}$ in 1993 ($T_C$ = 133 K) [3, 4]. However, at high pressures, new superconductors with record $T_C$ values have been predicted and then experimentally obtained, represented by $Im\bar{3}m$-$H_3S$ ($T_C$ = 203 K) [5, 6] and $LaH_{10}$ ($T_C$ = 250–260 K) [7, 8], binary polyhydrides with anomalously high hydrogen content. Searching for compounds superconducting at even higher temperatures requires studying ternary and more complex hydrides, which dramatically increases the variety of possible compounds (combinations of atoms) that is virtually impossible to enumerate in a blind experimental search.

Recent (2005–2015) advances in computational materials science and chemical process prediction at extreme pressures of tens and hundreds of gigapascals (GPa) have changed approaches to finding new superconductors. Evolutionary algorithms have now reached a high level of predictive accuracy in determining new crystal structures of inorganic compounds and are less costly than "blind" experimental enumeration. One of the best and most powerful methods for predicting thermodynamically stable compounds is the USPEX algorithm [9-12]. It has been used to achieve important results in obtaining new superhard materials [13], the first high-temperature superconducting hydrides ($H_3S$ [5, 14] and $Si_2H_6$ [15, 16]), and magnetic and electronic materials. In the case of polyhydrides synthesized at high pressures, ab initio methods make it possible to establish the structure of the hydrogen sublattice which cannot be done using X-ray methods. The results of a structural search can be verified by measuring the critical temperature of superconductivity in hydrides because high $T_C$ > 100–200 K is usually associated with a highly symmetrical hydrogen sublattice to achieve the desired parameters of the electron–phonon interaction.

After the discovery of superconductivity in sulfur hydride $H_3S$ in 2015 [5], the next research milestone was the experimental work by Z. Geballe et al. (2018) [17] in which the authors had managed to synthesize the previously predicted $LaH_{10}$ superhydride [18, 19] at a pressure of 175 GPa. A year later, superconductivity at 250 K has been revealed in the newly found $LaH_{10}$ [7], which has thus surpassed in critical temperature and critical magnetic field all cuprate-based compounds found in the previous 33 years since 1986. Only two years had passed between the prediction and discovery of the new record superconductor. This minimal time gap illustrates the progress (Figure 1a) that has been made in computational materials science and experimental techniques working with ultrahigh pressures using diamond anvils. By now, lanthanum hydrides still have not been sufficiently investigated. Higher lanthanum hydrides: $LaH_{10}$ ($C2/m$, $R\bar{3}m$, $Fm\bar{3}m$, $P6_3/mmc$) and $P4/nmm$-$LaH_{11}$, have been obtained in experiment, as well as $Pm\bar{3}m$-$LaH_{12}$ (at 167 GPa); many superconducting transitions in lower lanthanum polyhydrides $LaH_x$ ($x < 10$) have been experimentally observed [20, 21].

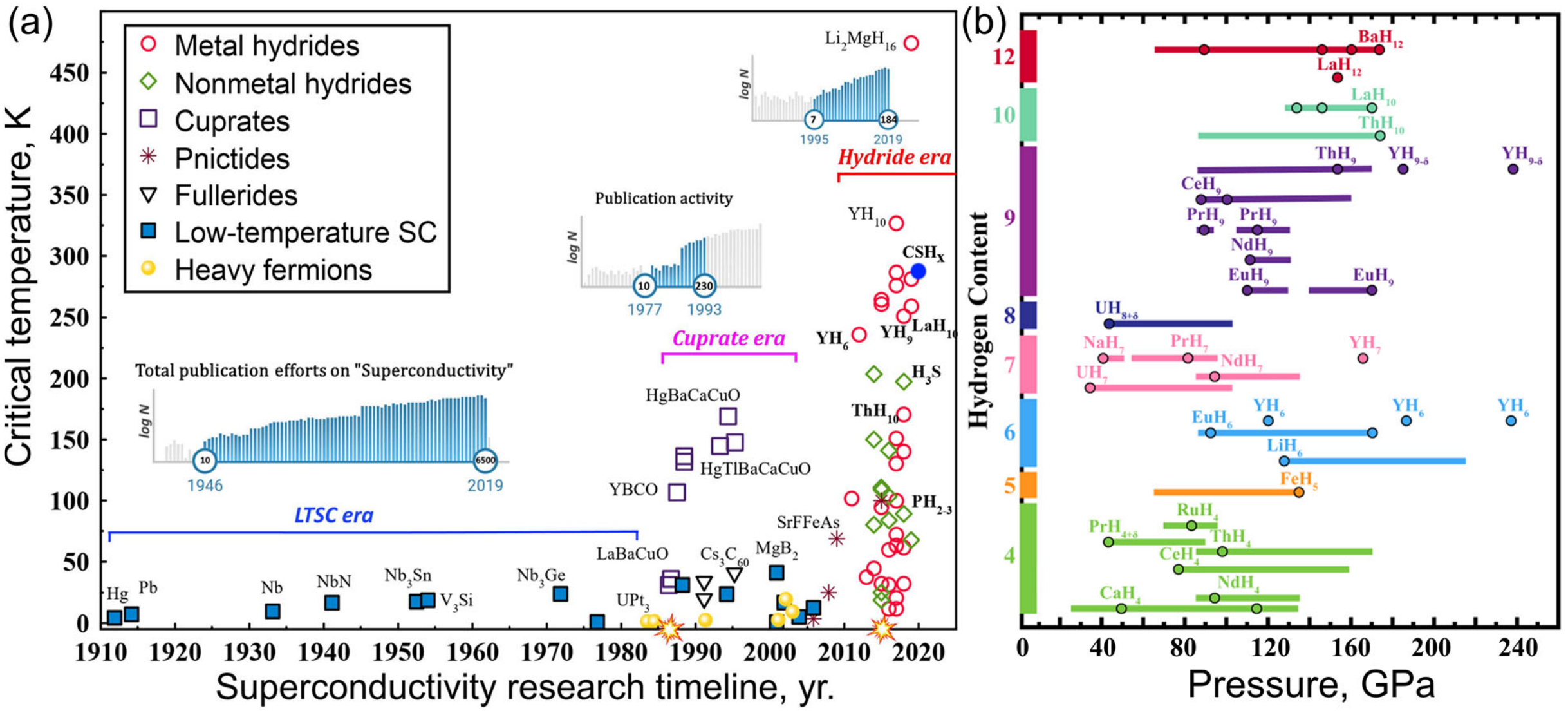

**Figure 1.** (a) Time scale of superconductors discovery and their critical temperature of superconducting transitions. Metal hydrides are shown by red circles, covalent hydrides of nonmetals — by green rhombuses, iron-containing pnictides — by asterisks, cuprate compounds — by hollow squares, low-temperature superconductors described mainly by the Bardeen–Cooper–Shrieffer–Migdal–Eliashberg theory — by filled squares, systems with heavy fermions — by yellow shaded circles, and fullerides — by triangles. The insets show the number of publications on the corresponding topics on a logarithmic scale (according to Semantic Scholar). (b) Stability regions (in GPa) and hydrogen content per 1 atom of metal (up to $x = 12$) in the best-known superhydrides investigated over 2015–2021.

Over the past six years (2015–2021), many different superhydrides have been synthesized (Figure 1b), both non-superconducting ($FeH_5$ [22, 23], magnetic neodymium superhydrides $NdH_7$ and $NdH_9$ [24], cubic and hexagonal praseodymium hydrides $PrH_9$ [25]) and superconducting (uranium hydrides $UH_7$, $UH_8$, and $UH_9$ stable at record-low pressures [26], thorium polyhydrides $ThH_9$ and $ThH_{10}$ ($T_C = 161$ K [27]), cerium hydrides $CeH_9$ and $CeH_{10}$ ($T_C \sim 110$ K [28, 29]), and yttrium hydrides $YH_6$ and $YH_9$ ($T_C = 224$ [30] and 243 K [31])). Most of these compounds were first predicted theoretically and then obtained experimentally, which proves the efficiency of the evolutionary search for thermodynamically stable compounds based on the methods of the density functional theory.

Binary hydrides have already been studied quite extensively as possible high-temperature superconductors, therefore the most intensive research in the field is now focused on ternary systems. The calculations using artificial intelligence algorithms, in particular neural networks, point to ternary hydrides as more promising in terms of both critical temperature and reduction of the synthesis pressure [32-36]. For example, it has been discovered in 2020 that at a pressure of several gigapascals a mixture of methane and sulfur hydride $H_2S$ photochemically forms a molecular compound which under further compression transforms into a strange material — either carbon-doped or methane doped $H_3S$ - which at 270 GPa demonstrates a sharp resistance drop at 15 ºC [37].

Computer simulations of metal superhydrides show that at pressures of up to 200–300 GPa (at higher pressures the experimental study of superconductivity is now difficult), the maximum critical temperatures of superconductivity are achieved in hydrides of elements of groups 2 and 3, such as Ca, Sr, Sc, Y, La, Ac, Hf, Zr, Th, Ce, and Mg, containing 6–10 hydrogen atoms per metal atom. At the moment, the experimental search for promising hydride superconductors is mainly limited to this set of elements and their combinations [36]. For example, calculations show that in the Li–Mg–H system, clathrate hydride $Li_2MgH_{16}$ with $T_C$ above 400 K may exist [32] (although it would be metastable and unlikely to be obtained in experiment, moreover, taking into account the melting of the hydrogen sublattice leads to a decrease in $T_C$); for the Ca–Y–H and Ca–Mg–H systems, cubic hexahydrides $Pm\overline{3}m$-$CaYH_{12}$ and $CaMgH_{12}$ with critical temperatures of 240–260 K have been predicted. Ternary hydrides of lanthanum–cerium, lanthanum–thorium, lanthanum–boron [38], and potassium–boron [39] are expected to demonstrate exceptionally high stability (stabilization pressure from 12 to 60 GPa) and critical temperatures of superconductivity above 100 K, bringing us closer to a discovery of a new class of hydrogen-bearing compounds which would be stable under normal conditions. Thus, a lot of work is to be done to synthesize ternary superhydrides and investigate their properties.

## 1.2 Classes of polyhydrides

In hydrides, hydrogen can be present in different forms: molecular (e.g., $LiH_6$), ionic (KH), and atomic ($YH_6$). By the type of the element–hydrogen bond, hydrides can be divided into covalent ($H_3S$, $SnH_4$), ionic ($AlH_3$), metallic ($LaH_{10}$), and mixed (molecular metal $BaH_{12}$) [40-42]. Also, a subclass of magnetic polyhydrides can be defined. For example, magnetic ordering is expected in the hydrides of neodymium $NdH_9$ [24], europium $EuH_9$ [43], samarium $SmH_9$, and many other lanthanides. The simultaneous realization of superconductivity in the hydrogen sublattice of hexagonal (e.g., $NdH_9$) or layered (such as $FeH_5$) hydrides and antiferromagnetic ordering in the metal sublattice can in principle lead to some exotic physical effects typical of cuprates and iron pnictides (e.g., Ref. [44]).

The interest in molecular and mixed superhydrides with a high hydrogen content (pseudo-tetragonal $SrH_{22}$ [45] and $BaH_{21-23}$ [46]) is due to the similarity of their hydrogen sublattices to the structure and properties of some crystal modifications of pure hydrogen (phases II, IV, V). However, the formation of these superhydrides (or hydrogen doped with 4–6% of Sr or Ba) is observed at much lower pressures (100–170 GPa) than those required to obtain the corresponding modifications of pure hydrogen (350–500 GPa). In molecular strontium superhydrides, gradual metallization and a change in optical transparency with a pressure increase from 90 to 160 GPa can be observed [47], whereas barium hydride $BaH_{12}$ [46] shows the emergence of superconductivity and an increase in the critical temperature, just as it has been predicted and partially confirmed experimentally for semiconducting and metallic hydrogen [48-51].

Properties of ionic and mixed metal hydrides will probably allow their use as ionic conductors and electrolytes for electrochemical synthesis of hydrides at high pressures [52]. Indeed, calculations show that the hydrogen diffusion rate ($\sim 6 \times 10^{-6}$ cm$^2$/s[53]) at high pressures in hydrides — in $Li_2MgH_{16}$ in particular — is much higher than in known ionic conductors.

Covalent polyhydrides are the most enigmatic class. Because of strong element–hydrogen interactions, formation of extended polymer chains and various organic groups is possible. The best studied covalent system is sulfur–carbon–hydrogen. Recent studies of $CS_2$ compression on diamond anvils indicate the formation of complex branched polymers with semiconducting properties [54]. In the H–S system, which initially seemed quite simple [15], even five years after the discovery of superconductivity in $H_3S$ more and more hydrides continue to be found [55], often with a very complex structure, such as $H_6S_5$ [56]. The observation of a sharp drop in the resistance in sulfur–carbon hydride (C,S)$H_x$ [37], interpreted by the authors as superconductivity, has attracted universal attention and represents an even more difficult task in terms of establishing the structure of this compound. However, serious doubts have been expressed recently that the effect detected in the C–S–H system is superconductivity [57-59].

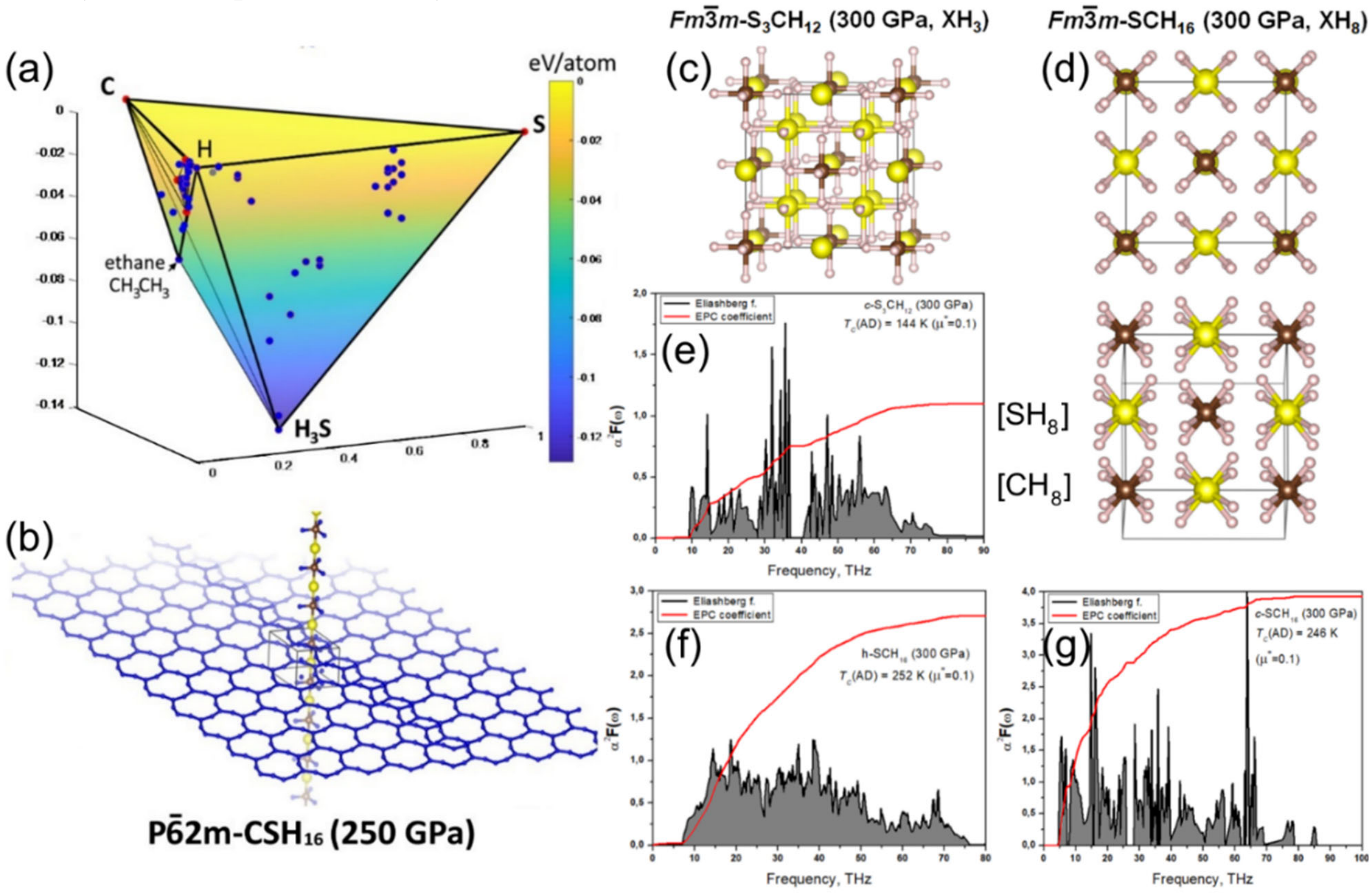

**Figure 2.** (a) Thermodynamic stability diagram of the carbon–sulfur–hydrogen system at 250 GPa shows the absence of stable ternary compounds. (b, c, d) Structures of unstable C–S–H ternary superhydrides for which high-temperature superconductivity is theoretically possible. (e, f, g) Eliashberg spectral functions and superconducting transition temperatures $T_C$ calculated using the Allen–Dynes formula (AD [60], see also Appendix).

Large-scale theoretical studies of 2020–2021 [58, 61-63] showed the absence of thermodynamically stable and high-$T_C$ superconducting ternary phases in the C–S–H system up to pressures of 300–350 GPa. Those rather rare phases that could show room temperature superconductivity due to the strong electron–phonon interaction (e.g., $P\bar{6}2m$-$CSH_{16}$ (Figure 2) or hypothetical $Pn\bar{3}m$-$CH_7$) appear to be essentially metastable and should decompose on heating to form previously studied $Im\bar{3}m$-$H_3S$. Almost all organic compounds are metastable with respect to decomposition into simple molecules ($CO_2$, $H_2O$, $N_2$). However, being situated in the local minima of the potential energy surface, they appear to be dynamically stable, exist for a long time and are formed by chemical reactions with kinetic control. A similar situation may occur as a result of photochemical synthesis at high pressures in the C–S–H system. In this regard, changing the criterion of structure selection in the evolutionary search from the minimum enthalpy to other parameters, such as the best agreement with the experimental X-ray diffraction pattern, different spectra, or the electron–phonon interaction parameters, becomes a relevant suggestion.

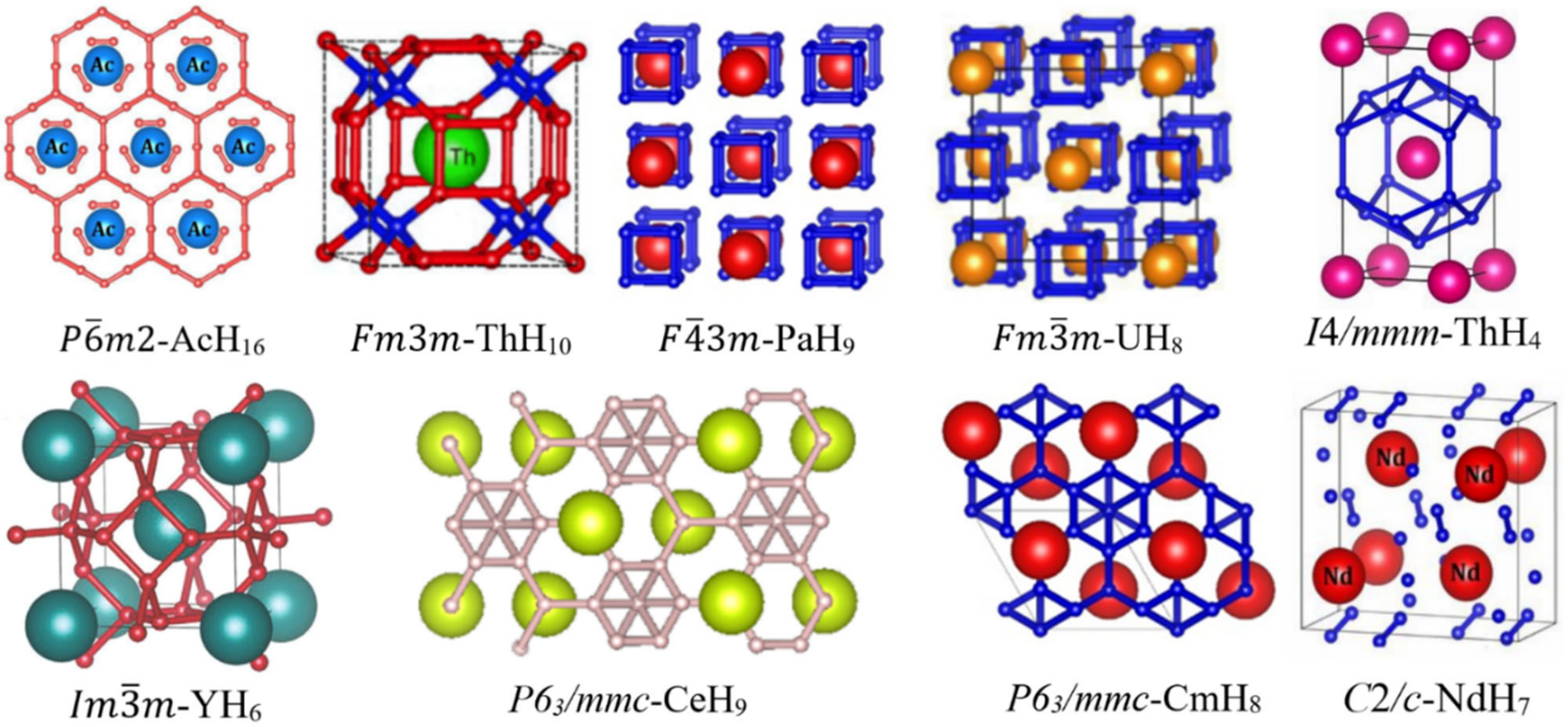

**Figure 3.** Structural motifs of actinide and lanthanide superhydrides with a hydrogen atomic sublattice. At high pressures, hexagonal and cubic dense packings of heavy atoms predominate.

Finally, the simplest and at the same time one of the most important classes are metal superhydrides with an atomic hydrogen sublattice (Figure 3). Such structures are typical metals, with the behavior of electrical resistance in the normal state described by the Bloch–Grüneisen formula [64, 65]

$$R(T) = R_0 + A\left(\frac{T}{\theta_D}\right)^5 \int_0^{\frac{\theta_D}{T}} \frac{x^5}{(e^x-1)(1-e^{-x})}\,dx, \quad (1)$$

where $A$, $\theta_D$, and $R_0$ were found using the least squares method. Below the critical temperature they exhibit the properties of high-temperature superconductors, have a high density of electronic states at the Fermi level, sometimes even a Van Hove singularity near it [66-68]. As we have shown previously [36], in this class of compounds there is a certain optimal number of hydrogen atoms per metal atom for achieving the highest critical temperature, $n$ = 6–10 in $XH_n$ hydrides, which formally corresponds to the transfer of ~0.33 electrons per hydrogen atom [19]. Such hydrides are formed at high pressures in reactions of $d^0$–$d^2$ metals with hydrogen and usually have cubic and hexagonal densely packed structures. We have recently found [36] that from a thermodynamical point of view, it is often more favorable for these structures to be slightly distorted, which, however, is difficult to detect using modern experimental methods. As the pressure decreases, at first more and more deviations from the high-symmetry structure are observed, the critical temperature decreases smoothly, then the loss of some hydrogen, change of the lattice composition and symmetry, and a sharp drop in $T_C$ occur. Thus, the $T_C(P)$ diagram usually has the form of a bell or an asymmetrical parabola [5, 7]: $T_C$ decreases in both directions from the maximum both at a pressure increase (with a decrease in the electron–phonon interaction constant λ due to "quenching" of the phonon modes) and at its decrease (with distortion of the lattice and decomposition of the compound).

Highly symmetric superhydrides formed by f-block elements Pr, Nd, Sm, U, Pu, Am, and so forth (e.g., $PrH_9$ [25], $EuH_9$ [43], $NdH_9$ [24], $UH_7$ [26, 69]), do not possess pronounced superconducting properties due to the Cooper pair scattering with spin reversal at paramagnetic centers [70]. Moreover, small additions of f-block elements effectively suppress superconductivity [47] in hydrides of $d^0$–$d^2$ elements ($LaH_{10}$, $YH_6$) almost without changing their structure, which can be used to study the magnetic phase diagram of superhydrides down to the lowest temperatures. It is important for understanding the mechanism of superconductivity in hydrides that small additions of nonmagnetic elements (such as C, B, N, Al) have almost no effect on the critical temperature of hydrides, whereas introduction of paramagnetic centers (e.g., Nd) dramatically reduces $T_C$.

Let us discuss here how the results of experimental structural studies of polyhydrides correspond to theoretical predictions. Structural information that can be obtained for polyhydrides in experiment concerns primarily the parameters of the unit cell. Experimental information about the interatomic distances can be obtained directly from the X-ray absorption fine structure (XAFS) for heavy atoms (e.g., *d*(Y–Y) in $YH_3$ [71]). Precise comparison cannot be made because of possible variations in the hydrogen content of synthesized polyhydrides in different experiments. In addition, DFT calculations performed with different pseudopotentials (using Quantum ESPRESSO or VASP) often give slightly different results. For $LaH_{10}$ at 150 GPa, the experimental and theoretical cell volume $V = 33.7 \pm 0.62$ Å$^3$/La, which can be used to find the minimum distance between the hydrogen atoms *d*(H–H) = 1.15 ± 0.007 Å (0.6%) [72].

## 1.3 Metal polyhydrides research methods

X-ray powder diffraction analysis in high-pressure diamond cells using synchrotron radiation remains the main method of determining the structure of hydrides [73]. The technique of focusing the synchrotron radiation has reached submicron resolution, which makes it possible to examine samples of several microns or even a few hundred nanometers in size sandwiched between diamond anvils [74]. The single-crystal diffraction [75, 76] at megabar pressures is becoming increasingly popular method. It requires the use of diamond anvils of a special shape (Boehler-type) with a wide aperture (70–80°) and is applicable if researchers manage to grow microcrystals of sufficient size (0.05–0.5 μm or larger) in a series of laser "annealing" cycles. Despite the fact that intense X-ray radiation is dangerous for diamond cells at pressures above 200 GPa because of the risk of anvil cracking [77], other instrumental methods cannot provide an amount of information comparable in value.

Recently (2018–2022), the group of Profs. L. Dubrovinsky and N. Dubrovinskaya (University of Bayreuth, Germany) developed a promising method of single micro- and nanocrystal diffraction (SCXRD) in diamond anvil cells at pressures above 100 GPa, which makes it possible to accurately determine the positions of heavy atoms in polyhydrides. There have even been reports about the possibility of establishing the positions of hydrogen atoms and molecules, at least some of them [78]. In particular, sulfur[56], sulfur-carbon[79, 80], and lanthanum polyhydrides [81] have been studied using this method. In the latter case, SCXRD led to the discovery of a new phase, *Pm*-3*n*-$La_4H_{23}$, not previously detected using the powder diffraction. I believe that this method could revolutionize the high-pressure research in the next five years.

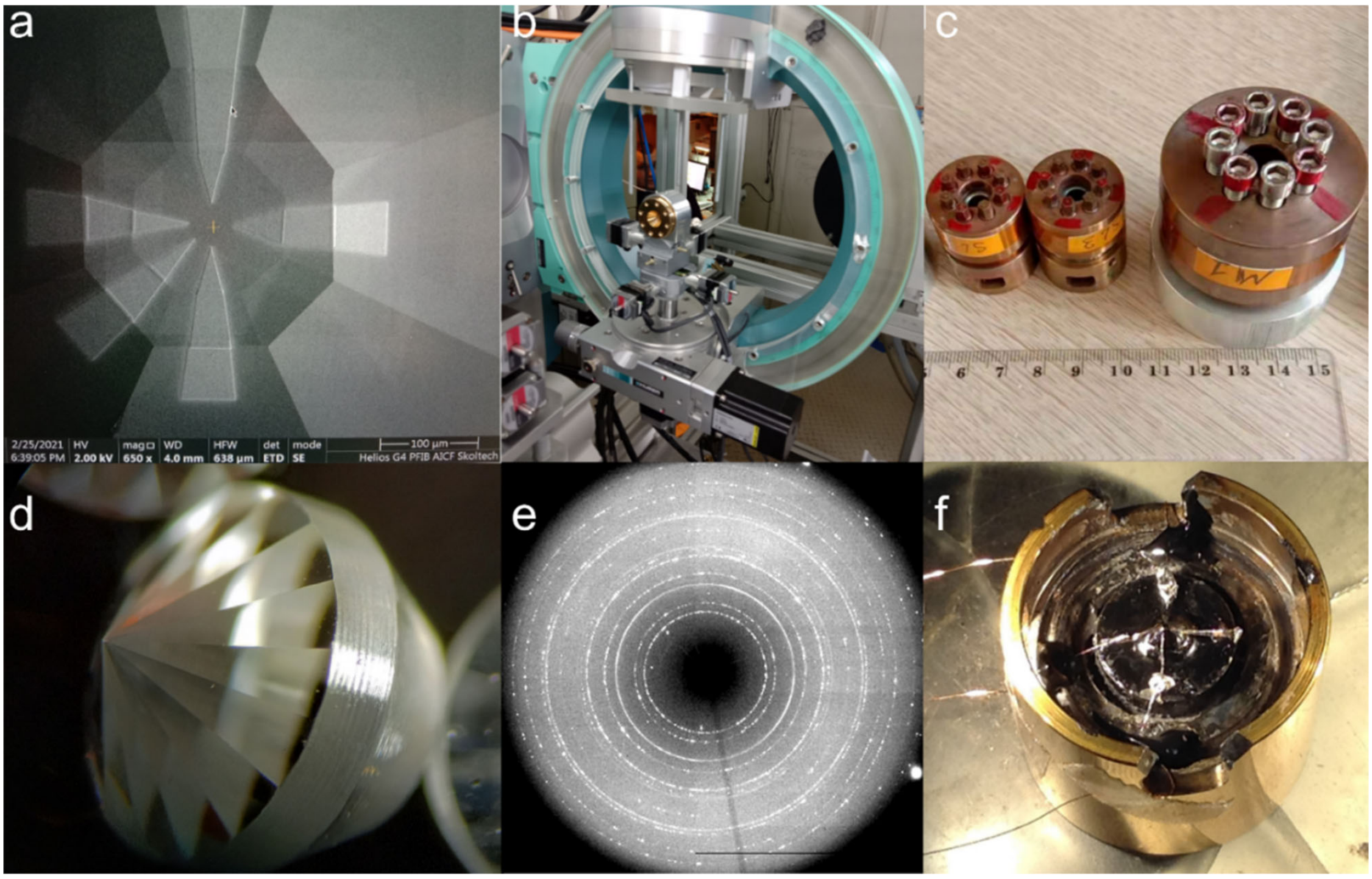

**Figure 4**. Elements of high-pressure diamond chambers and the experiment technique. (a) System of five platinum electrodes applied by a focused ion beam to the surface of a diamond anvil. (b) High-pressure chamber mounted on a goniometer for synchrotron imaging (Kurchatov synchrotron radiation source). (c) High-pressure bronze chambers of various types and sizes. (d) Diamond anvil with a conical base made by TISNCM (Moscow). (e) X-ray diffraction pattern of thorium superhydride $ThH_{10}$ obtained at ID27 beamline of the ESRF in Grenoble. (f) Open high-pressure chamber with a four-electrode system and copper wires to study the electrical transport characteristics of hydrides.

X-ray diffraction (Figure 4b,e) makes it possible to determine the hydrogen content indirectly, using the equation of state of the matter investigated in a certain pressure range, and to make assumptions about the structure of the hydrogen sublattice. For this purpose, the results are compared with the most thermodynamically advantageous structures found using evolutionary algorithms based on the density functional theory (DFT) calculations. Diffraction spectra can be calculated easily and accurately from known crystal structures. On the other hand, calculations of the critical temperature of superconductivity, Raman spectra, or sample reflectance are time-consuming and highly dependent on the technical parameters such as pseudopotential; they are more suitable for confirming a presumed structure but not for finding it from scratch.

The main methods for investigating the physical properties of hydrides at pressures above 250–300 GPa are the Raman and infrared spectroscopy of semiconductor phases, optical reflection, and measurements of the electrical transport properties. These methods, which have been successfully developed to study hydrogen metallization [48-51], provide limited information and require time-consuming calculations. Recently, attempts have been made to compare the Raman signals detected in some cases from samples in diamond cells with the expected spectra of metallic superhydrides [82]. This is a rather risky undertaking because calculations of the resonance Raman spectra for metals are complex, whereas dielectric micro-impurities and nanofilms of oxides, hydroxides, organic resins, and other compounds used in the design of diamond cells give a comparable or stronger Raman response than the expected signal from metal hydrides. To date, no systematic studies have been published of the correspondence between Raman spectra and X-ray diffraction results for metal hydrides at high pressures.

Reflection/transmission spectroscopy in the infrared and visible regions makes it possible to determine the bandgap of compounds, the magnitude and temperature dependence of the superconducting gap, and to compare the calculated electronic band structure with experimental data [50, 83, 84]. Important requirements for realization of these methods are the purity of diamonds — low content of nitrogen, defects, and impurities — and their low luminescence. These requirements are fulfilled for synthetic diamonds produced by the HPHT and CVD methods. To reduce absorption, diamond anvils can be partially drilled [85].

A study of the electrical transport characteristics (Figure 4a,f) makes it possible to determine the hydride's conductivity type, critical parameters of the superconducting state ($T_C$, $I_C$, $B_{C2}$), and electrical resistivity in the normal state. In some cases, the Debye temperature can be estimated using the Bloch–Grüneisen formula [86, 87]. The experimental compressibility $V^{-1}dV/dP$ calculated from the equation of state can be used to obtain some mechanical parameters and, using theoretical models, also estimate the Debye temperature.

Polyhydrides can be considered intermetallides formed by metals and metallic hydrogen. An effective approach to finding their structures would be to compare them with the known binary intermetallides formed by atoms with significantly different radii. Such an approach has been successfully applied in the theoretical study of clathrate $Li_2MgH_{16}$ (predicted $T_C$ is up to 473 K [32]) and in the experimental discovery of $Eu_8H_{46}$ [43] and $Ba_8H_{46}$ [88], which have a large number of prototypes such as $Ba_4Si_{23}$, $Ba_4Ge_{23}$, $Cs_4Sn_{23}$, and so forth.

The fully mathematical approach to the study of the structures of inorganic compounds under pressure proposed by R. Koshoji and colleagues is of interest [89, 90]. The researchers studied the densest packing of spheres of different radii in a three-dimensional Euclidean space. It is known that at high pressures, the packing density of atoms plays one of the crucial roles in the stabilization of chemical compounds. The researchers found that for packages with two and three types of spheres ($A$ and $B$), when the ratio of the radii $r_A/r_B$ is large, the space is most optimally filled with clathrate structures when the ratio $A:B$ = 1:12, 1:10, 1:9, and 1:6. Polyhydrides of many elements do have such stoichiometries. Unfortunately, the exact ratio of the effective radii of the atoms of hydrogen and the hydride forming element depends on both the pressure and the charges on the atoms, which prevents the results of these works from being directly applied. Nevertheless, they serve as a mathematical basis for the formation of polyhydrides at high pressures.

Among the main problems in the theoretical calculations of superhydrides, the first one is the theoretical prediction of the formation of certain polyhydrides at high pressures. It was believed that the set of thermodynamically stable phases lying on the convex hull (one-dimensional curve) is practically all that can be obtained in experiment. In 2010–2020, hundreds of publications on polyhydrides under pressure were done in the following design: calculating the DFT formation enthalpy of various phases using structural search algorithms (USPEX, CALYPSO, AIRSS, etc.) → building the convex hull → selecting the most stable phases → verifying their dynamic stability within the harmonic approximation → calculating the electronic and superconducting properties. This approach turned out to be erroneous, leading to correct results (coinciding with the experimental data) only in exceptional cases. Currently, the convex hull cannot be calculated accurately enough because of the presence of various subtle effects (ZPE and anharmonicity of the hydrogen sublattice, strong correlations in *d*- and *f*-shells, entropy factor, inaccuracy of pseudopotentials, etc.) and the limited size of the unit cell that can be studied using available supercomputer resources. In fact, instead of a one-dimensional convex hull line with a small number (~10) of candidate phases, a band (about 0.1 eV/atom wide) above the convex hull should be considered as a potential region containing the experimental phases. This band already contains 100–1000 candidate structures for which an exhaustive analysis of the physical properties cannot be performed. That is why in the last four years, in most studies the sublattice of heavy atoms is experimentally established at the beginning, and then the most favorable hydrogen sublattice is found theoretically.

## 1.4 Experimental techniques

In 1959, diamond anvils were first used to create ultrahigh pressures in special chambers [91]. The use of diamond, the hardest of the known materials, which is optically transparent up to 220 nm (bandgap 5.5 eV), opened up wide opportunities both for increasing the range of the investigated pressures and for applying optical methods of studying and changing the state of matter. A significant improvement of the diamond anvil was the addition in 1978 of a series of bevels in the vicinity of the culet to smooth its shape [92]. This improvement made it possible to systematically reach pressures of 100–200 GPa and perform routine experiments with a variety of materials. In particular, many works in 1970–2000 were devoted to studying the behavior of pure elements under pressure. In terms of superconductivity, the studies have revealed 22 elements transitioning to the superconducting state under pressure, in addition to the known ~31 elements superconducting at normal pressure [93]. These discoveries have led to the understanding that increasing pressure generally promotes the manifestation and enhancement of superconducting properties, the critical temperatures often increase at compression, and the behavior of $T_C(P)$ function is nonlinear and often surprising (e.g., in NbTi [94]).

The design of diamond anvil cells includes (see also Appendix):

(1) a gasket (*c*-BN, MgO, $CaF_2$, Re, W, Al, Be, etc.), which is a ceramic or metal plate with a hole serving as the walls of the chamber where high pressure is created;

(2) a medium transferring pressure to the sample ($H_2$, Ar, Ne, He, organic liquids, ammonia borane $NH_3BH_3$, etc.);

(3) a pressure sensor (ruby luminescence, X-ray diffraction from gold or platinum). The pressure can also be estimated from the edge of the Raman signal of the diamond;

(4) an insulating layer (e.g., $Al_2O_3$, thickness: 10–500 nm) deposited onto the anvils to protect them from aggressive media (hydrogen, helium, fluorine, etc.) and to thermally insulate the sample;

(5) an electrode system, usually multilayered, which is used to supply and read the electrical signal from the sample (Au, Mo, Au/Ta, B-alloyed diamond, etc.). Electrodes are formed using lithography, focused ion beam, magnetron sputtering or deposition from the gas phase (PVD);

(6) diamond anvils (usually synthetic, Figure 4d), in which the shape and size of the culet mainly determine the maximum pressure attainable in the diamond chamber. Special shapes of diamond anvils allow generating pressures of up to thousands of gigapascals in an area of several microns [95]. To improve the performance, the culet surface of diamond anvils can be modified using focused ion beam etching (Xe FIB). Because diamond anvils cannot be unloaded without partial cracking due to jamming effects when pressures reach 70–80 GPa, experiments with pressures above 1 megabar (100 GPa) almost always require their replacement or remaking;

(7) bases (seats) for diamond anvils, which transmit and distribute the force from the chamber to the anvils with minimal deformation. They are usually made of tungsten carbide and boron nitride. Seats with a conical anvil seat have the best characteristics [96];

(8) a diamond cell cylinder and piston, screws, and springs to create and smoothly transmit the force (Figure 4c,f). The cell material should be as hard as possible, nonmagnetic, and having minimum thermal expansion. Suitable materials are beryllium and titanium bronzes and NiCrAl alloys, which are very difficult for milling and lathing.

The most important step in obtaining superhydrides is the synthesis during laser heating. As a source of hydrogen, we systematically used the solid complex of ammonia with boron hydride $NH_3BH_3$ (ammonia borane, or AB) [97-99], which decomposes into hydrogen and amorphous polymer $[NBH_x]_n$ at temperatures above 200–250 ºC [100, 101]. In principle, hydride synthesis can be carried out at a low temperature of 250–400 ºC, although heating to 1000–1500 ºC is more common. The heating of the metal target accelerates its reaction with hydrogen to form a hydride that is stable at a given temperature and pressure. It is important to fix the sample between the anvils so that it does not touch them (making a sandwich structure, AB/sample/AB or AB/sample/electrodes): diamond has an extremely high thermal conductivity which makes impossible the effective laser heating of a sample pressed to an anvil.

## 1.5 Peculiarities of the superconducting properties of polyhydrides

High-temperature superconductivity in various hydrides under pressure, which was predicted by N. W. Ashcroft [102], was then discovered in many compounds using the DFT calculations. To date, more than 90–95% of the works on hydrides are still theoretical. It is important that in almost all cases, first-principles calculations resulted in an overestimation of the critical temperature of superconductivity (Table 1) due to the failure to take into account the anharmonic vibrations of the hydrogen sublattice as well as because of a possible increase in the effective Coulomb pseudopotential $\mu^*$ to 0.2 (usually, $\mu^* = 0.1$–$0.15$ is assumed in calculations, see below for a detailed explanation).

The superconducting properties of metallic and covalent hydrides differ in a number of aspects. One of the features of covalent hydrides is nonlinear temperature dependence of the electrical resistance in the normal state, observed for both $H_3S$ [5] and $(C,S)H_x$ [37]. This prevents even an approximate estimation of the Debye temperature using the electrical resistivity in the normal state and the Bloch–Grüneisen formula [86, 116]. Another feature of covalent hydrides is a relatively low upper critical magnetic field. Thus, for room-temperature superconductor $(C,S)H_x$ with $T_C = +13$–$15$ ºC, the extrapolated to 0 K value of $H_{C2}(0) \sim 70$ T [37], whereas for "weaker" superconductors $LaH_{10}$, $YH_6$, and $YH_9$ with $T_C < 250$ K it exceeds 120–160 T. For comparison, large values of $H_{C2}(0)$ — up to 300 T on extrapolation — can be achieved only in some iron-containing pnictides, for example in $NdFeAsO_{0.82}F_{0.18}$ ($T_C = 49$ K) [117].

**Table 1.** Highest critical temperatures obtained experimentally and theoretically in the harmonic approximation (at $\mu^* = 0.1$) of some hydride superconductors. The theoretical $T_C$ values presented have been obtained before the publication of experimental works. Because it is difficult to find data for the same pressure, the comparison is shown for illustration only.

| Compound | Experimental pressure, GPa | Estimated $T_C$, K | Experimental $T_C$, K |
|---|---|---|---|
| $Im\bar{3}m$-$H_3S$ | 150 | 200 [15] | 203 [5] |
| $Fm\bar{3}m$-$LaH_{10}$ | 160 | 286 [18, 19] | 250-260 [7, 8] |
| $P6_3/mmc$-$YH_9$ | 200 | 303 [19, 103] | 243 [31] |
| $Im\bar{3}m$-$YH_6$ | 170 | 270 [104] | 224 [30] |
| $Fm\bar{3}m$-$ThH_{10}$ | 170 | 160–193 [27] | 161 [27] |
| $P6_3/mmc$-$UH_7$ | 70 | 46 [26] | 8 [47] |
| $F\bar{4}3m$-$PrH_9$ | 150 | 56 [36] | 6 [25] |
| $P6_3/mmc$-$CeH_9$ | 110 | 117 [28, 29] | ~90 [105] |
| $Fm\bar{3}m$-$CeH_{10}$ | 100 | 168 [106] | ~115 [105] |
| *c*-$SnH_x$ | 190 | 81–97 [107] | 76 [108] |
| $PH_x$ | 200 | ~100 [109] | 100 [110] |
| $Pm\bar{3}n$-$AlH_3$ | 110 | >24 [111, 112] | <4 [112, 113] |
| $Im\bar{3}m$-$CaH_6$ | 170 | 220–235 [114] | 215 [115] |

### *1.5.1 Anharmonic effects*

The works of the groups of I. Errea and A. Bergara [113, 118-120] led to understanding of a large contribution of the anharmonic oscillations of the hydrogen sublattice to the thermodynamic stability and superconductivity of polyhydrides and, to a lesser extent, polydeuterides. Using the stochastic self-consistent harmonic approximation (SSCHA) method, the researchers answered many questions, including why $T_C$ of palladium deuteride PdD is higher than that of the corresponding hydride PdH, and the question about the unexpected stability of decahydride $LaH_{10}$, which should decay below 210 GPa in the harmonic approximation but exists at pressure drops to 140–145 GPa in experiment. The analysis of the anharmonic corrections shows that in many cases the critical temperature in hydrides lowers by 20–25 K, the electron–phonon interaction coefficient decreases by 20–25% (due to "quenching" of soft phonon modes), whereas the logarithmic frequency increases by 40–50% (300–350 K) in comparison with the harmonic approximation. This decrease in $T_C$ is critical for low-temperature superconductors (e.g., $AlH_3$ [112]), in which anharmonic effects practically suppress superconductivity.

**Table 2.** Comparison of superconducting state parameters of hydrides in harmonic and anharmonic approximations. The table illustrates the importance of taking into account anharmonicity when studying hydrides.

| **Compound (pressure, GPa)** | **λ (harm.)** | **λ (anharm.)** | **$\omega_{log}$, K (harm.)** | **$\omega_{log}$, K (anharm.)** | **$T_C$, K (harm.)** | **$T_C$, K (anharm.)** | **$T_C$, K (exp.)** |
|---|---|---|---|---|---|---|---|
| $Im\bar{3}m$-$H_3S$ (200) [118] | 2.64 | 1.84 | 1049 | 1078 | 250 | 194 | 190 |
| $Fm\bar{3}m$-$LaH_{10}$ (214) [119] | 3.42 | 2.06 | 851 | 1340 | 249 | 238 | 245 |
| $Im\bar{3}m$-$YH_6$ (165) [30] | 2.24 | 1.71 | 929 | 1333 | 272 | 247 | 224 |
| $Pm\bar{3}n$-$AlH_3$ (110) [113] | 0.95 | 0.52 | 485 | 1050 | 31 | 15 | <4 |
| PdH (0) [120] | 1.55 | 0.4 | 205 | 405 | 47 | 5 | 9 |

The main disadvantage of the SSCHA method is the computational complexity. The calculations of the anharmonic Eliashberg function for a single compound can take up to several months. In several recent works, researchers have implemented another approach to take into account anharmonic corrections, which is based on the use of machine-learning potentials and molecular dynamics of polyhydride supercells containing ~1000 atoms [43, 72, 121-124]. This method makes it possible to calculate the anharmonic spectral densities of the phonon states at any given temperature in several days and therefore to correct the phase diagram of compounds, the phonon spectrum, and the high-frequency part of the Eliashberg spectral function $\alpha^2F(\omega)$.

### *1.5.2 Role of the Coulomb pseudopotential*

The Migdal–Eliashberg theory [125, 126] has one uncertain parameter responsible for the effective Coulomb interaction — the so-called Coulomb pseudopotential $\mu^*$ whose values are usually 0.1–0.15. Because anharmonic effects are often insufficient to explain the overestimation of $T_C$ in theoretical calculations, it has been suggested in several works [30, 127, 128] that in pressurized hydrides $\mu^*$ can take much higher values of 0.2–0.5. The exact value of $\mu^*$ affects the critical temperature and other parameters of a superconductor, therefore calculating this parameter correctly is quite important.

In 2022, F. Marsiglio[129] pointed to increasing $\mu^*$ in polyhydrides caused by a significant decrease in the Tolmachev logarithm ln ($E_F/\omega_D$) in formulas like the Morel-Anderson formula [130] for the screened Coulomb potential ("Coulomb pseudopotential")

$$\mu^* = \frac{\mu}{1+\mu\cdot\ln(E_F/\omega_D)}, \quad (2)$$

where $\mu$ – is the average potential of the Coulomb interaction of electrons in a metal (usually ~0.3-0.4 eV), $E_F$ – is the Fermi energy, $\omega_D$ – is the characteristic phonon energy (e.g., the Debye frequency). Similar formulas are obtained when using the Holstein-Hubbard model and are quite universal, related to the very nature of the Coulomb interaction [131].

Given that the properties and density of polyhydrides under pressure are very close to those of ordinary metals at ambient pressure, there is no reason to expect any anomalous values for μ. However, the electron-phonon interaction in polyhydrides affects much deeper layers below the Fermi surface, which leads to a stronger contribution of the electron-electron repulsion and an increase in μ* (Appendix Table A7).

Currently, the most common method for taking into account the effect of the Coulomb interaction on superconductivity is the so-called DFT method for superconducting compounds (SCDFT), which is based on solving the Kohn–Sham equations for the order parameter [132, 133]. Successfully applied to many superconductors (Nb [134], $MgB_2$ [135], $V_3Si$ [136], $H_3S$ [137]), this method nevertheless gives underestimated $T_C$ (and thereby overestimated μ*) values for many superhydrides, for example, $YH_6$ [30], $YH_9$ [47], $LaH_{16}$ [127]. There is some progress due to the recent introduction of a new exchange–correlation functional SPG2020 [136], which approximates the experimental values better (Table 3). However, only a systematic application of the fully anisotropic SCDFT method and comparison of calculation results with experimental values of $T_C$ can lead us in the future to an understanding of what values the Coulomb pseudopotential can really take in hydrides at high pressures.

**Table 3.** Critical temperature calculated within the DFT theory for superconductivity (SCDFT) compared with the experimental values. The theoretical $T_C$ values are mostly underestimated.

| Compound (pressure in GPa) | $T_C$, K (LM 2005) | $T_C$, K (SPG 2020) | $T_C$, K (exp.) |
|---|---|---|---|
| $Im\bar{3}m$-$H_3S$ (200) [137] | 180 | - | 190 |
| $Im\bar{3}m$-$YH_6$ (165) [30] | 156 | 181 | 224 |
| $P6_3/mmc$-$YH_9$ (200) [31, 47] | 179 | 246 | 243 |
| $Fm\bar{3}m$-$LaH_{10}$ (214) [119] | 210 | - | 245 |
| $P6/mmm$-$LaH_{16}$ (200) [127] | 156 | - | 241* |

* Harmonic approximation, calculations using the isotropic Eliashberg equations.

### *1.5.3 Anisotropic effects*

The reason for another important correction to the superconducting transition temperature is the anisotropy of the superconducting gap. In most of the early works from 2010–2018, the Migdal–Eliashberg equations were solved in the isotropic approximation, without taking into account anharmonicity, and with empirical values of μ* = 0.1–0.15. However, it has been found in 2015 that factoring in the anisotropy of the Fermi surface, electron–phonon and electron–electron interactions in the energy space in many hydrides leads to a ~20–30 K increase in the superconducting transition temperature [104, 137-139] compared to isotropic calculations. It has been shown that $Fm\bar{3}m$-$YH_{10}$ exhibits a significant gap anisotropy of Δ ± 5 meV, whereas $Im\bar{3}m$-$YH_6$ has two superconducting gaps of 32 and 50 meV [104]. In the more recent work, Wang et al. [139] have found that $Fm\bar{3}m$-$LaH_{10}$ also has a significantly anisotropic main superconducting gap of 46 ± 5 meV and a small additional gap $\Delta_2 \approx 6.2$ meV. The anisotropy of the electron–phonon interaction is now considered necessary to be taken into account in all cases regardless of the superhydride structure. For many hydrides, solution of the anisotropic Migdal–Eliashberg equations adds approximately 20–30 K to $T_C$ found within the isotropic theory.

An important feature of superconductivity is the existence of the critical electric current density $J_C$. As was first shown by me in several works [27, 30, 72], in hydrides the extrapolated critical current density $J_C(0)$ reaches very large values, from 10 to 100 kA/mm$^2$, which are comparable with or exceed those of all currently known types of superconductors. When estimating the critical current density, attention must be paid to the thickness of the sample placed between the diamond anvils. This thickness does not exceed the distance between the diamond culets, determined by interference of visible light. At pressures above 100 GPa, the distance between the anvils is about 1 μm with a sample diameter of about 20–40 μm and a current of several amperes at a liquid helium temperature. The possibly labyrinthine nature of the current flow can be additionally considered on the basis of theoretical calculations of the normal resistance of hydrides using the EPW package [140-143]. The calculations show that due to the strong electron–phonon interactions, the electrical resistivity of hydrides is relatively high, being at the level of such materials as mercury, constantan, and Nichrome. Using the van der Pauw formula [144, 145] for estimation, the effective thickness of the $YH_6$ or $LaYH_{20}$ hydride samples of 0.5–0.75 μm can be obtained, which further increases the critical current density estimate in superhydrides.

### *1.5.4 Reproducibility of polyhydride studies*

In recent years, studies of hydrides increasingly shift to ternary systems such as C–S–H [55, 146] and Y–Pd–H [82]; let us touch upon superconductivity in such systems. As we have recently shown for the La–Y–H system [72], during the synthesis of a hydride from the La–Y alloy both atoms are randomly distributed in the metal sublattice. In the X-ray diffraction spectrum, a set of lines characteristic of pure binary hydride of one of the components (e.g., $LaH_{10}$) with an altered volume of the sublattice is observed and no additional lines are detected from the sublattice of the second component. Because of this disorder, the width of the superconducting transition of compounds increases significantly — up to 10–50 K. This broadening is an expected effect for all complex systems with a large number of atoms, which will narrow the field of potential applications of multicomponent hydride superconductors in the future.

Table 4 shows the reproducibility of the measurements of the temperature of resistive (and sometimes magnetic) superconducting transitions in different hydrides. These experiments were performed using different initial materials containing different impurities, cells were loaded in an inert atmosphere and in the air, and the hydrogen source was both ammonia borane and hydrogen gas. Different authors used different equipment for laser heating (and even synthesis by keeping samples in the hydrogen atmosphere for a long time), different cryostats and thermometers, and so forth. Resistive transitions in hydrides are reproduced with a good accuracy of 10–15 K (~5%). At the same time, these results show the implausibility of theories that insignificant impurities of carbon, boron, and nitrogen within narrow concentration limits can dramatically increase the critical temperature of superconductivity in hydrides [61, 62]. The insignificant effect of nonmagnetic impurities on superconductivity can be investigated directly; such a study has been performed, for example, in the group of W. Chen, X. Huang, and Tian Cui [147] for carbon-doped $LaH_{10}$/C or aluminum-doped $LaH_{10}$/Al. They showed that in this case $T_C$ decreases by only ~5–10 K. Polyhydrides are also known to have nonstoichiometric hydrogen content. Despite the fact that different groups obtained La superhydride with different hydrogen content ($LaH_{10\pm2}$), their critical temperature of superconductivity remained ~250 K, in accordance with Anderson's theorem [148].

**Table 4.** Reproducibility of the superconducting transition temperature measurements by the drop in the electrical resistance (in some cases, by the jump in the magnetic susceptibility) by different scientific groups.

| Compound | Maximum experimental $T_C$, K | Institution / Scientific group |
|---|---|---|
| $Im\bar{3}m$-$H_3S$ | 204 | Max Planck Institute for Chemistry in Mainz (A. Drozdov, ... M. Eremets) [5] |
| | 190 | Osaka University (M. Einaga, ... K. Shimizu) [149] |
| | 183 | Jilin University (X. Huang et al. ) [150] |
| $Fm\bar{3}m$-$LaH_{10}$ | 260 | University of Illinois Chicago (M. Somayazulu, ... R. J. Hemley) [8] |
| | 250 | Max Planck Institute for Chemistry in Mainz (A. Drozdov, ... M. Eremets) [7]] |
| | 250 | Institute of Physics CAS (Fang Hong et al. [151]) |
| | 245 | Jilin University (W. Chen, ... X. Huang) |
| $Im\bar{3}m$-$YH_6$ | 224 | Institute of Crystallography RAS (I. Troyan et al. [30]) |
| | 227 | Max Planck Institute for Chemistry in Mainz (P. Kong, ... M. Eremets [31]) |
| | 211 | Jilin University (W. Chen, ... X. Huang) |
| $c$-$SnH_{4+x}$ | 75 | Institute of Physics CAS (Fang Hong et al.) [108] |
| | 71-74 | Institute of Crystallography RAS (I. Troyan et al.) |
| $P6_3/mmc$-$YH_9$ | 243 | Max Planck Institute for Chemistry in Mainz (P. Kong, ... M. Eremets [31]) |
| | 237 | Institute of Crystallography RAS (I. Troyan, D. Semenok et al.) |
| | 230 | University of Bristol (J. Buhot et al. [152]) |
| $Im\bar{3}m$-$CaH_6$ | 215 | Jilin University (L. Ma, ... Y. Ma) [115] |
| | 195–210 | Institute of Physics CAS (Z. W. Li et al.) [153] |

## 1.6 Critique of hydride superconductivity

After the discovery of a large number of superconducting hydrides ($H_3S$, $LaH_{10}$, $ThH_{10}$, $YH_6$, $YH_9$, and $CSH_x$), this area attracted the attention of researchers from wider fields. J. E. Hirsch, F. Marsiglio, M. Dogan, and M. L. Cohen [87, 154-157] have expressed certain doubts about the existence of superconductivity in hydrides and that it can be described within the framework of the electron–phonon coupling mechanism. The authors' main arguments relate to the small width of superconducting transitions in hydrides, the insufficiently large broadening of superconducting transitions in the applied magnetic field, and the lack of clear evidence of diamagnetic shielding and the Meissner–Ochsenfeld effect [158] for hydrides. The latter is not surprising because the available instrumental techniques for studying microscopic samples at ultrahigh pressures are limited to spectroscopic and X-ray diffraction methods. Electrode sputtering techniques and electrical measurements in diamond cells proved relatively well developed. The sensitivity of detecting a superconducting transition using the van der Pauw four-contact method [144, 145] is proportional to $L/S$ [$m^{-1}$], where $L$ is the characteristic size (diameter) of the sample, $S$ is the average cross-sectional area, whereas the magnetic field change in the vicinity of the sample is proportional to its volume $L \times S$ [$m^3$]. Therefore, resistive measurements lend themselves well to miniaturization, whereas magnetic measurements of micron-sized samples present a very challenging technical problem. Recently, new promising methods for studying diamagnetic shielding have been developed, based on detecting the fluorescence of nitrogen NV centers created on the surface of the diamond anvil (Figure 5c) and on using high-frequency current in single coils sputtered on the culet of the diamond anvil in the immediate vicinity of the sample (Figure 5a,b).

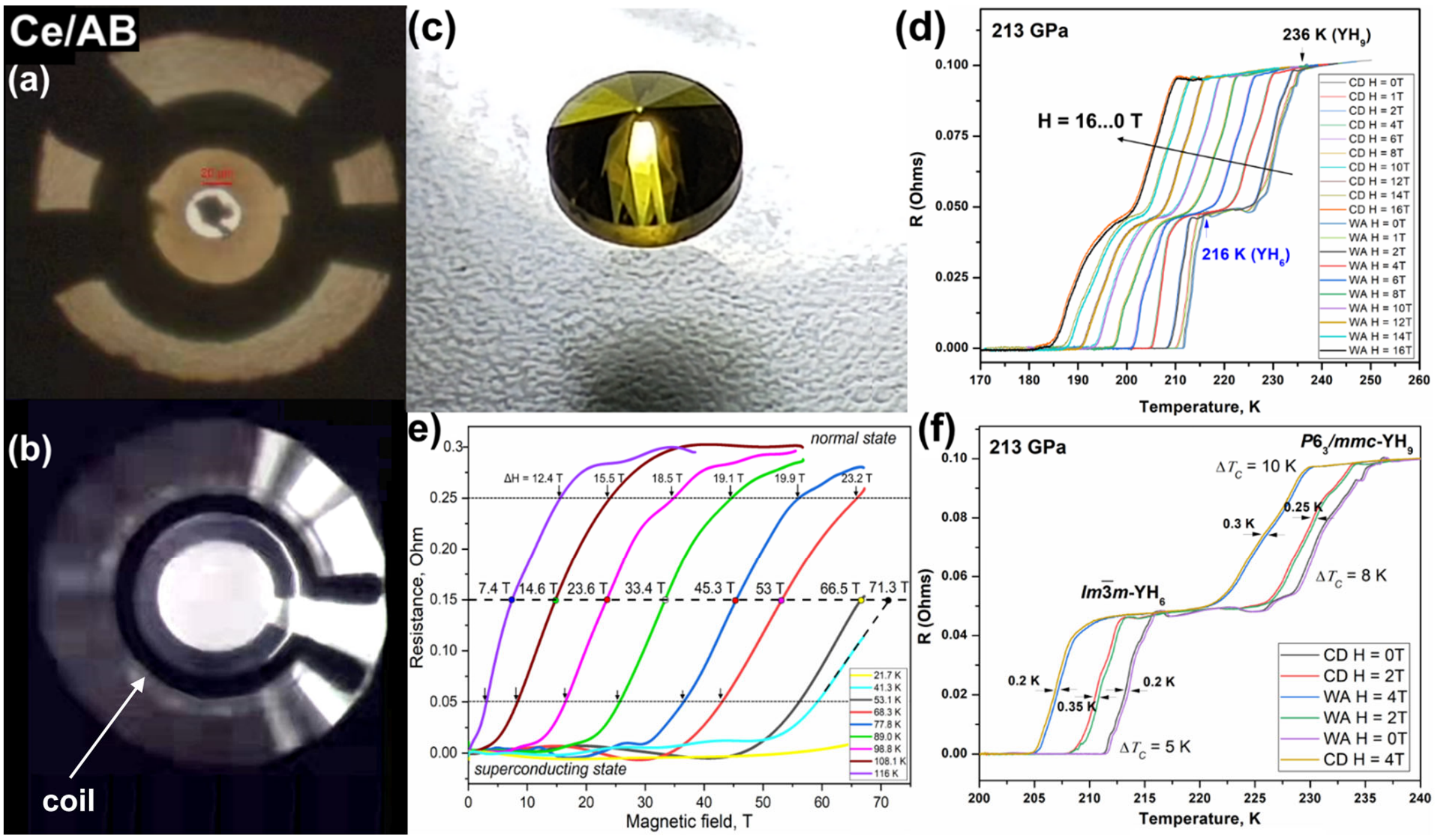

**Figure 5.** (a, b) Single coils sputtered on the diamond anvils in the immediate vicinity of the sample, (a) after loading and (b) before loading. (c) Nitrogen-doped diamond anvil with NV centers. (d) Broadening of the superconducting transitions in the magnetic field in a two-phase $YH_6$ + $YH_9$ sample at 213 GPa. (e) Broadening of superconducting transitions in a (La,Nd)$H_{10}$ sample containing 7–9 mol % of neodymium in strong pulsed magnetic fields of up to 70 T [47]. (f) Hysteresis of the superconducting transitions in a two-phase $YH_6$ + $YH_9$ sample at 213 GPa during cooling (CD) and heating (WA).

The critique of the small width of superconducting transitions [57] primarily addresses the results of the measurements in $CSH_x$ by Snider et al. [37] where the superconducting transition indeed shifted by 20 K practically without broadening in a magnetic field of 9 T. The doubts of J. Hirsch and F. Marsiglio concerning the width of superconducting transitions can be partially resolved if we take into account that metal superhydrides have very high values of the upper critical magnetic field ($B_{C2}(0)$ over 100–150 T), whereas most studies in the permanent magnetic field consider weak fields $B_{C2}(T)/B_{C2}(0) \sim 0.1$, in which the broadening of the superconducting transitions is insignificant. In addition, lower hydride impurities are often present in the samples, leading to multistep transitions to the superconducting state. A more detailed analysis of the data for $YH_6$, $YH_9$, and (La, Nd)$H_{10}$ shows (Figure 5d,e,f) that the superconducting transitions in the hydrides we studied are significantly broadened in magnetic fields and their initial width is determined by the quality of the single crystals lying on the electrodes. The width of the superconducting transitions in hydrides can be significantly reduced by repetition of laser heating and cooling cycles ("annealing"). This is especially true for ternary hydrides, where this broadening is due to a disorderly arrangement of heavy atoms in the structure. The small broadening of the superconducting transitions cannot be called an exclusive property of superconducting hydrides. Many iron-containing superconductors, in particular those of class 11, show an extremely insignificant broadening down to the lowest temperatures [159-161].

One of the most important arguments in favor of the electron–phonon nature of superconductivity in hydrides is the isotope effect manifested by a decrease in the superconducting transition temperature when hydrogen is replaced with deuterium in the structure of a compound. This effect has been observed in $H_3S$ [5], $LaH_{10}$ [7], $YH_6$ [30], $YH_9$ [31], $CeH_{9-10}$ [105], (Pd,Y)$H_x$ [82], and several other compounds. In all these cases, the isotope coefficient $\alpha = -\ln(T_C)/\ln(M)$, where $M$ is the mass of the atom, was in the range of 0.3–0.6, in reasonable agreement with the theoretical prediction. A certain difficulty is introduced in the analysis by the fact that the chemistry of deuterides does not fully coincide with that of hydrides, and the stability limits and distortion regions of hydride and deuteride structures on the pressure scale do not coincide to an even greater extent. For this reason, comparing $T_C$ for hydrides and deuterides at the same pressure is sometimes incorrect because their structures may be different. Another factor is the significantly smaller influence of anharmonicity on superconductivity in deuterides.

In general, deuterides exhibit the same properties as hydrides: the superconducting transition shifts depending on the applied magnetic field; their critical thermodynamic magnetic field $H_C(0)$ can be estimated from the interpolation formulas found by Carbotte and is usually significantly smaller than in hydrides; there is a critical current whose value also depends on the magnetic field strength. When the pressure decreases, first the crystal structure of superdeuterides distorts, the critical temperature of the superconducting transition markedly decreases, then the compound decomposes with the formation of lower deuterides and $D_2$.

Of particular interest in recent years are experiments on diamagnetic screening in hydrides and their criticism [59, 150, 154, 162-165]. Hot discussions are caused by the unexpectedly high ability of polyhydrides to shield an external magnetic field. In particular, in one of the first papers by J. Hirsch [154], the existence of superconductivity in $H_3S$ was called into question. However, as D. M. Gokhfeld noticed, it is necessary to correctly take into account the penetration of the magnetic flux into a sample. In a type-II superconductor ($H_3S$), Abrikosov vortices and a magnetic field at the sample center are absent as long as the external field is weaker than the total penetration field $H_p$. In the critical state model [166]

$$H_p \sim J_c \times a, \qquad (3)$$

where $J_c$ – is the critical current, $2a$ – is the sample size in the direction, perpendicular to the external magnetic field. The field $H_p$ for this sample can be larger than the external magnetic field $H_{ext}$ = 0.68 T (from experiment [162]), as well as the field $H_{edge}$ ~ 4.3 T (expected at the edges of the sample due to the demagnetizing factor $H_{edg}=H_{ext}/(1\text{-}N)$). Assuming that the depth of magnetic flux penetration in the experiment [162] was less than 5 μm, estimation of $J_c$ from the depth of screened area $J_c = H_{edge}/5\mu m$ leads to a reasonable value $J_c \sim 6.8\times10^7$A/cm$^2$, comparable with the critical density of intragranular currents in cuprate HTSC [167].

At the same time, the lower critical field $H_{c1}$ in $H_3S$ appears to be much smaller than 0.68 T. In this range of fields ($H_{c1} < H < H_p$) the distribution of the Abrikosov vortices in the sample should be inhomogeneous, as in all superconductors of the type II, and the magnetic flux density decreases from the edges to the center of the sample. Therefore, the formulas for homogeneous field obtained in [162] are inapplicable for this experiment. When estimating the critical current density, we should also consider the current circulation in a layer with a thickness equal to the sample thickness of 5 μm (along the field direction) and a depth equal to the vortex penetration depth [168] (in the plane perpendicular to the field), rather than the magnetic field penetration depth λ. Using a 5×5 μm$^2$ layer cross section, we obtain $J_C = 6.4\times10^7$ A/cm$^2$, which agrees with the results of recent measurements [30, 72], and is more than an order of magnitude lower than the estimate in [162]. Thus, for the screening effect established in [162], the values of $H_{c1}$ and $J_C$ for $H_3S$ are quite consistent with those for other superconductors.

When analyzing the recent preprint of V. Minkov et al. [164], where the magnetic field expulsion from $LaH_{10}$ and $H_3S$ samples has been investigated using a SQUID magnetometer, we should bear in mind that hydride samples are probably porous and consist of microscopic grains (~0.05-0.5 μm). In this case, the demagnetization coefficient should be calculated for a random packing of spherical particles and is between 0.33 and 0.5 [169, 170]. The magnetic field penetrates the sample between the individual grains and, therefore, there is no change in the sample magnetization at temperatures near $T_C$ during field cooling (FC). Thus, the penetration fields $H_p(0)$ = 96 mT for $H_3S$, and 41 mT for $LaH_{10}$ found by the researchers are the lower limit of $H_{c1}(0)$, and a more realistic estimate gives $H_{c1}(0) \sim H_p(0) / (1\text{-}N) = (1.5\text{-}2)\times H_p(0)$.

## 1.7 Main experimental properties of superconducting hydrides

Experimental studies [93, 171] have revealed the following properties of superhydrides:

**(1)** the isotope effect at replacement of hydrogen by deuterium, $\alpha = 0.3–0.6$ (see Appendix Table A8);

**(2)** the presence of a sharp drop in the electrical resistance ($10^3–10^5$ times to several micro-ohms within a few kelvins) at a certain temperature ($T_C$), the same in heating and cooling cycles;

**(3)** the dependence of the critical temperature $T_C$ on the applied magnetic field, linear at low fields $H_{C2} \sim |T\text{-}T_C|$;

**(4)** the presence of the critical current $I_C$ that depends on the applied magnetic field and temperature;

**(5)** the bell-shaped form of the critical temperature–pressure dependence, corresponding at low pressures to a distortion of the highly symmetric structure, at high pressures — to a decrease in the electron–phonon interaction constant due to the high-frequency shift of the soft phonon modes;

**(6)** broadening of the superconducting transitions in a magnetic field (especially in strong pulsed fields);

**(7)** broadening of the superconducting transitions in ternary hydrides (transition width up to 30–50 K) due to the disordered structure;

**(8)** significant suppression of $T_C$ in hydrides by paramagnetic impurities (e.g., 1 atom % of Nd leads to $\Delta T_C \approx -10$ K) and insignificant influence of nonmagnetic impurities (C, Al, Be, O, H) on $T_C$;

**(9)** diamagnetic screening, registered in several experiments in $H_3S$ and $LaH_{10}$ (see the recent paper [164];

**(10)** behavior of the reflectivity in the infrared range at incident radiation energies ~2Δ. Here it is necessary to consider the criticism of this experiment in Ref. [172];

**(11)** trapping of magnetic flux in hydrogen-rich high-temperature superconductors [178].

The most consistent explanation for all these properties is superconductivity, which is also expected from the first-principles calculations. To date, only isolated deviations are known in the behavior of hydrides from the Bardeen–Cooper–Schrieffer–Migdal–Eliashberg theory (e.g., the linear dependence of $H_{C2}(T)$ over the entire temperature range or anomalously low $T_C$ for $YH_6$), whereas no alternative interpretations explaining the entire set of the observed phenomena have been proposed by the authors of critical articles.

## 1.8 Future research directions

What is rational in the critique by J. Hirsch and F. Marsiglio is that hydrides should be investigated in more detail. At the moment, the basic parameters of the electron–phonon interaction in these compounds (electron–phonon coupling (EPC) constant $\lambda$, superconducting gap $\Delta$, Eliashberg function $\alpha^2F(\omega)$, logarithmic frequency $\omega_{log}$, and Coulomb pseudopotential $\mu^*$) are known mainly from first-principles calculations. Obviously, future studies will have to fill this void. There are several promising approaches for experimental investigation of the superconducting state parameters of polyhydrides that can be realized in high-pressure diamond cells.

**A.** Femtosecond reflection spectroscopy can allow direct experimental determination of the electron–phonon coupling constant via the relaxation rate of the electron temperature, as has been done for simple metals and intermetallics [173]. The difficulty in this approach is the nonlinear optical characteristic of diamonds, which leads to defocusing of the femtosecond pulse.

**B.** Infrared reflection spectroscopy in a wide energy range at low temperatures allows direct determination of the superconducting gap $\Delta$ and its temperature dependence $\Delta(T)$. This method was efficiently used in the study of $H_3S$ in 2017 [84]. Its disadvantage is the need for large samples, 70–150 μm, which leads to limitations on the maximum pressures (<175 GPa) in the cells. Nevertheless, $LaH_{10}$, $YH_6$, $ThH_{10}$, and $CeH_{9–10}$ are the primary targets for this method.

**C.** Superstrong pulsed magnetic fields (up to 70–80 T, and much more in magnetic flux compressors) have started to be applied to diamond anvil cells relatively recently (see the works of 2019–2020 on $H_3S$ and $LaH_{10}$ [174, 175]). The method makes it possible to significantly reduce the uncertainty in extrapolating the upper critical magnetic field $H_{C2}(0)$, plot a magnetic phase diagram to the lowest temperatures, and verify which model the $H_{C2}(T)$ dependence follows. The difficulty in this case is the necessity to perform measurements at high frequencies (1–50 kHz) in miniature diamond cells ($d$ = 15 mm) made of special steel. The method imposes serious requirements of minimization of the parasitic capacitance and inductance in the electrode system of the diamond cell.

Unfortunately, the currently available pulsed magnetic fields (70–80 T) are still not strong enough to completely suppress superconductivity in the most interesting superhydrides, for which $H_{C2}(0)$ exceeds 120–140 T. Therefore, this technique is most effective for compounds with low $T_C \sim$ 100–150 K, whereas for $H_3S$ and $LaH_{10}$, the obtained $H_{C2}(T)$ dependences remain linear even in the strongest available magnetic fields.

**D.** Andreev reflection spectroscopy and microcontact spectroscopy [176] are promising research methods that allow determining not only the value of the superconducting gap $\Delta(T)$ but also its anisotropy when there are several gaps simultaneously. This method has been successfully used recently to establish the anisotropic nature of the superconducting gap in metallic yttrium at a high pressure[177]. The researchers found that at high pressures yttrium has two superconducting gaps, around 3.6 and 0.5 meV, with $2\Delta/kT_C$ ratio reaching 8.2, which is in favor of a superconductivity mechanism with strong coupling. The difficulty of this study is that the Andreev contact must be nanoscale, which is hard to monitor during compressing and heating in DACs.

In 2022, V. Minkov et al. [178] proposed to use a SQUID magnetometer and generation of a trapped magnetic flux to study the superconducting properties of polyhydrides in diamond anvil cells. This method uses cooling of a polyhydride below $T_C$ in a uniform magnetic field followed by switching off the field. Because of the highly defective topology of hydride samples obtained after the laser heating of metals in a hydrogen environment in DACs, there are always many defects in them that allow magnetic field to penetrate the samples. As a result, polyhydrides under pressure displace the external magnetic field very weakly and do not exhibit a pronounced jump in the magnetic susceptibility upon cooling in a magnetic field (FC) [179]. However, in contrast with pure superconductors, polyhydrides under pressure capture the magnetic flux very well when the external magnetic field is turned off in the superconducting state of the sample. This makes it possible to create a small (~100–150 μm) but extremely strong permanent magnet from a hydride sample, in the vicinity of which the magnetic field induction can reach several Tesla at low temperatures. In turn, this allows confident detection of the frozen magnetic flux using the SQUID magnetometer. This gives us the temperature dependence of the critical current $J_C(T)/J_C(0)$ in hydride superconductors [178], and makes it possible to confirm the extremely slow damping of eddy currents in polyhydrides, as well as to find the lower critical magnetic field $H_{C1}(T)$. We can also determine the superconducting gap from the critical current data obtained over a wide temperature range [180]. This powerful and convenient method makes it possible to study the superconducting properties without sputtering electrodes and to combine such measurements with diffraction studies, for which the presence of electrodes is undesirable.

Thus, the design of future experimental studies of hydride superconductivity should include single crystal diffraction in high pressure diamond cells. Reflectance IR/UV/vis spectroscopy with determination of the superconducting gap value $\Delta(T)$; resistive measurements over a wide frequency range up to 10–100 kHz in constant and strong pulsed magnetic fields with detection of $H_{C2}(T)$, $J_C(T)$, magnetoresistance, Hall effect, and Andreev reflection; and measurements of magnetic susceptibility, $H_{C1}(T)$, and magnetic ordering character using X-ray magnetic circular dichroism (XMCD) in lanthanide superhydrides (Nd, Sm, Gd, Eu, etc.). From the theoretical point of view, future works will include thermodynamic calculations at finite temperatures with machine-trained interatomic interaction potentials, considering anharmonicity for supercells containing 100–150 atoms; and calculations of superconductivity and resistivity in normal state factoring in anharmonicity and electron–phonon interaction anisotropy and including first-principles calculations of the Coulomb interaction contribution using the SCDFT method.

More complex experiments using superconducting hydrides include creating conductive structures on the surface of the diamond anvil, such as creating S–N–S interfaces with a 1–10 nm dielectric gap between the superhydride electrodes; SQUID magnetometers; multilayer interfaces using layer-by-layer deposition of different metals and oxides; placing microthermometers and microheaters on diamond to measure the jump in the heat capacity; and creating microrings from superhydrides to study the flux pinning. It is also important to determine the position of hydrogen (deuterium) atoms in at least a small number of superhydrides stable at low pressures ($ThD_4$, $UD_{5–8}$, $CeD_{8–10}$) using neutron diffraction to verify the results of theoretical calculations.

# Chapter 2. Synthesis and superconductivity of thorium polyhydrides

## 2.1 Thorium and its hydrides at low pressures. Studies of actinide hydrides.

This chapter is based on the results of two publications: a theoretical analysis of the formation of new phases in the Th-H system [181] and an experimental verification of the obtained results [27] performed at the European Synchrotron Source (ESRF) in 2018.

Study of the Th-H system was preceded by a theoretical investigation of uranium hydrides by Ivan Kruglov, Alexander Kvashnin, and subsequent experimental verification of the predictions by the group of Prof. Alexander Goncharov (Geophysical Laboratory, Carnegie institution of Washington) [26]. An interest in uranium hydrides arose from computational work in the field of ensuring the safety of nuclear reactors, more specifically, in the process of calculating the diffusion ability of hydrogen in various steels, performed by Dr. I. Kruglov at the VNIIA named after N.L. Dukhov (Moscow).

Thorium is a weakly radioactive alpha-emitter $^{232}Th \rightarrow {}^{228}Ra + {}^{4}He$ with a half-life of 14.05 billion years, slowly oxidizing in the air, easily reacting with hydrogen to form *I*4/*mmm*-$ThH_2$ and *I*-43*d*-$Th_4H_{15}$. The latter hydride has the highest hydrogen content (1: 3.75) among metal hydrides, stable at ambient pressure. This hydride also exhibits pronounced superconducting properties ($T_C$ = 7.5-8 K). Metallic thorium is also a superconductor at normal pressure and temperature below 1.6 K. When the pressure rises to 15 GPa, the critical temperature in thorium drops below 1 K [182].

Although the properties of thorium are similar to those of Zr and Hf, it differs from them in the higher chemical activity of both the metal itself and the carbides and nitrides [183]. Further studies have shown that Zr and Hf form superconducting hydrides *I*-43*d*-$X_4H_{15}$ [184, 185] similar to $Th_4H_{15}$ when the pressure is increased to 30-50 GPa. Their superconductivity temperatures are also close: 4.5-6 K. Moreover, uranium also forms the same phase $U_4H_{15}$ at a pressure of about 5-7 GPa. This compound was found back in Prof. A. Goncharov's first work on U-H when unloading one of the DACs, but was misinterpreted (Figure 6). At ambient pressure, $U_4H_{15}$ is metastable, can be stored at liquid nitrogen temperature, but decomposes into hydrogen and $UH_3$ around -70 °C [186]. Increasing the pressure to megabar levels leads to the formation of better superconductors, for example, in the Zr-H system a new unknown compound with $T_C$ > 70 K is formed at a pressure of about 200 GPa [187]. We note that $T_C(Th_4H_{15}) > T_C(Zr_4H_{15}) > T_C(Hf_4H_{15})$, which suggests that the same sequence will be observed for higher polyhydrides: $T_C(ThH_x) > T_C(ZrH_x) > T_C(HfH_x)$. Current experimental data support this conclusion.

New $Th_4H_{15}$ phase modifications (high-pressure (*HP*) and low-pressure (*LP*) phases, *LP* is well-known cI16), recently synthesized at ~5 GPa, show slightly lower critical temperature ($T_C$ = 6 K for *HP*-$Th_4H_{15}$ and 8 K for *LP*-$Th_4H_{15}$) than the modification obtained at ambient pressure [188]. We performed an experiment to synthesize thorium hydrides at a low pressure of about 30 GPa (Th and $NH_3BH_3$ were used) and found two cubic phases of $ThH_{3.75}$ with different diffraction patterns (Figure 6). The aim of this experiment was to synthesize *I*4/*mmm*-$ThH_4$ at a pressure of about 30 GPa. However, instead of $ThH_4$, a new modification, *HP*-$Th_4H_{15}$, was obtained, the XRD pattern of which resembles that of compounds with symmetry group *Im*-3*m* (Figure 6f–g). The comparison with the other experiment (Th + $H_2$, Dr. M. Kuzovnikov, Figure 7) confirms our conclusion about the probable synthesis of a new high-pressure $Th_4H_{15}$ modification at 30 GPa. Unfortunately, the samples of the obtained hydrides were contaminated with thorium dioxide (*fcc* $ThO_2$), which was formed during the prolonged contact of metallic thorium with air oxygen. Since thorium is to the left of iron in the series of electrochemical activity, its oxide cannot be reduced by hydrogen.

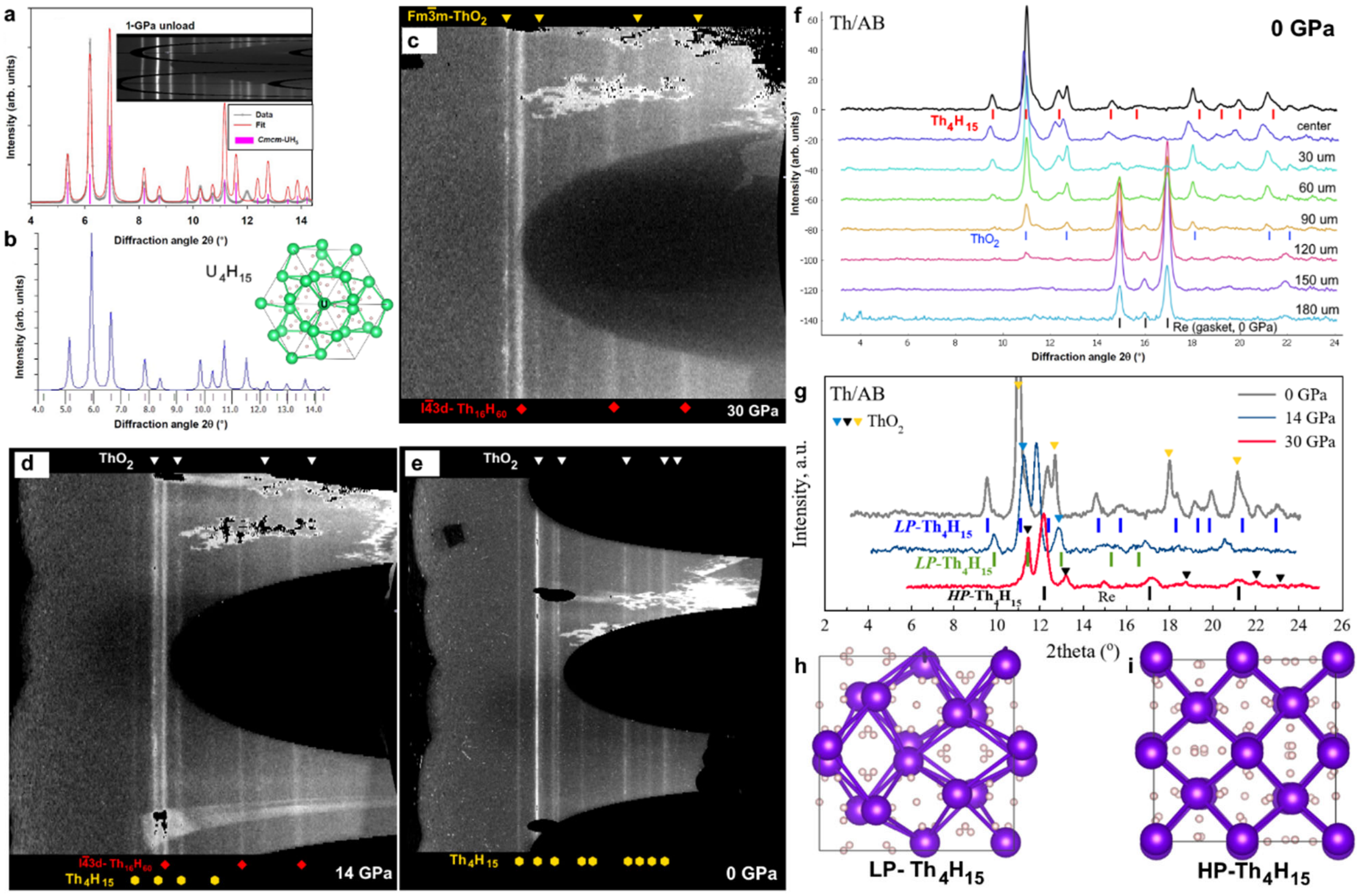

**Figure 6**. Formation of lower hydrides of uranium and thorium. (a) Formation of *I*-43*d*-$U_4H_{15}$ during decompression of the diamond anvil cell to 1 GPa in the experiment [26] and an explanation of the observed diffraction pattern proposed by I. Kruglov et al. (*Cmcm*-$UH_5$, $\lambda = 0.3344$ Å). (b) A much better interpretation of the experiment based on *I*-43*d*-$U_4H_{15}$ ($a = 8.7574$ Å at 0 GPa). (c–e) Diffraction patterns ($\lambda = 0.62$ Å) at 30, 14, and 0 GPa in the Th–H system. The hydrides were synthesized by laser-heating a mixture of Th and ammonia borane (AB), a rhenium gasket was used. There is a significant $ThO_2$ admixture due to air oxidation of thorium. (f, g) Integrated diffractograms of thorium hydride samples. The right side of panel (f) shows the shift of the beam center relative to the center of the sample (in micrometers). (h, i) Presumed structures of the two-phase modifications *LP*-$Th_4H_{15}$ and *HP*-$Th_4H_{15}$.

**Table 5.** Unit cell parameters of two modifications of $Th_4H_{15}$ (primitive cell is $Th_{16}H_{60}$) synthesized from Th and AB at 30 GPa, and thorium dioxide (admixture, unit cell is $Th_4O_8$).

| **Pressure, GPa** | **a, Å (*HP*)** | **V, Å$^3$/Th-atom (*HP*)** | **a, Å (*LP*)** | **V, Å$^3$/Th-atom (*HP*)** | **a, Å ($ThO_2$)** | **V, Å$^3$/Th-atom ($ThO_2$)** |
|---|---|---|---|---|---|---|
| 30 | 8.279 | 35.46 | - | - | 5.382 | 38.97 |
| 14 | 8.389 | 38.31 | 8.795 | 42.5 | 5.492 | 41.41 |
| 0 | - | - | 9.1054 | 47.18 | 5.604 | 44.00 |

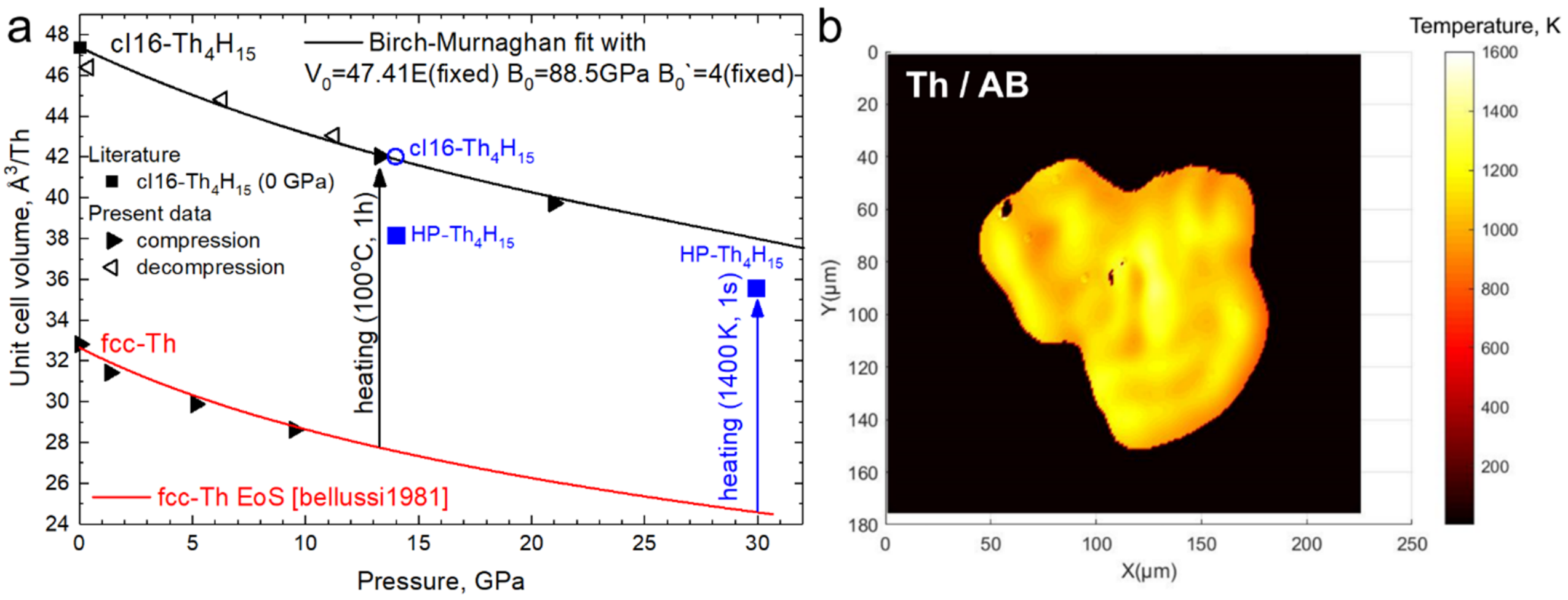

**Figure 7.** Equations of state of lower thorium hydrides. The data marked by black symbols were obtained by Dr. M. Kuzovnikov [186]; those marked in blue were obtained in this work when heating the sample at a higher temperature and pressure. (b) Heat map of a laser-heated thorium sample in an ammonia borane medium, obtained using an acousto-optical filter in the laboratory of Prof. P. Zinin (STC UI RAS, Moscow) [189, 190].

## 2.2 Thorium polyhydrides at high pressures

Theoretical studies of the possibility of formation of higher polyhydrides were first performed using the USPEX code [9-12] at pressures of 0, 50, 100, and 200 GPa (Figure 8) [181]. The calculations show that starting at low pressures of 0-50 GPa the formation of different polyhydrides is possible: $ThH_4$, $ThH_6$, and even $ThH_{16}$, which is likely to lose stability when the ZPE is considered. Unfortunately, our calculations were limited by the number of atoms in the unit cell (no more than 36), so we could not include $Th_4H_{15}$, which has a primitive unit cell of $Th_{16}H_{60}$ with 76 atoms. At 0 GPa and 0 K this phase has an enthalpy of -4.684 eV/atom and an enthalpy of formation of -0.556 eV/atom. Inclusion of this hydride to the convex hull at 0 GPa results in displacement and destabilization of $Th_3H_{10}$ and $ThH_3$; however, $ThH_4$ and $ThH_{16}$ remain stable under these conditions (without the ZPE), which is not consistent with the experimental data. It is possible that considering the ZPE will restore agreement with the experimental observations. These results were generally confirmed in an independent theoretical study using the CALYPSO code [191]. With further increases in the pressure up to 100 GPa, new stable thorium polyhydrides appear: $Fm\bar{3}m$-$ThH_{10}$, $P6_3/mmc$-$ThH_9$ (metastable without the ZPE), and $C2/m$-$ThH_7$. The latter hydride is analogous to the compounds that were later found in the Nd-H [24] and Eu-H [43] systems. A similar low-symmetric molecular hydride $XH_7$ may also exist for other lanthanides (e.g., Sm, Gd [192]). The detailed calculations of the phonon and electronic structure of these phases indicate their dynamic stability and metallic properties at pressures above 100 GPa [181].

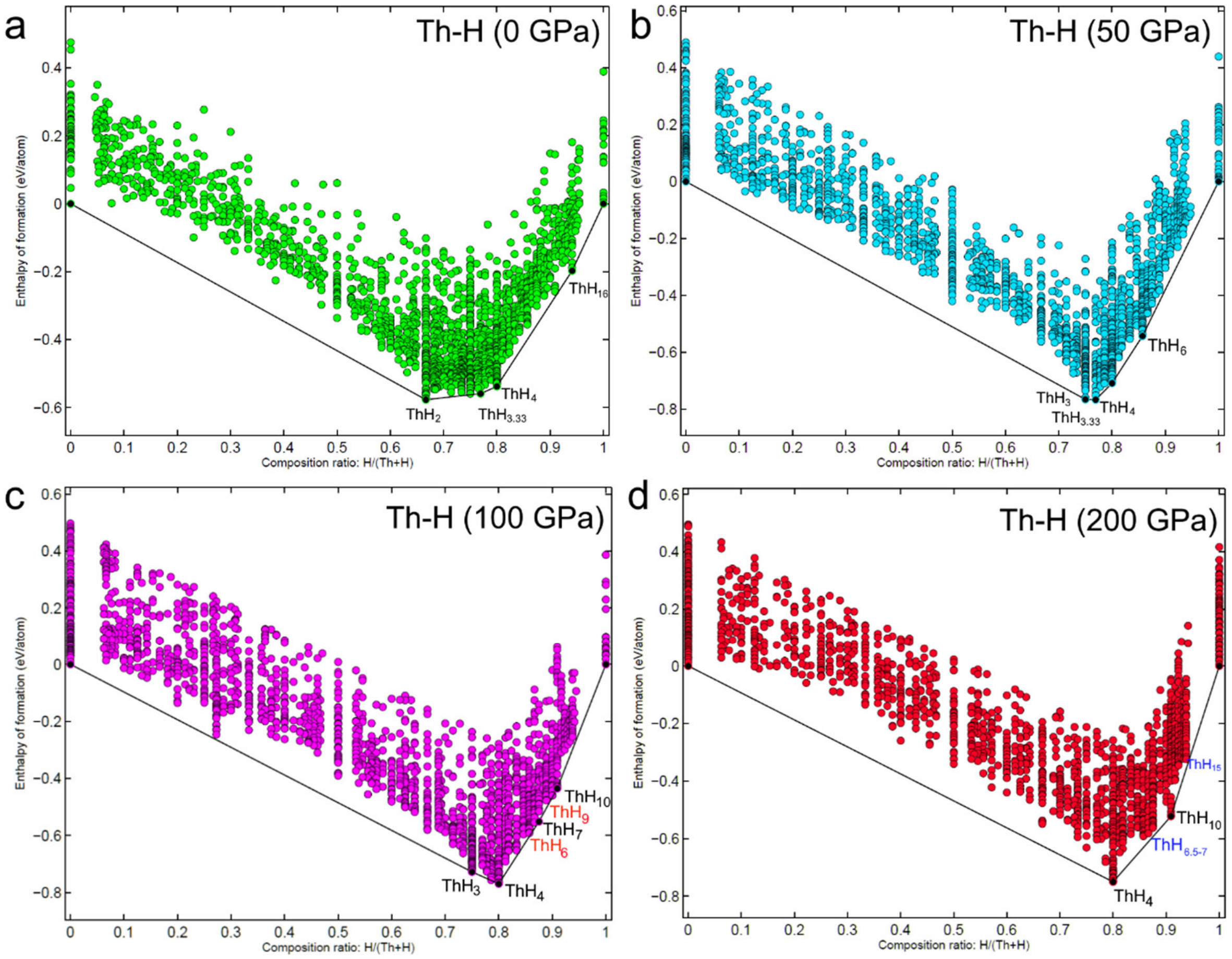

**Figure 8**. Thermodynamic convex hulls (enthalpy of formation - composition) of thorium hydrides at pressures 0 - 200 GPa calculated at 0 K without considering the zero-vibration energy. The VASP code with the PAW PBE functional was used in the calculations [193, 194].

Tetragonal *I*4/*mmm*-$ThH_4$ ($Th_4H_{16}$), theoretically stable and experimentally detected in a wide range of pressures above 80 GPa, has only a slightly higher hydrogen content than $Th_4H_{15}$ but significantly lower symmetry. As a result, the calculated superconducting properties of $ThH_4$ are significantly weaker, $T_C$ ~ 3 K at 85 GPa [181]. In the experiment at 30 GPa (Figure 6), we did not observe the formation of $ThH_4$. At all pressures, the right "hydrogen" part of the phase diagram is much more saturated with metastable phases than the left "metallic" part. This speaks in favor of the possibility of formation of a large variety of as yet unknown higher thorium polyhydrides.

We note an increasing interest in tetrahydrides for their superconducting properties. Despite relatively low $T_C$, their stabilization pressures are often about 50–100 GPa, which makes it easy to synthesize large tetrahydride samples and study them in detail. For example, in a recently published theoretical paper, room superconductivity has been predicted in $MgH_4$ at 280 GPa [195]. Tetrahydrides of $YH_4$ [196], which exhibits superconducting properties with maximum $T_C$ = 88 K at 150 GPa, as well as $LaH_4$, $ZrH_4$, and $ScH_4$ [197] have been experimentally and theoretically investigated. In general, tetrahydrides are a promising subclass of superconducting hydrides whose synthesis does not require pressures above 1 Mbar.

Among the compounds we found in calculations, the most interesting because of their superconducting properties are highly symmetrical polyhydrides: hexagonal *P*6₃/*mmc*-$ThH_9$ and cubic $Fm\bar{3}m$-$ThH_{10}$, stabilizing above 100 GPa (Figure 9). In this publication, we already noted that in the hydrogen sublattice of thorium decahydride (as well as all $Fm\bar{3}m$-$XH_{10}$ superhydrides), there are two significantly different types of hydrogen atoms (H1, H2) with different contributions to superconductivity. This was later investigated in detail in the theoretical papers on the electronic properties of $LaH_{10}$ [66] and $CeH_{9–10}$ [198]. In this regard, it seems promising to use the 1H NMR method in high-pressure diamond anvil cells [199] to confirm this theoretical prediction. It is also interesting that the band structure of $ThH_{10}$ lacks any features near the Fermi level (for example, the Van Hove singularities) that are usually considered important factors in the emergence of high-temperature superconductivity [66, 68, 200]. Nevertheless, as we show below, thorium decahydride exhibits high-temperature superconductivity.

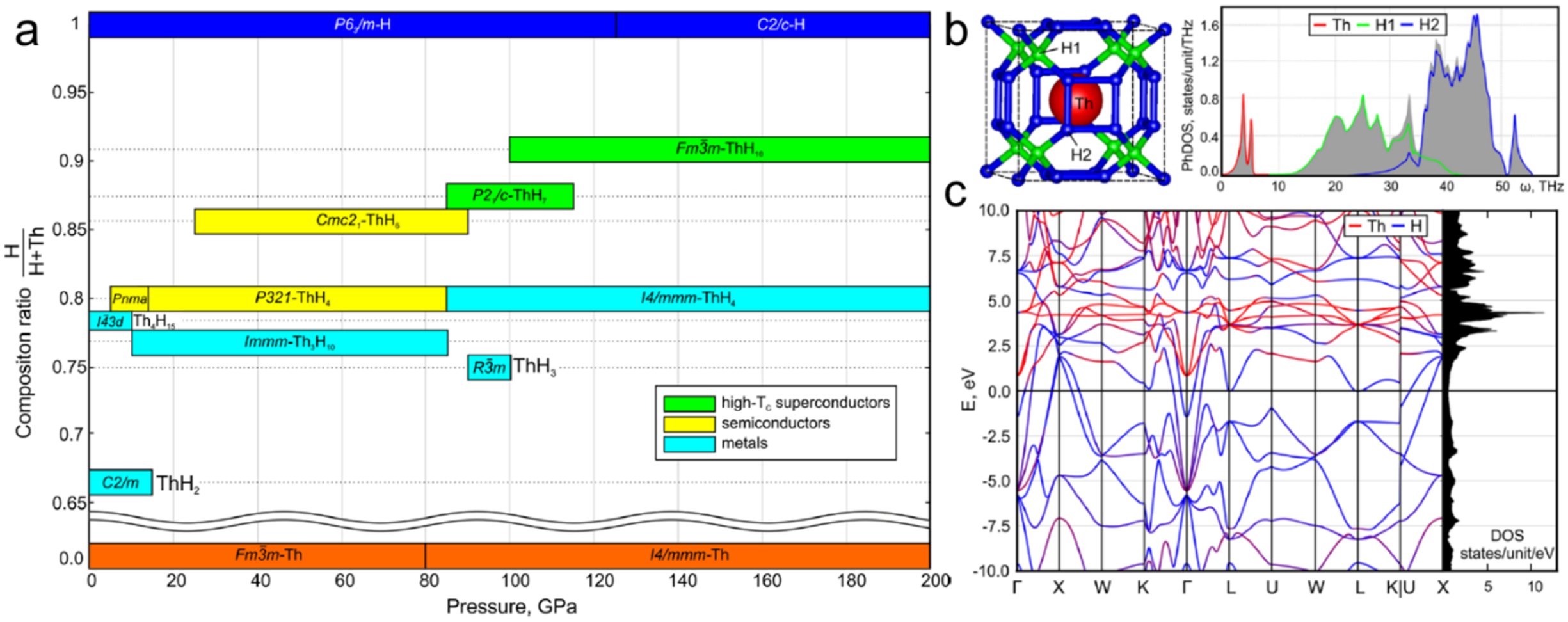

**Figure 9.** (a) Composition–pressure phase diagram for the Th–H system. Semiconducting, metallic, and high-temperature superconducting thorium hydrides are shown in yellow, blue, and green, respectively. (b) Crystal structure and the phonon density of states of $Fm\bar{3}m$-$ThH_{10}$ at 100 GPa. Colors indicate the contribution from different types of atoms (Th, H1, and H2). (c) Electronic band structure of $ThH_{10}$ at 100 GPa.

The calculations of the electron–phonon interaction within the isotropic harmonic approximation using the Quantum ESPRESSO code and the algorithm for solving the Eliashberg equations developed by the Polish group of R. Szczęśniak (Czestochowa University of Technology) [201] showed that $ThH_{10}$ should exhibit very high critical temperature $T_C$ = 200–240 K (Table 6), which decreases with increasing pressure. This is expected because the structure of $ThH_{10}$ is completely similar to $LaH_{10}$. However, as we will see soon, the use of theoretical calculations leads to a significant error in maximum $T_C$, which is 161 K in experiment.

**Table 6.** Calculated parameters of the electron-phonon interaction in $Fm\bar{3}m$-$ThH_{10}$ at different pressures, with the Coulomb pseudopotential $\mu^*$ = 0.1 (0.15)

| Pressure, GPa | λ | $N_F$, states/Å$^3$/Ry | $\omega_{log}$, K | $T_C$(McM), K | $T_C$(A-D), K | $T_C$ (E), K |
|---|---|---|---|---|---|---|
| 100 | 2.50 | 0.215 | 1073 | 176.8 (160.3) | 221.1 (193.9) | 241.2 (220) |
| 170* | 1.91 | - | 1209 | 165 | 197 | 214 |
| 200 | 1.35 | 0.227 | 1627 | 166.3 (139.4) | 182.6 (150.5) | 228 (205) |
| 300 | 1.11 | 0.23 | 1775 | 144.2 (114.2) | 155.4 (121.4) | 201 (174) |

*Separate calculation, $\Delta(0)$ = 46 meV, $\omega_2$ = 1433 K.

During 2020–2022, the reason for the discrepancy have been mostly clarified. First, this is the anharmonicity of the hydrogen sublattice vibrations, as a result of which the electron–phonon interaction parameter decreases, the logarithmically averaged phonon frequency increases, and the resulting critical temperature decreases by approximately 20–25 K [30, 119]. The second reason is the spin fluctuations. The introduction of impurities of magnetic atoms with *d, f* electrons into hydrides leads to the scattering of Cooper pairs with a change in the spin of the electrons. As a result, as has been shown in the experiments with the introduction of Nd [202], Ce [105, 203], and Pd [204] into the known hydride superconductors, magnetic impurities significantly suppress the superconducting properties of hydrides. The $d^2$ electrons of thorium atoms can partially fill the *f*-orbitals characteristic of all actinides and thus suppress superconductivity in $ThH_{10}$ within this mechanism.

More accurate (k-mesh: 24×24×24, q-mesh: 4×4×4) calculations of the Eliashberg function performed for $Fm\bar{3}m$-$ThH_{10}$ give approximately the same results (Figures 10-11, Table 6 (170 GPa)) as the previous calculations, confirming the need for corrections for anharmonicity and spin fluctuations to match the experiment. Another way to achieve this correspondence is to change (increase) the Coulomb pseudopotential $\mu^*$, which was discussed in many papers, for instance, for the cases of $LaH_{10}$ [127] and $YH_6$ [30]. However, there are no physical reasons for the anomalous growth of the Coulomb pseudopotential in polyhydrides. Despite the fact that the physical density of electrons in matter increases with increasing pressure, the importance of their electrostatic interaction with respect to kinetic energy decreases in the same way that the ideality of plasma increases with its density. As we will see below, the density of charge carriers and the electrical resistivity of hydrides in the normal state take values typical for metals, where $\mu^*$ = 0.1-0.13. Therefore, there are no reasons to expect any anomalous values of $\mu^*$ for superhydrides either.

The superconducting properties of $ThH_9$ and $ThH_{10}$ can be calculated without using the empirical value of the Coulomb pseudopotential $\mu^*$ within the superconducting DFT approach (SCDFT) [132, 133]. These calculations were carried out by Prof. R. Akashi (Tokyo University) using ultrasoft (USPP) and norm-conserving (NCPP) pseudopotentials for Th and H and the Sanna-Pellegrini-Gross functional (SPG2020) [136]. As a result, the following values for the critical temperature were found: $T_C$(USPP) = 226 K and $T_C$(NCPP) = 207 K for $ThH_{10}$ (170 GPa); $T_C$(USPP) = 185 K and $T_C$(NCPP) = 184 K for $ThH_9$ (150 GPa). The critical temperatures obtained are also significantly higher than those observed experimentally.

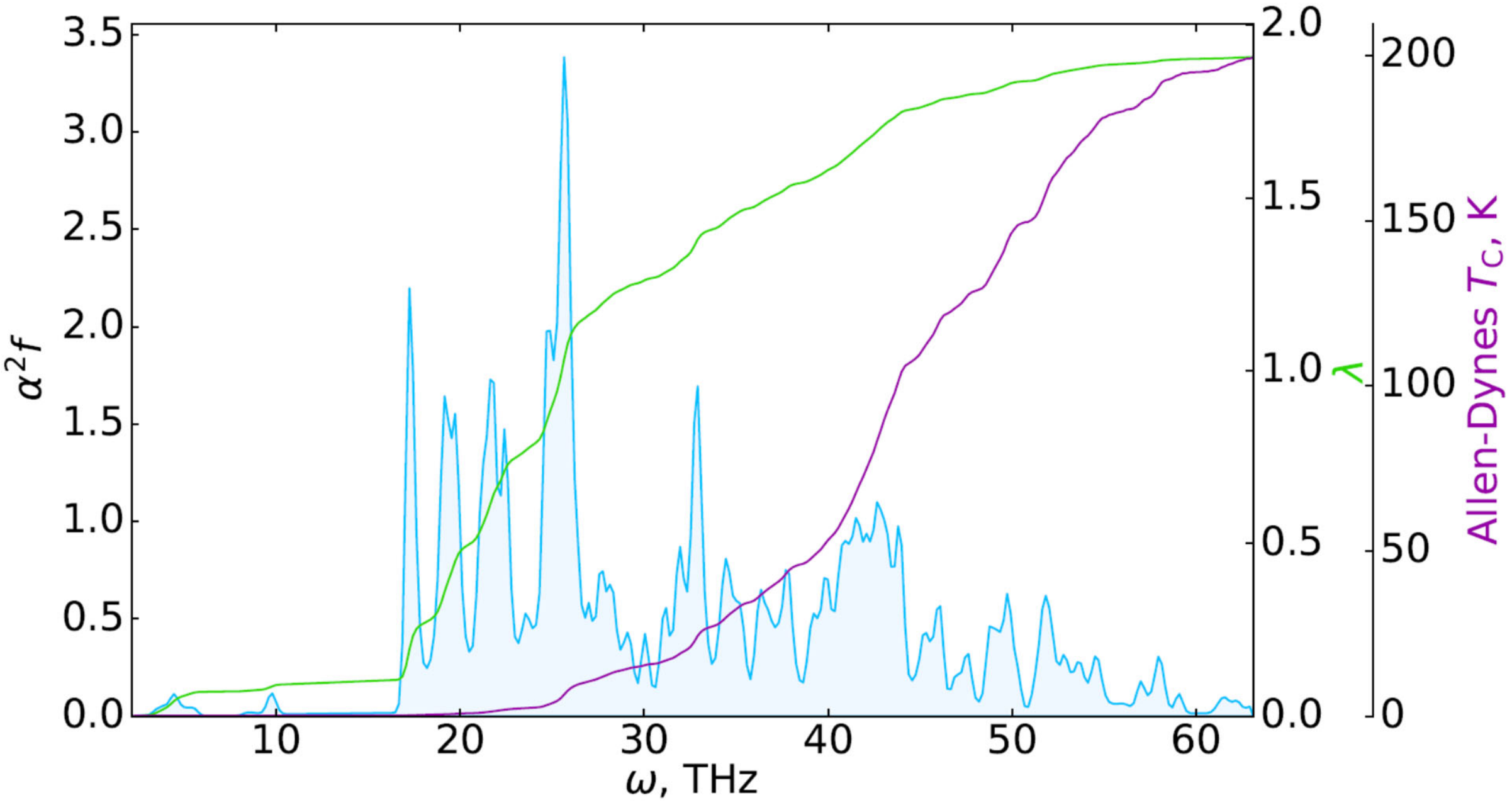

**Figure 10.** Harmonic Eliashberg function $Fm\overline{3}m$-$ThH_{10}$ calculated at 170 GPa using Quantum ESPRESSO [205]. Additional scales show the electron-phonon interaction parameter (λ), and the critical temperature of superconductivity calculated using the Allen-Dynes formula [60] (see also Appendix).

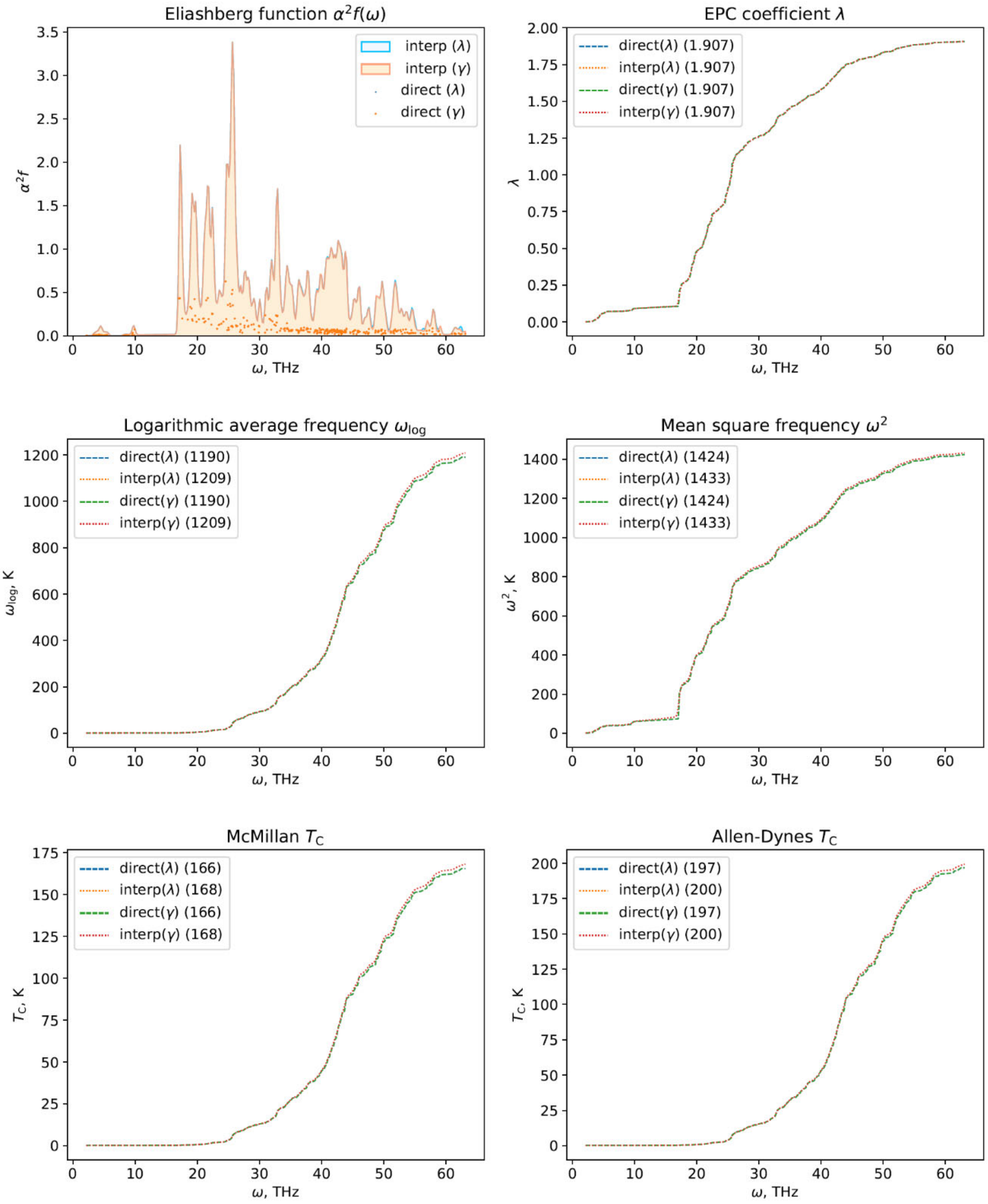

**Figure 11.** Eliashberg function and the integral $\lambda(\omega)$, $\omega_{log}(\omega)$, $\omega^2(\omega)$, and $T_C(\omega)$ parameters of the electron-phonon interaction calculated using different methods (interpolation over the first Brillouin zone and direct summation of the partial EPC coefficients) for $ThH_{10}$ at 170 GPa, showing good convergence of the obtained results. The calculations were performed using the script [206].

## 2.3 Experimental study of thorium polyhydrides at pressures above 1 Mbar

Synthesis of thorium superhydrides [27] was performed in a series of diamond anvil cells with a 50-micron culet and tungsten gaskets loaded with thorium and ammonia borane. The first stage of laser heating was performed at 88 GPa and resulted in only two $ThH_4$ modifications (*I*4/*mmm* and *P*321) predicted theoretically (Figure 12). The next laser heating was performed at a higher pressure, 152 GPa, and led to the formation of a small amount of *P*6₃/*mmc*-$ThH_9$, which is actually surprising because the stable phase should be $ThH_{10}$. The reason for this behavior is probably the lack of hydrogen (or ammonia borane, AB) in this cell, as verified by subsequent cycles of laser heating and pressure increase, which did not lead to a change in the set of the synthesis products (Figure 12). As we will see later, $ThH_9$ almost always accompanies $ThH_{10}$. The problem is to obtain these compounds separately.

The synthesis of almost pure $ThH_{10}$ was carried out in the other diamond anvil cell at a pressure of about 170 GPa (Figure 13). A very small $ThH_9$ impurity was also present in this DAC. Reducing the pressure showed that $Fm\bar{3}m$-$ThH_{10}$ decomposes with a distortion of the cubic structure between 101 and 85 GPa, and with a simultaneous increase in the $ThH_9$ content. This behavior is a common scenario for metal polyhydrides. When the pressure decreases, their structure first undergoes a distortion, a loss of hydrogen, and a transformation into a phase with a smaller hydrogen content: $Fm\bar{3}m$-$ThH_{10}$ → *Immm*-$ThH_{10}$ → $P6_3/mmc$-$ThH_9$ → $Cmc2_1$-$ThH_6$ [27]. As was shown later, $LaH_{10}$ undergoes a similar metamorphosis with decreasing pressure ($Fm\bar{3}m$ → *R*-3*m* → *C*2/*m* → *P*1) [207].

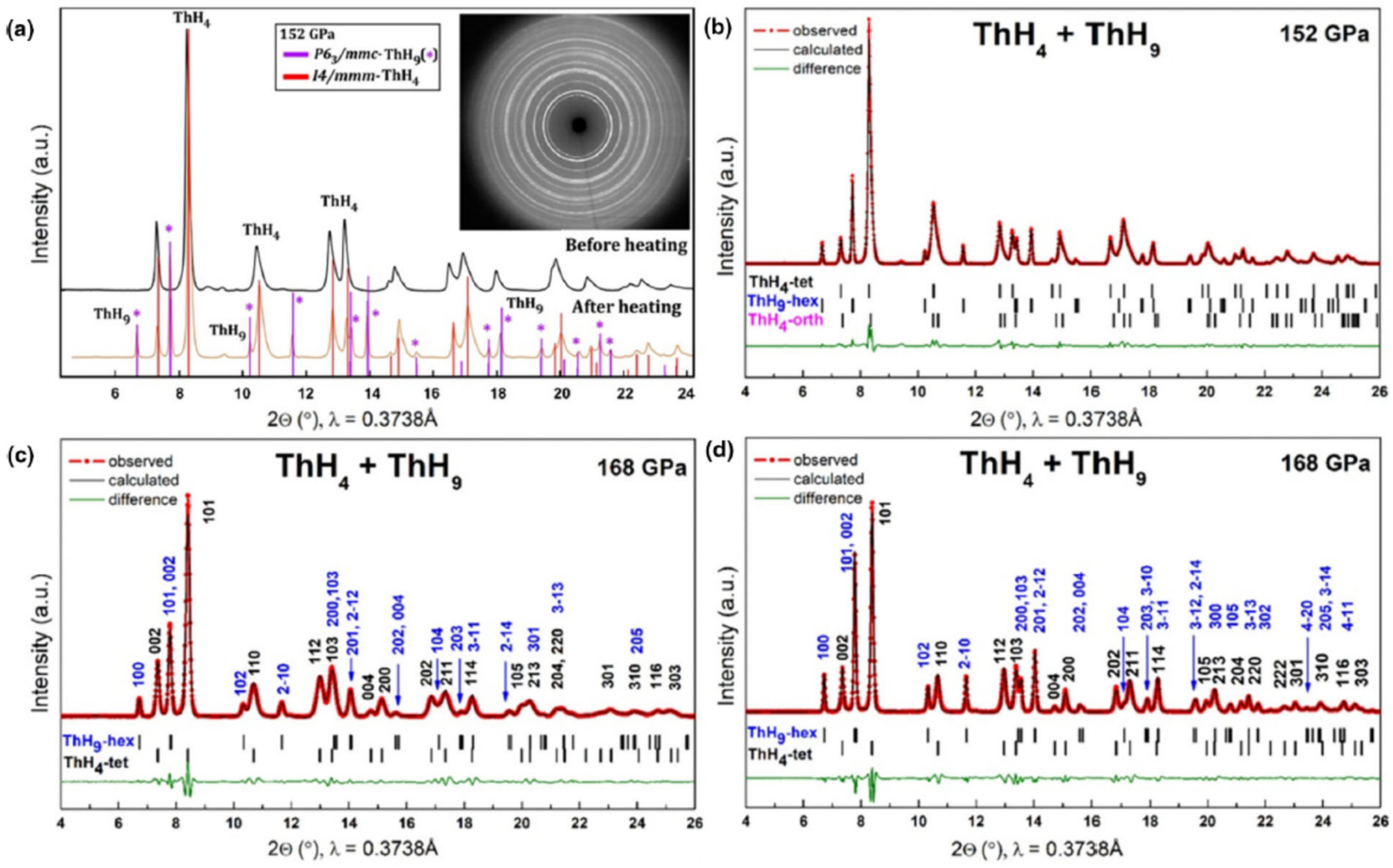

**Figure 12.** (a) Comparison of the XRD patterns of sample M2 at 152 GPa before and after the laser heating. The formation of $ThH_9$ is observed. New peaks are marked by asterisks. The inset shows the experimental XRD pattern obtained with the incident X-ray wavelength of 0.3738 Å; (b) Le Bail refinements of $P6_3/mmc$-$ThH_9$ and $ThH_4$ at 152 GPa and (c) after the third and (d) fourth cycles of heating at 168 GPa (sample M2). The experimental data, fit, and residues are shown in red, black, and green, respectively.

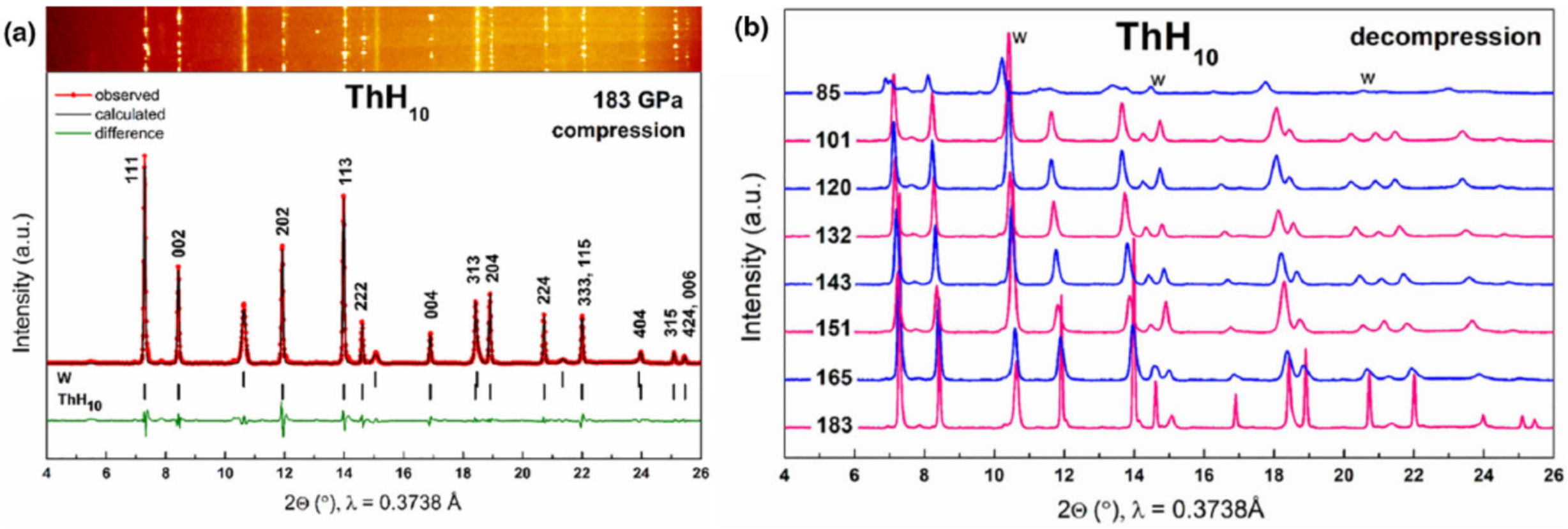

**Figure 13.** (a) Le Bail refinement of $Fm\bar{3}m$-$ThH_{10}$ and *bcc*-W at 183 GPa. The experimental data, fit, and residues are shown in red, black, and green, respectively. (b) Experimental XRD patterns of $Fm\bar{3}m$-$ThH_{10}$ in the pressure range of 183–85 GPa.

*P*6$_3$/*mmc*-$ThH_9$ is the first discovered member of the family of hexagonal superconducting polyhydrides, which now also includes $PrH_9$ [25] $NdH_9$ [24], $EuH_9$ [43], $CeH_9$ [28, 106], and $YH_9$ [31]; we investigated this compound's properties in detail. We noticed the difference in the unit cell volume, which is expected to be smaller by 1.2–1.3 Å$^3$/Th (>150 GPa) than for $ThH_{10}$ at high pressures (Figure 14). This agrees with the idea that each hydrogen atom corresponds to a certain volume in the unit cell $V_H(P)$ of polyhydrides [21] (Figure 14). This calculation scheme allows us to estimate the hydrogen content in $ThH_{10}$ from the experimental data as $10 \pm 0.5$. However, for the hexagonal hydride $ThH_9$, the situation is more complicated. By this method of estimation, its composition is closer to $Th_2H_{19}$, which is also confirmed by the equation of state at pressures below 150 GPa (see Supporting Information to Ref. [27]): the unit cell volume of the cubic $ThH_{10}$ phase increases much slower during the decompression of the DAC than that of the hexagonal $ThH_9$ phase, indicating that both hydrides are possibly nonstoichiometric. This could also be a reason for the decrease in $T_C$ in the system with the decreasing pressure (Figure 18).

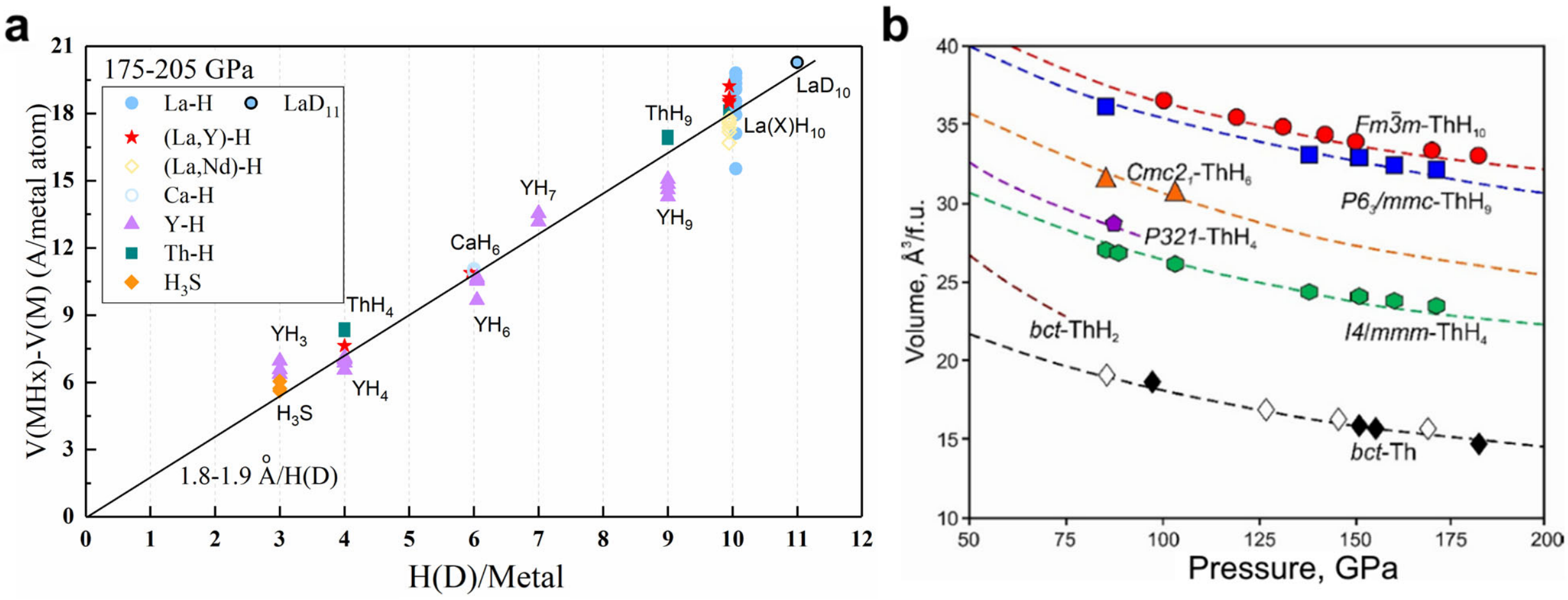

**Figure 14.** (a) Dependence of the volume per hydrogen atom in the unit cell of polyhydrides (deuterides) proposed by Dr. M. Kuzovnikov. (b) Equations of state for the synthesized thorium polyhydrides (filled symbols) in comparison with theoretical calculations (lines) and the literature [208] (hollow symbols).

The analysis of the band structure for an ideal $P6_3/mmc$-$ThH_9$ crystal at 100 and 150 GPa (Figure 15) shows the absence of any features (for example, van Hove singularities), as for $ThH_{10}$. The superconducting properties of $ThH_9$, according to calculations (Table 7), are less pronounced than in $ThH_{10}$: $T_C$(A–D) at 150 GPa reaches 145 K and decreases as the pressure decreases to 100 GPa. More accurate calculations give a similar result: $T_C$ = 148 K (150 GPa, Figure 17). This is quite close to the experimental $T_C$ = 146 K at 170 GPa (Figure 16). For both $ThH_9$ and $ThH_{10}$, the electrical resistance of the samples decreases upon cooling, in agreement with the Bloch–Grüneisen formula [64, 65], from 0.06–0.08 Ω at 170 GPa (corresponding to $\rho \sim 2–3 \times 10^{-7}$ Ω·m at a sample thickness of 3–4 μm) to 10–14 μΩ ($\rho \sim 4 \times 10^{-11}$ Ω·m), which is about 10,000 times smaller than the resistance in the normal state and 100 times smaller than in the best normal metals (e.g., Ag) at 150 K (Figure 17). The calculation of the electrical resistivity of $ThH_9$ in the normal state at 150 GPa and 200 K, carried out using the EPW code [140-143], yields 4–4.5 × $10^{-7}$ Ω·m, which is quite close to the experimental values. The critical temperature of superconductivity depends linearly on the applied external magnetic field (0–16 T, Figure 16). A linear extrapolation gives the upper critical field $\mu_0 H_{C2}(0) \approx 65$ T. As has been recently shown [202], the linear extrapolation of the upper critical magnetic field for superhydrides is more accurate than the Werthamer–Helfand–Hohenberg model (WHH), which gives underestimated values of $\mu_0 H_{C2}$ at low temperatures (see also Appendix, "Details of the upper critical magnetic field calculations").

In recent years, after the works of E. Talantsev [86, 209, 210], the analysis of the temperature–resistance curve $R(T)$ in the normal state using the Bloch–Grüneisen formula has become popular [64, 65]. Using previously unpublished data for $ThH_{10}$ at 170 GPa (Figure 17c, see also [209]) we found $\lambda = 1.65$, $\omega_{\log} \approx 1116$ K, and $\theta_D = 1350$ K at $\mu^* = 0.1$, which is quite close to the calculated values. Thus, the transport properties of thorium polyhydrides in the normal and superconducting states can be obtained with good accuracy by modern computational methods within the framework of the known models of electron–phonon interaction in metals.

**Table 7.** Parameters of the superconducting state of $P6_3/mmc$-$ThH_9$ at 100 and 150 GPa calculated using $\mu^* = 0.1$ (0.15). $\alpha$ is the isotope coefficient.

| **Parameters** | **$P6_3/mmc$-$ThH_9$** | |
|---|---|---|
| | **100 GPa** | **150 GPa** |
| $\lambda$ | 2.15 | 1.73 |
| $\omega_{log}$, K | 728 | 957 |
| $\alpha$ | 0.48 (0.47) | 0.48 (0.47) |
| $T_C$ (A-D), K | 138 (118) | 145 (123) |
| $T_C$ (E), K | 156 (142) | 161 (145) |
| $T_C$ (E, $ThD_9$), K | 112 (102) | 115 (105) |
| $\Delta(0)$, meV | 35.2 (31.2) | 33.9 (29.6) |
| $\mu_0 H_{C2}(0)$, T (eq. A8) | 41 (37) | 37 (33) |
| $R_\Delta = 2\Delta(0)/k_B T_C$ | 5.24 (5.11) | 4.89 (4.74) |

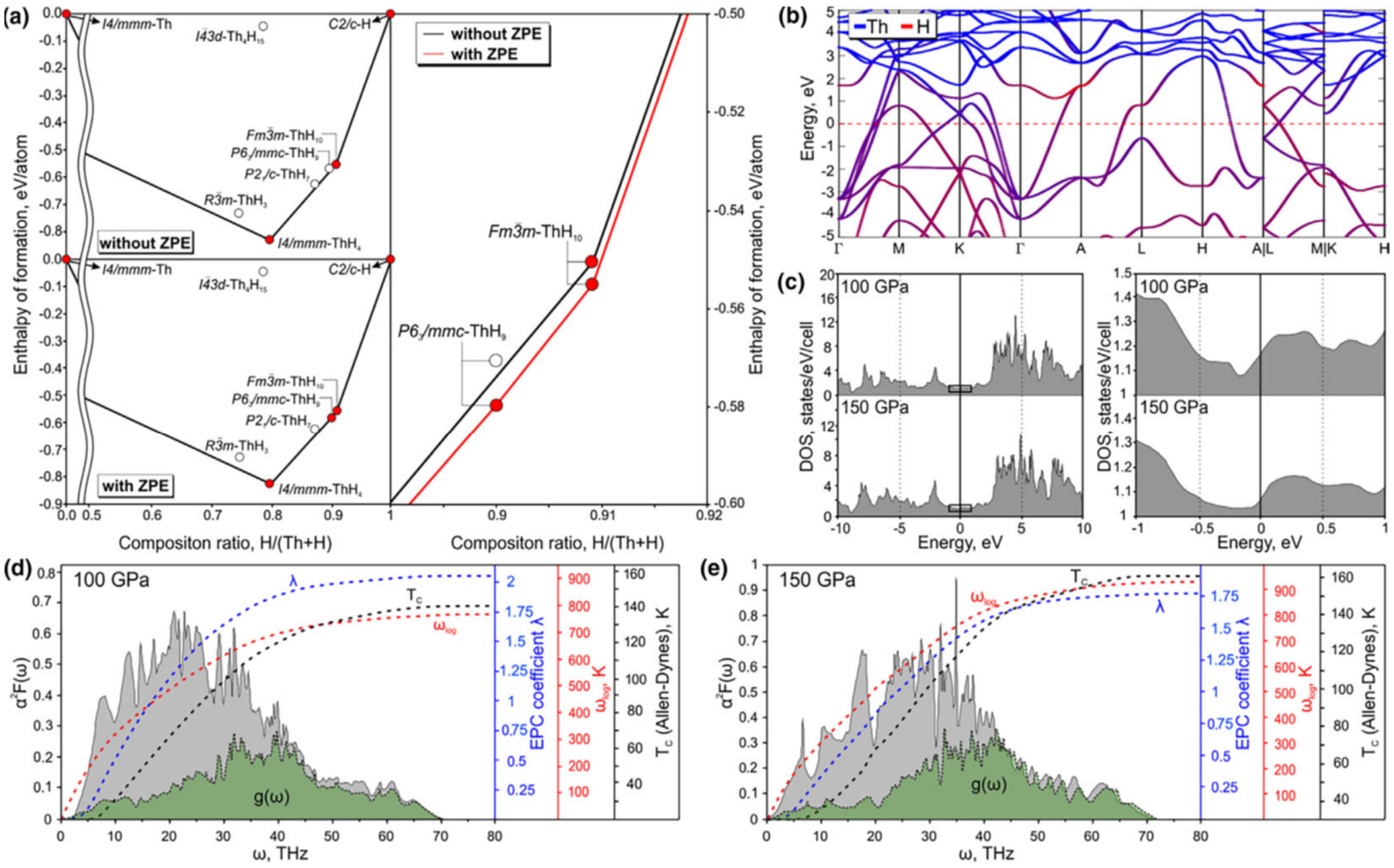

**Figure 15.** (a) (left) Thermodynamic convex hull of the Th–H system with and without the zero-point energy (ZPE) contribution at 150 GPa. (right) Magnified region near $ThH_9$ and $ThH_{10}$ showing that $ThH_9$ is stabilized by the zero-point energy. Hereinafter, the hollow circles correspond to metastable phases, and the filled ones – to stable phases. (b) Band structure of $P6_3/mmc$-$ThH_9$ at 150 GPa. Contributions of hydrogen and thorium are shown in red and blue, respectively. (c) Electronic DOS of $P6_3/mmc$-$ThH_9$ at 100 and 150 GPa in the energy ranges of ±10 eV and ±1 eV around the Fermi level. (d, e) Phonon density of states (green, in a.u.), the Eliashberg function $\alpha^2F(\omega)$ (gray), and superconducting parameters of $P6_3/mmc$-$ThH_9$ at 100 and 150 GPa.

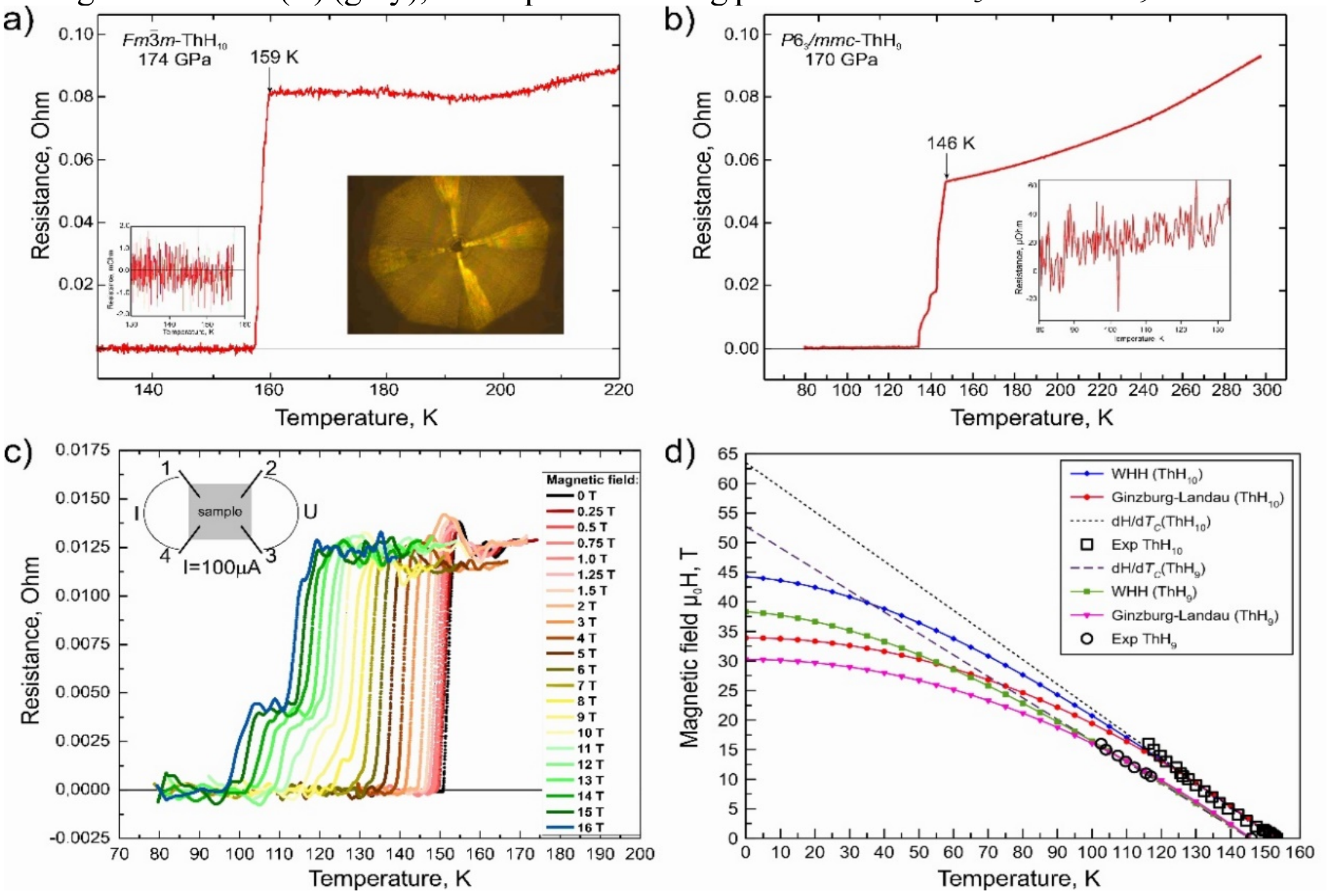

**Figure 16.** Observation of superconductivity in (a) $ThH_{10}$ and (b) $ThH_9$. The temperature dependence of the resistance $R$ of thorium superhydride was determined in a sample synthesized from Th and $NH_3BH_3$. The resistance was measured with four electrodes deposited on a diamond anvil on which the sample was placed (the right inset in panel (a)), at an excitation current of 100 μA. In the other insets, the resistance in the superconducting state is shown in a smaller scale. (c) Temperature dependence of the electrical resistance in an external magnetic field at 170 GPa. (d) Dependence of the critical temperature $T_C$ of $ThH_{10}$ and $ThH_9$ on the magnetic field.

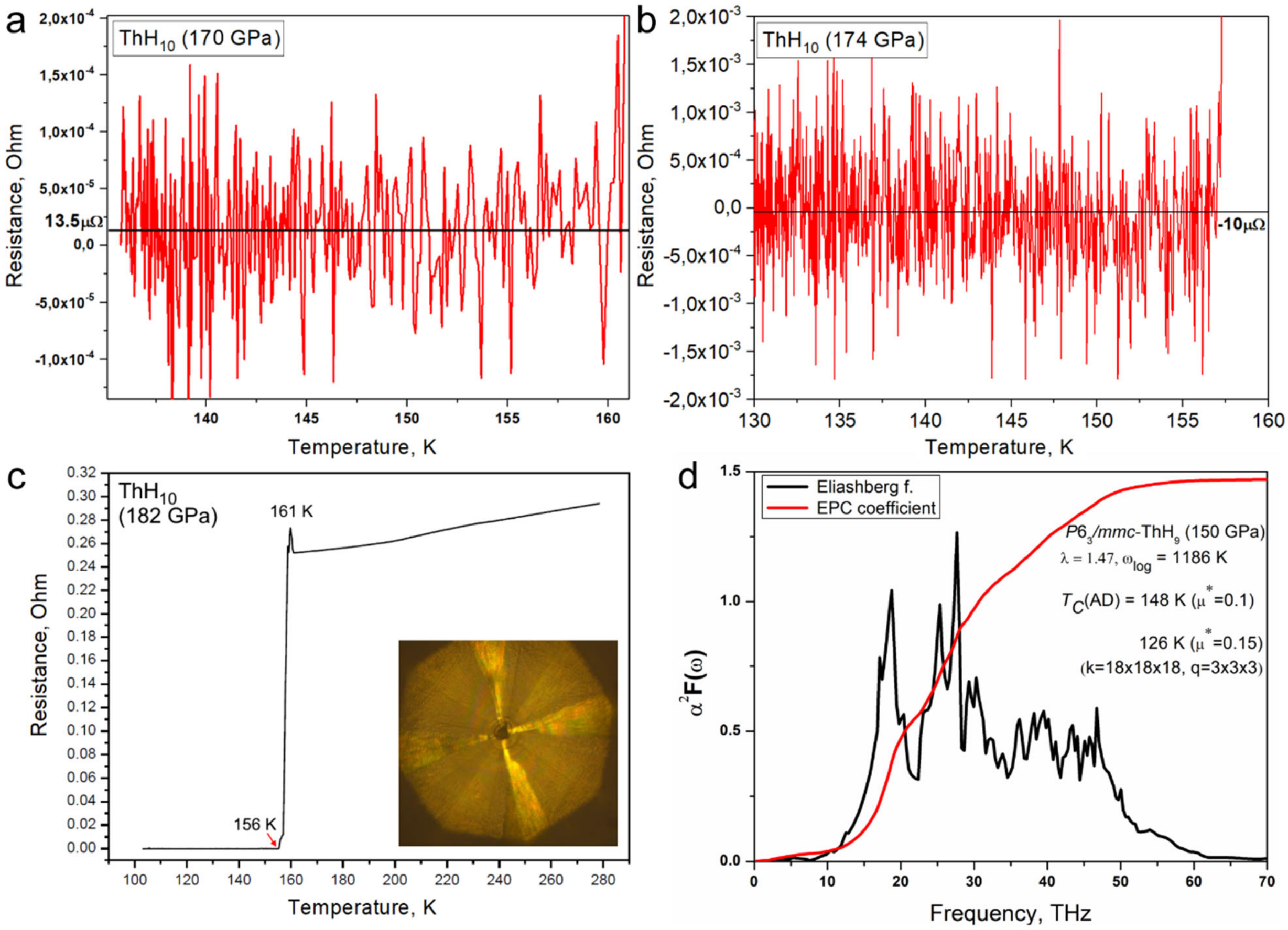

**Figure 17.** Superconducting properties of thorium polyhydrides. (a, b) Residual resistance of $ThH_{10}$ in the superconducting state at 170–174 GPa. (c) Superconducting transition width $\Delta T_C = 5$ K for $ThH_{10}$ at 161 K. Inset: thorium polyhydride sample lying on a diamond anvil with electrodes placed on it. The culet size is 50 µm. (d) Refined Eliashberg function of $P6_3/mmc$-$ThH_9$ calculated at 150 GPa.

In the article [27], we did not discuss some details of the superconducting behavior of thorium hydrides, in particular the pressure dependence of the critical temperature. In subsequent years, this experiment was performed and led to ambiguous results. In contrast to theoretical predictions, the slope $dT_C/dP \approx +0.3$ K/GPa (150 GPa) is positive and $T_C$ decreases with the pressure. The XRD analysis of the sample in this DAC shows the presence of two thorium polyhydrides, $ThH_9$ and $ThH_{10}$ (Figure 18), in which $dT_C/dP$ gradient has a different sign. Thus, one of the reasons of the deviation from the theoretical calculations may lie in the phase composition of the studied sample. Another reason could be the anharmonic effects.

Another aspect not discussed in the article [27] is the Hall effect in $ThH_{10}$ at 165 GPa in low magnetic fields of about 3 T ($I = 1$ mA, Figure 19). After some initial growth (up to 1 T), the Hall voltage goes to a linear mode: $V_H/B_z = I/e{\times}n_e{\times}t = 5–7 \times 10^{-8}$ V/T, then the density of charge carriers is equal to $n_e = 3–4 \times 10^{28}$ electrons/m$^3$, which does not differ in the order of magnitude from the values for metals and corresponds to one charge carrier per $ThH_{10}$ unit cell. Thus, in the normal state, $ThH_{10}$ demonstrates the behavior typical for ordinary metals. Considering the ratio $R(300\text{ K})/R(T_C)$, which is in the range of 1–1.5 for most hydrides, a conclusion can be made on the basis of Matthiessen's rule that the sample has many impurities and defects. Matthiessen's rule says that the total resistivity ρ can be approximated by adding up several different terms: $\rho_{total} = \rho_{thermal} + \rho_{impurity}$ $+ \rho_{deformation}$, where the components stand for different types of electron scattering: $\rho_{thermal}$ – on phonons, $\rho_{impurity}$ – on impurities (which does not depend on the temperature), $\rho_{deformation}$ – on deformation-induced defects. For pure metals, the resistance ratio $R(300\ K)/R(T_C)$ reaches 10–1000.

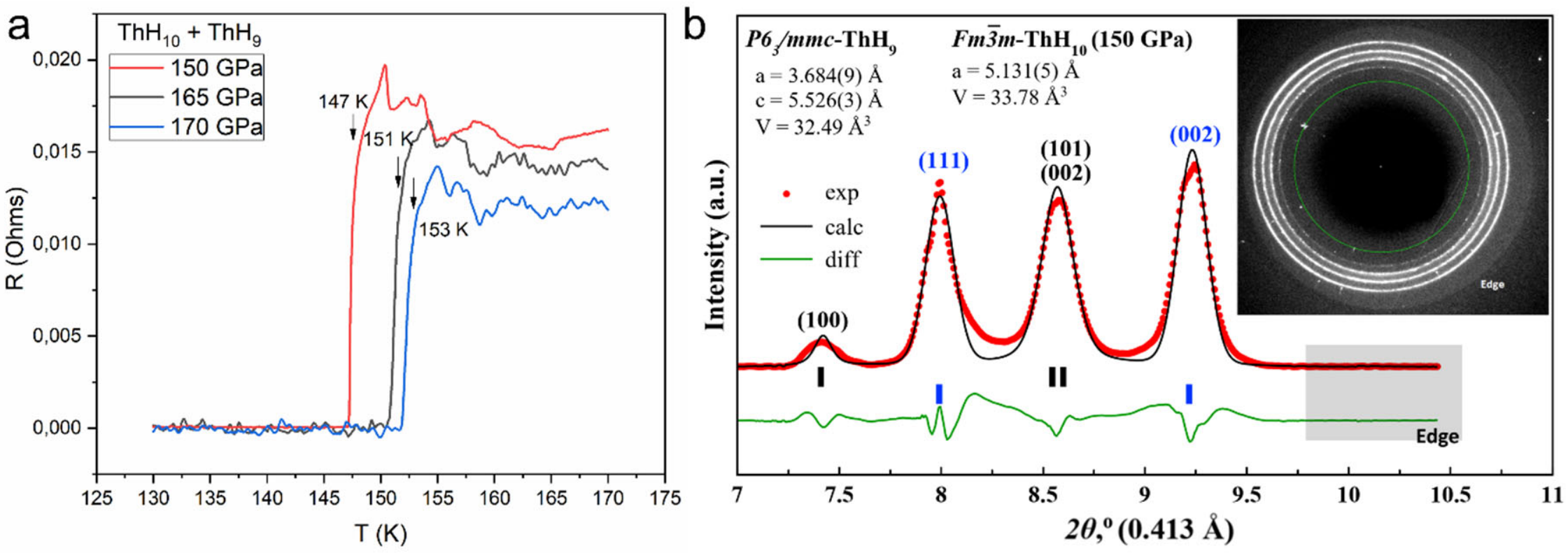

**Figure 18**. (a) Pressure dependence of the critical temperature for the $ThH_9$ + $ThH_{10}$ sample (ratio approximately 1:1). (b) Diffraction pattern of the aperture-limited electrical cell and the Le Bail refinement of the $ThH_9$ and $ThH_{10}$ unit cell parameters at 150 GPa. Inset: XRD pattern from a sample confined to the diamond anvil cell aperture. The oscillations before the transition in panel (a) are due to an unstable operation of the electrode system of the DAC because of its damage during decompression.

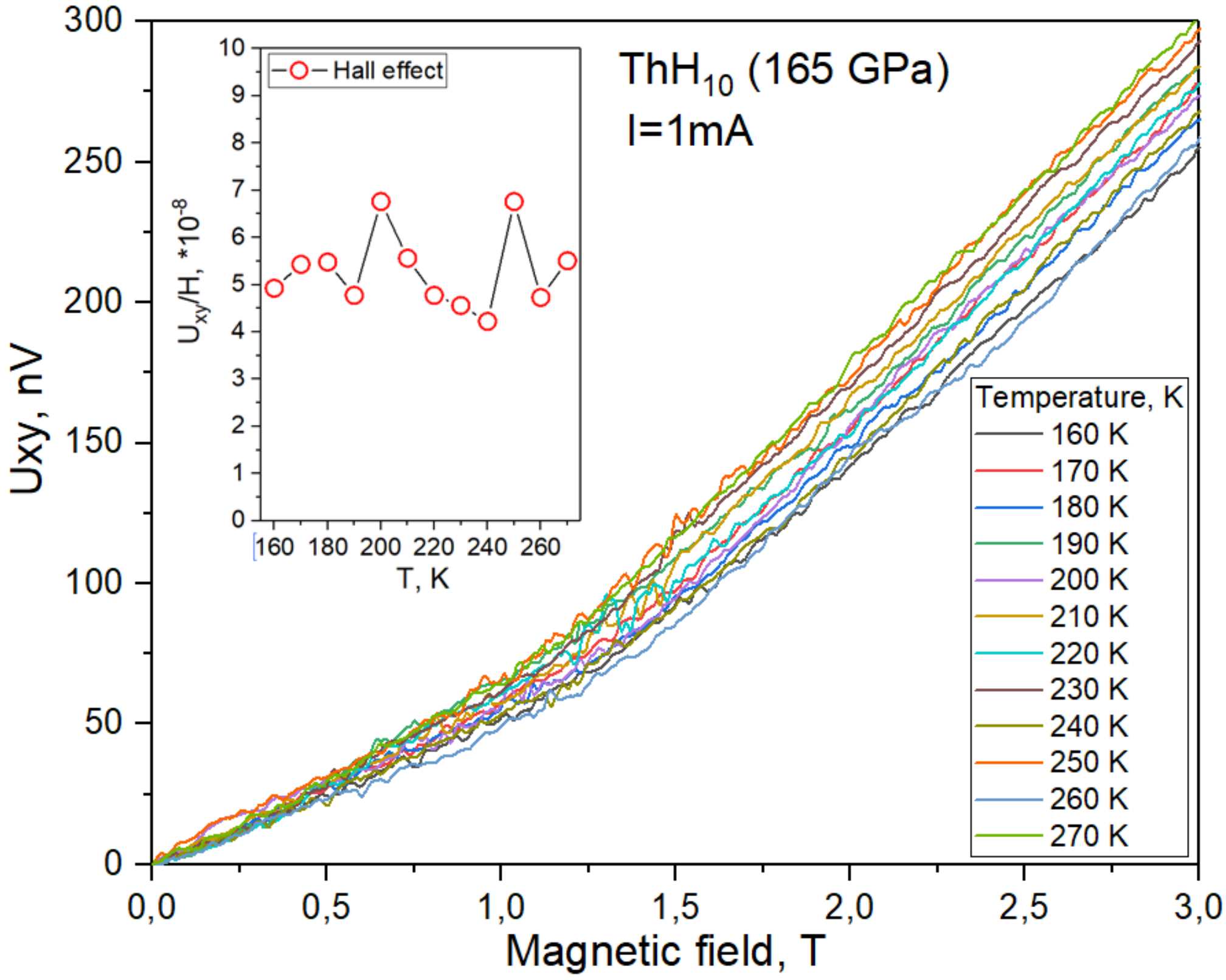

**Figure 19.** Hall effect in $ThH_{10}$. Dependence of the Hall voltage ($U_{xy}$, nV) on the applied magnetic field at different temperatures from 160 to 270 K. Inset: temperature dependence of the Hall voltage divided by the value of the external magnetic field for the linear part of the diagram (H > 1.2 T).

## 2.4 Plans for continuing research on thorium polyhydrides

Unresolved tasks in the field of thorium hydride research are the measurement of the isotope effect and study of superconductivity in $ThD_{10}$ and $ThD_9$, including their investigation in pulsed magnetic fields to determine the upper critical magnetic field (the expected value at 0 K is ~65 T) as accurately as possible. A detailed Raman study of thorium hydrides has also never been performed. From the chemical point of view, synthesis at pressures around 50 GPa is of interest. A theoretical study of the effect of anharmonicity on superconductivity in $ThH_{10}$ and $ThH_9$ is planned in the near future.

Given the similarity of the Th-H system with the Zr-H, Hf-H, and Ce-H systems, the synthesis of such nona- and decahydrides as $ZrH_9$, $ZrH_{10}$, $HfH_9$, and $HfH_{10}$ predicted in the theoretical papers [211], should be expected in the future. Indeed, studies of 2020-2021 have shown the formation of high-temperature superconductors $Fm\bar{3}m$-$CeH_{10}$ and $P6_3/mmc$-$CeH_9$ at moderate pressures around 80-120 GPa. The superconducting properties of cerium hydrides are weaker (i.e., have a lower max $T_C$ and the EPC parameter λ) than those of thorium hydrides, probably due to the greater probability of the Cooper pairs decay by scattering on Ce atoms, whose valence electrons have a more pronounced *f*-character than those of Th. After the recent discovery of hexagonal ternary hydrides (La, Ce)$H_{9\text{-}10}$ [203, 212], which possess the high-$T_C$ superconducting properties at moderate pressures of about 100 GPa, synthesis of La-Th alloys (1:1, 4:1) and superhydrides based on them will also be a promising direction for further research. Because Th suppresses superconductivity less than Ce, the critical temperature in such compounds is expected to be even higher at a comparable pressure of about 100 GPa.

Increasing the critical temperature of superconductivity of polyhydrides is another avenue of research. One approach to this problem relies on the search for systems with significant structural anisotropy. The superconducting properties of cuprates and $MgB_2$ are strongly anisotropic, and the value of the superconducting gap Δ(0) along the crystallographic planes is much higher than in the perpendicular direction. The same idea was used in nano-optics to create hyperlenses: deterioration of the target material properties in one direction can lead to multiple improvements of the same properties in the other direction. Hexagonal binary and ternary hydrides containing flat hydrogen layers are of particular interest from this point of view. Indeed, in the recent study of the Ce–La–H system, the introduction of Ce allowed the stabilization of the hexagonal ternary phases $P6_3/mmc$-(La,Ce)$H_{9\text{–}10}$ [203] which have better superconducting properties at low pressures than lanthanum polyhydrides.

It is interesting that the $T_C$ of hexagonal $ThH_9$ is lower than that of cubic $ThH_{10}$. This is also observed in other systems: $T_C(CeH_{10}) > T_C(CeH_9)$ [105] and probably $T_C(YH_{10}) > T_C(YH_9)$ [31]. There are currently no examples of hexagonal polyhydrides with a higher critical temperature than the nearest cubic polyhydrides of the same metal with similar hydrogen content.

## 2.5 Conclusions from the study of thorium polyhydrides

The study of thorium hydrides showed that formation of high-$T_C$ superconducting compounds at high pressures is not a unique property of the H–S system [5] and is inherent in a wide class of polyhydrides. Some of the first representatives of cubic clathrate [19] decahydrides ($ThH_{10}$), hexagonal clathrate nonahydrides ($ThH_9$), and tetragonal tetrahydrides ($I4/mmm$-$ThH_4$) were experimentally discovered. In a series of consequent papers [27, 181], high efficiency of computational methods in predicting new superconducting hydrides at high pressures was demonstrated for the first time. Additionally, we would like to note the following:

- Thorium tetrahydride $I4/mmm$-$ThH_4$ cannot be obtained at 30 GPa and below despite the theoretically predicted thermodynamic and dynamic stability of this compound.
- Equation of state of "$P6_3/mmc$-$ThH_9$" indicates a more complex structure and non-stoichiometric composition, close to $Th_2H_{19}$.
- There is a significant deviation of the experimental $T_C(ThH_{10})$ and $dT_C/dP$ from the results of theoretical calculations due to anharmonicity of the hydrogen sublattice and possible spin fluctuations that arise because of the partial *f*-character of the valence electrons of thorium.

# Chapter 3. High-temperature superconductivity in yttrium hydrides

## 3.1 Prediction of high-temperature superconductivity in yttrium hydrides

This chapter is based on the results of the theoretical-experimental publication [30] devoted to the properties of yttrium hexahydride $YH_6$ and hexadeuteride $YD_6$. All studies of polyhydrides by our group are of a comprehensive theoretical-experimental nature, and we are convinced that this approach is correct and effective.

Interest in yttrium hydrides arose as a result of a series of theoretical papers published in 2017 [18, 19, 213], which predicted the outstanding superconducting properties of $LaH_{10}$, $YH_6$, and $YH_{10}$ cubic superhydrides at pressures of 150–300 GPa. The first of them was experimentally obtained a year later and showed a record-high stability: $Fm\bar{3}m$-$LaH_{10}$ can be synthesized at 140–150 GPa. The critical temperature of $LaH_{10}$ is slightly lower than predicted: $T_C$(exp) = 250 K, $T_C$(theory) = 286 K. Moreover, the hexagonal modification of $LaH_{9–10}$ occurred only sporadically in the experimental diffraction patterns [72]. As the main product, $P6_3/mmc$-$LaH_{9-10}$ can be obtained by stabilization with cerium as part of the ternary hydride $P6_3/mmc$-(La,Ce)$H_{9–10}$ [203, 212]. In addition, various phase modifications of $LaH_{10}$ (*R*-3*m*, *C*2/*m*, *P*1) were also obtained. In the case of yttrium stabilization, *Im*-3*m*-$LaH_6$ in the ternary polyhydride (La,Y)$H_6$ can also be obtained [72]. In the binary La–H system, *Im*-3*m*-$LaH_6$ as an individual compound has not been obtained so far.

Surprisingly, the phase diagram of yttrium hydrides differs significantly from that of lanthanum hydrides, despite the very similar chemistry of these elements. Like the lanthanum hydrides, according to the calculations [18, 19, 104], $YH_6$, $YH_9$, and $YH_{10}$ are room-temperature superconductors with maximum $T_C$ > 270 K. For a long time, yttrium superhydrides could not be obtained experimentally. Our group succeeded in synthesizing $YH_6$ for the first time in 2019 [214]. The electrical measurements showed that the critical temperature of superconductivity in this compound (224 K) is far from the theoretical predictions (>270 K [104], Figure 20).

Very soon, at pressures above 200 GPa, the $P6_3/mmc$-$YH_9$ was synthesized [31], whose critical temperature (243 K) also differed significantly from predictions (303 K, [19]). A further pressure increase to 400 GPa and intense heating to 2000-3000 K failed to detect any trace of the formation of the $Fm\bar{3}m$-$YH_{10}$ cubic phase. An attempt was also made to stabilize $YH_{10}$ by lanthanum [72]. Indeed, at lanthanum concentrations > 50 at%, it was possible to obtain cubic $Fm\bar{3}m$-(La,Y)$H_{10}$, with a critical temperature only slightly higher (253 K) than that of pure $LaH_{10}$. However, when the yttrium concentration is increased to 75 at%, the critical temperature decreases and many synthesis byproducts are found in the mixture [72]. This indicates the thermodynamic and possibly dynamic instability of $YH_{10}$ at pressures < 200 GPa. Unlike $YH_{10}$, other yttrium hydrides — $I4/mmm$-$YH_4$, $YH_6$ and $YH_9$ — are stable. In comparison, in the La-H system $LaH_{10}$, $LaH_{11}$ [21, 119] and $I4/mmm$-$LaH_4$ are stable, as well as a series of compounds not yet identified. Note that all yttrium hydrides, including $I4/mmm$-$YH_4$ [196] (the Eliashberg function is shown in Figure 20b) are high-temperature superconductors with a $T_C$ > 77 K.

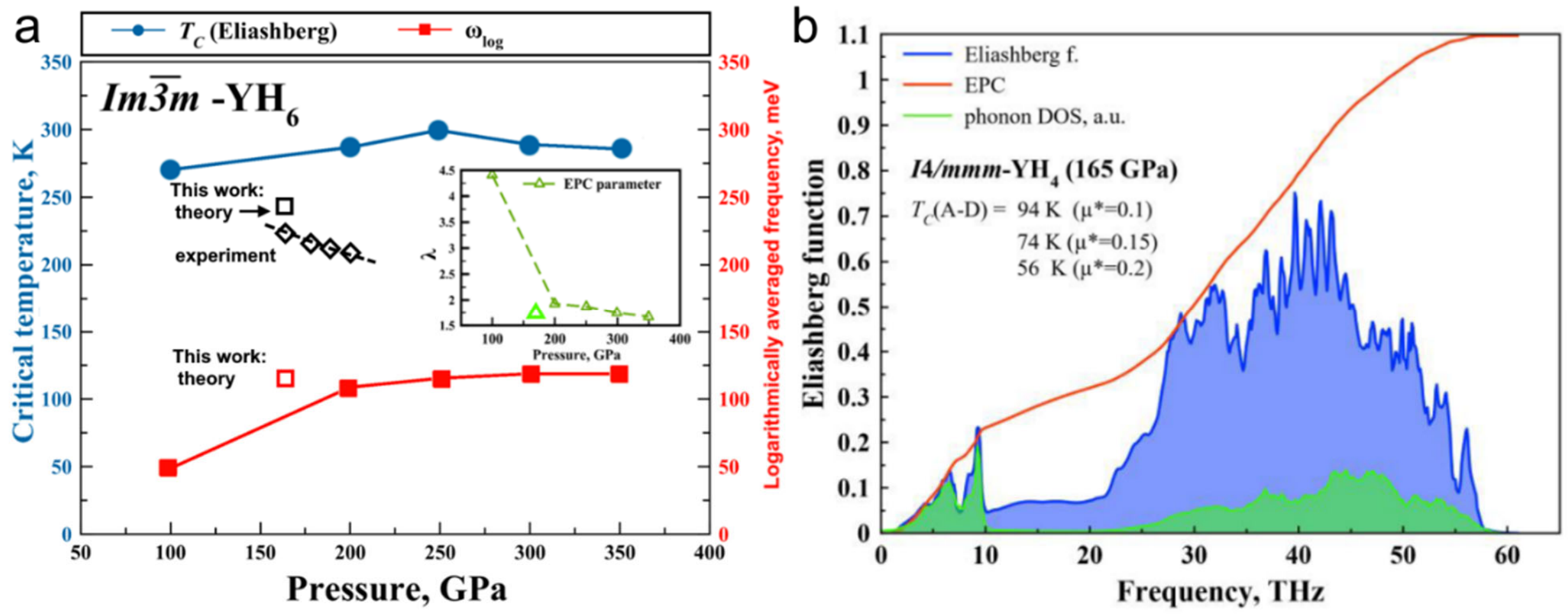

**Figure 20.** (a) Yttrium hexahydride properties from previous theoretical studies: pressure dependence of the superconducting critical temperature $T_C$, electron–phonon coupling coefficient λ, and logarithmically averaged frequency $\omega_{log}$ [104]. (b) Harmonic Eliashberg functions and superconducting properties of for *I*4/*mmm*-$YH_4$ calculated using QE with a 16×16×16 k-grid, 2×2×2 q-grid, and σ-smearing of 0.01 Ry at 165 GPa. EPC means the electron-phonon coupling strength.

The cherished goal of researchers remains yttrium decahydride *Fm*$\bar{3}$*m*-$YH_{10}$, which according to theoretical predictions [18, 104] should exhibit room-temperature superconductivity. However, other yttrium polyhydrides, *Im*-3*m*-$YH_6$ and *P*63/*mmc*-$YH_9$, show significantly lower $T_C$ than was predicted by theoretical methods. Moreover, we know that cubic hydrides exhibit maximum $T_C$'s only 10% higher than their hexagonal neighbors with similar hydrogen content: $ThH_9$ – 146 K, $ThH_{10}$ – 161 K (+10%) [27]; $CeH_9$ – 105 K, $CeH_{10}$ – 115 K (+10 %) [105]. This means that we can expect $T_C(YH_{10})$ = 267 K, considering that $T_C(YH_9)$ = 243 K. This is not a bad result, though it is much lower than the predicted critical temperature.

In this regard, it is interesting to discuss the recent ambiguous paper by E. Snider and R. Dias et al. [82], where the researchers reported that sputtering of a 10 nm palladium layer on yttrium provides a hydride with superconducting transition at 244-262 K at a pressure of 144-182 GPa. At such low pressures, only $YH_4$ [196], $YH_6$ [30, 31] and $YH_9$ (metastable) [31] can be produced from Y or $YH_3$. The authors of this article indicate that they obtained "$YH_9$" but provide no evidence for this. The introduction of a Pd impurity into any yttrium hydride, according to Anderson's theorem [148], cannot lead to an increase in $T_C$. We also performed a verification experiment to rule out the formation of Y-Pd-H ternary superhydrides. Using arc melting, we synthesized $Y_3Pd$ intermetallic with the highest yttrium content known from the literature (Figures 21 and 22), and used it as a starting material, together with AB, for the hydride synthesis at 177-184 GPa. No resistive transitions were observed in the sample between 120 and 300 K despite three stages of the laser heating. Using the Bloch-Grüneisen formula [64, 65] to interpolate the temperature dependence of resistance R(T) leads to fairly large Debye temperature, typical for polyhydrides. This can be considered as an indirect evidence of relatively high hydrogen content in the sample, but the superconductivity seems to be suppressed by palladium. Thus, the reproducibility of this work by E. Snider and R. Dias et al. [82] is questionable.

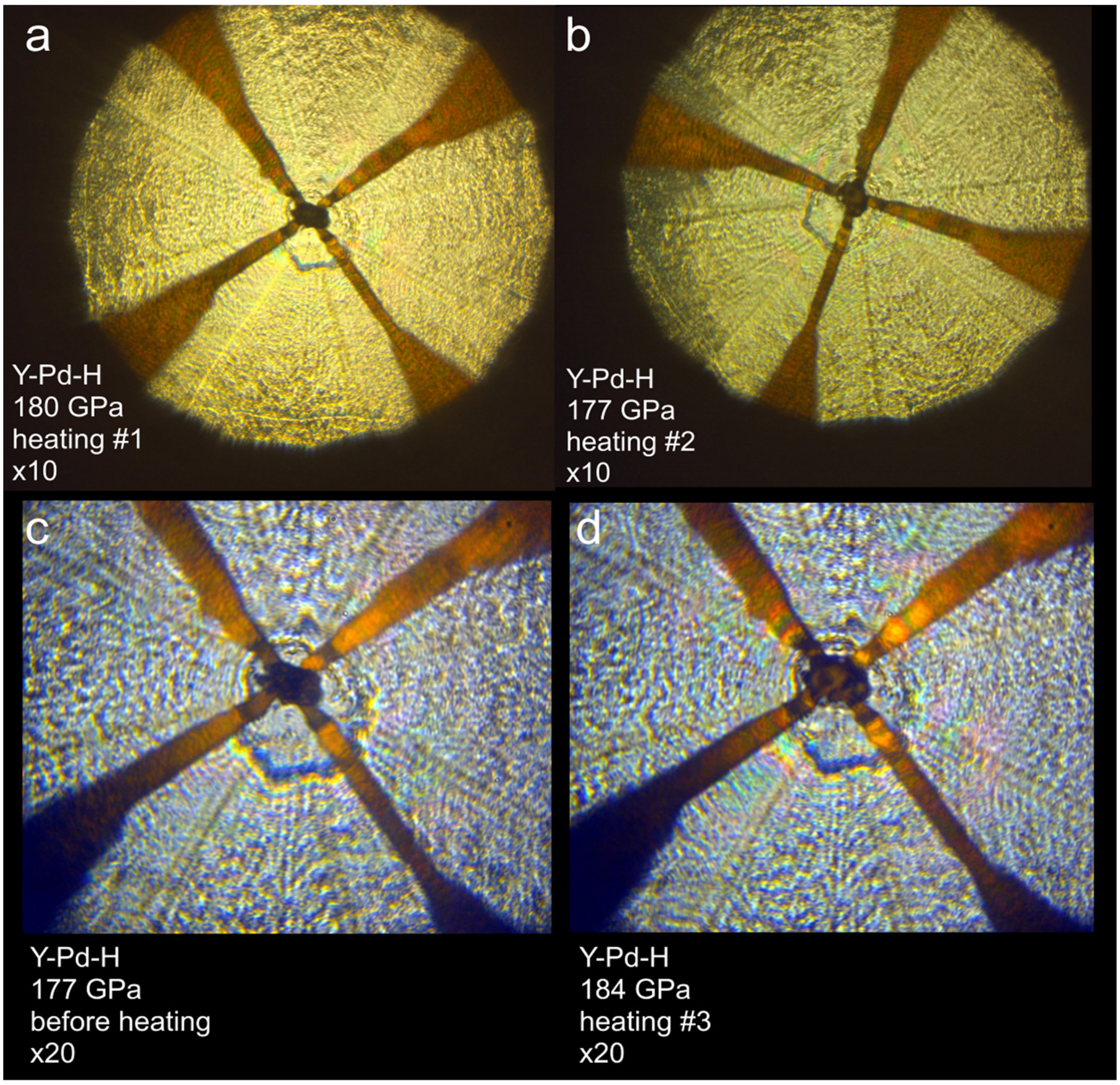

**Figure 21**. Optical photographs of the electrode system and $(Y_3Pd)H_x$ sample at different stages of the laser heating. Gasket consists of $CaF_2$/epoxy.

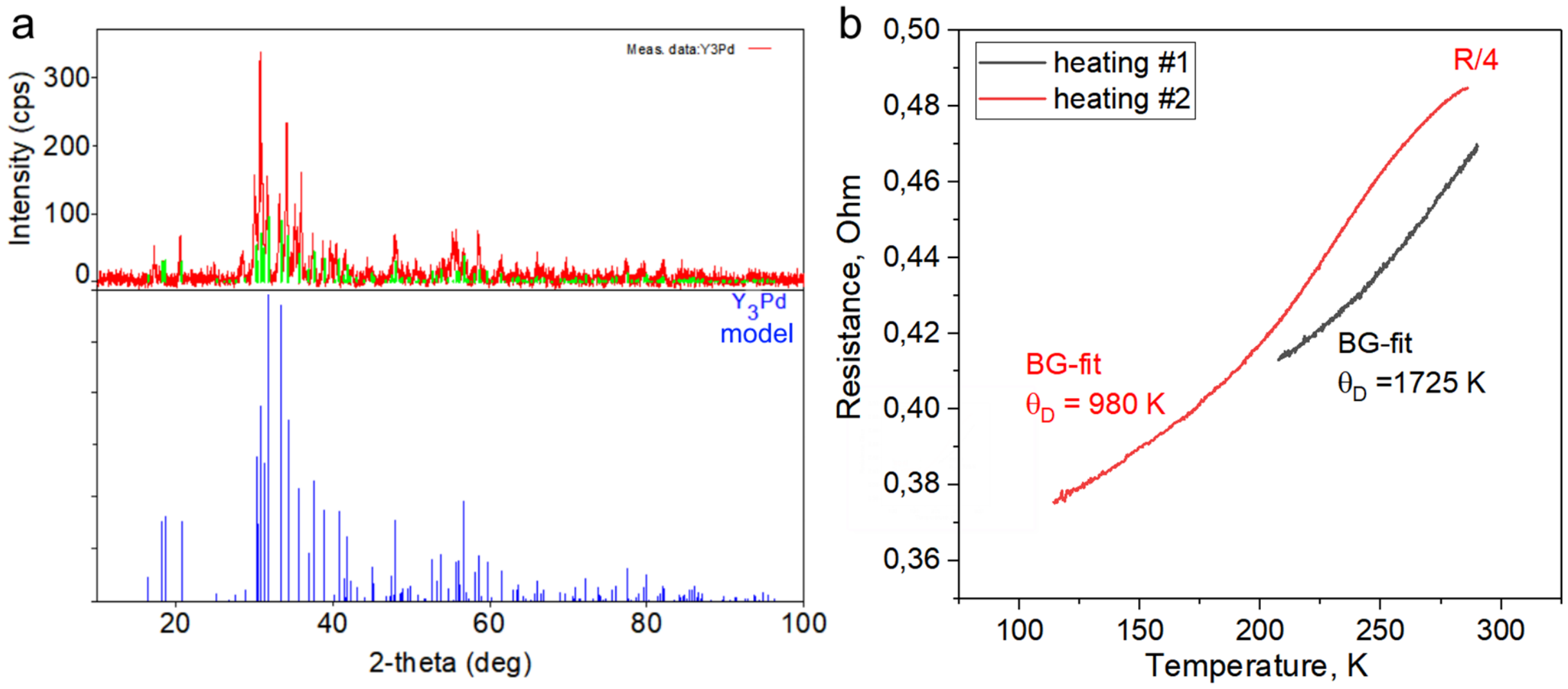

**Figure 22.** (a) X-ray diffraction pattern (wavelength: $CuK_a$) of an initial $Y_3Pd$ intermetallic prepared by arc melting (K. Pervakov, LPI) and used for loading a high-pressure diamond anvil cell. (b) Temperature dependence of the electrical resistance of the sample $(Y_3Pd)H_x$ after the first and second laser heating.

## 3.2 Synthesis and study of yttrium hexahydride $YH_6$

In this section, the description of the experimental synthesis and superconducting properties of yttrium hexahydride $YH_6$ is preceded by a theoretical analysis of the behavior of various phases of the Y-H system at high pressures. Conducting a theoretical analysis with simultaneous validation and comparison with experimental results is extremely useful because it allows us to build a reliable theoretical model for predicting the superconducting properties of polyhydrides as well as a set of stable phases, the use of which saves time and material resources (diamond anvils).

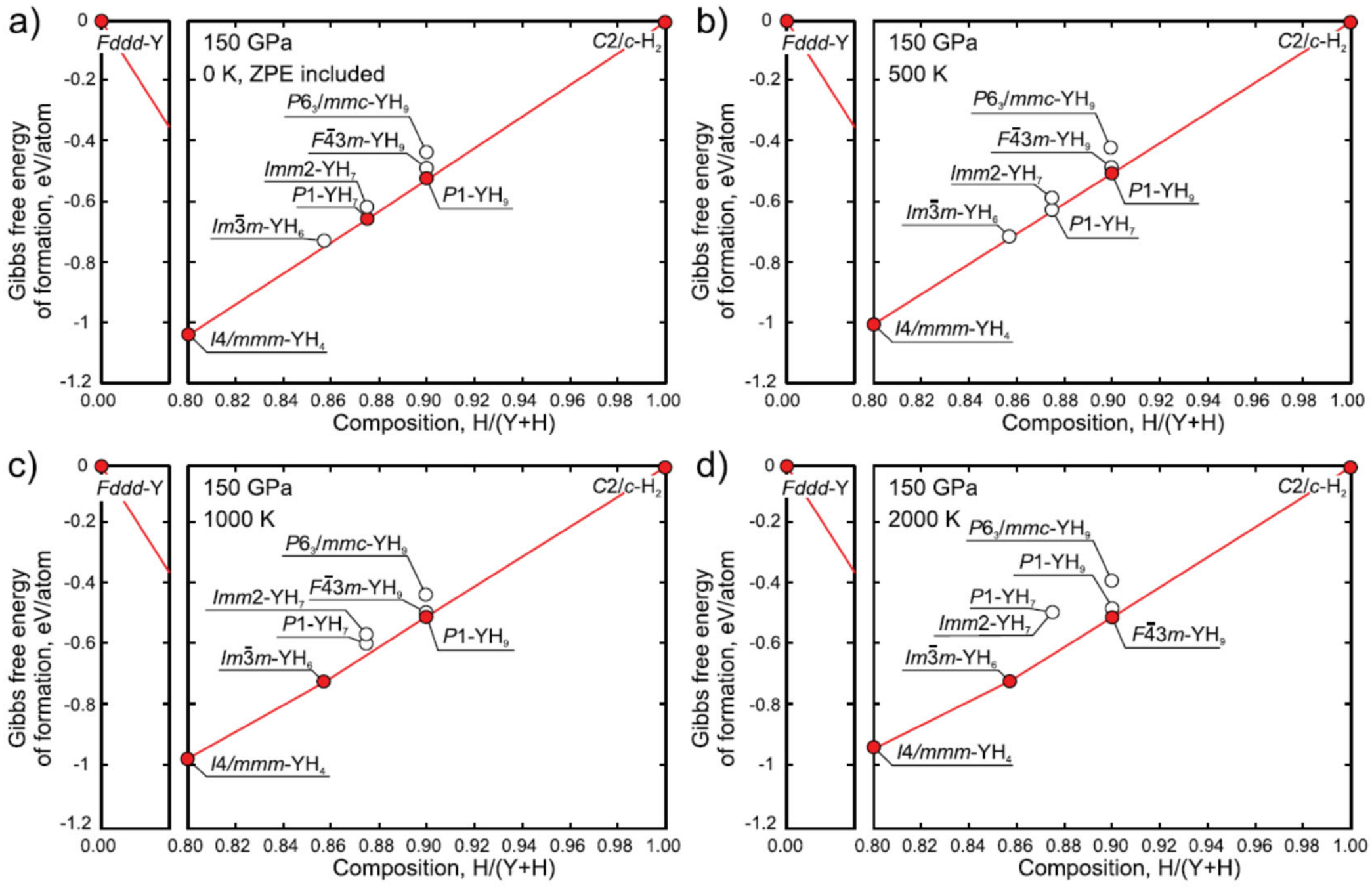

**Figure 23.** Calculated convex hulls of the Y–H system at 150 GPa and (a) 0, (b) 500, (c) 1000, and (d) 2000 K under the assumption of the crystalline state of all phases. Thermodynamically metastable and stable phases are shown by hollow circles and red circles, respectively. Red line marks the thermodynamic convex hull of the Y–H system.

We searched for stable phases in the Y-H system using the USPEX code [9-12] at 150 GPa at various temperatures, taking into account the ZPE and the vibrational entropy (-*TS*) contribution to the Gibbs energy of formation (Figure 23). The temperature consideration is due to the physical possibility of "freezing" of metastable phases during laser heating, whereby cooling of the sample from temperatures of 1000-1500 K to 300 K can be quite rapid (1-10 ms). As a result, some of the hydride phases can be in a metastable state.

Studying the Y-H system at 150 GPa, we found that in the harmonic approximation at 0 K, yttrium hexahydride *Im*-3*m*-$YH_6$ can be thermodynamically unstable at low temperatures with respect to decay to *I*4/*mmm*-$YH_4$ and low-symmetry phases *P*1-$YH_7$ and *P*-1-$YH_9$, but stabilizes under the laser heating conditions at 1000-2000 K. It is interesting that at these temperatures cubic *F*-43*m*-$YH_9$ appears at the thermodynamic convex hull, which should have a higher $T_C$ than *P*63/*mmc*-$YH_9$ and could, in principle, be a candidate to explain the results of E. Snider and R. Dias et al. [82]. Thus, the theory predicts that high temperatures during the laser heating promote the formation of yttrium hexahydride $YH_6$. As was later shown, a similar situation is observed for *Im*-3*m*-$CaH_6$, which can be obtained only at high temperature and very intense laser heating [115, 153].

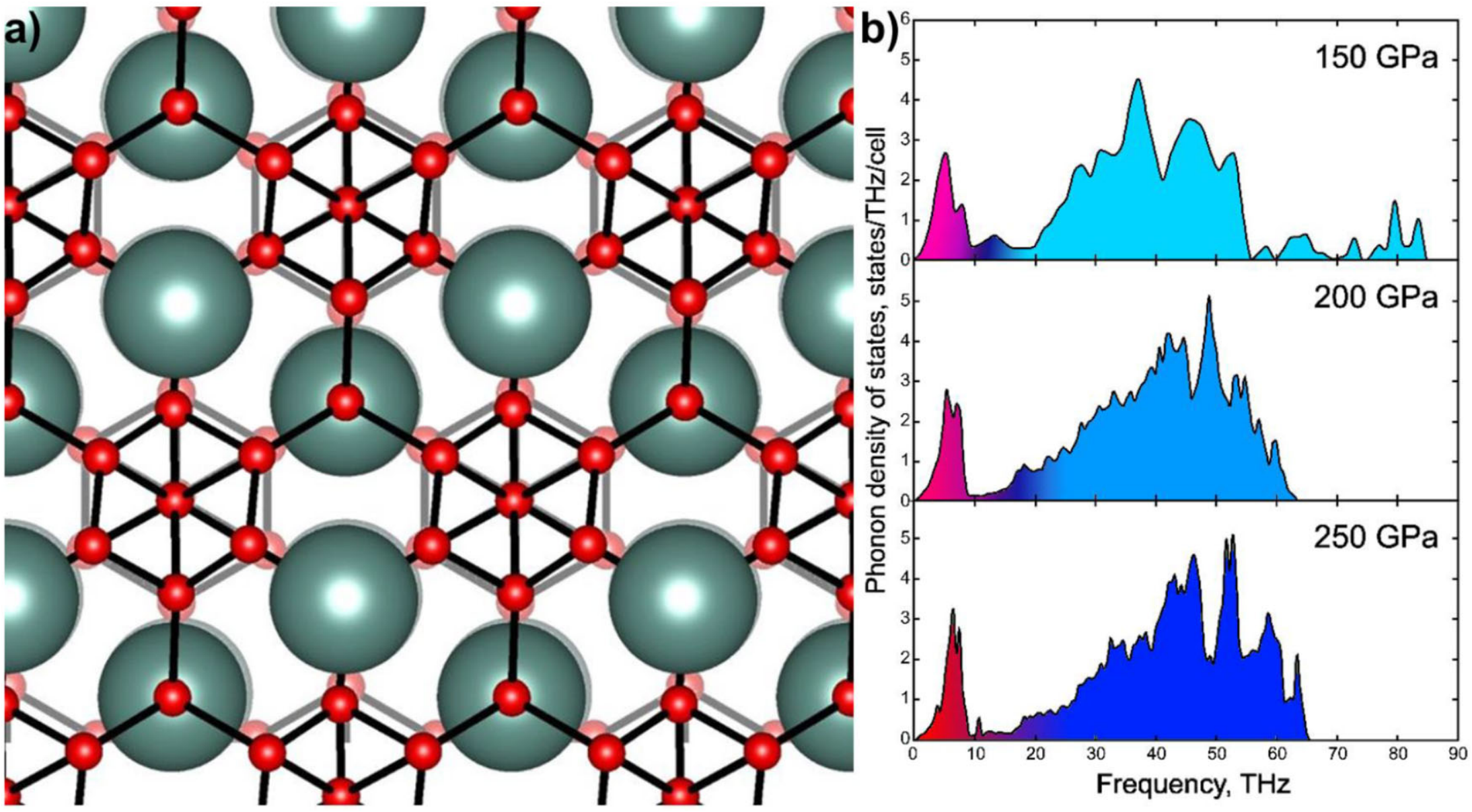

**Figure 24.** (a) Crystal structure of $P$-1-$YH_9$ (= $Y_4H_{36}$) with the distorted hydrogen sublattice in comparison with ideal symmetric $P6_3/mmc$-$YH_9$ (shaded). The hydrogen atoms are shown in red. (b) Phonon density of states of $P$-1-$YH_9$ at 150, 200, and 250 GPa.

Examining the structure of the low-symmetry $P$-1-$YH_9$ phase, which is a small distortion of $P6_3/mmc$-$YH_9$ (Figure 24), we conclude that it is dynamically and thermodynamically stable in the harmonic approximation. In other words, a small distortion of the highly symmetric $P6_3/mmc$ structure is thermodynamically advantageous. At the same time, this distortion is very difficult to observe experimentally given the real width of the XRD spectral lines. This is not an isolated case: we observed an absolutely similar picture for $BaH_{12}$ [46] and $Sr_8H_{48}$ [45], where the highly symmetric phases lose in enthalpy of formation to distorted structures. However, the experiment does not give an unambiguous answer whether the known hydrides have a distorted lattice or not. This requires the single-crystal diffraction analysis for polyhydrides, which is a matter of the near future.

The study of the thermodynamic stability of yttrium hydrides at a higher pressure of 200 GPa (Figure 25) leads to an understanding of possible reasons of the failure of $YH_{10}$ synthesis. At temperatures up to 500 K, pseudohexagonal $P$-1-$YH_9$ displaces $YH_{10}$ from the convex hull. To obtain $YH_{10}$ under such conditions, an impurity must be added to destabilize the hexagonal hydrides. In other words, we should add an element for which hexagonal hydrides are not characteristic, such as lanthanum. At temperatures of 1000–2000 K, $P1$-$YH_7$ becomes stable, which also prevents the formation of $YH_{10}$. Thus, two phases, $YH_7$ and $YH_9$, thermodynamically prevent the formation of yttrium decahydride at both high and low temperatures. At 200 GPa, yttrium hexahydride $Im$-3$m$-$YH_6$ and tetrahydride $YH_4$ appear stable throughout the temperature range. The experiment supports this conclusion (Figure 26).

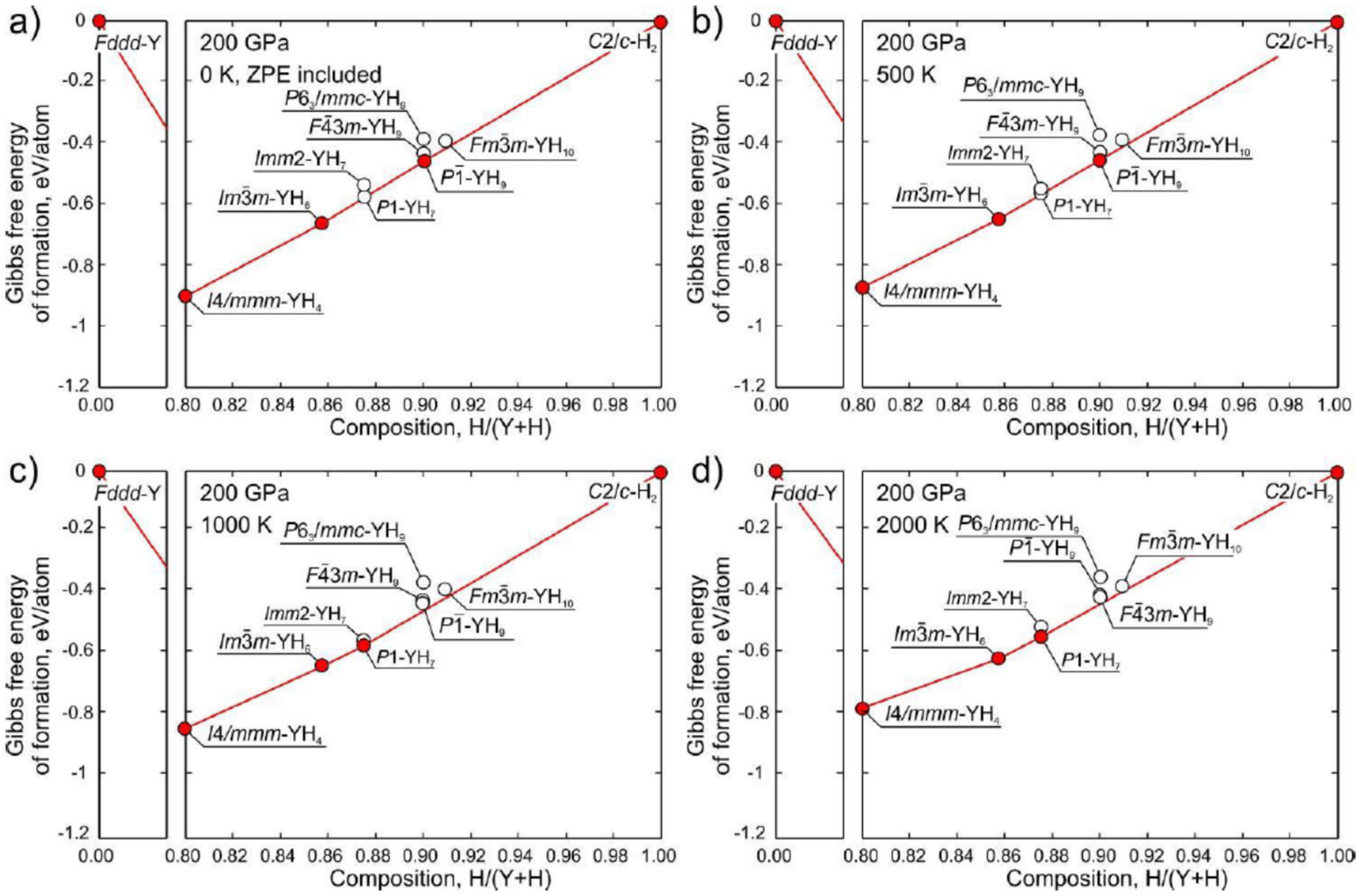

**Figure 25.** Calculated convex hulls of the Y–H system at 200 GPa and (a) 0 K, (b) 500 K, (c) 1000 K, and (d) 2000 K under the assumption of the crystalline state of all phases.

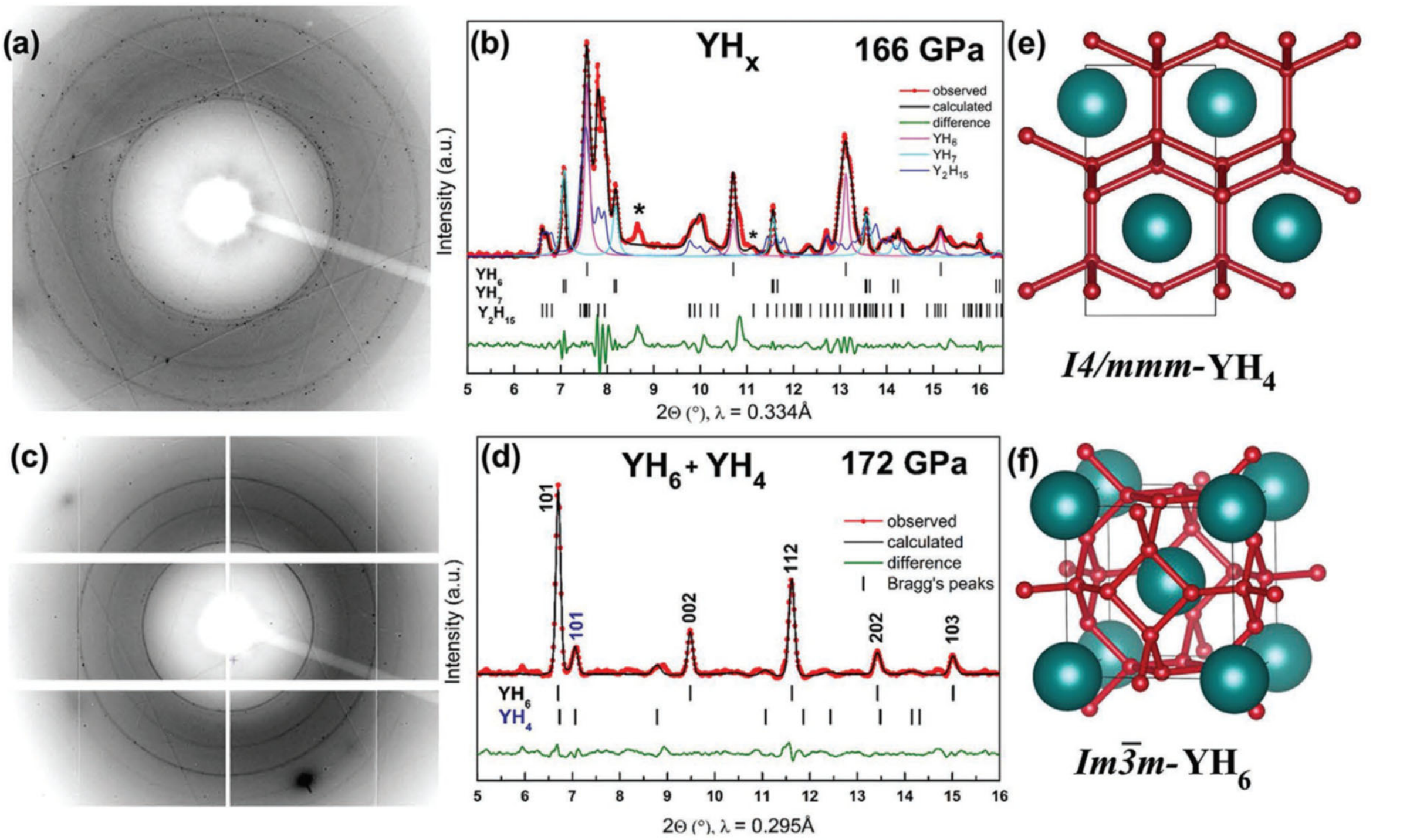

**Figure 26.** (a) XRD pattern of the sample in DAC K1 at 166 GPa obtained at a wavelength λ = 0.334 Å at ambient temperature (25 ºC). (b) Le Bail refinements of *Im*-3*m*-$YH_6$, *Imm*2-$YH_7$, and *P*1-$YH_{7+z}$ (*z* = 0–0.5) at 166 GPa. Unidentified reflections are marked by asterisks. (c) XRD pattern of sample M3 at 172 GPa obtained with λ = 0.295 Å. (d) Le Bail refinements of *Im*-3*m*-$YH_6$ and *I4/mmm*-$YH_4$. The experimental data, fitted line, and residues are shown in red, black, and green, respectively. (e, f) Crystal structures of $YH_4$ and $YH_6$.

Figure 26 shows the experimental XRD patterns of yttrium hydride samples obtained in diamond anvil cells after the laser heating of the yttrium particles in an ammonia borane medium at 166 and 172 GPa. The X-ray pattern of the sample at 172 GPa is relatively simple and shows the presence of two phases: *I*4/*mmm*-$YH_4$ and *Im*-3*m*-$YH_6$. The sample at 166 GPa is more difficult to interpret: in addition to the fine-crystalline $YH_6$ fraction, there is also a coarse-crystalline phase which approximately corresponds to *P*1-$YH_7$ or *P*1-$Y_2H_{15}$ with a similar structure. This phase also occurs in the synthesis of La–Y ternary hydrides [72].

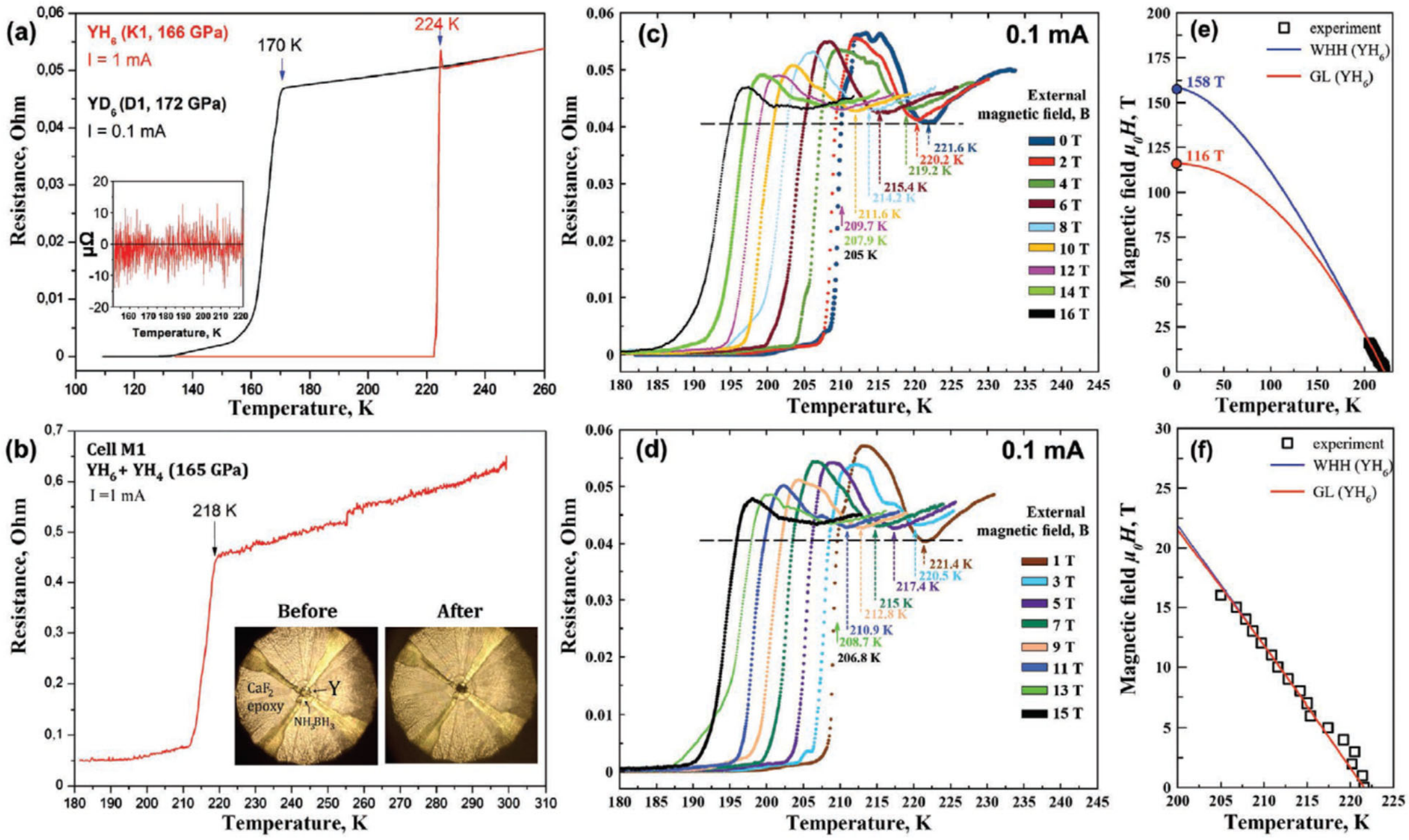

**Figure 27.** Superconducting transitions in *Im*3*m*-YH6: (a) temperature dependence of the electrical resistance *R*(*T*) in $YH_6$ (DAC K1) and $YD_6$ (DAC D1). Inset: the resistance drops to zero after cooling below $T_C$; (b) temperature dependence of the electrical resistance in DAC M1. A ninefold decrease is observed. Inset: chamber of DAC M1 with Y sample and electrodes before and after the laser heating. (c, d) Dependence of the electrical resistance on the external magnetic field (0–16 T) at 183 GPa and a current of 0.1 mA for (c) even and (d) odd values of the magnetic field. Because of the presence of several hydride phases in the sample, the superconducting transition in $YH_6$ can be observed as an upward feature of the *R*(*T*, *H*) curves due to the shunting effect in the fine-grained samples. The critical temperatures were determined at the onset of the resistance jump. (e) The upper critical magnetic field was extrapolated using the Werthamer–Helfand–Hohenberg theory [215] and the Ginzburg–Landau [216] theory. (f) Dependence of the critical temperature $T_C(YH_6)$ on the applied magnetic field.

Yttrium hexahydride exhibits pronounced superconducting properties (Figure 27) at temperatures below 224 K (166 GPa). The detected resistive transition shifts in magnetic fields to a lower temperature region, which allows us to extrapolate the value of the upper critical field $\mu_0 H_{C2}(0)$ = 158 T (WHH model). However, given the large number of defects in the sample, a linear extrapolation seems a more correct solution [188]: $\mu_0 H_{C2}(0)$ = 205 T (166 GPa). The resistance jump before the superconducting transition can be caused by the fine-dispersed structure of the sample with a large number of S/N contacts [217-219].

Here we formulate an algorithm for analyzing the superconducting properties of polyhydrides, which we will continue to follow.

**1.** Analysis of R(300 K)/R($T_C$) shows how many defects and impurities are in the sample. An insignificant change in the resistance with decreasing temperature indicates that phonon scattering plays a minor role compared to scattering on impurities and defects.

**$YH_6$:** R(300)/R($T_C$) ~ 1.2 at 166 GPa. It means that the electrical resistance is determined by defects, not phonons.

**2.** Analysis of R(T) in terms of the Bloch-Grüneisen formula [64, 65] often yields the Debye temperature $\theta_D$. For compressed polyhydrides, $\theta_D$ is usually in the range of 1000 ± 500 K. This allows us to distinguish lower hydrides from higher ones and from metals, and to see if the laser heating was successful. This analysis was first widely applied to polyhydrides by E. Talantsev [86].

**$YH_6$:** $\theta_D$ = 1200 ± 25 K at 166 GPa in close agreement both with calculations based on the elastic constants and with calculations in Quantum ESPRESSO [205] when the anharmonic effects are considered, see [30] for additional information.

**3.** Electrical resistance, as well as its difference R(300 K) - R($T_C$) can be used to estimate the thickness of the sample as $\rho \sim R_{VdP} \times h$, where $R_{VdP}$ – is the resistance of the sample measured using the van der Pauw method, assuming that the sample is an ideal disk of thickness h. This estimate is supplemented by first-principles calculations of resistivity using the EPW code [140-143]

**$YH_6$:** calculations in EPW at 165 GPa give ρ(300) - ρ(230) = 36 μΩ·cm. Then, the sample thickness is h = [ρ(300 K) - ρ(230 K)] / [R(300 K) - R(235 K)] ~ 36 μm, which is highly unlikely because it is a large thickness for this kind of sample (very small resistance < 0.1Ω).

**4.** The Debye temperature obtained at stage 2 can be applied to the estimation of the electron–phonon interaction constant λ via the approximate transformation of $\theta_D$ into the logarithmically averaged frequency $\omega_{log} = 0.827 \times \theta_D$ [220] and using the well-known Allen–Dynes formula (A–D) [60]. Despite the fact that for high-$T_C$ hydrides the Allen–Dynes formula gives essentially underestimated values of $T_C$(A–D) compared to the results of solving the Eliashberg equations $T_C$(E), considering the anharmonic effects reduces the critical temperature and brings us back from $T_C$(E) to $T_C$(A–D). Thus, due to fortunate circumstances, the simple Allen–Dynes formula gives a very good lower bound estimate for the experimental $T_C$.

**$YH_6$:** $\lambda_{BG}$ = 2.71 (BG – means λ, obtained using the Bloch-Grüneisen and Allen-Dynes formulas), which is substantially larger than the theoretical analysis gives. This may be caused by the interpolation temperature interval being too short. A wider temperature interval for $YD_6$ gives λ = 1.96 and $\theta_D$ = 1200 ± 5 K at 172 GPa.

**5.** Analysis of the residual resistance, the hysteresis loop in the cooling and heating cycles, and the width of the superconducting transition are important in terms of distinguishing between superconductivity and insulator-to-metal transition. For example, in the case of hydrogen sublattice melting (probably [221]), there will also be a sharp increase in the electrical resistivity. The width of the transition and the residual resistance allow us to evaluate the potential applications of a given hydride as a sensor and to compare the resistivity of the hydride in the superconducting state with the best metals.

**$YH_6$:** the width of the superconducting transition is about $\Delta T_C$ = 3-4 K (166 GPa), there is an initial resistance growth region at 226 K, which should be attributed to the beginning of the superconducting transition (see also [31]). The average residual resistance is 21 μΩ, the initial resistance is 0.05 Ω. Thus, there is a resistance drop by a factor of 2380. We investigated the hysteresis in a separate cell at 200 GPa (Figure 28) and found $\Delta T_{C\text{-}H}$ = 0.04-0.05 K at a heating/cooling rate of 0.5 K/min, which is close to the experimental temperature sensor error in the cryostat. At the same time, this sample exhibited a smaller extrapolated critical field $\mu_0 H_{C2}(0)$ = 143 T (linear extrapolation) and a significantly larger R(300 K)/R($T_C$) = 1.5. It is likely that $\mu_0 H_{C2}(0)$ depends on concentration of defects in the sample.

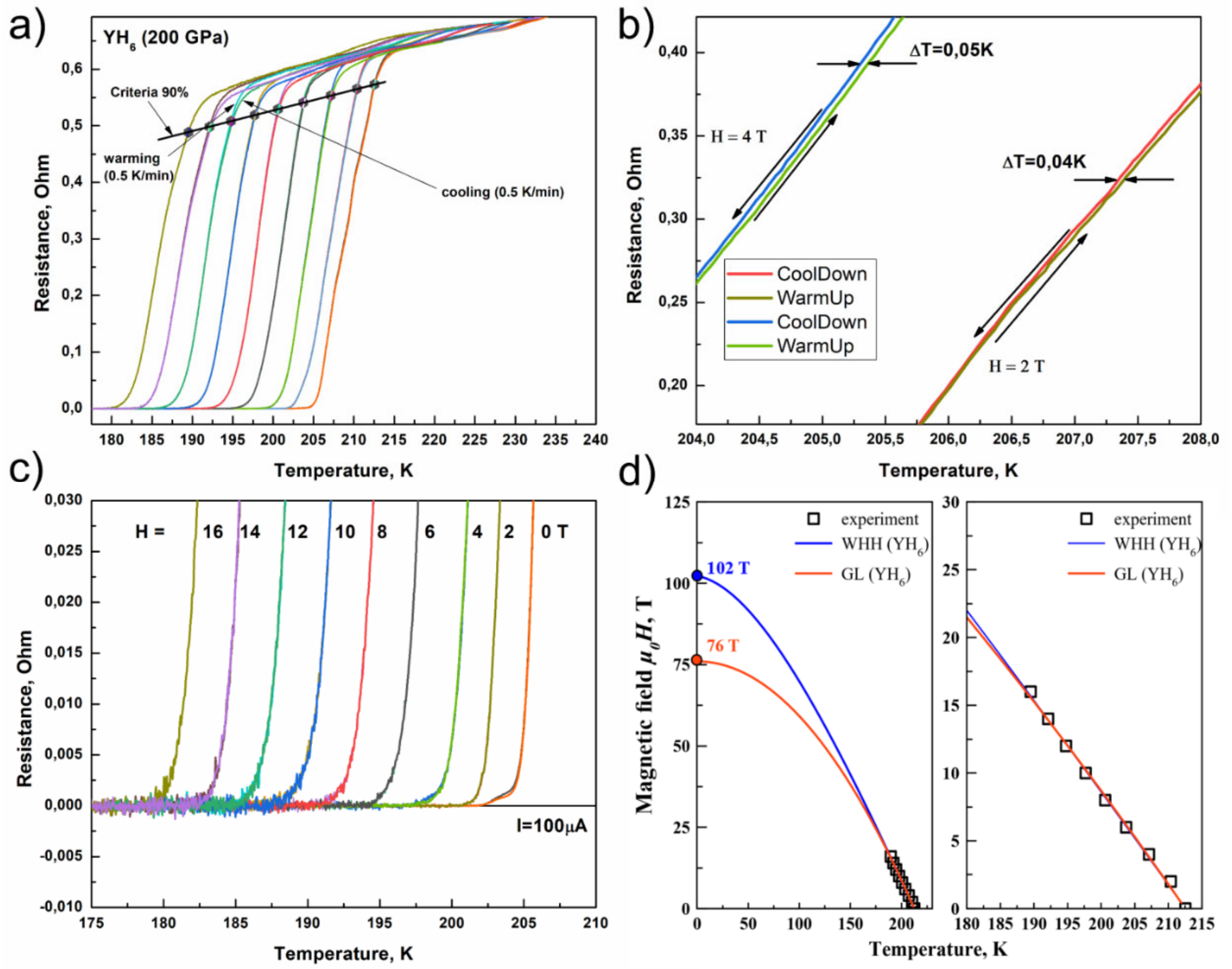

**Figure 28.** Superconducting transitions in $Im$-$3m$-$YH_6$ at 200 GPa: (a) dependence of the electrical resistance and critical temperature of $Im\bar{3}m$-$YH_6$ (90% criteria) on the external magnetic field (0–16 T) at 200 GPa within the cooling and warming cycles (0.5 K/min); (b) hysteresis of the electrical resistance of the $YH_6$ sample upon cooling and warming in magnetic fields; (c) reduction of the electrical resistance in the final section of the superconducting transitions in magnetic fields (different colored lines correspond to magnetic fields 0–16 T); (d) extrapolation of the upper critical magnetic field using the Werthamer–Helfand–Hohenberg (WHH) theory [215] and the Ginzburg–Landau (GL) theory [216]. The coherence length calculated from the experimental data $\xi_{BCS}(YH_6, 200\ \text{GPa}) = 0.5(h/\pi e H_{c2})^{0.5}$ = 18 Å.

**6.** Broadening of superconducting transitions in a magnetic field. This type of analysis is proposed by J. Hirsch [157].

**$YH_6$:** there is no significant broadening of the superconducting transitions in the fields up to 16 T in $YH_6$ (Figure 28a). Exactly the same situation is observed for many other superhydrides [72, 202].

For $YH_6$ we investigated the isotope effect by synthesizing $YD_6$ at 172 GPa from yttrium and fully deuterated ammonium borane $ND_3BD_3$. The resulting sample (Figure 29) exhibited a superconducting transition at 172 K, shifting to the low temperature region when an external magnetic field was applied. Thus, the isotope coefficient $\alpha = 0.4$. The upper critical magnetic field $\mu_0H_{C2}(0) = 156$ T, found by linear extrapolation, is not much inferior to the critical field of $YH_6$.

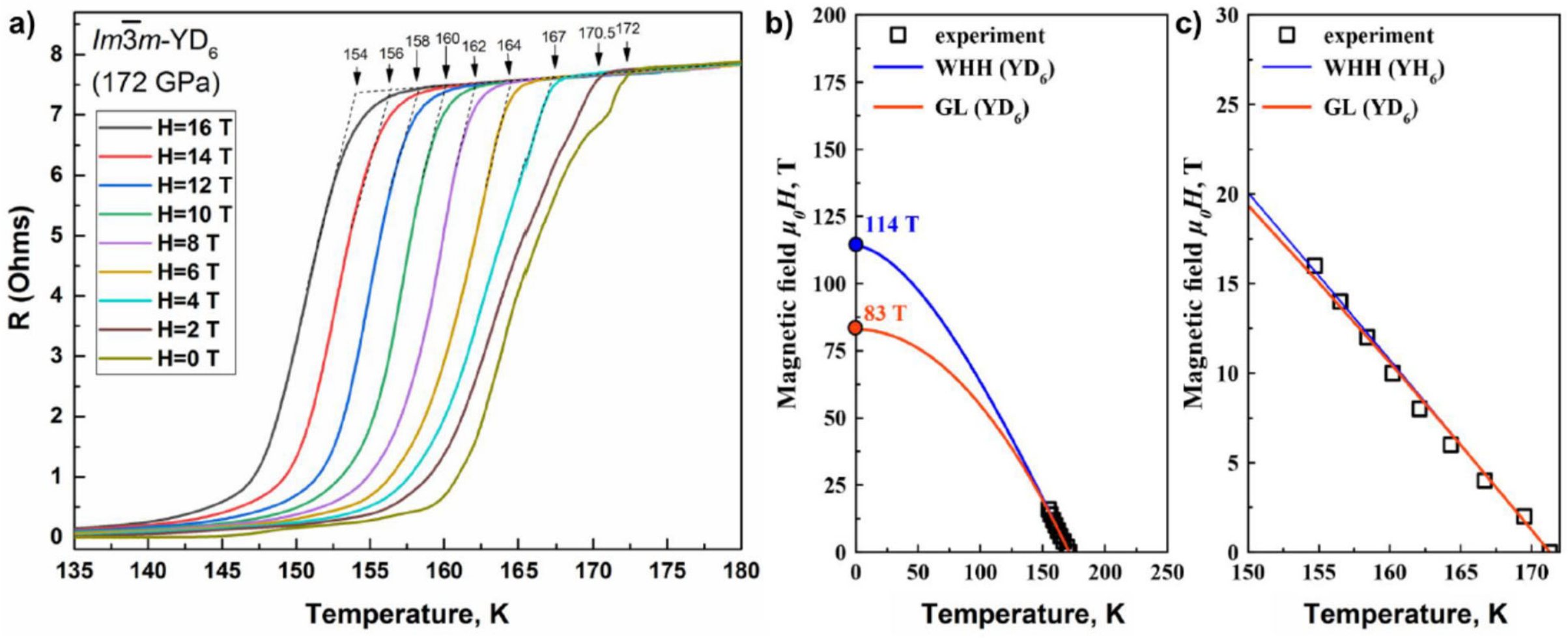

**Figure 29.** Superconducting transitions in $Im\bar{3}m$-$YD_6$: (a) dependence of the electrical resistance and critical temperature on the external magnetic field (0–16 T) at 172 GPa; (b) extrapolation of the upper critical magnetic field using the Werthamer–Helfand–Hohenberg (WHH) theory [215] and the Ginzburg–Landau (GL) theory [216]; (c) dependence of the critical temperature $T_C$ ($YD_6$) on the applied magnetic field. The coherence length calculated from the experimental data $\xi_{BCS}(YD_6) = 0.5(h/\pi eH_{c2})^{0.5} = 17$ Å.

### 3.3 Anomalies of superconducting properties of yttrium hexahydride $YH_6$

As was mentioned above, even before the experimental discovery of superconductivity in $YH_6$, first-principle calculations of the electron-phonon interaction were performed, considering the anisotropy of the superconducting gap in this compound [104]. These calculations resulted in $T_C$ of about 270 K, more than 50 K higher than the experimental critical temperature. It has recently been observed that a similar problem may occur for $CaH_6$ [222].

Immediately after discovering such a serious discrepancy between theory and experiment, we performed independent calculations of the anharmonic corrections to the superconducting properties of $YH_6$ [30]. As a result, we found the anharmonic Eliashberg function for $YH_6$ at 165 GPa (Figure 30) and the values of the critical temperature (Table 8).

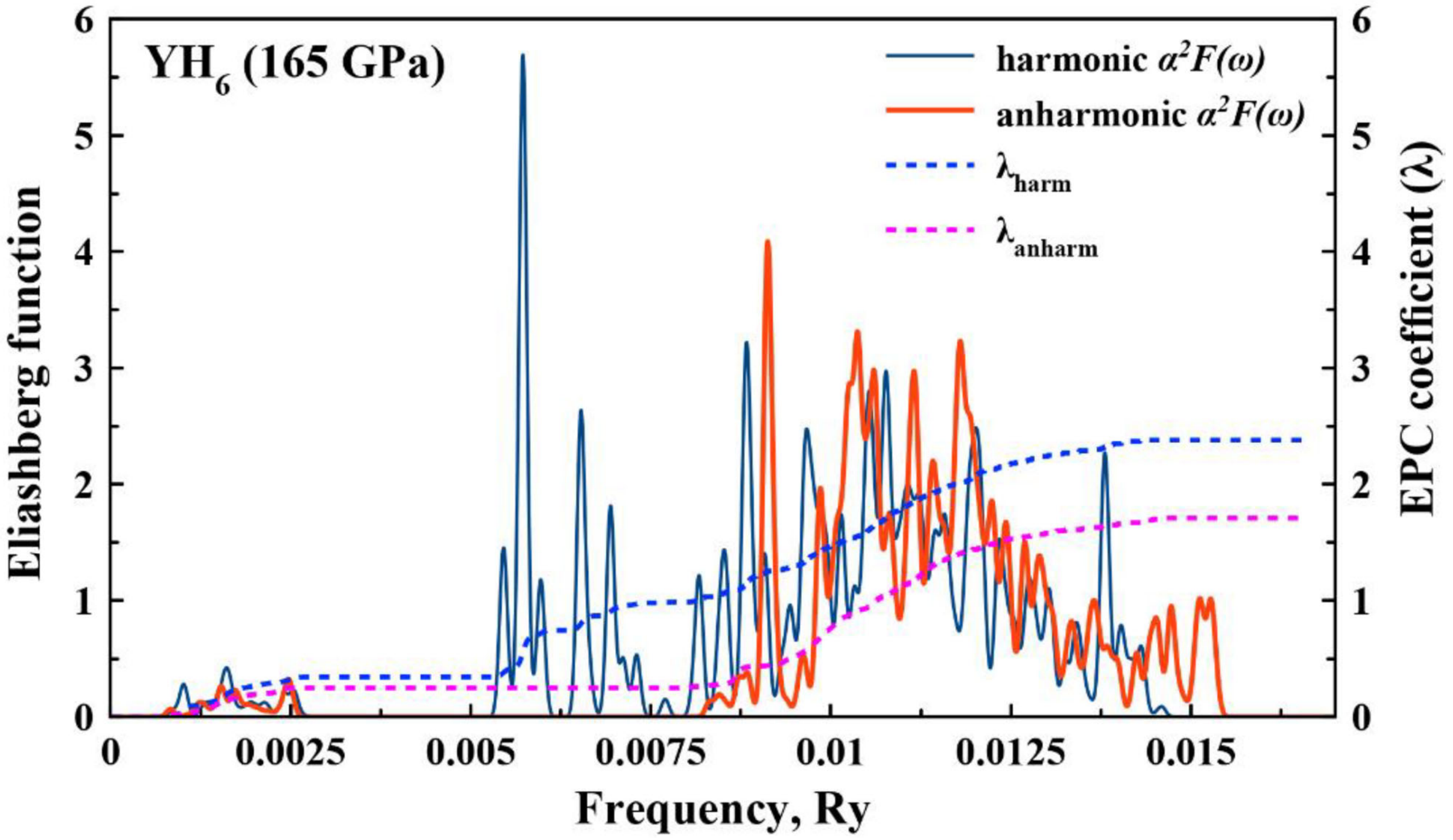

**Figure 30.** Harmonic and anharmonic Eliashberg functions of *Im*-3*m*-$YH_6$ at 165 GPa. A 60×60×60 grid of *k*-points and 6×6×6 grid of *q*-points was used in the calculations.

**Table 8.** Numerical solution of the Migdal–Eliashberg equations for harmonic and anharmonic $\alpha^2F(\omega)$ of $YH_6$ at 165 GPa for different Coulomb pseudopotentials.

| **Coulomb pseudopotential μ*** | **0** | **0.1** | **0.13** | **0.15** | **0.195** |
|---|---|---|---|---|---|
| $T_C$, K (anharmonic) | 269 [a] | 247 | 240 | 236 | 226 |
| $T_C$, K (harmonic) | 294 [a] | 272 | 265 | 261 | 251 |

[a] Linear extrapolation of $T_C(\mu^*)$ to $\mu^*=0$.

As can be seen, in the harmonic approximation within the generally accepted Coulomb pseudopotential interval $\mu^*$ = 0.1-0.15, the solutions of the Eliashberg equations lead to overestimated $T_C$ values. The anharmonic corrections calculated within the SSCHA method [223] significantly reduce the λ (electron-phonon interaction coefficient) from 2.24 to 1.71, resulting in $T_C(YH_6)$ also decreasing by 25 K. However, this is not sufficient to explain the experimental results. The calculation still gives $T_C$ overestimated by 20 K. Therefore, there must be another effect that reduces experimental $T_C$, which can be represented as an effective increase in the Coulomb pseudopotential $\mu^*$ to 0.195.

To exclude the influence of the semiempirical parameter $\mu^*$, we calculated the critical temperature of superconductivity of $YH_6$ within the SCDFT theory, which considers electron-electron interactions and does not need the introduction of any empirical parameters. Such calculations have been previously performed for $LaH_{10}$ [119, 127] and showed good convergence with the experimental data. We did our calculations using two exchange-correlation functionals, LM2005 [132] and SPG2020 [136], and the Eliashberg anharmonic function from Figure 30, as a result we received: $T_C$ = 156 K (LM2005), $T_C$ = 181 K (SPG2020). The new functional (SPG2020) shows better results than the old one, but the deviation from the experiment is rather large (- 43 K) and has the opposite sign than the calculations within the framework of the Migdal-Eliashberg theory give [126, 142].

Thus, there is currently no theoretical model that satisfactorily explains the unexpectedly low value of the critical temperature of superconductivity in $YH_6$. There must be some additional effect that suppresses superconductivity in yttrium hydrides but is absent in lanthanum hydrides. It is possible that this yet unexplored effect is related to spin fluctuations.

## 3.4 Synthesis and study of superconducting properties of yttrium nonahydride $YH_9$

Studies of yttrium hexagonal hydride $YH_9$ were performed almost simultaneously with the discovery of $YH_6$ in 2019 [31] by the group of M. Eremets in Mainz, Germany. In this section, I will present the results of the theoretical calculations we performed for this compound in 2020-2021, and of our experimental studies confirming the data obtained by P.P. Kong et al. [31].

*P*6$_3$/mmc-$YH_9$ crystallizes in the hexagonal space group and expectedly exhibits anisotropy in the superconducting gap as investigated by the EPW code [140-143] (k-mesh 12×12×12, q-mesh 4×4×4, Figure 31). Calculations in the harmonic approximation show that at 200 GPa, $YH_9$ has some small superconducting gap anisotropy ($\Delta_1$ = 50 meV, $\Delta_2$ = 55 meV) and should exhibit superconductivity at 257 K, which is 14 K higher than the experimental critical temperature (243 K [31]). At the same time, the isotropic approximation gives significantly lower $T_C$ = 236 K, which indicates the importance of a systematic study of the anisotropy.

The study of the electronic structure of $YH_9$ (Figure 32) shows the presence of the Van Hove singularity at the Fermi level (direction "M"). The same situation is observed for $YH_6$ [30]. The Fermi surface in $YH_9$ is mostly open, therefore we expect the manifestation of linear magnetoresistance [224, 225] in this compound in strong pulsed magnetic fields.

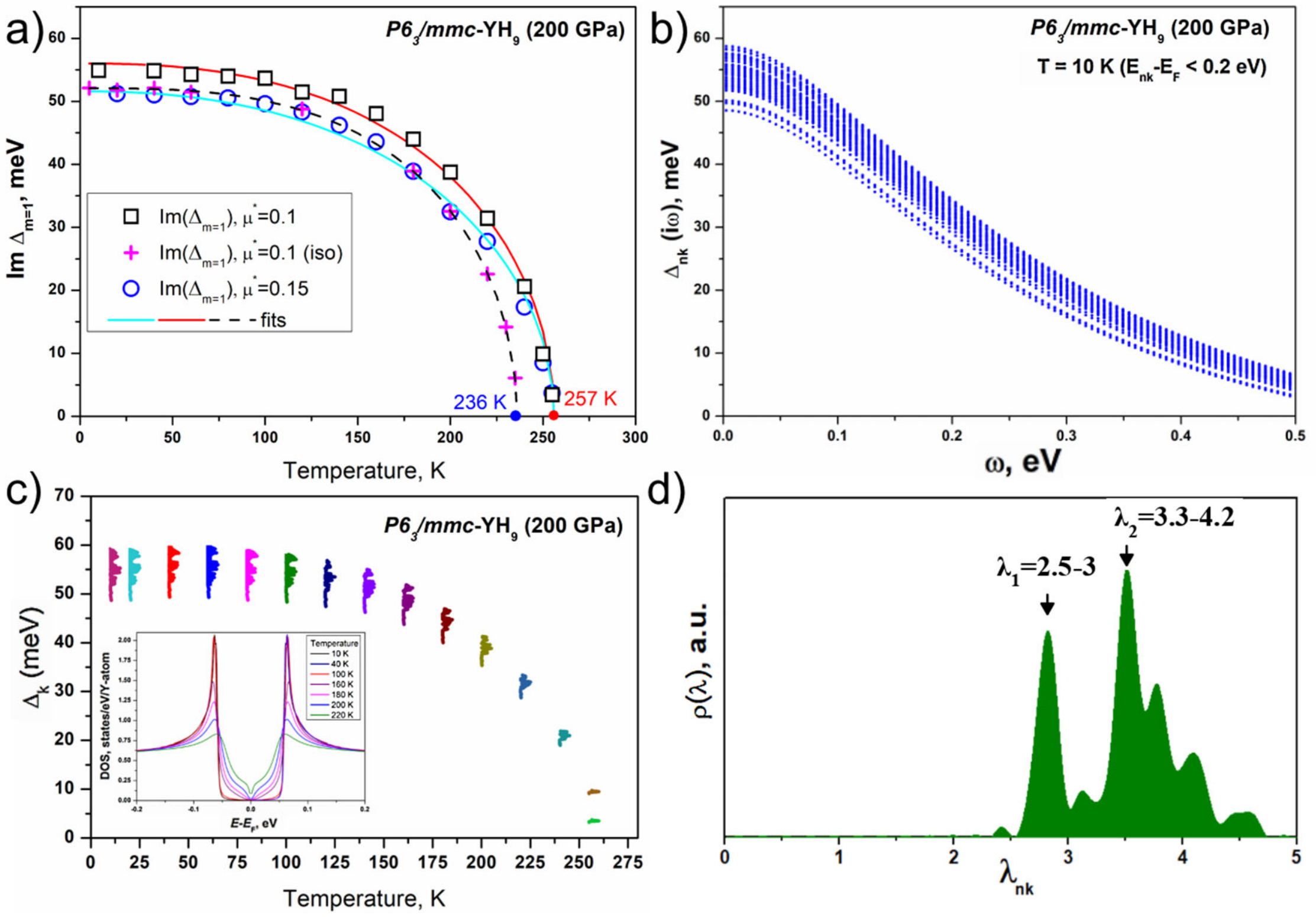

**Figure 31.** Anisotropy of the superconducting properties of $YH_9$ at 200 GPa. (a) Temperature dependence of the imaginary part of the superconducting gap for μ* = 0.1 and 0.15, compared with the isotropic case (iso). (b) Energy dependence of the superconducting gap along the imaginary axis at 10 K. Only the Kohn-Sham states *nk* with an energy within 0.2 eV from the Fermi level are shown. (c) Histogram of the contributions to the superconducting gap at different temperatures. Inset: dependence of the density of electronic states on the energy in the superconducting state at different temperatures. (d) Histogram of the contributions to the electron–phonon interaction coefficient from different zones of the Fermi surface (n = 0).

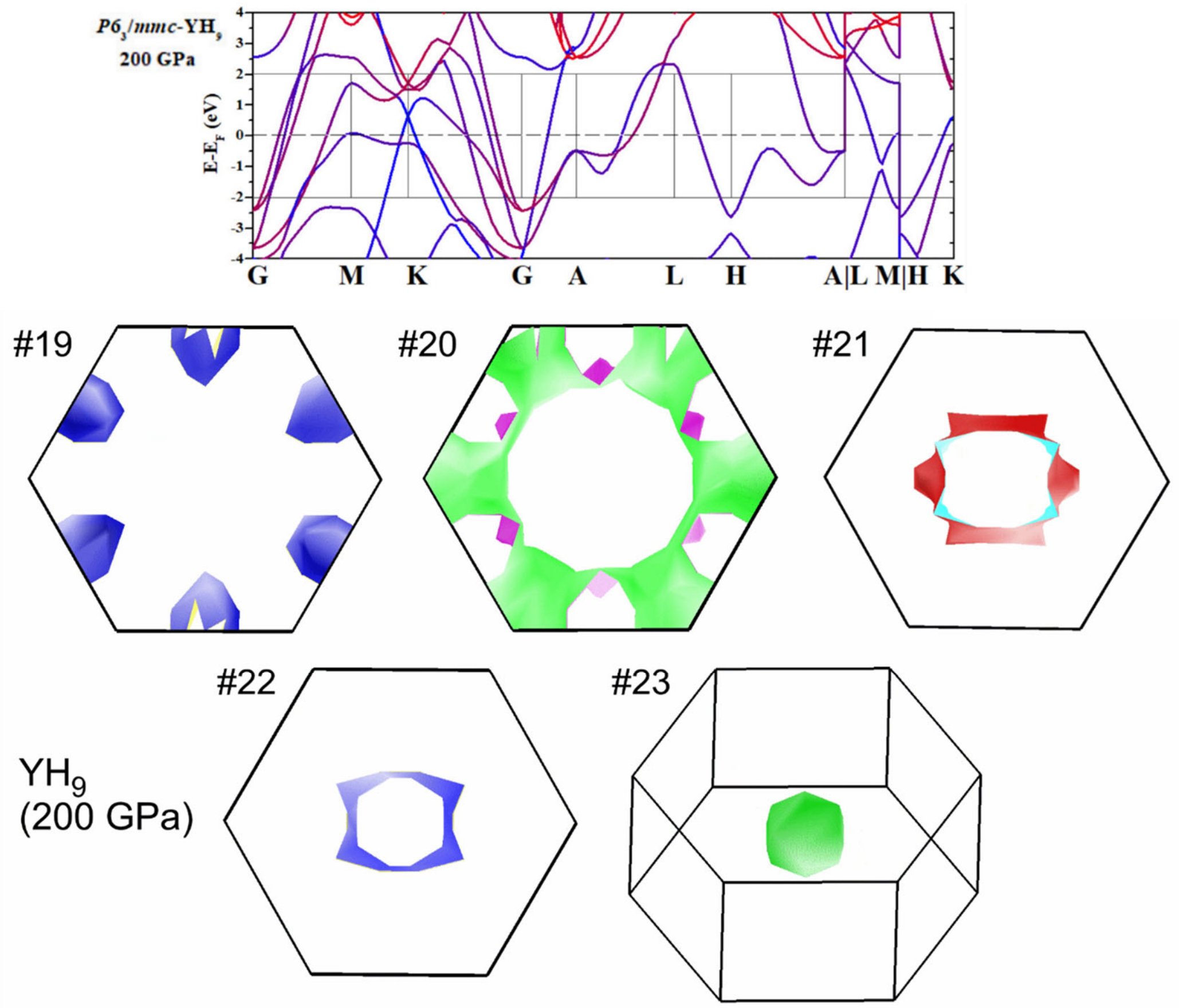

**Figure 32.** (a) Band structure of $YH_9$ at 200 GPa near the Fermi level. (b) Three-dimensional visualization of the Fermi surface for the electron levels crossing it (bands #19-23).

The first-principles calculations of the electron-phonon interaction in $YH_9$ show good convergence with respect to the number of k-points in the grid used for calculations (Figure 33), yielding $\lambda = 2.72$, $\omega_{log} = 804$ K. These values are very close to those obtained from the experiment and the Allen-Dynes formula [60] (see below, $\lambda = 2.66$, $\omega_{log} = 990$-$1093$ K). Despite good agreement with the experimental data, this approach is not entirely correct, since we would have to proceed from the solution of the anisotropic Eliashberg equations (Figure 31) and introduce the anharmonic corrections to it. At the moment, such an approach has not been implemented. The numerical solution [72] of the isotropic equations for the Eliashberg function of $YH_9$ (Figure 33) leads to a somewhat underestimated value of $T_C$(E) = 227 K. Considering the anisotropy of the superconducting gap will increase calculated $T_C$ and bring it closer to the experimental values.

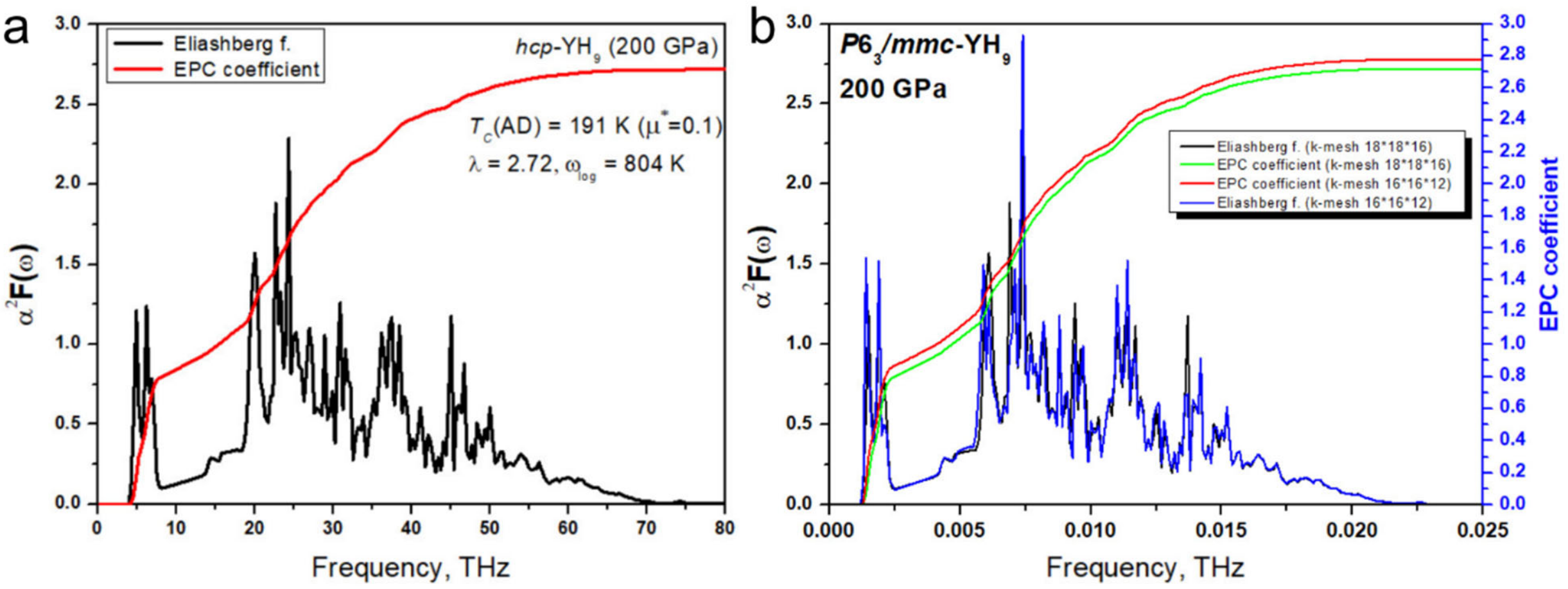

**Figure 33.** Eliashberg functions for $YH_9$ calculated by tetrahedron integration over the Brillouin zone [226] with different k, q-grids. (a) k-mesh 18×18×16, q-mesh 4×4×2. (b) Comparison of the results obtained in (a) with calculations using the 16×16×12 k-mesh. There are virtually no differences.

We performed the SCDFT calculations for $P6_3/mmc$-$YH_9$ (200 GPa) using LM2005 and SPG2020 functionals (the Eliashberg function was taken from Figure 33), resulting in $T_C$ = 179 K (LM2005) and 246 K (SPG2020). Thus, for $YH_9$ the critical temperature can be calculated with good accuracy using the new SPG2020 functional, whereas the previously developed functional LM2005 gives a very large error and is practically unusable. We do not observe the anomaly that was found for *Im*-3*m*-$YH_6$ — in the case of $YH_9$, all computational approaches and the experiment yield close results.

Anharmonic effects in $YH_9$ at 300 K and a pressure of 200 GPa were factored in using molecular dynamics. Machine-learning potentials (MTP) of interatomic interaction for Y and H were created, which allowed a dynamic simulation in a supercell of ~ 1000 atoms within tens of picoseconds. Analysis of the atomic velocities made it possible to construct a phonon density of states spectrum (Figure 34), which is not very different from the harmonic one. As a result of a simplified rearrangement of the Eliashberg function based on the idea of isolating the contribution from the electron-phonon interaction matrices $\alpha^2(\omega) = \alpha^2F(\omega)\ /\ g(\omega)$, where $g(\omega)$ – is the phonon density of states. $g^{anh}(\omega)$ is close to the harmonic phonon density of states, therefore we can obtain the anharmonic Eliashberg function for the hydrogen sublattice as $g^{anh}(\omega)\times[\alpha^2F(\omega)/g(\omega)]$ (Figure 34). For the metal sublattice, this approach gives a very large error because of the abrupt changes of $\alpha^2(\omega)$ on a short frequency interval. But this part of the Eliashberg function is practically unchanged. As the analysis of the obtained anharmonic $\alpha^2_{anh}F(\omega)$ shows that the considered effects at 300 K (which is near experimental $T_C$) even increase the critical temperature by 16 K due to the growth of $\omega_{log}$. It has previously been shown that anharmonic effects can not only worsen the superconducting properties, but also improve them [227, 228], but no confirmed examples have been found so far.

Experimental studies of $YH_9$ were performed for two samples at 205 GPa and 213 GPa. In the first case (Figure 35), a fairly pure, single-phase sample with a narrow transition at 235 K, the transition width $\Delta T_C$ = 5 K, was used to confirm the $YH_9$ → $YH_6$ transformation around 200 GPa.

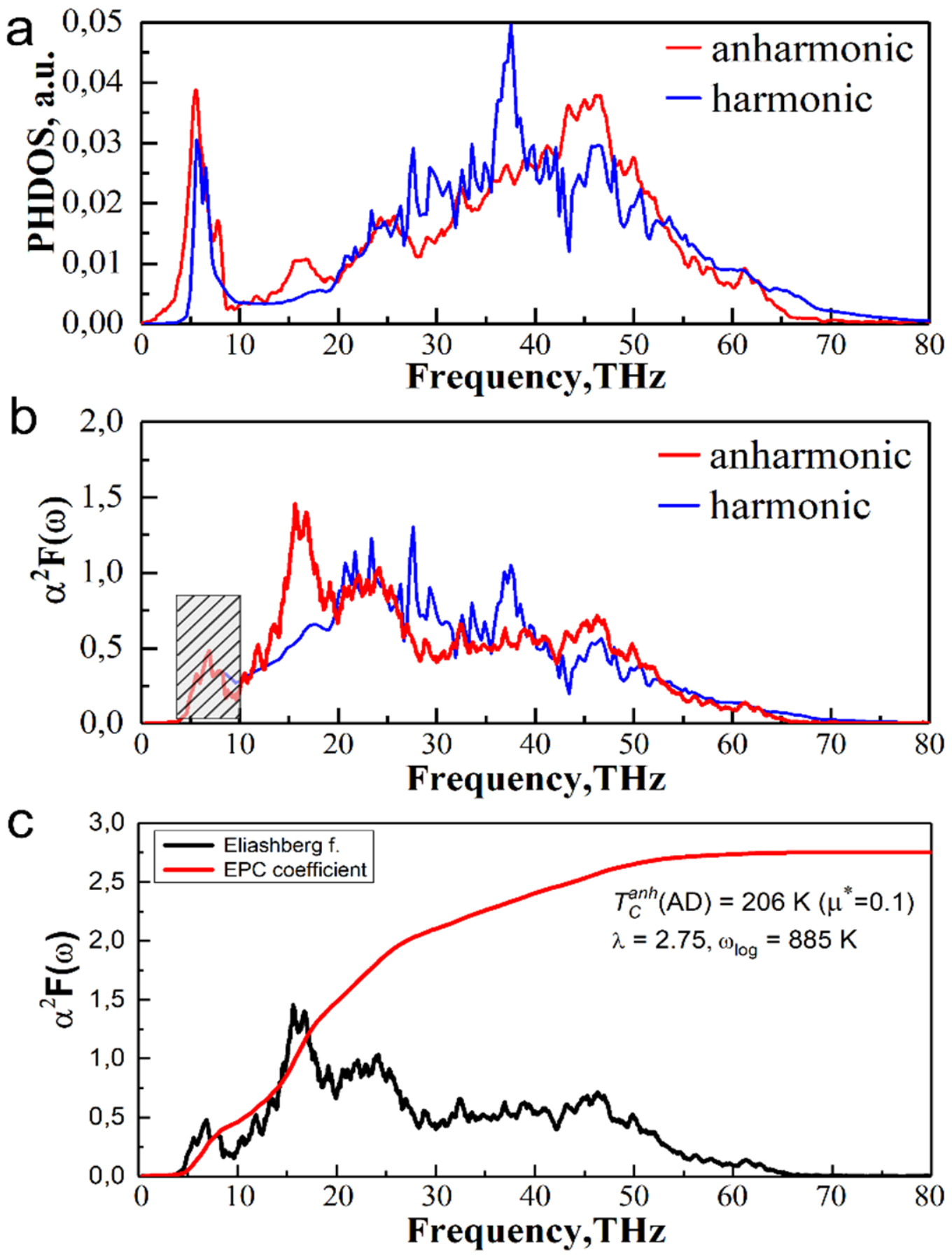

**Figure 34.** Anharmonic effects in *P*6₃/mmc-$YH_9$ at 200 GPa. (a) Comparison of harmonic and anharmonic phonon density of states. (b) Comparison of the harmonic and anharmonic Eliashberg function of $YH_9$ at 200 GPa calculated on a 18×18×12 k-point grid and a 6×6×3 q-point grid. The low-frequency part of the spectrum up to 10 THz was not changed. (c) Anharmonic Eliashberg function and the parameters of the electron-phonon interaction.

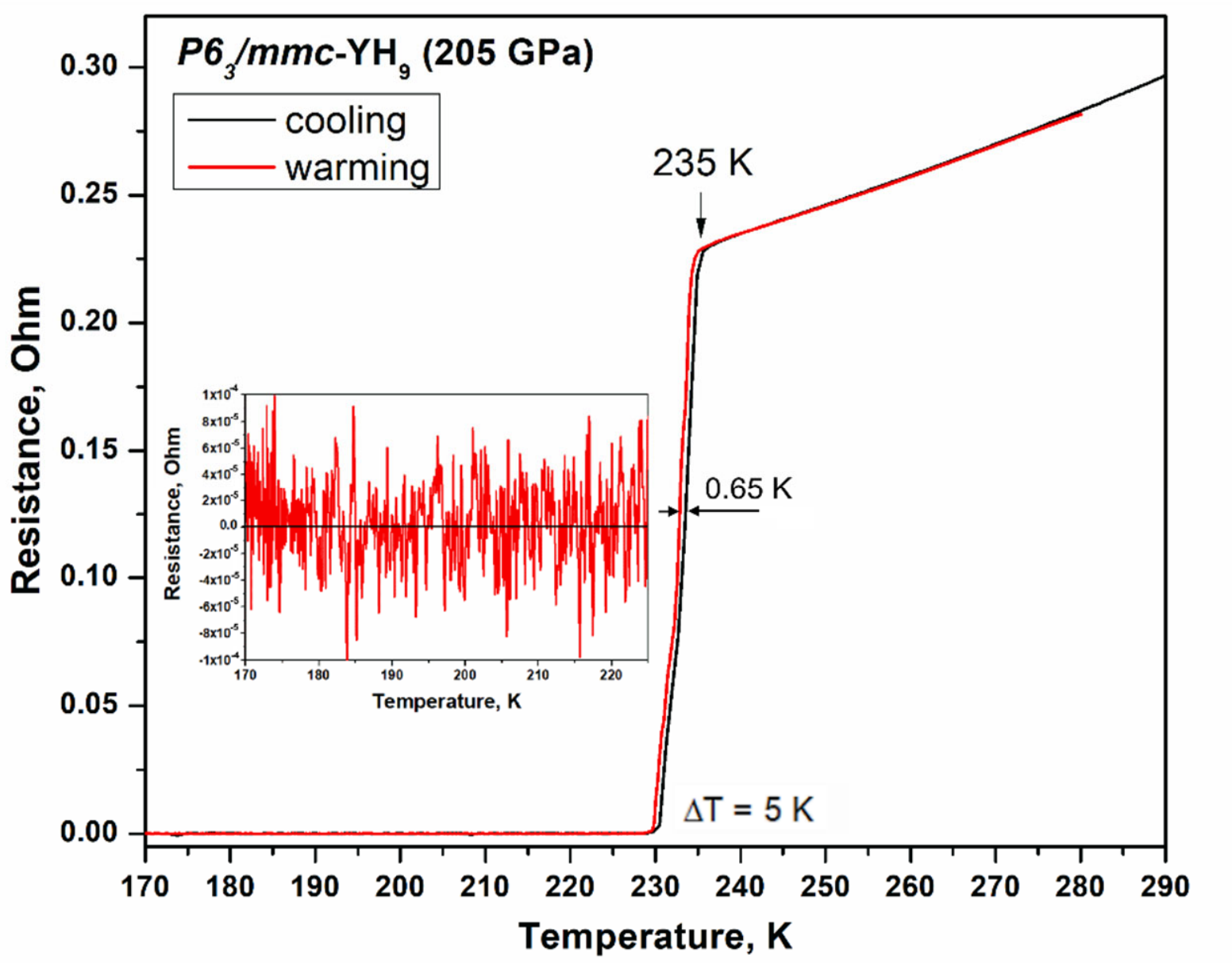

**Figure 35.** Temperature dependence of the electrical resistance of $YH_9$ at 205 GPa in the cooling and heating cycles. The hysteresis does not exceed 0.65 K. Inset: residual resistance in the superconducting state, not exceeding 16 μΩ.

The second sample examined at 213 GPa was found to be two-phase, containing $YH_6$ and $YH_9$ (Figure 36). Despite this, the residual resistance of the sample in the superconducting state does not exceed 20 μΩ. The superconducting transitions exhibit very small hysteresis of 0.2–0.3 K. Studying the behavior of both yttrium hydrides in magnetic fields up to 16 T allows us to linearly extrapolate the upper critical magnetic fields for both compounds: $\mu_0 H_{C2}(0)$ = 181 T for $YH_6$ and 143 T for $YH_9$.

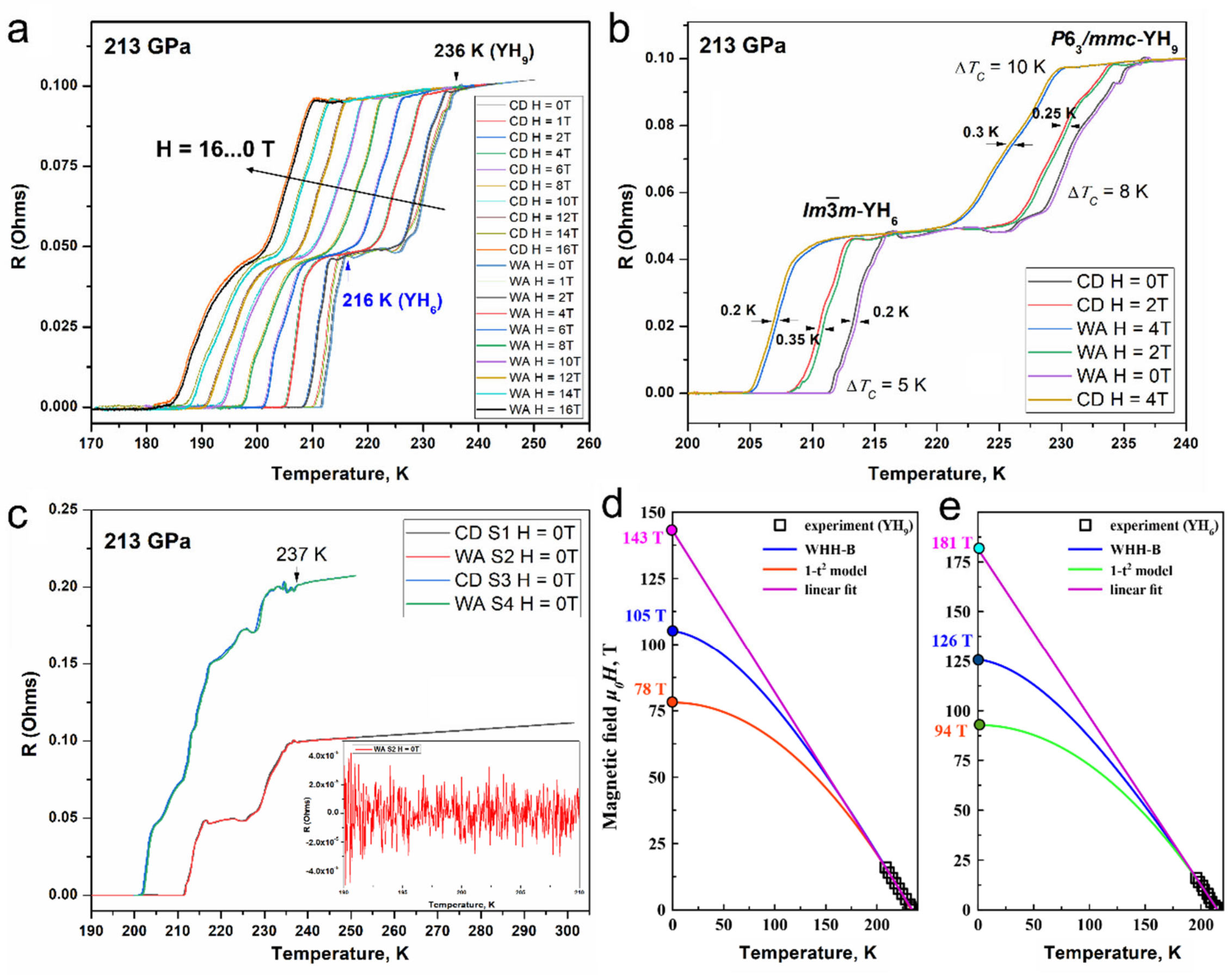

**Figure 36.** Superconducting properties of yttrium hydrides $YH_6$ and $YH_9$ at 213 GPa. (a) Dependence of the resistance of the two-phase sample $YH_6 + YH_9$ $R(T, H)$ on the temperature and applied magnetic field in the cooling (CD) and heating (WA) cycles. As the magnetic field increases, the critical temperatures for both hydrides converge as $\mu_0 H_{C2}(0)$ for $YH_9$ is smaller than for $YH_6$. (b) Transition widths (8–10 K and 5 K) and hysteresis values (0.2–0.3 K) for sample heating/cooling cycles. (c) Residual resistance and the onset of the superconducting transition (237 K) for different pairs of electrodes. (d, e) Extrapolations of the upper critical magnetic field for $YH_6$ and $YH_9$.

In the conclusion of this chapter, I briefly summarize the findings concerning *P*6₃/*mmc*-$YH_9$, using the algorithm of analysis described earlier and supplemented by several theoretical points (e – experiment, t -theory, K - Kelvins):

**1e.** Analysis of R(300)/R($T_C$)

**$YH_9$:** R(300)/R($T_C$) = 1.33 for the sample at 205 GPa.

**2e.** Analysis of R(T) in terms of the Bloch-Grüneisen formula [64, 65]

**$YH_9$:** for $YH_9$ (205 GPa) we got $\theta_D$ = 1275 ± 10 K ($\omega_{log} \approx$ 1093 K). For the mixture $YH_6$ + $YH_9$ (213 GPa) we obtained slightly lower $\theta_D$ = 1160 ± 10 K ($\omega_{log} \approx$ 990 K).

**3e.** Estimation of the sample thickness and resistivity using the EPW code [140-143].

**$YH_9$:** R(300 K) = 0.3 Ω, R(235 K) = 0.225 Ω. Calculations in EPW give the following values for the resistivity: ρ (300 K) = 19.5 μΩ·cm and ρ (235 K) = 12 μΩ·cm. This allows us to make an estimate of the thickness of the sample:
h = [ρ(300 K) - ρ(235 K)] / [R(300 K) - R(235 K)] ~ 1 μm, which is close to the results of the optical measurements.

**4e.** Estimation of the electron-phonon interaction constant (λ) using the well-known Allen-Dynes formula (A-D) [60].

**$YH_9$:** λ = 2.66 (μ*=0.1), $T_C$(exp) = 235 K at 205 GPa.

**5e.** Analysis of the residual resistance, hysteresis loops in cooling and heating cycles, and superconducting transition width.

**$YH_9$:** R(SC) = 16 μΩ, R($T_C$) = 0.225 Ω, the resistance drop by a factor of 14,000, transition width $\Delta T_C$ = 5 K, hysteresis up to 0.2 K. The residual resistivity ρ(SC) = 1.6 nΩ·cm.

**6e.** Broadening of superconducting transitions in a magnetic field.

**$YH_9$:** at 213 GPa in fields up to 16 T the superconducting transition practically does not broaden.

**1t**. Analysis of anisotropy $T_C^{aniso} - T_C^{iso}$ by solving the anisotropic Eliashberg equations using the EPW program.

**$YH_9$:** $T_C^{aniso} - T_C^{iso}$ = 21 K at 200 GPa, there is a significant influence of anisotropy.

**2t**. Analysis of the anharmonic contribution $\lambda_{anh}$, $\lambda_{harm}$, $T_C^{harm} - T_C^{anharm}$ performed using the SSCHA or molecular dynamics.

**$YH_9$:** $T_C^{harm} - T_C^{anharm}$ = - 16 K, unexpectedly, but anharmonic effects can enhance $T_C$ of $YH_9$.

**3t.** SCDFT analysis using the SPG2020 functional, which allows us to isolate the contribution of the empirical Coulomb potential μ* and obtain the $T_C$(SCDFT) value.

**$YH_9$:** $T_C$(SCDFT) = 246 K, close to the experimental value.

**4t**. Analysis within the framework of the Migdal-Eliashberg theory and Allen-Dynes formula in the isotropic harmonic approximation, allowing us to obtain the Eliashberg function, electron-phonon interaction coefficient, logarithmically averaged phonon frequency, etc.

**$YH_9$:** $T_C$(E) = 231 K (μ*=0.1), 238 K (μ*=0.08), 248 K (μ*=0.06).

λ = 2.75, $\omega_{log}$ = 884.5 K, $T_C$(A-D) = 206 K, Δ(0) = 54.6 meV

**5t**. EPW calculations of the electrical resistivity in the normal state.

**$YH_9$:** the EPW calculations give the following values for the resistivity: ρ(300 K) = 19.5 μΩ·cm and ρ(235 K) = 12 μΩ·cm.

## 3.5 Conclusions from studies of yttrium hydrides

1. High-pressure chemistry of the Y-H system differs significantly from that of the La-H system. The synthesized $YH_6$ and $YH_9$ polyhydrides exhibit high-temperature superconductivity at 224-243 K.

2. At present, there is no satisfactory theoretical explanation for the anomalous suppression of superconductivity in yttrium hexahydride $YH_6$.

3. The upper critical magnetic field of $YH_9$ is lower than that of $YH_6$. The latter has one of the highest upper critical magnetic fields (~180-200 T) among all polyhydrides, second only to the recently found *hcp*-(La, Ce)$H_9$ with the critical field up to 216 T [203].

4. For $YH_9$, we demonstrated the importance of considering the anisotropy of the superconducting gap, which leads to a noticeable (20 K) increase in the critical temperature in comparison with the isotropic calculation.

5. Predicted room-temperature superconductor $YH_{10}$ may have $T_C$ of about 267 K, significantly lower than expected, which follows from the systematic error in the predictions of the superconducting properties for yttrium polyhydrides $YH_6$ and $YH_9$ and the fact that the cubic superhydride modifications have a $T_C$ only 10% greater than that of hexagonal nonahydrides ("nona" means nine, Greek prefix) of the same metal. It also follows from the fact that $T_C$ of the ternary hydrides of (La,Y)$H_{10}$ differs only slightly from that of $LaH_{10}$.

# Chapter 4. Synthesis and magnetic properties of europium superhydrides

## 4.1 Studies of lanthanide polyhydrides

This chapter is based on a series of papers on the synthesis and transport properties of cerium [28, 105, 106], praseodymium [25], neodymium [24], and europium [43] polyhydrides.

After the discovery of record superconductivity in $LaH_{10}$, studies of polyhydrides at high pressures became mainstream, and during 2018-2020, lanthanide hydrides had been in the focus of high-pressure experiments. Soon it was found theoretically and experimentally that the superconducting properties of cerium hydrides were significantly inferior to those of lanthanide hydrides [105], whereas in hydrides of praseodymium, neodymium and europium, superconductivity was either completely absent or expressed very weakly ($T_C < 10$ K). The crystal structure of all lanthanide hydrides, their hydrogen content, volume of the unit cell, and hydrogen-hydrogen distance $d_{H\text{-}H}$ were found to be very close to those of $LaH_{10}$. For several years, there was no clear understanding of why lanthanide polyhydrides (Pr, Nd, Eu) are not superconductors in the presence of an atomic hydrogen sublattice. Studies of magnetic ordering of metal atoms in hydrides partially clarified this question: because the magnetic field in matter suppresses superconductivity, the emergence of a magnetic order in hydrides could explain the absence of superconductivity in lanthanide polyhydrides.

Chemically, La, Ce, Pr, Nd, and, to a lesser extent, Eu are similar to each other and, under ordinary pressure, form cubic hydrides of different stoichiometry, close to $XH_2$ and $XH_3$ [229]. Increasing the pressure to 100-130 GPa leads to the formation of cubic and hexagonal polyhydrides *Im*-3*m*-$XH_6$, *F*-43*m*-$XH_9$, *P*6$_3$/*mmc*-$XH_9$, and $Fm\bar{3}m$-$XH_{10}$. For instance, in the Ce-H system, the formation of *P*6$_3$/*mmc*-$CeH_9$, $Fm\bar{3}m$-$CeH_{10}$ [28, 203], and also, perhaps, *Im*-3*m*-$CeH_6$ [203], was experimentally confirmed. In the Pr-H system, *P*6$_3$/*mmc*-$PrH_9$ and *F*-43*m*-$PrH_9$ ($V$ = 31 Å$^3$/Pr at 110-113 GPa) were found [25]. In the Nd-H system, *P*6$_3$/*mmc*-$NdH_9$ and, probably, *Pm*-3*n*-$NdH_{6-x}$ were synthesized [24]. Finally, the Eu-H system was studied simultaneously and independently by two groups at Jilin University. Both groups came to complementary conclusions using the same scheme of synthesis (ammonium borane and Eu metal): at pressures of about 90-170 GPa they obtained *P*6$_3$/*mmc*-$EuH_9$, *Im*-3*m*-$EuH_6$, *Pm*-3*n*-$EuH_{6-x}$, *F*-43*m*-$EuH_9$, and *P*6$_3$/*mmc*-$EuH_6$ [43, 230] (doubtful, interpretation ambiguous). Syntheses of higher samarium and gadolinium polyhydrides (*F*-43*m*-$SmH_9$ and $GdH_x$) are also known from private communications (group of T. Cui and X. Huang, Jilin university) [192], but these studies have not yet been published.

There is particular interest in europium hydrides to not only due to europium being an anomalous element with the lowest density and the highest chemical reactivity among the lanthanides, but also because of the Mossbauer isotope $^{151}$Eu, which is present in a high concentration (50%) in the natural europium. This, in principle, makes it possible to study the magnetic structure of europium polyhydrides at moderate pressures using the resonant synchrotron radiation scattering (NRS). Such experiments have been successfully performed at the European synchrotron research facilities (ESRF) at the station ID18 using the iron isotope $^{57}$Fe [231].

Thus, it was found that at high pressures the lanthanides form various cubic and hexagonal polyhydrides; in the Pr-H system, these polyhydrides are nonsuperconducting [25], whereas in the case of Nd and Eu polyhydrides, the temperature dependence of the electrical resistance contains features that can be attributed to magnetic transitions [24, 230].

Another feature of lanthanide polyhydrides is the need to take into account the effect of a correlated f-electron behavior in the valence shells of atoms. It has been found in the studies of lanthanum polyhydrides that the existing pseudopotentials for DFT calculations give a large discrepancy in the unit cell volume of $LaH_{10}$ at high pressures, making it impossible to ascribe an exact chemical formula to this compound ($LaH_{10\pm1}$). In the case of praseodymium hydrides [25, 232], it has been immediately observed that the consideration of the spin-orbit interaction (SOC) leads to a significant change in the enthalpy of formation, thermodynamic convex hull, equation of state, and the phonon band structure [25]. The situation becomes even more complicated in the case of neodymium, where the correction for the spin-orbit interaction is insufficient, and to obtain the correct equation of state and consider the electronic correlations, it is necessary to introduce the Hubbard correction *U-J* [233, 234]. Nonzero values of magnetic moments on metal atoms also affect the enthalpy of Nd hydrides. As a result, a theoretical analysis of lanthanide hydrides must take into account (a) anharmonicity, (b) spin-orbit interaction (SOC), (c) ZPE and temperature effects, (d) magnetic ordering, and (e) the Hubbard *U-J* correction. All this makes the work with lanthanide polyhydrides very time-consuming and the determination of their H-content using the DFT methods very approximate.

Regarding calculations of the superconductivity parameters in lanthanide polyhydrides, as part of the study of $P6_3/mmc$-$NdH_9$, we developed a method to account for the spin-splitting energy using the UppSC code [235]. The result of this study is quite trivial. The coupling energy in the Cooper pairs is ~10-20 meV, whereas the typical energy separating bands with opposite electron spins in $NdH_9$ is ~1 eV [24]. Thus, conventional superconductivity cannot practically compete with magnetism in neodymium and europium polyhydrides: spin flip involves the consumption of substantially more energy than can be obtained through electron-phonon interaction. In the Pr-H system, the situation is somewhat different. Even before the experimental discovery of cubic and hexagonal modifications of $PrH_9$, ab initio calculations of the superconducting properties [19, 36] gave contradictory results with $T_C$ from 50 to 4 K. We repeated these calculations and found that their results significantly depend on the Pr pseudopotential used in Quantum ESPRESSO. In fact, almost any result for $T_C(PrH_9)$ can be obtained by changing the pseudopotential for Pr.

## 4.2 Europium superhydrides: theory and experiment

Following the standard scheme of investigation, to interpret the experimental data obtained, we conducted calculations using the USPEX code for the Eu-H system at pressures of 50-150 GPa (Figure 37). The variable composition and fixed composition functions were applied. As a structure-optimizer code we selected VASP, which makes it possible to quite simply introduce the spin-orbit interaction (SOC), magnetic ordering, and Hubbard *U-J* corrections.

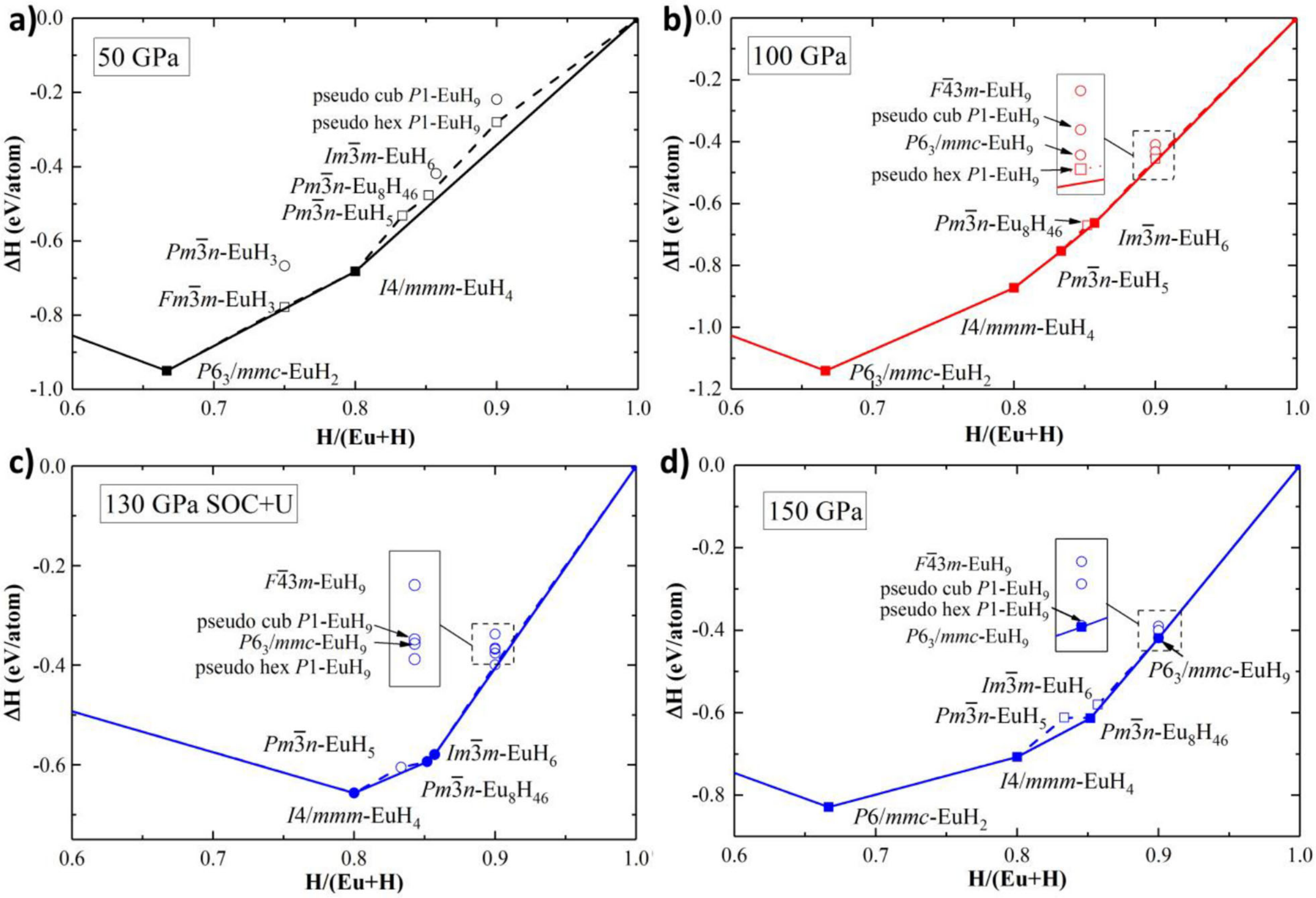

**Figure 37.** Calculated convex hulls of the Eu–H system at **(a)** 50 GPa, **(b)** 100 GPa, and **(d)** 150 GPa without SOC and *U-J* = 0, and **(c)** at 130 GPa with SOC and specific *U–J*. Thermodynamically metastable and stable phases are shown by hollow and filled circles, respectively.

The results of the computer search show that at low pressures around 50 GPa only $EuH_4$ formation is expected, whereas increasing the pressure to 100 GPa leads to stabilization of a large number of polyhydrides observed in the experiment: *Pm*-3*n*-$EuH_5$, *Pm*-3*n*-$Eu_8H_{46}$, and *Im*-3*m*-$EuH_6$. The disadvantage of this calculation is that the observed cubic and hexagonal $EuH_9$ phases are thermodynamically unstable even considering *U-J* and SOC. Taking into account the synthesis temperature (-$T\Delta S$ contribution) also does not change the situation [43]. The use of small structure distortions by analogy with $YH_9$, pseudohexagonal and pseudocubic (slightly less stable) *P*1-$Eu_4H_{36}$ (Figure 38), which have lower enthalpy of formation at pressures below 150 GPa and are dynamically stable, somewhat improves the situation.

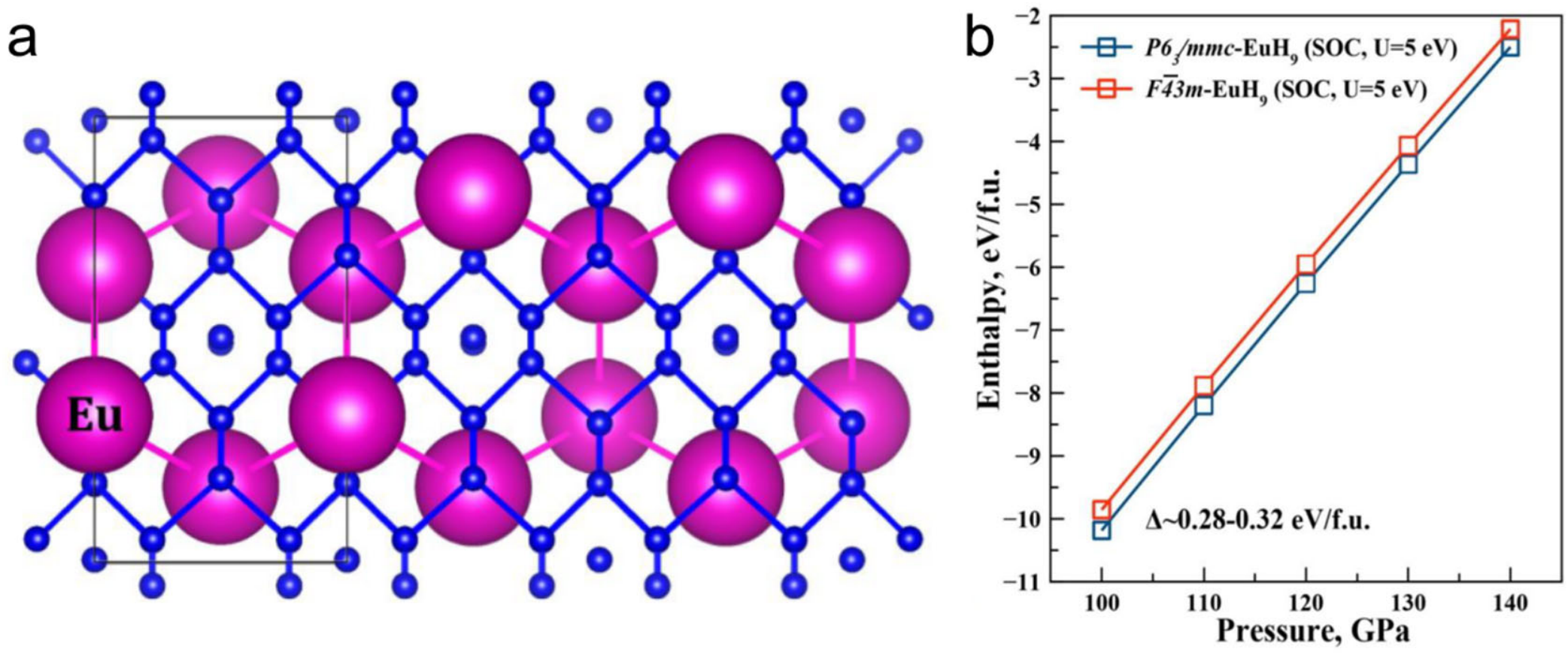

**Figure 38.** (a) Crystal structure of pseudohexagonal $P1$(hex)-$EuH_9$ (=$Eu_4H_{36}$) with the disordered hydrogen sublattice, visualized using VESTA software [236]. The hydrogen atoms are shown in blue. (b) Difference between the enthalpies of the cubic and hexagonal modifications of $EuH_9$.

Relatively new in this work is the use of the *Pm*-3*n*-$Eu_8H_{46}$ phase constructed by analogy with many Zintl phases of a similar structure, such as $K_4Ge_{23}$, $Rb_4Ge_{23}$, and $Ba_8Si_{46}$, which are also sometimes called Weaire-Phelan phases [88]. This was the first experience for us to search the structure using the databases of inorganic compounds. Once a prototype of the desired composition having between 5.5 and 6 hydrogen atoms per Eu atom and the desired symmetry (*Pm*-3*n*) was found, we performed the replacement of the atoms of the parent structure and its relaxation to the desired pressure (~100 GPa). We found that constructed *Pm*-3*n*-$Eu_8H_{46}$ satisfies the experimental equation of state much better than the alternative explanation via *Pm*-3*n*-$EuH_5$ (Figure 39) proposed in Ref. [230]. It is interesting to note a very large difference in the unit cell volume of both europium and neodymium polyhydrides calculated with and without the SOC and *U*-*J* corrections (Figure 39).

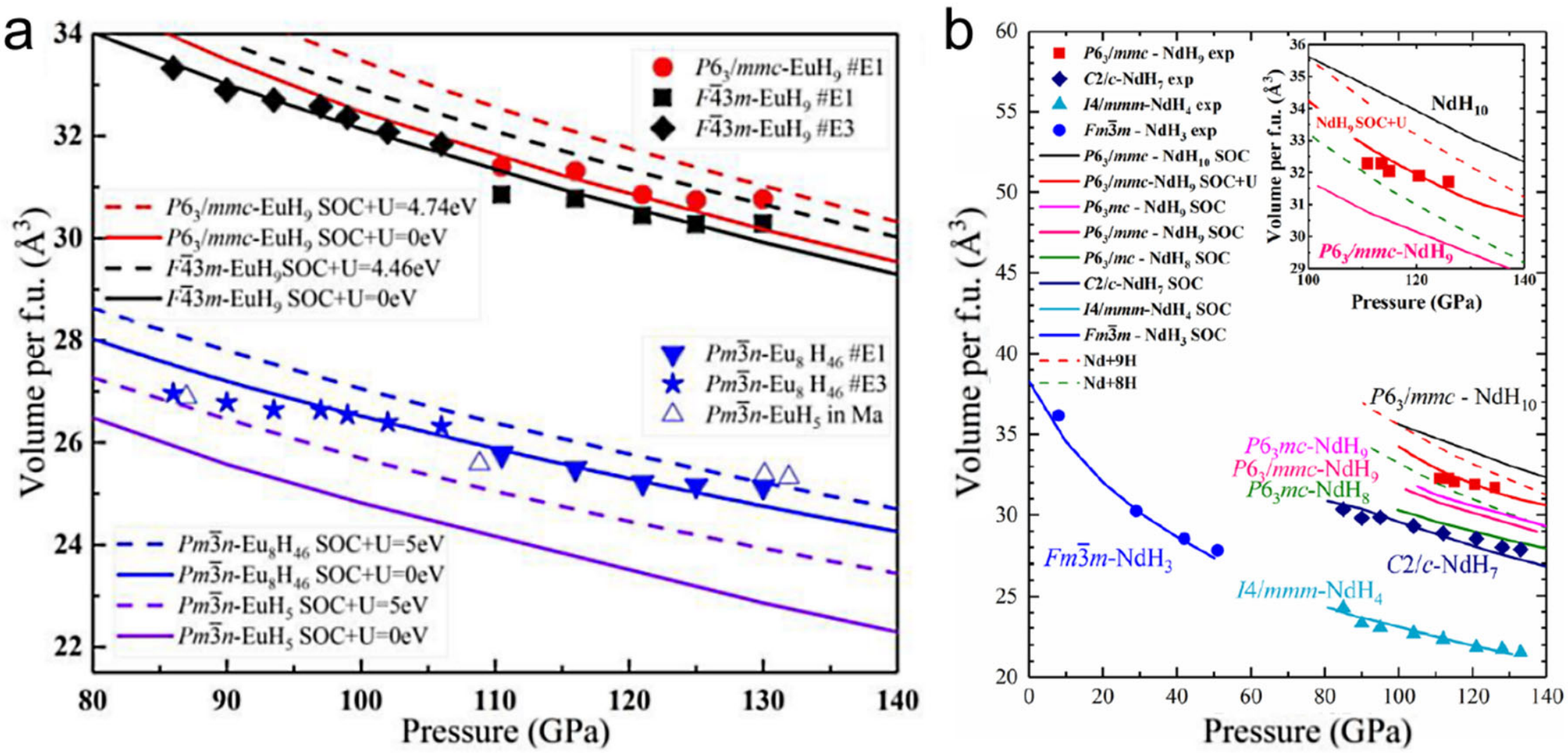

**Figure 39.** Equations of state of the synthesized Eu–H and Nd–H phases. (a) Pressure dependence of volumes of europium hydrides. The theoretical results considering SOC are shown by solid (*U*–*J* = 0 eV) and dash-dot (*U* – *J* = 5 eV) lines. The experimental data for $EuH_5$ from [230] are shown by hollow squares. (b) The equation of state of the synthesized Nd−H phases; theoretical results include spin−orbit coupling and magnetism. Inset: The distinction between the *P*6₃*mc*-$NdH_8$, *P*6₃/*mmc*- and *P*6₃*mc*-$NdH_9$, and *P*6₃/mmc-$NdH_{10}$ phases.

*Pm*-3*n*-$Eu_8H_{46}$ is also thermodynamically more stable at pressures of 130-150 GPa than *Pm*-3*n*-$EuH_5$. Furthermore, using molecular dynamics methods with machine-learning potentials (MTP) of interatomic interaction in a supercell of 960 Eu and H atoms, we showed that *Pm*-3*n*-$Eu_8H_{46}$ is dynamically stable at 300 K and 130 GPa for an interval of 40 picoseconds, which is equivalent to stability in the anharmonic approximation (see Supporting Information [43]). This was the first example where the calculation scheme with molecular dynamics and velocity autocorrelator was used to justify the dynamic stability of polyhydrides ($EuH_9$ and $Eu_8H_{46}$) in the anharmonic approximation at a high pressure and finite temperature (300 K). Subsequently, we have successfully used this calculation scheme in many other projects [72, 202]. The criterion of the dynamic stability within this approach is the preservation of the original structure of compounds when averaging the results of the dynamic simulation over a long-time interval.

The study of the magnetic and transport properties of europium hydrides in our work was mainly theoretical. Polyhydrides of praseodymium [25], neodymium [24], and europium [230] have been studied experimentally in electrical DACs. No resistive transitions above 10 K were found. In some cases, a sharp drop in resistance was observed below 10 K, but we cannot state unequivocally whether this refers to the sample or to superconductivity in the electrodes (Mo, $MoC_x$). A retrospective analysis of the Debye temperature for resistivity measurements of praseodymium hydrides (Figure 40a) [25] using the Bloch-Grüneisen formula [64, 65] gives the following results: $\theta_D = 303 \pm 5$ K (145 GPa), $\theta_D = 278 \pm 5$ K (138 GPa), $\theta_D = 260 \pm 5$ K (145 GPa), which is not typical for polyhydrides. Therefore, the electrical resistance drop at 6-8 K observed in our experiments most likely does not relate to praseodymium polyhydrides.

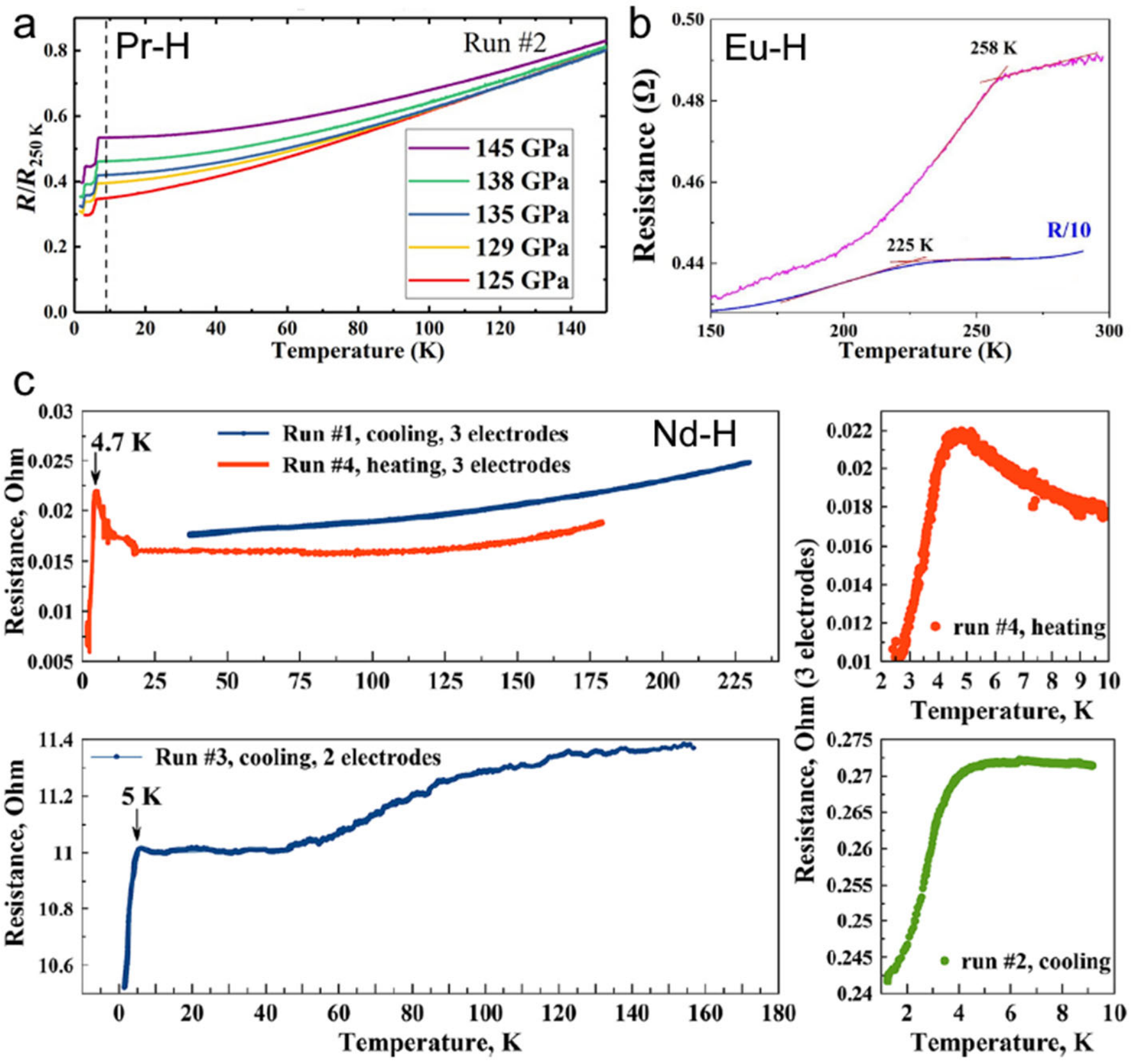

**Figure 40.** Temperature dependence of the electrical resistance R(T) for hydrides of different lanthanides: (a) praseodymium hydrides at 125-145 GPa, (b) europium hydrides at 140 GPa [230], and (c) neodymium hydrides at 110 GPa (runs 1-4).

A similar analysis for $NdH_x$ sample (run #1, Figure 40, [24]) gives $\theta_D$ = 760 ± 10 K (110 GPa), which corresponds to a typical Debye temperature for polyhydrides. For $EuH_x$ [230], such an analysis is not possible because of the nonmonotonic nature of the $R(T)$ dependence. In a recent work on ternary La-Nd-H hydrides [202], we have encountered a low-temperature anomaly of the electrical resistance at 6-8 K, which may also be relevant to the manifestation of both magnetic ordering and superconductivity in electrodes or impurities.

At normal pressure, europium shows antiferromagnetic ordering below $T_N$ = 90 K [237-240], accompanied by the tetragonal distortion of the cubic (*bcc*) lattice [241]. Measurements of the electrical resistivity and the position of the kink on the $R(T)$ curve indicate that when the pressure is increased to 15 GPa, $T_N$ decreases to 80 K. At higher pressures, new features appear on the $R(T)$ around 140 K, persisting up to 42 GPa [242]. The X-ray magnetic circular dichroism (XMCD) measurements indicate that this feature corresponds to a ferromagnetic ordering and up to about 50 GPa europium remains magnetic [243].

The first step in our study of the magnetic properties was to establish the absolute value of the equilibrium magnetic moments on the metal atoms in polyhydrides as a function of pressure (Figure 41). The calculations show that the magnetic moment on the europium atoms is weakly dependent on the pressure and is about 6-7 $\mu_B$ (Bohr magneton). The comparison with the Nd-H system shows that for $NdH_9$, this dependence is more significant and increasing the pressure and density of the Nd atoms suppresses the magnetization of this polyhydride (Figure 41b). This is also evident from the density of states plot for different electron spin directions. For europium hydrides, the positions of the electron state density maxima for the different spins are separated from each other by 7-8 eV, implying the emergence of a spontaneous magnetic order in the material at 0 K since only states with a certain spin will be filled.

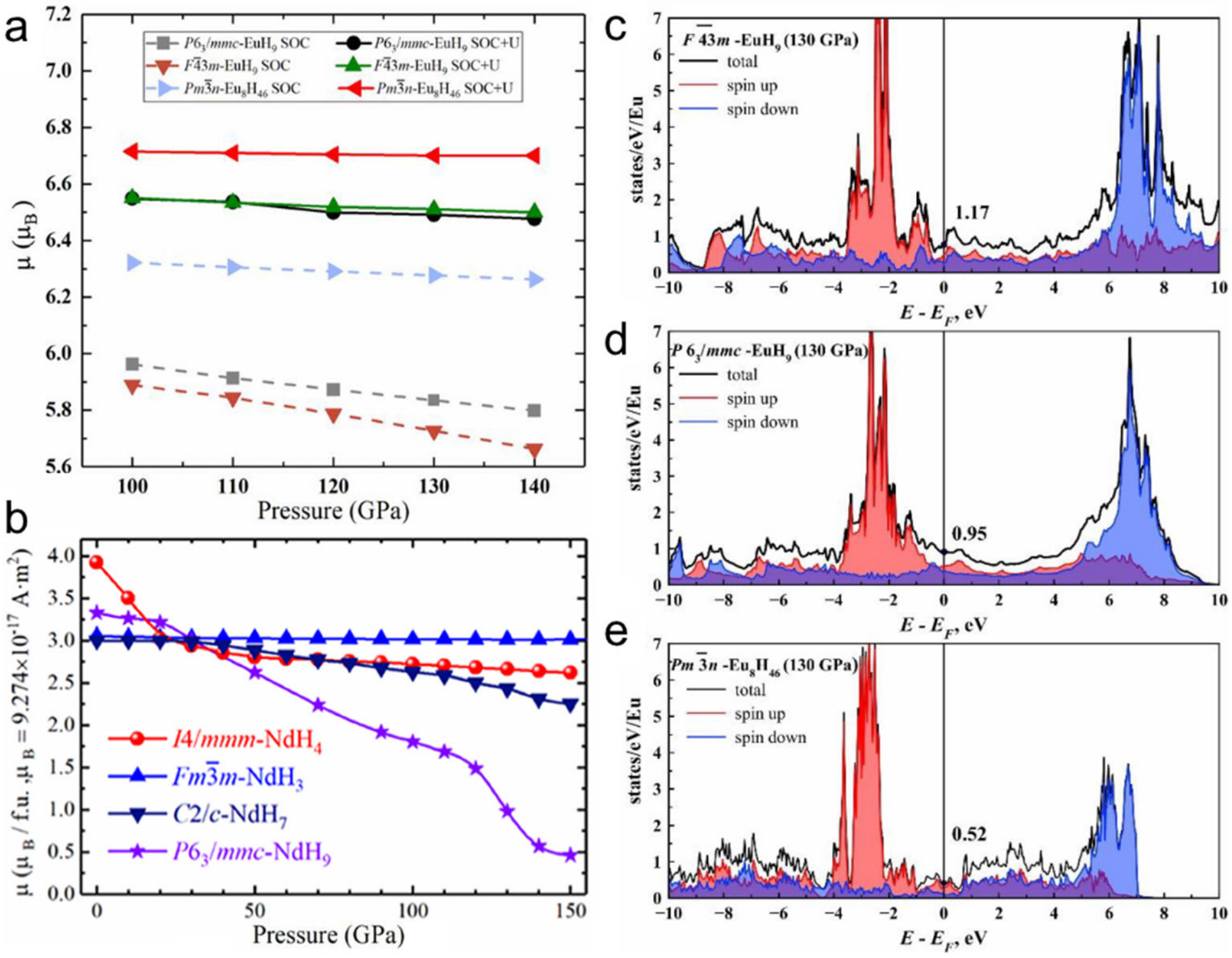

**Figure 41.** (a) Magnetic moments ($\mu_B$ per 1 Eu atom) of all synthesized Eu-H compounds with and without *U–J* and SOC. (b) Same for the Nd-H phases with *U-J* and SOC. (c–e) Electron density of states (DOS) of different europium hydrides at 130 GPa, with contributions of different spin orientations to the total DOS.

After establishing the absolute value of the magnetic moments, we considered various options of their direction distribution in supercells of europium polyhydrides up to 8 Eu atoms (hydrogen excluded). Analyzing the enthalpy of antiferromagnetic and ferromagnetic structures, we found that cubic $EuH_9$ is antiferromagnetic at temperatures below 24 K, whereas hexagonal $EuH_9$ and cubic $Eu_8H_{46}$ are ferromagnetic at temperatures below 137 and 336 K, respectively (Figure 42). Of course, the calculation was limited by the size of the supercell, and we did not consider the possibility of a complex (e.g., spiral [244]) magnetic ordering. All of the calculations performed are illustrative and estimative, because no experimental investigations of the magnetic order in polyhydrides at pressures of 100 GPa and higher are known at present time. Only recently the first experiments on the Mossbauer effect have been made for lower europium $^{151}EuH_{2+x}$ hydrides at pressures up to 15 GPa [245]. Therefore, studies of the magnetic structure of polyhydrides, primarily europium and iron, are a matter of the future.

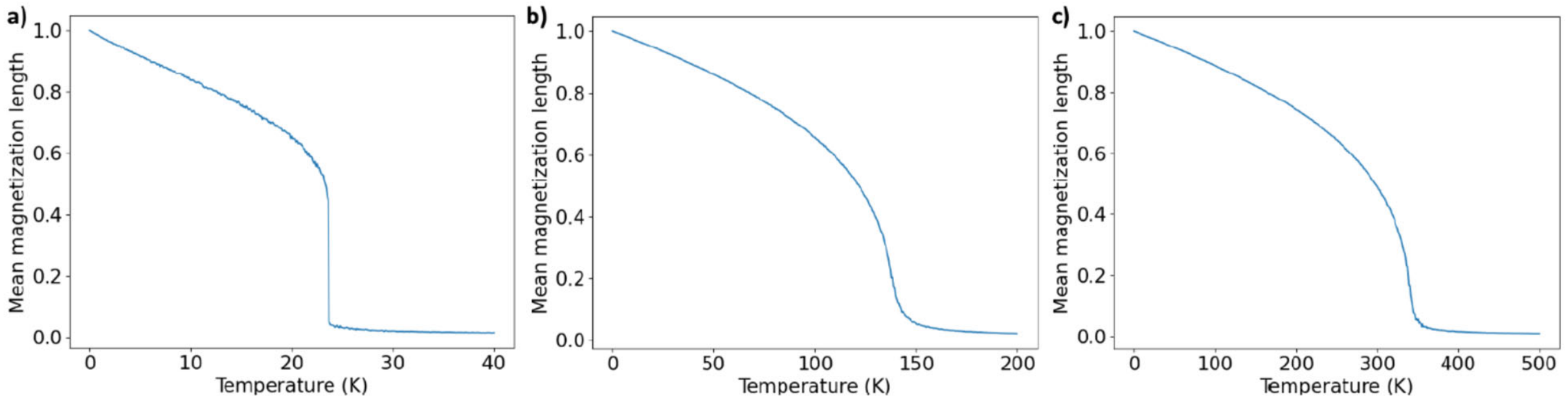

**Figure 42.** Normalized mean magnetization lengths with respect to temperature in (a) $F\bar{4}3m$-$EuH_9$, (b) $P63/mmc$-$EuH_9$, and (c) $Pm\bar{3}n$-$Eu_8H_{46}$ from the Monte Carlo simulations. For antiferromagnetic $F\bar{4}3m$-EuH9 the mean magnetization length of the spin-up channel is displayed.

In 2021, several theoretical papers have been published predicting high-temperature superconductivity in polyhydrides of heavy lanthanides, such as Yb and Lu. This refers primarily to *Im*-3*m*-$LuH_6$ and *Im*-3*m*-$YbH_6$ hexahydrides [246]. Because of the occupancy of the *f*-electron shell, the Lu atom should behave as an analog of Y, La, Ac, in other words, as a $d^1$-element, and form symmetric clathrate nonmagnetic polyhydrides $LuH_6$, $LuH_9$, and $LuH_{10}$ [211], with the critical temperature close to 270 K. Indeed, Lu itself is a superconductor with $T_C$ growing as the pressure increases (Figure 45). In a recent experimental work, our colleagues from Jilin University registered the superconducting transition in $Fm\bar{3}m$-$LuH_3$ at a pressure of 120 GPa and temperature $T_C$ = 12.4 K [247]. This, in principle, proves the possibility of superconductivity in other lutetium hydrides as well.

To test this hypothesis, we assembled a high-pressure diamond anvil cell loaded with Lu and ammonium borane at about 161 GPa (Figure 43). As a result of the first laser heating, the pressure decreased to 158 GPa. The examination of the temperature dependence of the electrical resistance (Figure 44) showed a smooth metallic $R(T)$ from 310 to 155 K without any indication of a superconducting transition, $R$(310K)/$R$(155K) = 1.4, which is typical for polyhydrides. Interpolation using the Bloch-Grüneisen formula gives the Debye temperature $\theta_D = 1278 \pm 5$ K (158 GPa), which is common for polyhydrides at megabar pressures. Thus, the electrical measurements are in favor of the successful synthesis of lutetium polyhydrides.

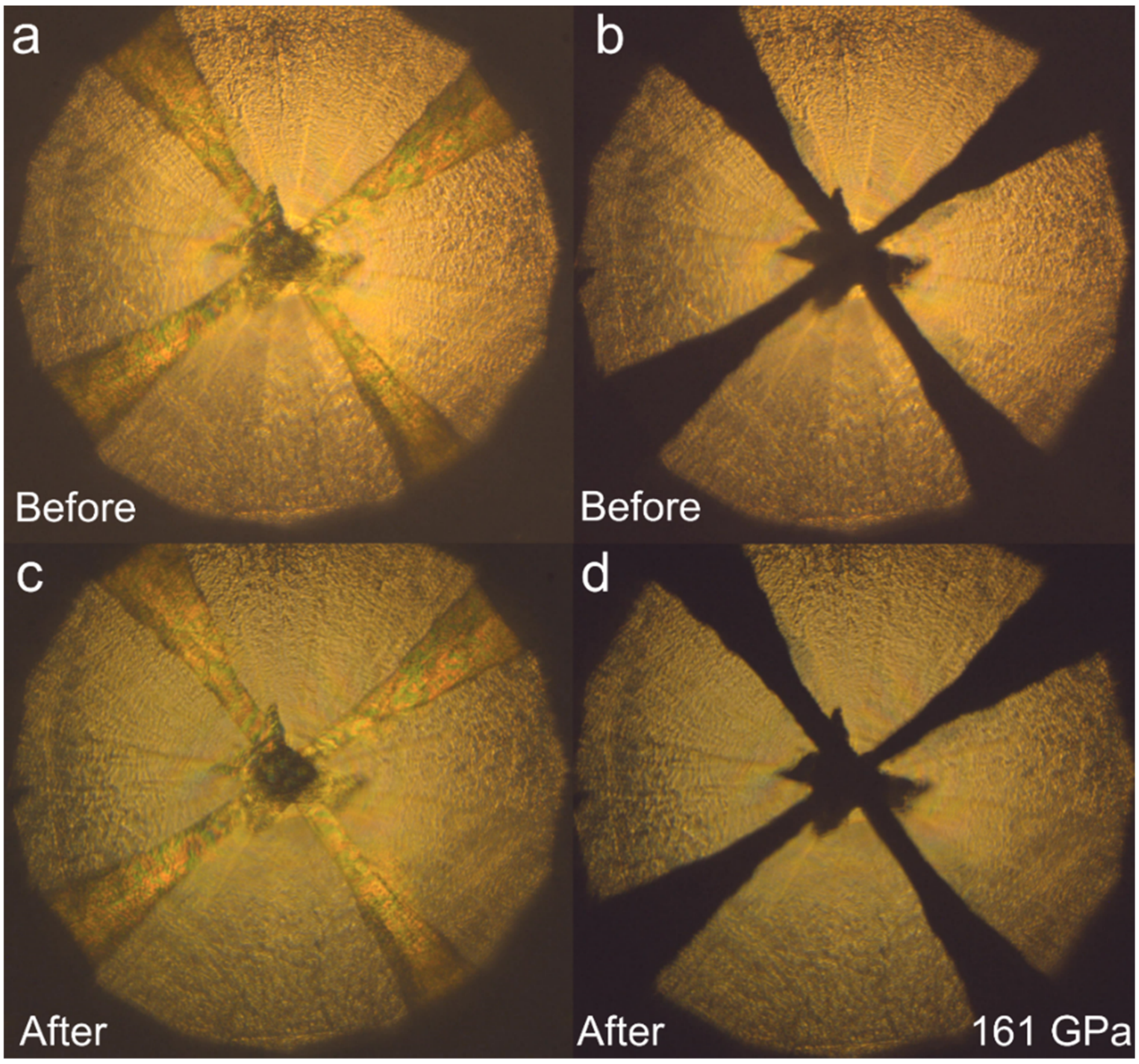

**Figure 43**. Photos of the Lu/AB sample (a, b) before and (c, d) after the laser heating at 161 GPa. Gasket consists of $CaF_2$/epoxy.

The obtained data shows that superconductivity in lutetium hydrides is not observed at temperatures above 90 K, which contradicts the published theoretical predictions [211, 246]. In fact, back in the article on praseodymium hydrides (Supporting Information to [25]), we noticed that the results of calculations of the superconducting properties for $PrH_x$, giving $T_C$ from 50 to ~0 K, depend significantly on the chosen pseudopotential for Pr. We believe that the same situation will be observed for other lanthanides because of impossibility to accurately factor in correlation interactions of *f*-electrons. This factor fundamentally limits the predictive ability of standard DFT approximations of the electron-phonon interactions for polyhydrides of lanthanides, including Lu. Also, a simple way to test $LuH_{10}$ for high-temperature superconductivity is to synthesize a lanthanum-lutetium decahydride $(La,Lu)H_{10}$ with a small lutetium concentration of 5-15 atom %. If an appreciable suppression of superconductivity in $LaH_{10}$ is found in this experiment, it would be a direct indication that $T_C(LuH_{10})$ is significantly lower than in theoretical predictions.

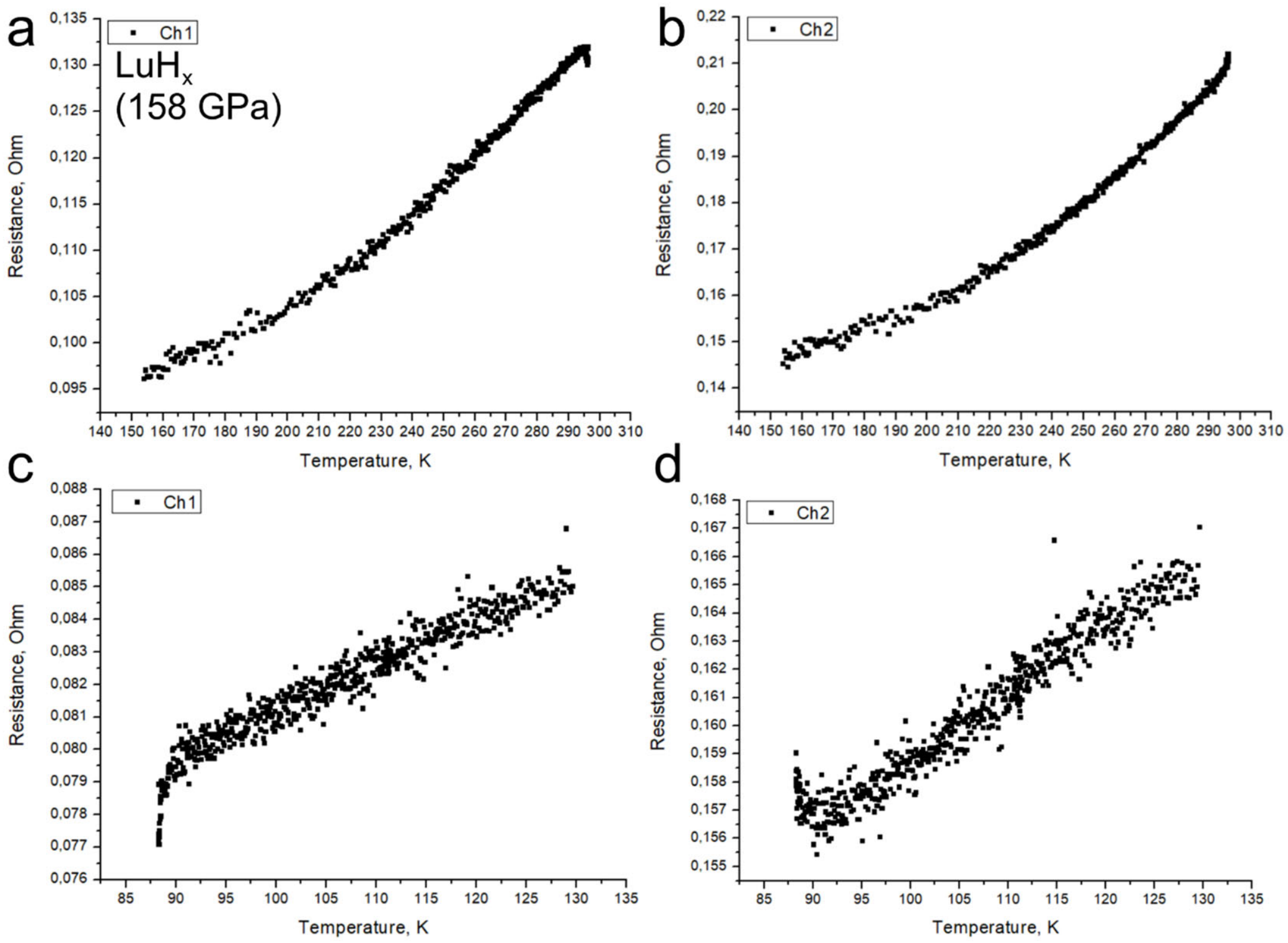

**Figure 44.** Temperature dependence of the electrical resistivity of the $LuH_x$ sample synthesized at 158-161 GPa for 2 channels (ch1, ch2, cooling). In the interval from 155 to 130 K, there is a gap in the data because of an error of a recording device. A possible superconducting transition begins around 90 K.

## 4.3 Analogy of properties of polyhydrides and pure metals

In this section, we will describe an approach to the qualitative description of the properties of both magnetic and superconducting metal polyhydrides based on an analogy with the properties of hydride-forming metals. The physical density of matter itself is related to the unit cell volume and the metal-to-metal distance in polyhydrides. For most of the hydrogen-rich compounds studied, the contribution of hydrogen to the physical density is insignificant, but its presence in the unit cell significantly increases the volume and the metal-to-metal distance in the crystal structure. If we neglect the redistribution of the electron density and change in the valence state of metal atoms in hydrides, we can find certain parallels between the properties of pure metals at normal or low pressures and polyhydrides at megabar pressures. In particular, the hydrogen sublattice itself is nonmagnetic (see also [248]), which we verified by direct DFT calculations: even at a pressure of several TPa, structures with a magnetic moment on hydrogen atoms are much higher in energy than nonmagnetic structures. Therefore, when considering the magnetic properties of polyhydrides, we can neglect the “invisible” hydrogen sublattice, which only changes the distance between the metal atoms.

The density of praseodymium nonahydrides is 8.06 g/cm$^3$ at 110 GPa, whereas the density of the metal under ambient conditions is 6.77 g/cm$^3$, and even at a small pressure of 5-10 GPa the distances *d*(Pr-Pr) in pure Pr become similar to those in $PrH_9$ at 110 GPa. Considering that praseodymium remains paramagnetic up to the lowest temperatures, and the known anomalies in its electrical properties are associated only with phase transitions [249, 250], we do not expect the manifestation of magnetic ordering in praseodymium polyhydrides at megabar pressures.

The density of neodymium nonahydride $NdH_9$ is 7.93 g/cm$^3$, whereas that of Nd metal under ambient conditions is 7.0 g/cm$^3$. Therefore, it is expected that the magnetic ordering of pure neodymium at low temperatures can also be observed in neodymium polyhydrides at high pressures. Indeed, as the results of the rather old experimental work [251] show, under a pressure of up to 1.6 GPa the magnetic properties of neodymium are enhanced, the Neel temperature $T_N$ and Curie temperature $T_c$ increase. Hexagonal Nd sites show antiferromagnetic ordering, whereas cubic Nd sites show ferromagnetic ordering, which also corresponds to our calculations for $P6_3/mmc$-$NdH_9$ at 110-130 GPa.

The density of europium is 5.24 g/cm$^3$, that of europium nonahydrides is 8.65 g/cm$^3$ at 115 GPa. Increasing the pressure for metallic europium leads to an increase in the density, which reaches a value of $\rho(EuH_9)$ at about 10 GPa [252]. At these values of pressure and density, metallic europium (II) exhibits pronounced antiferromagnetic properties with $T_N \approx 85$ K [253]. Thus, for europium polyhydrides we can also expect the manifestation of pronounced magnetism in this range of physical densities.

Let us now consider superconducting polyhydrides. There is an idea that the maximum $T_C$ in metal polyhydrides is reached approximately in the same density range as for pure hydride-forming metals. This idea is based on the fact that because of the large difference in the atomic masses of hydrogen and metal atoms, the vibrations (optical and acoustic phonons) of their sublattices practically do not interact, occupying separate regions in the phonon spectrum. This means that the Eliashberg function of metal polyhydrides can be represented as the sum of the acoustic and optical components coming from metal and hydrogen: $\alpha^2F = \alpha^2F_{metal}(\omega < 10 \text{ THz}) + \alpha^2F_H(H, \omega > 10 \text{ THz})$. Moreover, the metallic sublattice provides a very significant contribution to the electron-phonon interaction coefficient $\lambda$, whereas the contribution of the hydrogen sublattice is the high Debye temperature and the logarithmically averaged phonon frequency $\omega_{log}$. Both contributions are critical to achieve high-temperature superconductivity.

It is currently known that in superhydrides at megabar pressures, the contribution of metal and hydrogen to the total density of electron states at the Fermi level (DOS) is comparable. It is believed that the greater the contribution to the DOS from the hydrogen sublattice, the more promising this polyhydride is as a superconductor. Importantly, the introduction of hydrogen into the metal sublattice in many cases does not lead to any dramatic changes in the electronic band structure: the density of polyhydrides at megabar pressures and metal-metal distance, as we have seen, are very close to the density and $d_{Me\text{-}Me}$ in pure metals. This, of course, does not apply to the alkali and alkaline earth elements, where the introduction of hydrogen is accompanied by a loss of metallic properties due to a very strong charge transfer. But for $d^1$-$d^2$ elements (Sc, Ti, Zr, Y, La, Hf, Th, U), the metal contribution to the DOS remains significant even after formation of a polyhydride. This allows us to hope for a correlation between the behavior of superconductivity in pure metals and superhydrides under pressure. An illustration of this idea is shown below (Figure 45). For example, lanthanum has one of the highest $T_C$ at pressures up to 20 GPa. Li, Y, V, Sc, S, Ca, Lu, and Zr are also promising elements for superhydride synthesis. Indeed, it has been recently shown experimentally that $H_3S$ [5], $LaH_{10}$ [7], $YH_6$ [30, 31], $CaH_6$ [115, 153], $YH_9$ [31] are high-$T_C$ superconductors; $Li_2MgH_{16}$ [32] and $ScH_9$ [19, 254-256] have been predicted to be so.

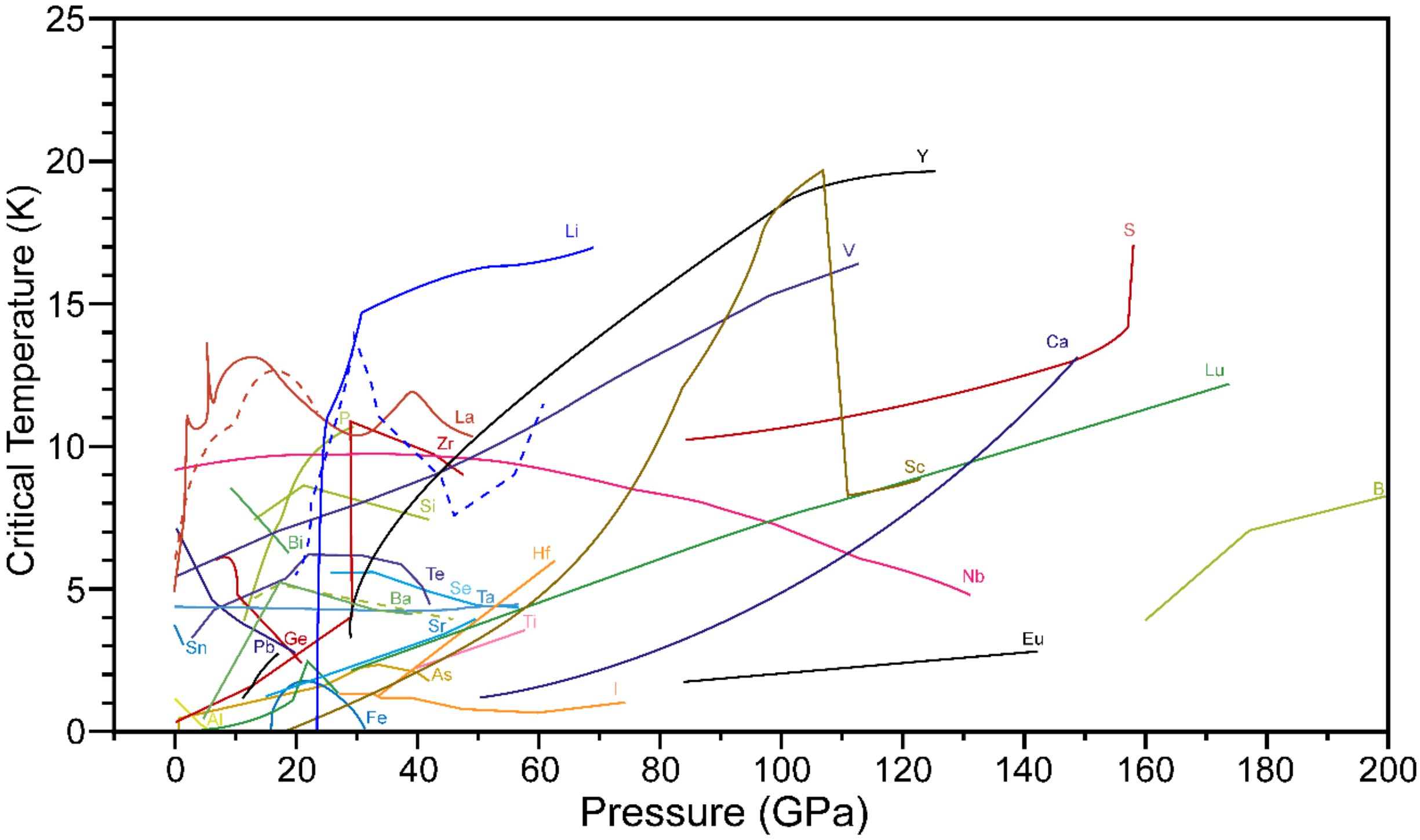

**Figure 45.** Pressure dependence of the critical temperature of superconductivity for various metals [257, 258].

It is known that superconductivity in metallic thorium disappears with increasing pressure (see Chapter 2). Formally, maximum $T_C$ for pure Th is in the “negative pressure” region. Thorium density at 0 GPa is 11.78 g/cm$^3$, at the same time $ThH_{10}$ at 125 GPa has a unit cell volume of 35 Å$^3$/f.u. and a density of 11.52 g/cm$^3$. As the pressure rises, the density of $ThH_{10}$ will grow and, according to our DFT calculations, $T_C(ThH_{10})$ will decrease along with $T_C$ of pure metal.

The density of La at ambient conditions is 6.16–6.18 g/cm$^3$, whereas that of $LaH_{10}$ at 150 GPa is 6.97 g/cm$^3$, which corresponds to the position of the maximum $T_C$ in pure La at 5-10 GPa [259]. Indeed, the physical density of lanthanum increases rapidly during compression and reaches 7 g/cm$^3$ just in this pressure range.

The density of $YH_6$ is 6.91 g/cm$^3$ at 165 GPa, which is significantly higher than that of yttrium at normal pressure – 4.47 g/cm$^3$. To achieve a similar density, yttrium must be compressed to about 65 GPa [260]. Indeed, as the pressure rises, $T_C$(Y) increases very rapidly and exceeds maximum $T_C$(La) at this pressure. As a rule, increasing the hydrogen content leads to decreasing the density. However, in the case of yttrium we need to increase the density. This leads to the need to use higher and higher pressures to stabilize higher yttrium polyhydrides and achieve high-temperature superconductivity in them. The maximum $T_C$ for the metal is observed at 100-120 GPa. At this pressure, the density of metallic yttrium is 10.6 g/cm$^3$ [260, 261]. This density is very difficult to achieve in Y hydrides. Thus, we believe that in the Y-H system greater $T_C$ can be achieved by increasing the pressure and physical density of the hydrogen-rich compounds.

## 4.4 Conclusions from the studies of lanthanide hydrides

1. At present, the magnetic structure of polyhydrides at megabar pressures has not been practically investigated in experiment. There are only several theoretical works, including those of the author, which lead to a conclusion that polyhydrides of most lanthanides are magnetic.

2. All lanthanide polyhydrides have similar structures at high pressures. Clathrate structures, such as $F\text{-}43m$-$XH_9$, $P6_3/mmc$-$XH_9$, $Fm\bar{3}m$-$XH_{10}$, $Im\text{-}3m$-$XH_6$, $Pm\text{-}3n$-$X_8H_{46}$, $Pm\text{-}3n$-$XH_5$, and $I4/mmm$-$XH_4$, predominate.

3. Calculations of the physical properties of lanthanide polyhydrides are extremely difficult because of the need to take into account strong correlations of *f*-electrons. The simultaneous consideration of the spin-orbit interaction, magnetism, the Hubbard *U-J* correction (which is individual for each phase and should be calculated separately by the linear response method), ZPE, and the entropy factor leads to a large amount of computations and low accuracy in predicting the stable structures of polyhydrides and their equations of state.

4. A supposed crossover of magnetism and superconductivity can be realized in lanthanide polyhydrides at low temperatures, where unexpected sharp drops in the electrical resistance are often observed.

5. Suppression of conventional BCS superconductivity in lanthanide hydrides is not due to magnetic ordering, which does not appear in all lanthanides, but caused by the scattering of Cooper pairs on magnetic atoms with spin flipping of one of the electrons and the destruction of the pair.

6. Despite the difficulty of theoretical calculations, polyhydrides of heavier lanthanides represent an extremely tempting target for experimental research, because almost all of these metals form higher hydrides at moderate pressures of about 100-150 GPa.

# Chapter 5. Lanthanum-yttrium ternary polyhydrides. Synthesis and properties.

## 5.1 Studies of ternary polyhydrides

This chapter is based on the studies of the structure and superconductivity of the La-Y-H [72], La-Nd-H [202], and La-Ce-H [203, 212] ternary hydride systems.

Immediately after the discovery of outstanding superconducting properties of binary polyhydrides of sulfur, lanthanum, and yttrium, the desire to improve their superconducting properties by doping arose. The simplest approach was based on selecting such a dopant that would increase the density of electronic states in the vicinity of the Fermi level, which should proportionally increase the electron-phonon interaction strength ($\lambda$) without changing any other properties. This idea owes its origin to the considerable experience gained with unconventional superconductivity in Fe-containing pnictides and cuprates, for which the degree and type of doping is one of the most important characteristics [262, 263].

A large number of additives have been proposed for $LaH_{10}$ [264], including Al and Si. Phosphorus has been suggested for doping of $H_3S$ [61, 265]. After the resonant publication by E. Snider et al. on room-temperature superconductivity in the C-S-H system [37], several attempts have been made to explain such high $T_C$ through the idea of doping $H_3S$ with carbon. It would have to lead to a shift of the van Hove singularity, present in the theoretically calculated $H_3S$ band structure, exactly to the Fermi level [266]. Subsequently, more accurate calculations have completely disproved the possibility of raising the critical temperature in $H_3S$ by carbon doping [267-271].

Experimental observations very soon have changed the concept of doping of known superhydrides. It has been found that in La-Y polyhydrides, no increase in the critical temperature was observed despite the presence of yttrium in a chemical environment characteristic of the predicted room-temperature superconductor $YH_{10}$. Similarly, the introduction of aluminum and carbon in amounts of 3-5 atom % was found to have very little effect on the superconducting properties of $LaH_{10}$ [147]. However, the introduction of magnetic atoms, such as Ce and Nd, leads to significant suppression of superconductivity in ternary hydrides [202, 203, 212]: in the case of Nd, $T_C((La,Nd)H_{10})$ decreases by 10-11 K for each atomic percentage of Nd. This is fully consistent with Anderson's theorem [148] and Gorkov's theory of Cooper pair scattering on isotropic and anisotropic centers [70]. As early as 1959, it had been shown that doping cannot in principle increase the critical superconductivity temperature within the framework of the classical Bardeen-Cooper-Schrieffer (BCS) theory, although impurities may significantly increase the upper critical magnetic field (see also Appendix, "Details of the upper critical magnetic field calculations"). The pronounced suppression of superconductivity in $LaH_{10}$ by magnetic Nd atoms and the robustness of $T_C$ with respect to nonmagnetic impurities, such as Y, Al, C, $CH_4$ (comes from diamond anvils [272]) and O (comes from the air during the loading of DACs in atmosphere) and H (such impurities occur in non-stoichiometric polyhydrides) under Anderson's theorem confirm that polyhydrides are conventional BCS superconductors and their critical temperature cannot be increased by doping.

In parallel with the doping concept, researchers developed the theoretical approach to true ternary polyhydrides with an ordered arrangement of atoms of all types. In this case, the unit cell of compounds appears quite compact and accessible for ab initio calculations. Many such calculations have been made and led to encouraging results, for instance $CaYH_{12}$ [34], $MgCaH_{12}$ [35], $Li_2MgH_{16}$ [32], $ScYH_6$ [273]. Many of these ternary compounds have been predicted to have much higher critical temperatures than the corresponding binary hydrides. However, experiments made significant corrections. It has been found that the synthesis from alloys (A,B) of the corresponding metals always yields not true ternary hydrides but solid solutions with a disordered arrangement of A and B atoms in the metal hydride sublattice [72, 202, 203, 212]. In other words, we return to the case of doping of polyhydrides. Calculations of the migration enthalpy (~1-10 μeV/atom) of atom A to the position of atom B (Supporting Information to [188]) show that metal atoms can easily migrate along the metal sublattice even at room temperature. Thus, in most cases true ternary polyhydrides simply do not form, giving way to solid solutions of the $(A,B)H_x$ type, which are very difficult to calculate using ab initio methods because of the large size of the model supercells.

But there is still light at the end of the tunnel. The introduction of impurities can lead to several interesting effects in superhydrides. First of all, an impurity can destabilize one of the crystal modifications and favor the formation of another one. For example, it is likely that in the ternary system $(Ce,La)H_9$, the addition of cerium stabilizes the hexagonal modification of nonahydride $LaH_9$, whereas for pure lanthanum, the main product under similar conditions will be cubic $LaH_{10}$. Second, the large difference in the properties and atomic radii of the metals A and B makes the formation of the A-B solid solutions impossible, and instead true ternary polyhydrides are formed. We observed this in the Ca-Y-H system (Figures 54-56) at 177-119 GPa. Another approach is to use lower hydrides or intermetallides with ordered A and B sublattices as starting compounds. Then, if the synthesis proceeds without too much heating, hydrogen will fill the voids in the metal sublattices without mixing them, and we will come to true ternary polyhydrides. Impurities increase the resistance of hydrides in the normal state, widening the superconducting transition to several tens of degrees, but also increase the upper critical magnetic field.

According to the Ginzburg-Landau-Abrikosov-Gorkov (GLAG) theory [216], the upper critical magnetic field for dirty superconductors is related to the normal resistance:

$$-\left.\frac{dH_{C2}}{dT}\right|_{T_C} = 3565.1 \cdot \gamma \cdot \rho(T_C), \quad (4)$$

where $dH_{c2}/dT$ is in T/K, the Sommerfeld parameter $\gamma$ is in $J/m^3K^2$, the resistivity at the transition point $\rho(T_c)$ is in Ω×m (see also [274]). Using the definition

$$\gamma = 1/3\, \pi^2 k_B^2 (1 + \lambda) \times (N_F/V_{cell}), \quad (5)$$

where $N_F$ – is the total density of electronic states per unit cell (mostly, per metal atom) in states/eV/f.u., we can write

$$-\left.\frac{dH_{C2}}{dT}\right|_{T_C} = \frac{1.4 \cdot 10^7 \cdot (1+\lambda) \cdot \rho(T_C) \cdot N_F}{V_{cell}}, \quad (6)$$

where $V_{cell}$ – is the unit cell volume in $Å^3$. Then, within the WHH model, we obtain the upper critical field in Tesla

$$H_{C2}^{WHH}(0) = T_C \frac{0.97 \cdot 10^7 \cdot (1+\lambda) \cdot \rho(T_C) \cdot N_F}{V_{cell}}, \quad (7)$$

or, expressing in terms of sample resistance,

$$H_{C2}^{WHH}(0) = T_C \frac{0.97 \cdot 10^7 \cdot (1+\lambda) \cdot R(T_C) \cdot h \cdot N_F}{V_{cell}}. \quad (8)$$

In this case, the thickness of the sample (h) is most correctly calculated through the resistance drop at a certain temperature interval h = [R(300) – R($T_C$)] / [$\rho_{EPW}$(300) - $\rho_{EPW}$($T_C$)], because EPW only considers phonons as a scattering factor. Thus, by introducing more impurities into a sample and increasing its normal resistivity we can significantly increase the upper critical field of a polyhydride (see also Appendix, "Details of the upper critical magnetic field calculations"). From these formulas it also becomes clear why $YH_6$ has such a high $\mu_0 H_{C2}$: One factor in this GLAG-type model is the unit cell volume, which is significantly lower for $YH_6$ than for $YH_9$ or $LaH_{10}$, at a comparable critical temperature.

In conclusion, a few words about the difficulty of preparing precursors for the synthesis of ternary polyhydrides. For metals with similar properties (Sc-Y, La-Y, La-Ce, etc.), the most convenient precursors are the corresponding alloys. However, the hydrides synthesized from them usually belong to the class of solid solutions. The preparation of alloys becomes extremely difficult in the case of large differences in the melting temperatures and volatility for the two metals to be alloyed. Thus, the preparation of promising alloys of La and Y with magnesium, calcium, and strontium is hampered by the extremely high volatility of these metals at high temperatures, and their high chemical reactivity. For example, for the synthesis of ternary Mg-Y and Ca-Y polyhydrides, it is much more convenient to use mechanochemical process of grinding $MgH_2$, $CaH_2$, and $YH_3$ in a hydrogen atmosphere, followed by loading the resulting product in a high-pressure diamond anvil cell.

## 5.2 Synthesis of lanthanum- yttrium polyhydrides

The theoretical analysis of the La-Y-H system was performed using the USPEX code [9-12] at a pressure of 200 GPa and temperatures from 0 to 2000 K (Figure 46). Despite the fact that we considered only ordered ternary hydrides, thermodynamic calculations give a good reference point for the region of stability of $(La,Y)H_x$ solid solutions.

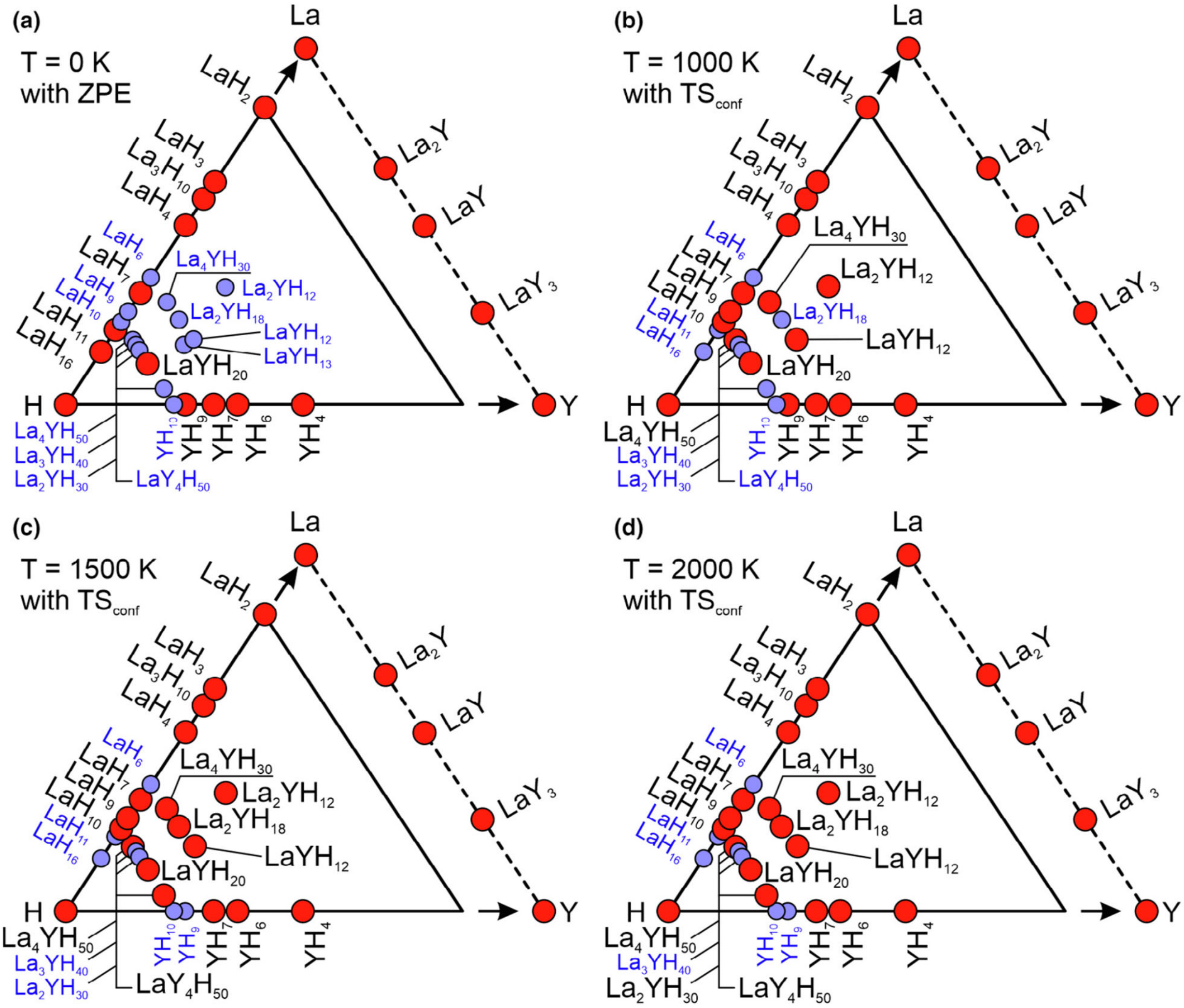

**Figure 46.** Ternary convex hulls of the La–Y–H system at a pressure of 200 GPa and temperatures of (a) 0 K, (b) 1000 K, (c) 1500 K, and (d) 2000 K, calculated with the contribution of the configurational entropy $TS_{conf}$ and the zero-point energy (ZPE) under the assumption of the crystalline state of all phases. The invisible Z axis, perpendicular to the plane of the figure, corresponds to the enthalpy of formation. Stable and metastable phases are shown in red and blue, respectively.

Calculations show that at a low temperature (0 K), there is only one stable ternary polyhydride: the high-$T_C$ superconductor $R$-3$m$-$LaYH_{20}$ [72]. At higher temperatures (1000 K), the set of stable ternary polyhydrides expands significantly: $LaYH_{12}$ and $La_4YH_{30}$ hexahydrides, $La_2YH_{12}$ tetrahydride, new $La_4YH_{50}$ and $Y_4LaH_{50}$ decahydrides form. The appearance of more and more of ternary hydrides indicates that laser heating promotes the formation of ternary compounds with ordered and disordered (solid solutions) structures. It should be noted that all calculations were made under the assumption of the crystalline state of all phases. However, melting of the hydrogen sublattice at 200 GPa is a very likely scenario at temperatures around 1000-1500 K [275, 276]. In ternary and quaternary hydrides with a large number of defects and impurities, the melting of the hydrogen sublattice can occur even earlier. In such a case, the thermodynamic calculations performed above become incorrect because the hydrogen atoms no longer have definite positions and the enthalpy cannot be calculated in the static approximation.

Here we would like to make a small digression and comment on the recent work of A.D. Grockowiak et al. [221], where the authors claim to have found superconductivity at 550 K in doped $LaH_{10}$ at pressure of 160-180 GPa. Molecular dynamics simulations with machine-learning potentials for La and H atoms at 150 and 200 GPa (NVT, 20-40 ps) show that complete melting of the lanthanum sublattice requires very high temperatures above 2000-2500 K (Figure 47), whereas the hydrogen sublattice melts completely already at 1500 K. However, even earlier, around 1000 K, the hydrogen atoms begin to migrate to neighboring positions and their diffusion capacity increases significantly. Impurities (B, N, C) in the $LaH_{10}$ structure and disorder will lower the melting temperature even further. At 150 GPa, the melting of the hydrogen sublattice occurs already at 800 K. It is known that molecular modeling overestimates the melting temperature because of the limitation of modeling time by tens and hundreds of picoseconds [277].

Melting or amorphization (formation of hydrogen glass) of the hydrogen sublattice will lead to a sharp increase in the electrical resistance of the samples, which will look similar to a superconducting transition [278]. Thus, we believe that the abovementioned study of the $LaH_{10}$ sample [221] represents the first detection of the hydrogen sublattice melting in superhydrides, which should be considered when doing thermodynamic calculations.

We also carried out calculations of the stable phases of the La-Y-H system with fixed compositions (fixcomp mode, Figure 48). We decided to search for stable phases of the $LaYH_{16}$ and $LaYH_{20}$ compositions, which are the formal combinations of well-known high-$T_C$ binary compounds $LaH_{10} + YH_6$ and $LaH_{10} + YH_{10}$, respectively. Even such a simple approach yielded two thermodynamically stable compounds with the ratio La:Y = 1:1, namely *P*4/*mmm*-$LaYH_{16}$ and *Amm*2-$LaYH_{20}$. $LaYH_{16}$ structure has a tetragonal unit cell, where each La and Y atom is surrounded by a cage of 32 hydrogen atoms (Figure 48B). $LaYH_{20}$ has an orthorhombic unit cell consisting of one La and one Y atoms with nine symmetrically inequivalent hydrogen atoms (Figure 48C). The calculations of the electron-phonon coupling in *P*4/*mmm*-$LaYH_{16}$ at 200 GPa indicate that this ternary hydride is a potential high-$T_C$ superconductor (Figure 48A) with $\lambda = 2.31$ and $T_C = 208$ K ($\mu^* = 0.1$).

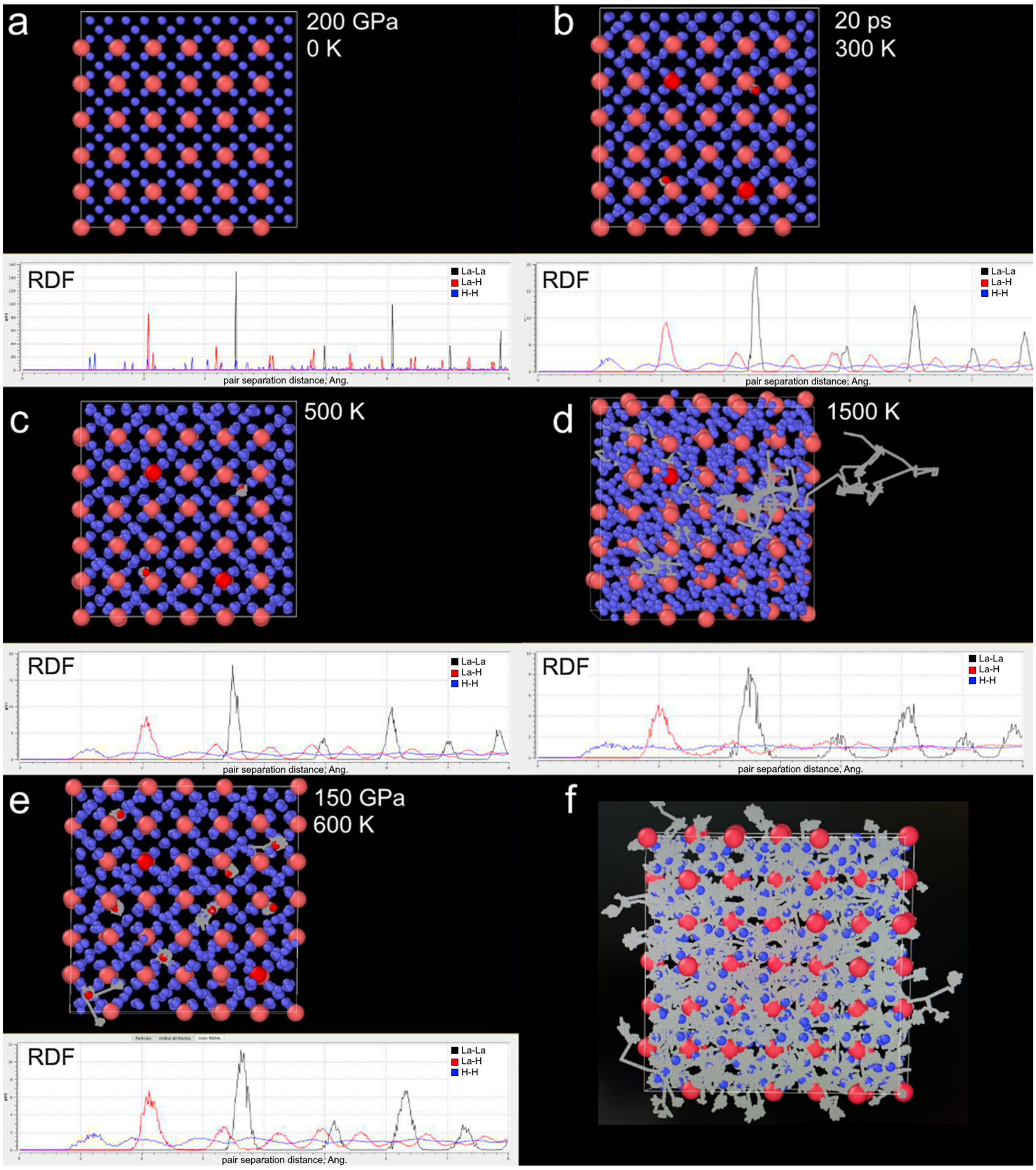

**Figure 47.** Molecular modeling of hydrogen behavior in the structure of $LaH_{10}$ at (a-d) 200 and (e-f) 150 GPa and temperatures (a-d) from 300 to 1500 K and (e-f) 600 K. Simulations were performed using machine-learning potentials (MTP) of the interatomic interaction for La and H for 20 ps in the group of Dr. I. Kruglov (VNIIA, Moscow). The figure shows the averaged structures, hydrogen trajectories, and radial distribution functions (RDF) for the La-La (black), La-H (red), and H-H (blue) bonds. The horizontal axis is the pair separation distance in Angstroms.

Another stable compound was predicted to have the $LaY_2H_{22}$ composition. Investigation of the ratio La:Y = 1:2 revealed the possibility of existence of tetragonal *I4/mmm*-$LaY_2H_{22}$ at 200 GPa. This structure has each La atom surrounded by an $H_{32}$ cage, and each Y atom — by an $H_{24}$ cage (Figure 48E). The compound has very high parameters of electron-phonon coupling (Figure 48D), with $\lambda = 2.6$ and $T_C = 243$ K ($\mu^* = 0.1$).

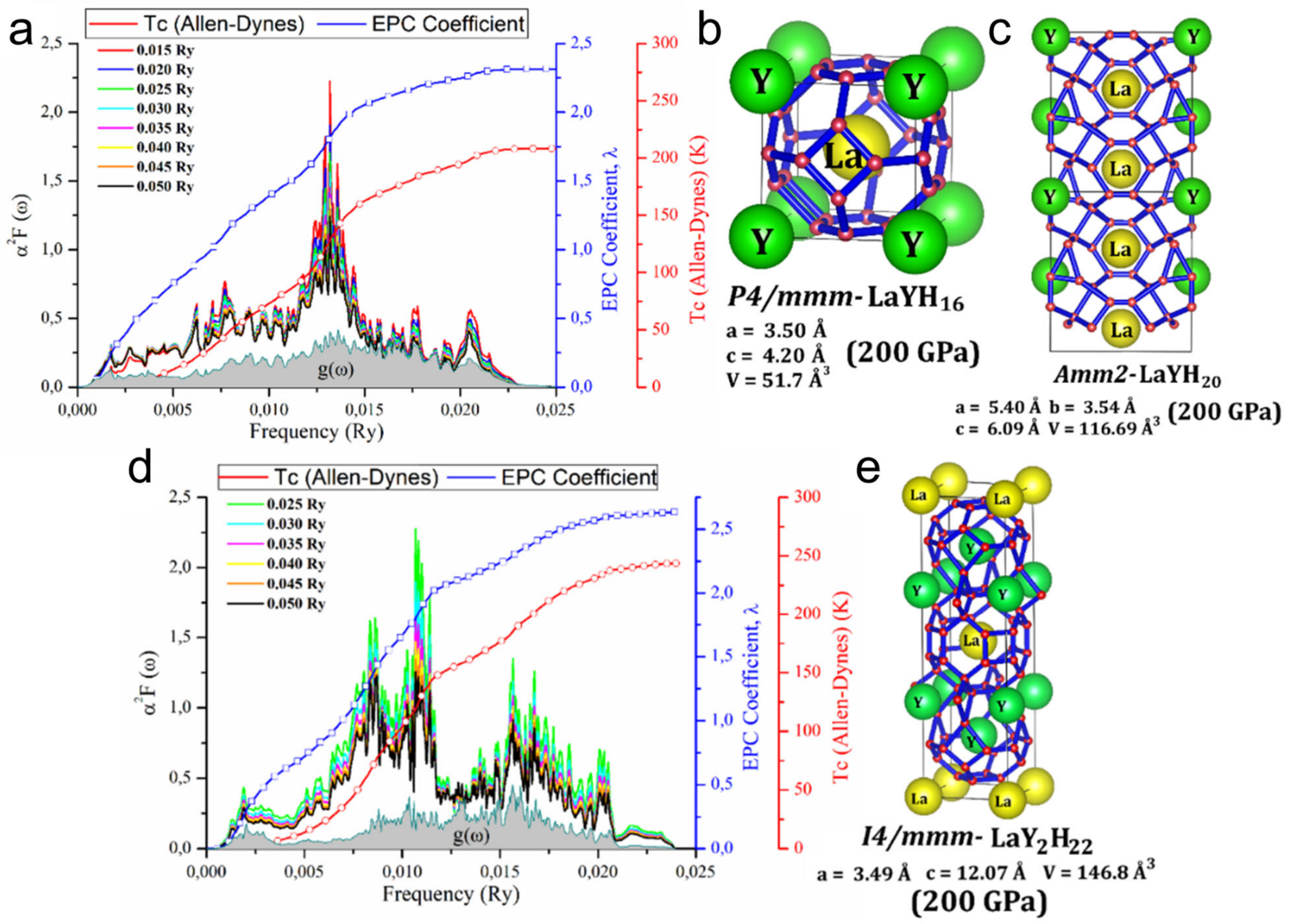

**Figure 48.** a) Eliashberg function $\alpha^2F$ of *P4/mmm*-$LaYH_{16}$ for different smearing values (σ = 0.015-0.05 Ry) calculated using Quantum ESPRESSO at 200 GPa. The phonon density of states is shown in gray; b) crystal structure and unit cell parameters of *P4/mmm*-$LaYH_{16}$ at 200 GPa; c) crystal structure of 1×1×2 supercell of *Amm*2-$LaYH_{20}$ superhydride at 200 GPa; d) Eliashberg function $\alpha^2F$ of *I*4/*mmm*-$LaY_2H_{22}$ for different smearing values (σ = 0.025-0.05 Ry); e) crystal structure and unit cell parameters of *I*4/*mmm*-$LaY_2H_{22}$ at 200 GPa. The atoms of lanthanum, yttrium and hydrogen are shown in yellow, green, and red, respectively.

Experimental synthesis of lanthanum-yttrium polyhydrides was performed using La-Y alloys prepared by alloying in a crucible and long annealing of La and Y at 1000 ºC, as well as by arc melting. In these experiments, ammonia borane was used as a hydrogen source. It has been shown that La-Y alloys at ambient pressure have hexagonal *dhcp* and *hcp* structures whose unit cell volume increases along with the lanthanum content [72]. Despite the hexagonal structure of the initial alloys, the main product of the reaction with hydrogen under pressure are cubic hydrides $(La,Y)H_{10}$ and $(La,Y)H_6$ with various concentrations of yttrium from 20 to 75 at%. (Figures 49, 50). No superstructural diffraction peaks were detected. Therefore, the obtained hydrides should have a metal sublattice of the La-Y solid solution type, existing in a wide concentration range of both components (from 20 to 75 atom % of yttrium).

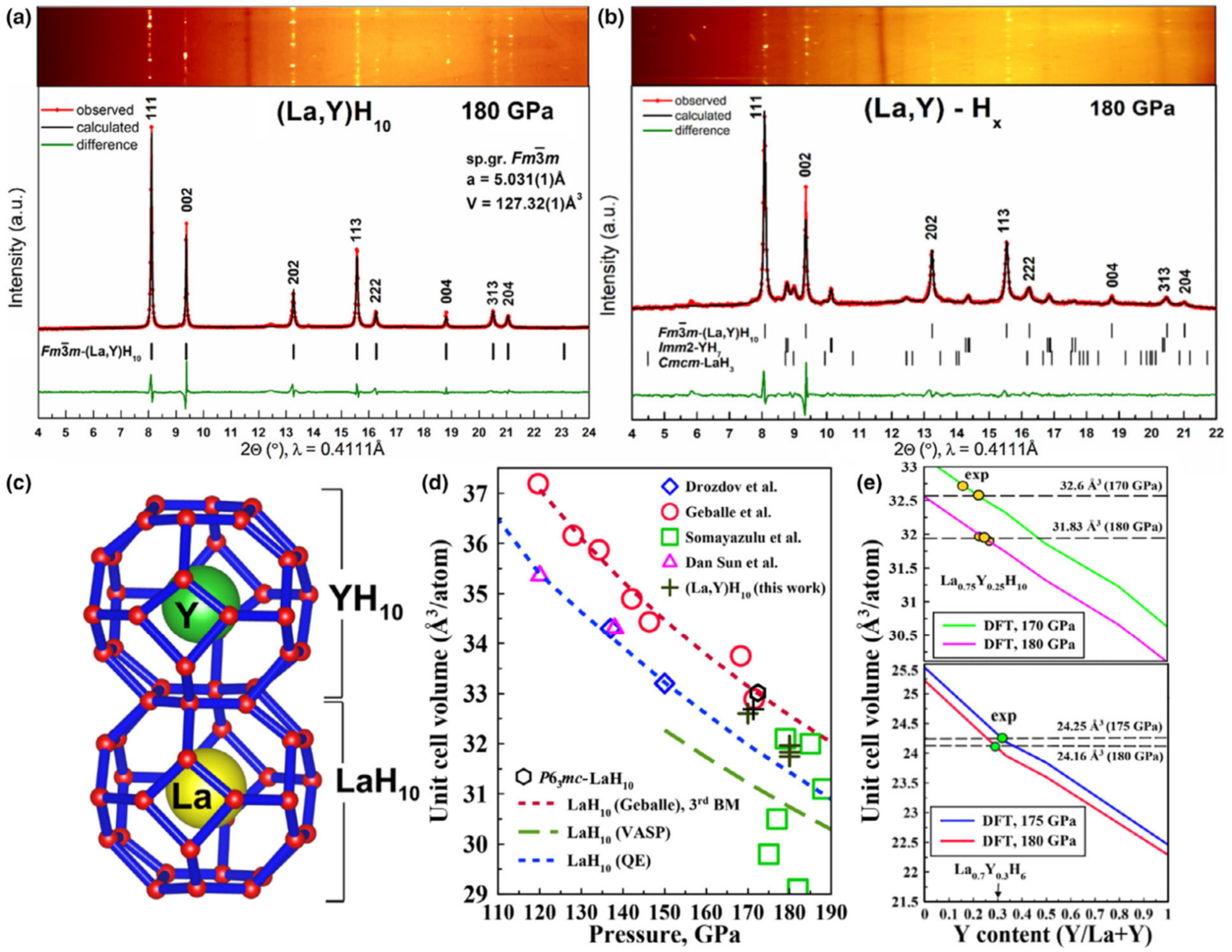

**Figure 49.** X-ray diffraction of La-Y hydrides. Experimental diffraction patterns and Le Bail refinements of the crystal unit cell parameters of (a) *Fm*3̄*m*-(La,Y)$H_{10}$ (DAC SL1) and (b) *Imm*2-$YH_7$ and *Cmcm*-$LaH_3$ (DAC M1). The experimental data, fit, and residues are shown in red, black, and green, respectively. (c) Fragment of crystal structure of (La,Y)$H_{10}$ where Y and La are neighbors (for illustrative purposes). (d) Pressure–unit cell volume diagram for *fcc* $LaH_{10}$: circles, squares, rhombuses, triangles, and crosses show the experimental data, lines depict the theoretical calculations. (e) Estimates of the Y content in (La,Y)$H_{10}$ and (La,Y)$H_6$ obtained using the experimental unit cell volumes.

In a series of other experiments with the $L_{0.66}Y_{0.33}$ alloy, another cubic hexahydride (La,Y)$H_6$ (Figure 50), whose structure is similar to $YH_6$, was obtained [30]. In addition to (La,Y)$H_6$, a lower tetrahydride (La,Y)$H_4$ was found in another region of the sample in the same run, indicating a probable lack of hydrogen in this DAC. Because the Y atom occupies a smaller volume in the space than La, the cell volume of the (La,Y)$H_x$ hydrides is between $LaH_x$ and $YH_x$ for all synthesized phases (Figure 50e). The volume of the (La,Y)$H_6$ phase is very close to that of *Im*-3*m*-$LaH_6$, which cannot be obtained by direct synthesis from lanthanum and hydrogen, but a small addition of yttrium helps to stabilize this compound.

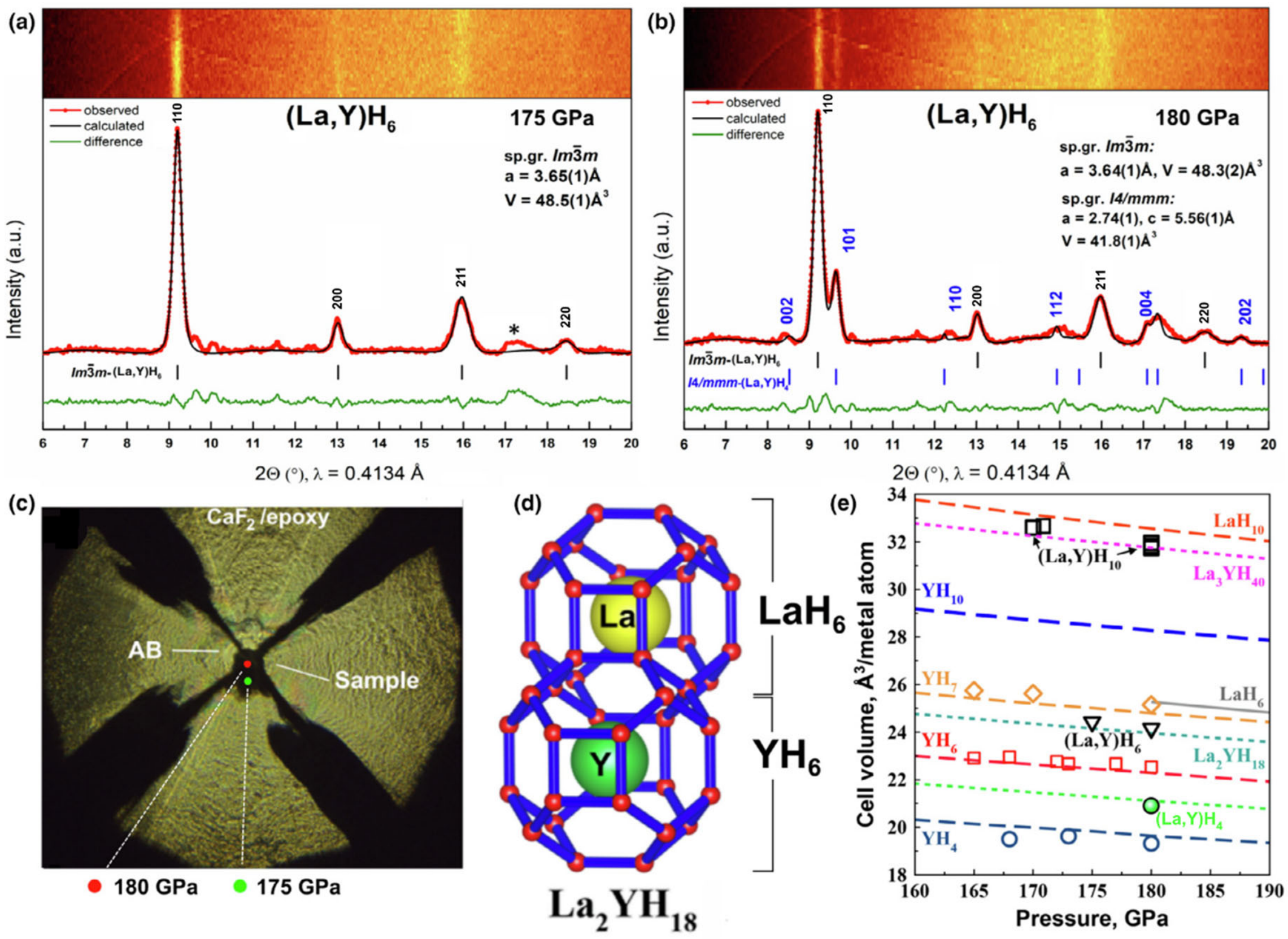

**Figure 50.** X-ray diffraction study of the La–Y hydrides in DAC M2_S. (a, b) Experimental diffraction patterns and the Le Bail refinements of the cell parameters of *Im*-3*m*-$(La,Y)H_6$ and tetragonal *I*4/*mmm*-$(La,Y)H_4$ at 175 and 180 GPa. The experimental data, fit, and residues are shown in red, black, and green, respectively. (c) Optical microscopy of the loaded DAC: sample, $NH_3BH_3$ medium, and four Ta/Au electrodes. (d) Fragment of the crystal structure of $(La,Y)H_6$ where Y and La are neighbors (for illustrative purposes). (e) Pressure–unit cell volume diagram of the studied La–Y–H phases. In the original paper [72], there is a mistake in the designation of the reflective planes of *Im*-3*m*-$(La,Y)H_6$ (panels a, b) pointed out by Prof. A. Abakumov (Skoltech). The mistake has been corrected in this thesis.

All synthesized $(La,Y)H_x$ phases show superconducting properties only slightly better or worse than $LaH_{10}$. One exception is $(La,Y)H_6$, whose critical temperature is 237 ± 5 K, noticeably higher than $T_C(YH_6)$. This is due to the fact that the lanthanum atoms in $(La,Y)H_6$ occupy an inherently unstable *Im*-3*m* structure in its pure form. A distinctive feature of superconductivity in ternary polyhydrides is a very broad superconducting transition $\Delta T_C$ = 15-20 K, which limits the use of these compounds in electronic devices because of the insufficient steepness of the current-voltage characteristic. This disadvantage is common to all ternary polyhydrides because of the increased concentration of defects and fluctuations in the concentration of the constituent elements. In other respects, La-Y hydrides exhibit practically the same properties as binary polyhydrides.

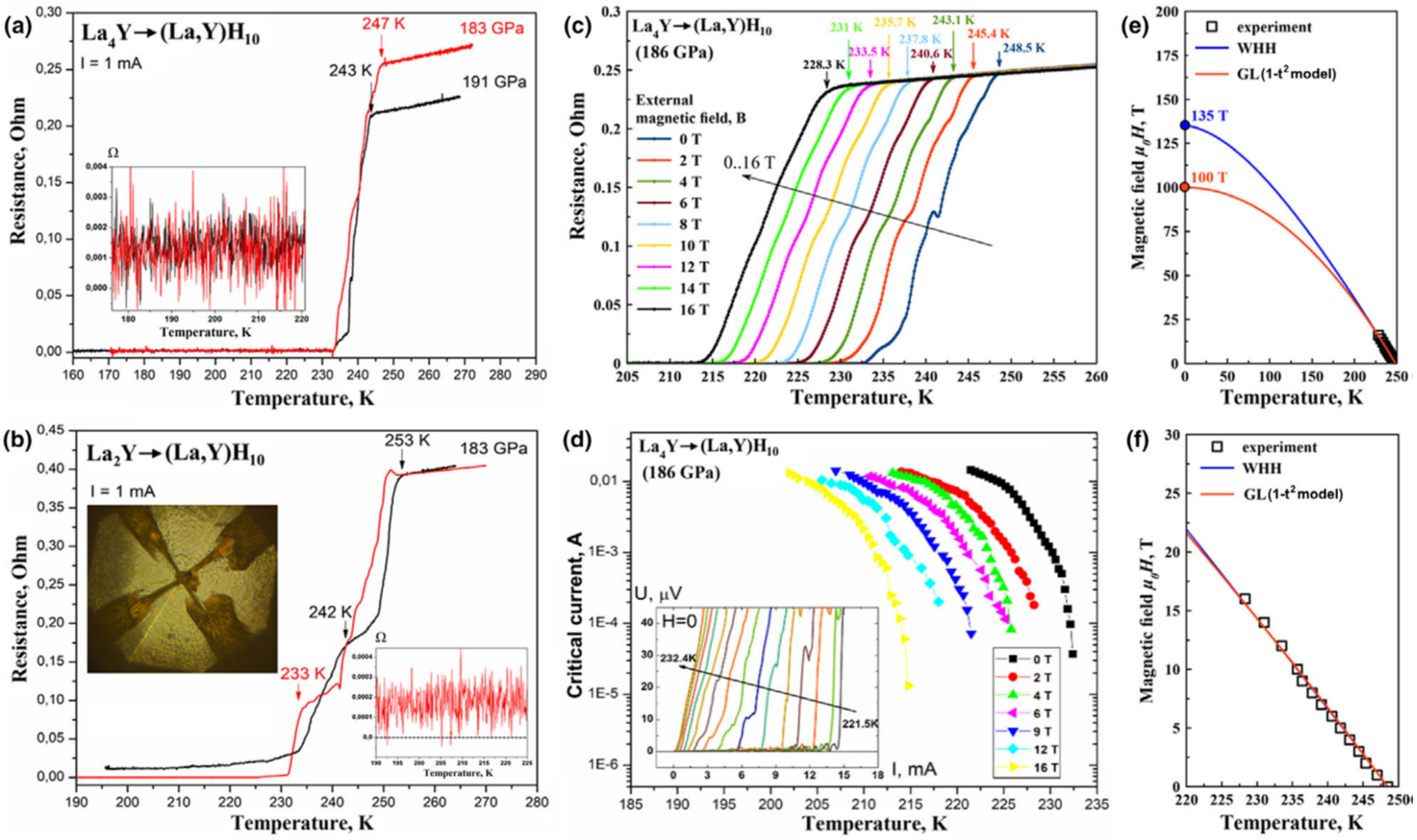

**Figure 51.** Superconducting transitions in $Fm\bar{3}m$-(La,Y)$H_{10}$: (a) Temperature dependence of the electrical resistance for the sample obtained from $La_4Y$. Inset: residual resistance after cooling below $T_C$. (b) Temperature dependence of the resistance for the sample obtained from $La_2Y$. Insets: residual resistance after cooling below $T_C$ and a photo of the DAC culet with electrodes. (c) Dependence of the electrical resistance of (La,Y)$H_{10}$ on the external magnetic field (0–16 T) at 186 GPa and 0.1 mA current. The critical temperatures were determined at the onset of the resistance drop. (d) Dependence of the critical current on the temperature and external magnetic field (0–16 T). The critical currents were measured near $T_C$. Inset: current–voltage characteristic near a superconducting transition. (e) Extrapolation of the upper critical magnetic field using the Werthamer–Helfand–Hohenberg theory [215] (WHH) and Ginzburg–Landau theory (GL) [216]. (f) Dependence of the critical temperature $T_C$ on the applied magnetic field.

We will analyze the superconducting properties of (La,Y)$H_{10}$ using the previously described algorithm with some simple additions. This algorithm will be used later to collect statistical data and to compare different hydrides with each other.

**1e.** Analysis of R(300)/R($T_C$)

**(La,Y)$H_{10}$:** R(300)/R($T_C$) ~ 1.05-1.08, typical for polyhydrides, the resistance of the sample is almost entirely due to structural defects and impurities.

**2e.** Analysis of R(T) in terms of the Bloch-Gruneisen formula [64, 65].

**(La,Y)$H_{10}$:** $\theta_D$ = 1022 ± 5 K for $La_{0.8}Y_{0.2}H_{10}$, 182-185 GPa.

**3e.** Estimation of the sample thickness and resistivity with calculations in EPW [140-143].

**(La,Y)$H_{10}$:** calculations in EPW were not carried out, however, the thickness of the sample, h ~ 1.2-2.4 µm, was investigated using optical methods (Figure 52, interference of light between diamond anvils). Given this sample thickness, we can make an estimate for the resistivity before and after the superconducting transition: $\rho_n$ = 48-96 μΩ·cm and $\rho_{SC}$ = 7.2-21.6 ×$10^{-4}$ μΩ·cm (averaged). In other words, the resistivity at the superconducting transition changes by a factor of about 70,000.

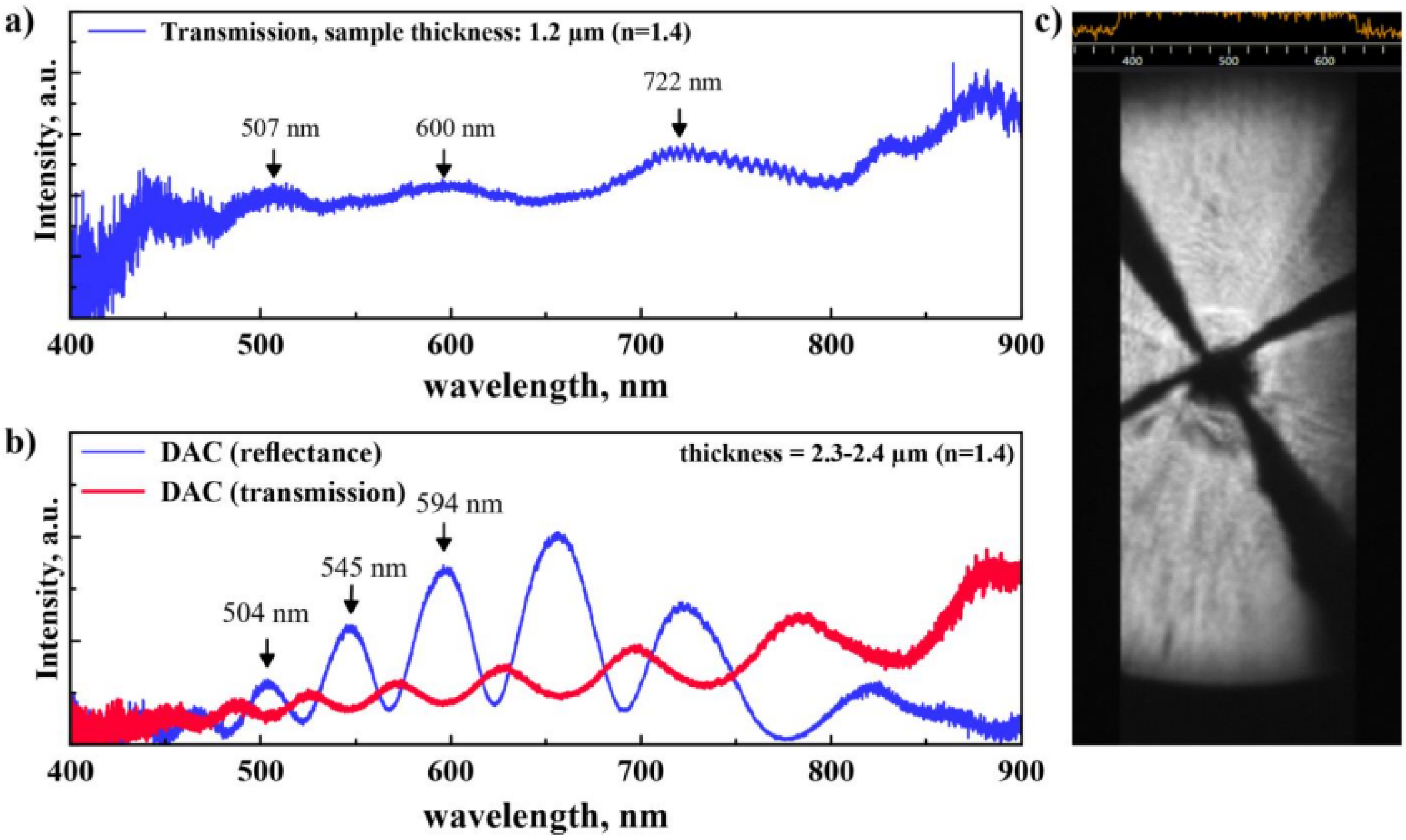

**Figure 52.** Light interference patterns for (La,Y) samples loaded in an electrical DAC: after (a) and before (b) compression. Taking into account the approximate refractive index of ammonia borane (n ≈ 1.4) the measurements of an anvil-anvil distance show that the thickness of the (La,Y) samples does not exceed 1– 1.5 μm. Similar values were obtained using the scanning electron microscopy of (La,Y) samples before loading DACs.

**4e.** Estimation of the electron-phonon interaction constant (λ) by using the well-known Allen-Dynes formula (A-D) [60].

**(La,Y)H10:** $\lambda$ = 3.63 ($\mu^*$=0.1), $\omega_{log}$ = 845 K, close enough to the calculated values. But the temperature range from 280 to 250 K for extraction of the Debye temperature is clearly insufficient. In the future, in high-temperature superconducting hydrides we need to perform the resistance measurements at up to 350-400 K.

**5e.** Analysis of the residual resistance, hysteresis loop in cooling and heating cycles, and superconducting transition widths.

**(La,Y)H10:** superconducting transitions are broadened to $\Delta T_C$ = 15-20 K, which is typical for ternary polyhydrides [202, 203, 212]. The average residual resistivity depends on the sample and is about 6-9 μΩ with an initial resistivity of about 0.2-0.4 Ω ($La_{0.8}Y_{0.2}H_{10}$).

**6e.** Broadening of superconducting transitions in magnetic fields.

**(La,Y)H10:** there is no broadening in the fields up to 16 T (Figure 51c). It is necessary to perform the pulse measurements in strong magnetic fields. Estimation of $\mu_0 H_{C2}(0)$ using the linear interpolation shows that superconductivity in (La,Y)$H_{10}$ will persist up to 180 T.

**7e.** Investigation of R(T) in a wide temperature range to search for low-temperature anomalies, possibly associated with the manifestation of magnetic ordering, and high-temperature anomalies associated with the melting of the hydrogen sublattice.

**(La,Y)H10:** future plans.

**8e**. Investigation of the resistance and cooling steps to identify anomalies in the measuring system.

**(La,Y)H$_{10}$:** there are sharp spikes in the resistance (ΔR) and temperature (ΔT) steps (Figure 53), but they are not associated with the region of the superconducting transition. No other anomalies were found in the electrical circuit or measuring system.

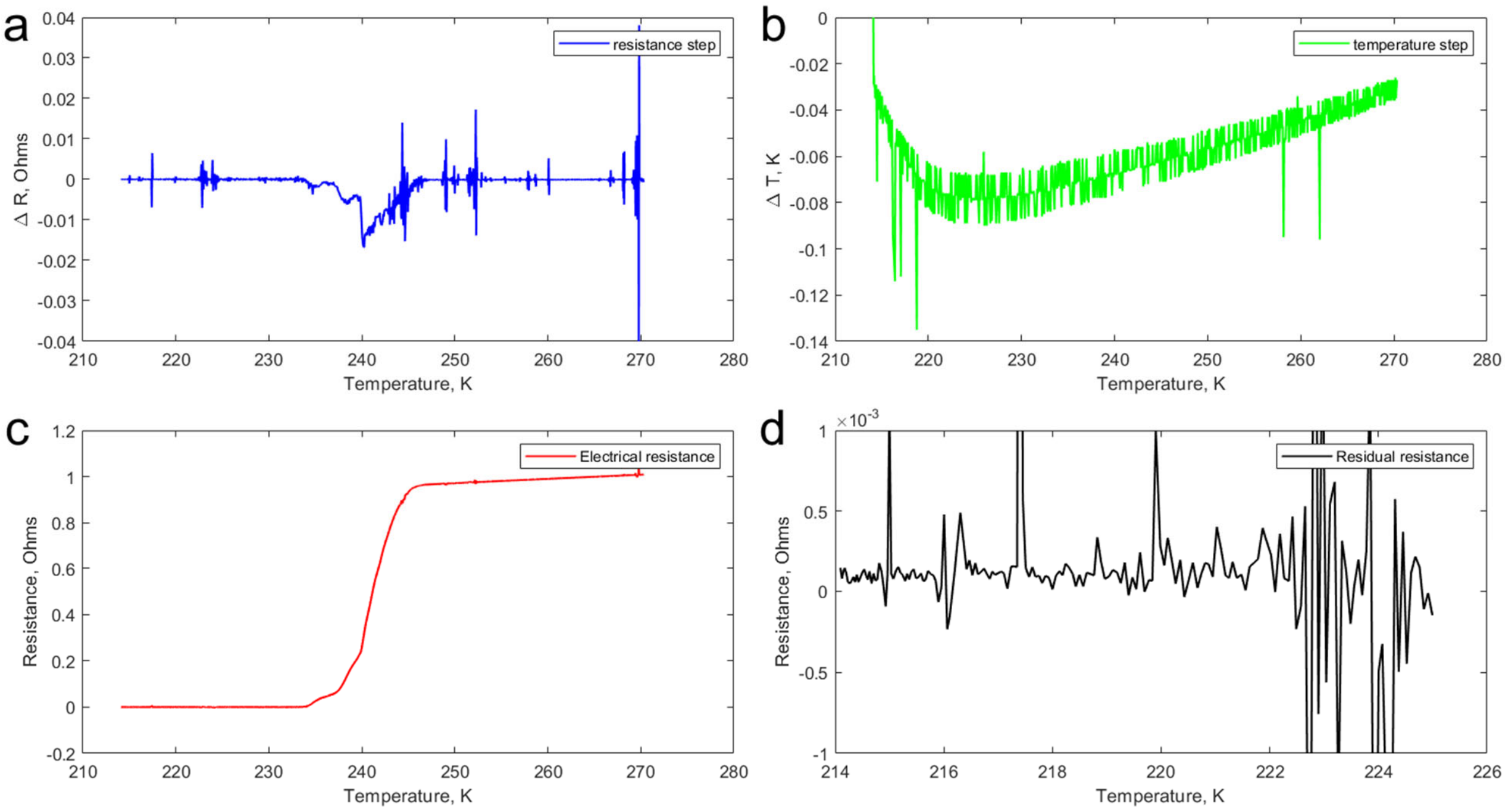

**Figure 53.** Searching for anomalies in the (La,Y)H$_{10}$ sample. (a) Resistance step vs. temperature; (b) temperature step in the cooling cycle vs. temperature; (c) temperature dependence of the sample resistance, and (d) residual resistance in the superconducting state for La$_{0.8}$Y$_{0.2}$H$_{10}$ at 182-185 GPa.

**1t**. Analysis of anisotropy $T_C^{aniso} - T_C^{iso}$ via solving the anisotropic Eliashberg equations in the EPW program.

**(La,Y)H$_{10}$:** future plans, the contribution of anisotropy is assumed to be similar to LaH$_{10}$ [66].

**2t**. Analysis of the anharmonic contribution $\lambda_{anh}$, $\lambda_{harm}$, $T_C^{harm} - T_C^{anharm}$ performed using SSCHA or molecular dynamics.

**(La,Y)H$_{10}$:** the contribution of anharmonic effects was analyzed using the molecular dynamics. Machine-learning potentials of interatomic interactions were created by dynamic modeling of the LaYH$_{20}$ structure at 180 GPa and 300 K. In the original paper, vibrations of the metal sublattice were also included in the calculation of the Eliashberg function, which leads to a very large increase in the electron-phonon interaction coefficient. I believe that it is more correct to exclude the low-frequency region from consideration in the case of dynamically unstable compounds (see Table 9).

**Table 9.** Comparison of results of harmonic and anharmonic approaches to the superconducting properties of *R*-3*m*-$LaYH_{20}$ at 180 GPa ($\mu^* = 0.1$). The low-frequency region was excluded from the calculations.

| Parameter | $\lambda$ | $\omega_{log}$, K | $\omega_2$, K | $T_C$(A-D), K | $T_C$(E), K |
|---|---|---|---|---|---|
| harmonic | 3.87 | 868 | 1208 | 266 | 300 |
| anharmonic | 3.76 | 804 | 1211 | 247 | 292 |

Table 9 shows that the anharmonic effects reduce the $T_C$(E) by 8 K and $T_C$(A-D) by 19 K. It is expected that the anharmonic corrections in the case of $LaYH_{20}$ will be of the same order of magnitude as for $LaH_{10}$ [119].

**3t.** SCDFT analysis, which allows us to isolate the contribution of the empirical Coulomb pseudopotential $\mu^*$ and obtain the critical temperature $T_C$(SCDFT).

**(La,Y)H$_{10}$:** detailed calculations were performed using the old LM2005 SCDFT functional for the most interesting ordered polyhydrides: the new phase *Im*-3*m*-$LaH_6$, *Pm*-3*m*-$LaYH_{12}$ and *R*-3*m*-$LaYH_{20}$. As a result, it was found that LM2005 gives a severely underestimated $T_C$(SCDFT) ~ 176 – 191 K for hexahydrides $LaH_6$, $LaYH_{12}$ and $YH_6$ [30]. Nevertheless, the introduction of La into hexahydrides increases $T_C$ by 16-32 K compared to $YH_6$. For La-Y decahydride the LM2005 gives (see also [119]) better result $T_C$(SCDFT) = 252 K, very close to experiment. For this reason, the Coulomb pseudopotential for this compound can be taken from the generally accepted range $\mu^*$ = 0.1-0.15.

**4t**. Analysis within the Migdal-Eliashberg theory and the Allen-Dynes formula in the isotropic harmonic approximation, allowing one to obtain the Eliashberg function, electron-phonon interaction coefficients, logarithmically averaged phonon frequency, etc.

**(La,Y)H$_{10}$:** ab initio calculations of the electron-phonon interaction parameter for La-Y decahydrides and hexahydrides give $\lambda(LaYH_{20})$ up to 3.87, whereas $\lambda$ = 2.68-2.82 for $(La,Y)H_6$. The logarithmically averaged phonon frequencies are in the range of 850-1050 K. The solution of the isotropic Eliashberg equations for decahydrides gives a somewhat overestimated result (as for $LaH_{10}$): $T_C$(E) = 281-300 K ($\mu^*$ = 0.1-0.15), whereas the Allen-Dynes formula fits the experimental data better $T_C$(A-D) = 232-266 K. A similar situation is observed for hexahydrides La-Y: $T_C$(E) = 223-241 K is close to the experimental data.

**5t**. EPW calculations of electrical resistivity in normal state.

**(La,Y)H$_{10}$:** future plans

This pioneering work on the synthesis of La-Y superhydrides outlined the scheme for their future research. We showed the need for a detailed study of the structure and composition of the initial La-Y alloys for the subsequent preparation of polyhydrides. It is very important that the alloy sample is homogeneous, because only a microscopic amount of the prepared substance is loaded into a diamond anvil cell. Before loading the metal particle, we also performed its preliminary compression on diamond anvils to obtain the desired thickness (~1 μm).

## 5.3 Is it possible to synthesize ordered ternary hydrides?

This natural question arises at the moment for all researchers dealing with hydride superconductivity, and the answer should be positive. Such a situation is observed if the properties of two metal atoms (electronegativity and atomic radius) significantly differ from each other. Recently we found an example of this, investigating the formation of calcium-yttrium polyhydrides using the lowest cubic hydride $(Ca,Y)_2H_5$ as a starting material. $(Ca,Y)_2H_5$ or $(Ca,Y)H_{2.5}$ was synthesized using the mechanochemical process of grinding a 1:1 mixture of $CaH_2$ and $YH_3$ powders at normal pressure in a hydrogen atmosphere in a ball mill. A powder X-ray diffraction pattern of the synthesis product is shown in Figure 54. In this compound, the metal sublattice appears to be a solid solution of Ca and Y.

The prepared precursor $(Ca,Y)_2H_5$ was loaded into a high-pressure diamond cell along with ammonium borane, compressed to 177 GPa, and heated by IR laser. The product is a mixture of several coarse-grained phases (Figure 55). All phases were found to be stable in the investigated pressure range of 119-177 GPa. The analysis of the diffraction patterns shows that the main product of the reaction of cubic $(Ca,Y)_2H_5$ with hydrogen is a new polyhydride $R\bar{3}m$-$(Ca,Y)H_5$ ($V$ = 19.16 Å$^3$/metal atom at 177 GPa), whereas the impurity phases are $Im\bar{3}m$-$CaH_6$ (additional reflection intensity at $(2\bar{1}0)$ in Figure 56a, $V$ = 20.02 Å$^3$/Ca at 177 GPa) and $Cc$-$YH_7$ (diffraction peaks at 8-10°, Figure 56a, Table 10), which we have previously discovered in the study of the Y-H system [30]. The experimental parameters of the unit cell are given in Tables 10-11. The cell volume of the impurity, cubic $CaH_6$, is in excellent agreement with the previously published data [115] (Figure 57).

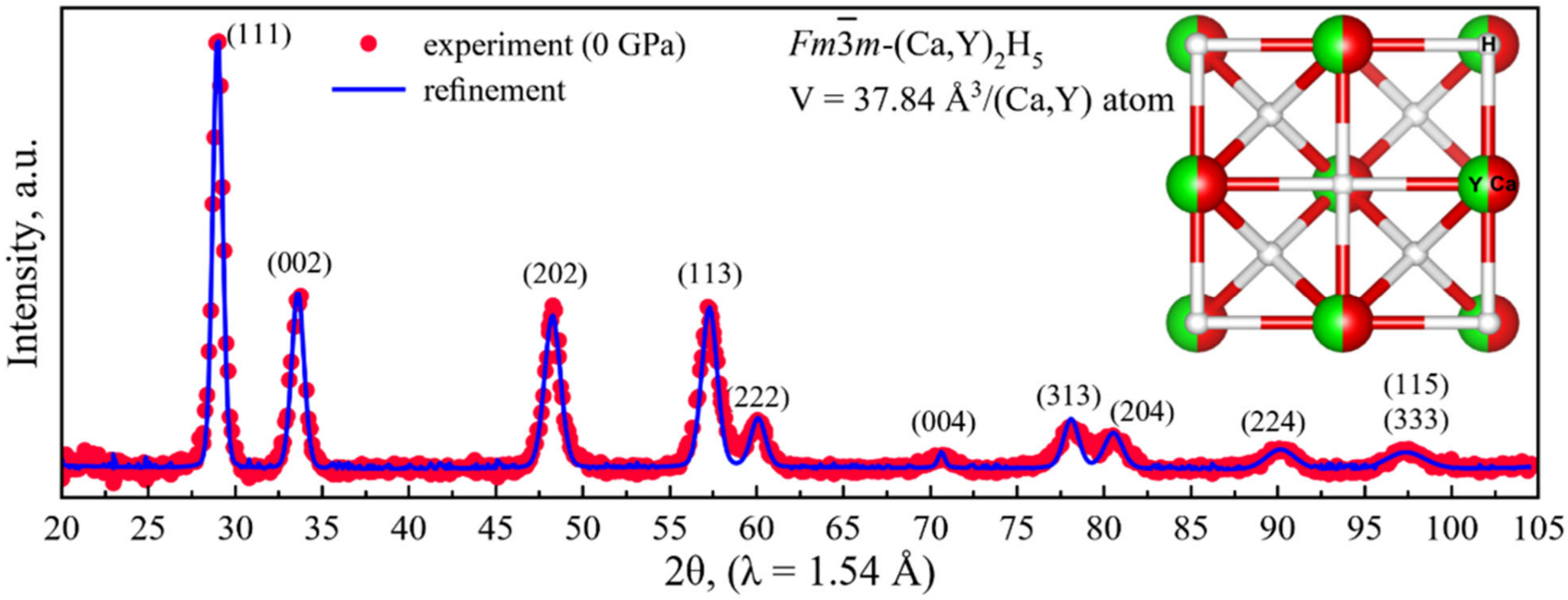

**Figure 54**. Experimental X-ray diffraction patterns of cubic $(Ca,Y)_2H_5$ precursor at 0 GPa. The experimental data and fitted line are shown in red and blue, respectively. Inset: proposed candidate structure $Fm\bar{3}m$-$(Ca,Y)_2H_5$.

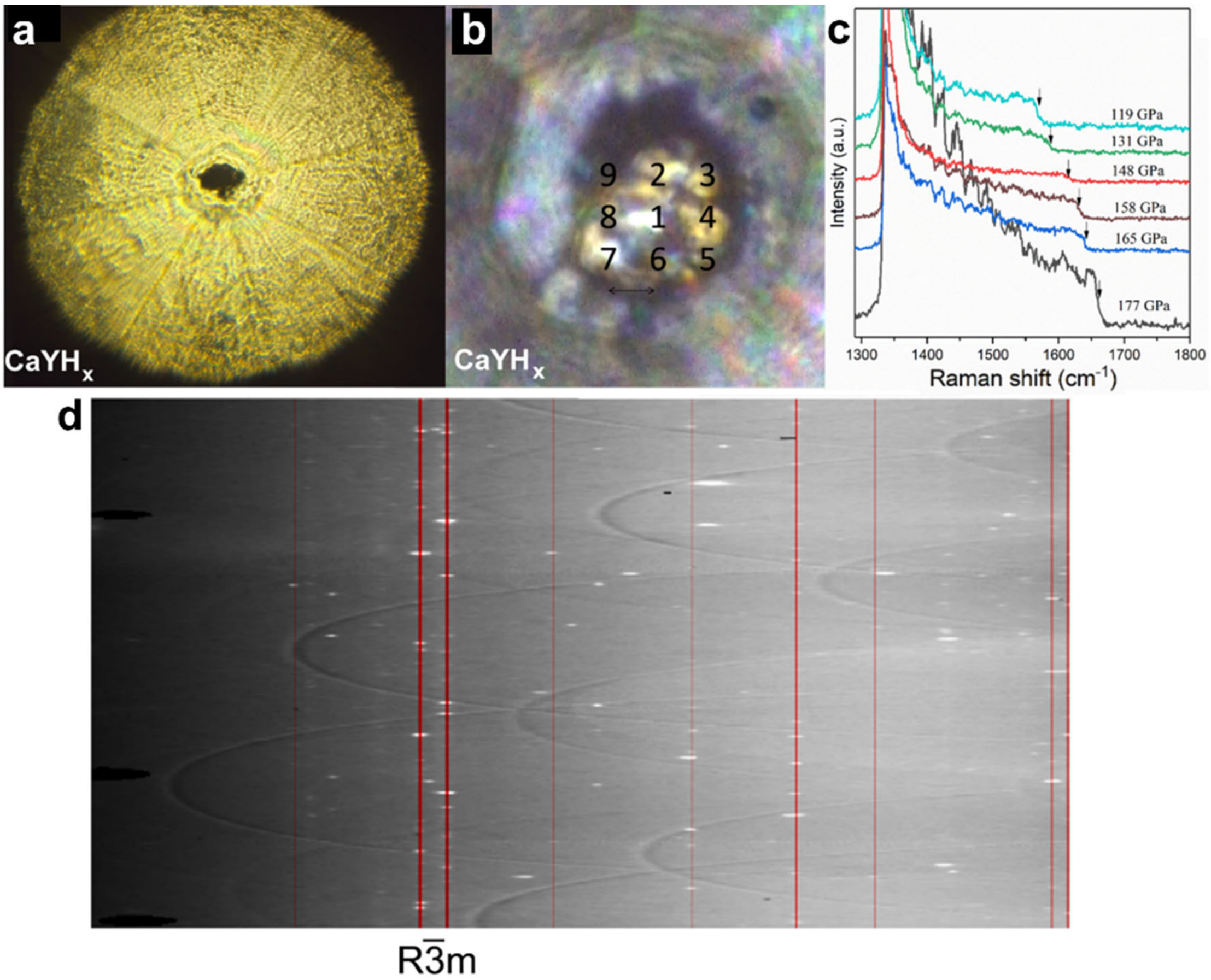

**Figure 55.** Optical photography of diamond anvils of (a) DAC MgY at 175 GPa and (b) DAC CaY at 177 GPa. Numbers indicate the area of the X-ray diffraction analysis of the $CaYH_x$ sample. (c) Raman spectra of the DAC CaY recorded during decompression. Arrows indicate the frequency points that were used to calculate the pressure. (d) Typical diffraction pattern ("cake") of a $CaYH_x$ sample at 177 GPa. Red lines correspond to the XRD pattern of $CaYH_{10}$.

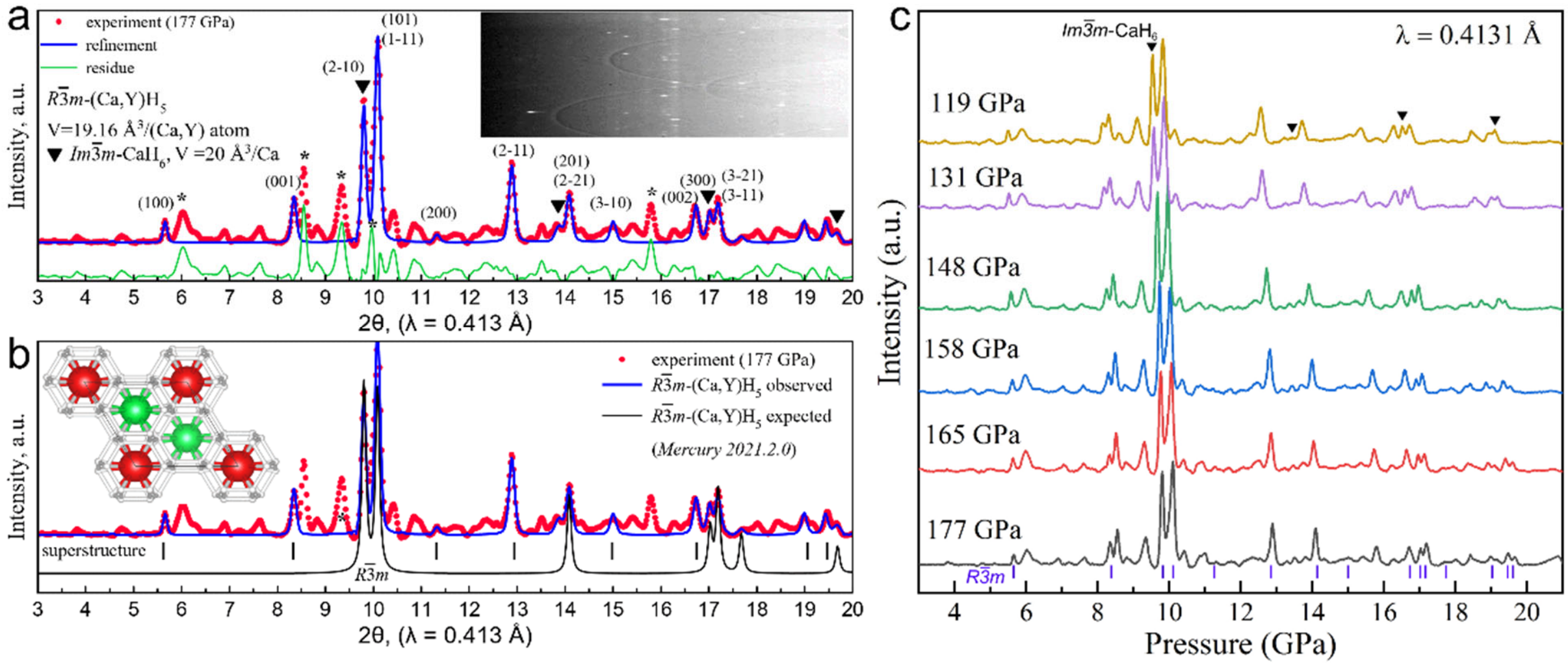

**Figure 56.** XRD patterns of sample $CaYH_x$ obtained from DAC CaY at 177 GPa (a, b) and in 119-177 GPa pressure range (c) during decompression of the cell. The experimental data and fitted line are shown in red and blue, respectively. In panel (a), the refinement was carried out only for the phases $CaH_6$ and $(Ca,Y)H_5$. In panel (b) an example of the structure of $CaYH_{10}$ is shown, which could explain the appearance of superstructure reflections in addition to the expected reflections of the $R\bar{3}m$ phase.

**Table 10.** Experimental cell parameters of $Im\bar{3}m$-$CaH_6$ (Z=2) and $R\bar{3}m$-(Ca,Y)$H_5$ (Z=3).

| Pressure, GPa | a, Å | V, Å³ | a, Å | c, Å | V, Å³ |
|---|---|---|---|---|---|
| 177 | 3.421 | 40.04 | 4.841 | 2.832 | 57.48 |
| 165 | 3.433 | 40.46 | 4.857 | 2.841 | 58.05 |
| 158 | 3.443 | 40.81 | 4.873 | 2.851 | 58.63 |
| 148 | 3.467 | 41.67 | 4.905 | 2.869 | 59.78 |
| 131 | 3.505* | 43.06 | 4.957 | 2.900 | 61.69 |
| 119 | 3.515 | 43.80 | 4.972 | 2.909 | 62.27 |
| | (c = 3.545)* | | | | |

* Tetragonal distortion

The main reaction product needs to be discussed. The hydrogen content in it is below the level of hexahydrides ($CaH_6$ and $YH_6$), and in the first approximation can be described as ~5 H atoms per metal atom. Usually metal atoms in ternary hydrides are randomly mixed in the metal sublattice of such compounds (e.g., La-Y-H [[72], La-Nd-H[202], Ca-Nd-Zr-H [279]). A black line marked $R\bar{3}m$ in Figure 56b shows how the $R\bar{3}m$-(Ca,Y)$H_5$ phase should look if it was a case of a solid solution of Ca and Y atoms. Only several broad reflections (such as $2\bar{1}0$, 101, $1\bar{1}1$, etc.) would have to appear, as was observed for the initial hydride (Ca,Y)$_2H_5$. However, we see well-defined, narrow, superstructural reflections (100), (001), (200), ($2\bar{1}1$), ($3\bar{1}0$), and (002), which correspond to single crystal point reflections (Figures 55d, 56a).

All these facts indicate that the main phase can be a true ternary polyhydride $CaYH_{10}$ (instead of (Ca,Y)$H_5$, Ca:Y ratio is expected to be close to 1:1) with certain positions of Ca and Y atoms and their individual sublattices, reflections from which we observe in the XRD patterns. To the best of our knowledge, this compound is the first example of such true ternary polyhydride synthesized at a high pressure.

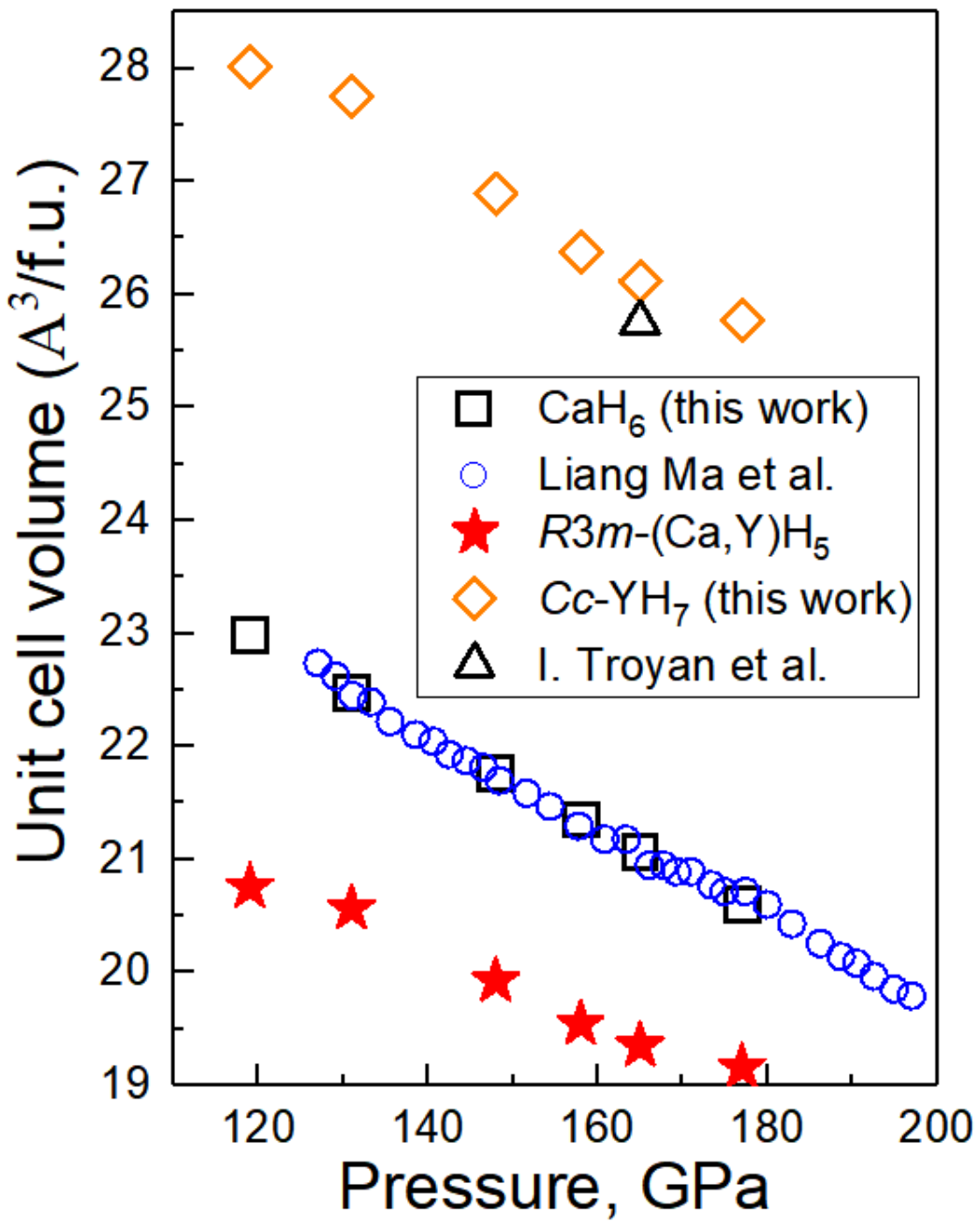

**Figure 57.** Pressure–unit cell volume diagram for Ca,Y polyhydrides: blue circles - $CaH_6$ [115], black squares – $CaH_6$ in this work, red stars - $R\bar{3}m$-(Ca,Y)$H_5$,orange diamonds – *Cc*-$YH_7$.

**Table 11.** Experimental unit cell parameters of *Cc*-$YH_7$ (Z=4, for simplicity, β = 90°).

| **Pressure, GPa** | **a, Å** | **b, Å³** | **c, Å** | **V, Å³** |
|---|---|---|---|---|
| 177 | 3.264 | 5.693 | 5.547 | 103.08 |
| 165 | 3.279 | 5.718 | 5.571 | 104.45 |
| 158 | 3.290 | 5.737 | 5.590 | 105.50 |
| 148 | 3.311 | 5.774 | 5.626 | 107.56 |
| 131 | 3.346 | 5.835 | 5.685 | 111.0 |
| 119 | 3.356 | 5.853 | 5.703 | 112.04 |

## 5.4 Conclusions from studies of ternary metal polyhydrides

1. Synthesis of ternary polyhydrides from alloys of metals with similar properties leads to formation of compounds with a solid solution type of metal sublattice. Anderson's theorem for dirty and non-stoichiometric superconductors, which states the impossibility of increasing the critical temperature of superconductivity in binary hydrides by doping them, is applicable to such compounds.

2. The introduction of magnetic atoms, such as Nd and Ce, significantly reduces the superconducting properties of the corresponding ternary hydrides because of scattering of Cooper pairs with spin flipping.

3. Doping can lead to stabilization of new crystal structures that are not stable themselves in the parent binary system. In other words, when element A is introduced into the B-H system, crystal structures specific only for the A-H system, but not for the B-H system, appear among the reaction products (A,B)$H_x$. Examples: stabilization of *Im*-3*m*-(La,Y)$H_6$, although *Im*-3*m*-$LaH_6$ does not exist [72]; appearance of *P*6$_3$/*mmc*-(Ce,La)$H_9$, whereas *P*6$_3$/*mmc*-$LaH_9$ is unstable (according to calculations, see also [81]) [203].

4. Last, atoms having different properties can form true ternary hydrides with ordered sublattices of each type of atoms, which are characterized by the appearance of superstructural peaks on X-ray diffraction patterns. Example: calcium-yttrium polyhydrides.

5. At the moment, the effect of the initial precursor structure and synthesis temperature on resulting ternary polyhydrides is an open question.

# Chapter 6. Distribution of superconducting properties in binary and ternary metal polyhydrides

## 6.1 Distribution of superconducting properties in binary polyhydrides

This chapter is based on the data from Ref. [36], where we have investigated the distribution of superconductivity in binary metal-hydrogen systems under pressure.

The distribution of the superconducting properties in a certain class of materials is key information for a materials scientist. It is well known that the Bardeen-Cooper-Schrieffer (BCS) theory of conventional phonon-mediated superconductivity, developed in 1957, does not provide an answer to the question of which materials will be good superconductors and which will be bad ones. If we know the structure of the compound, then by doing some calculations within the framework of the BCS theory we can say approximately whether the material is promising or not. The BCS theory cannot tell us the kind and number of atoms that should be taken and how they should be arranged in space to provide good superconducting properties. This can be illustrated by several examples.

The first example is superconductivity with $T_C$ = 39 K in *P*6/*mmm*-$MgB_2$ [280], which remained unknown until 2001, and a recent discovery of hexagonal $MoB_2$, superconducting below 32 K at a pressure of 100 GPa [281]. None of the widely known theories and models were able to predict superconductivity in these borides before their experimental discovery, although the DFT calculations almost immediately after their discovery confirmed the electron-phonon mechanism of coupling in these compounds. Surprisingly, the theoretically predicted $T_C$ is in excellent agreement with experimental data for $MgB_2$ and $MoB_2$.

Second, the BCS theory does give some (ambiguous) indications of where to look for good superconductors. For example, light elements are preferred because of the higher frequency of vibrations in the crystal lattice and the phonon energy associated with them. J. Hirsch [282] has repeatedly used this predictive tool to criticize the BCS theory because at ambient pressure among simple metals and their alloys (intermetallics), the best superconducting properties belong to rather heavy elements (Nb, Pb, Pb-Bi, $Nb_3Sn$, $V_3Ge$, etc.). Of course, in the BCS theory, the critical temperature is determined by two factors: 1) the electronic band structure and the electron-phonon interaction matrices, and 2) the phonon averaged frequency. For simple metals, $T_C \sim (1/m_A)^{0.5}$, where $m_A$ is the mass of an atom. One way to change the electronic structure of matter is to compress it. Indeed, as Figure 45 shows, many light elements (Li, Ca, Ti, B, S, etc.) exhibit record superconductivity at high pressures in accordance with the ideas of the BCS theory. The superhydrides described in this thesis are a prime example of the application of the atomic mass reduction principle for the purpose of increasing the critical temperature of matter.

From the Matthias rules [283, 284], we also know that crystal symmetry plays an important role for superconductivity. For cubic and hexagonal crystals, the energy levels of phonons are degenerate in the k-space, so several phonon modes have the same energy at once. If this energy coincides with the energy of electronic transitions near the Fermi level, we would expect superconducting properties in the compound. If the symmetry of a crystal is low, then even if some phonons have a suitable energy, other phonons will have energy outside the resonant band due to anisotropy. This leads to a deterioration of the superconducting properties with a decrease in the crystal symmetry. This rule is actively used nowadays for the design of new superconductors using structural templates. For instance, the well-known cubic superconductor $Nb_3Sn$ ($T_C$ = 18 K) of type $AX_3$ (or A15) is the basis on which many other superconductors have been obtained by atom substitution, such as $V_3Si$ and $V_3Ge$. The following templates are popular for hydrides: *I*4/*mmm*-$XH_4$ (for low pressures [197]), *Pm*-3*n*-$X_4H_{23}$, *Im*-3*m*-$XH_6$, *P*6$_3$/*mmc*-$XH_9$, and $Fm\overline{3}m$-$XH_{10}$. There are numerous papers on this topic [19, 36]. In this approach, the superconducting properties of such clathrate polyhydrides are analyzed separately from their thermodynamic and dynamic stability, the establishment of which is a much more difficult task.

Indeed, this approach proved to be quite fruitful and many outstanding hydride superconductors were predicted before the experimental discovery: $H_3S$ [5], $YH_6$ [30, 31], $YH_9$ [31], $LaH_{10}$ [7, 8], $ThH_{10}$ [27], $YH_4$ [182], etc. In most cases, prediction and discovery are separated by 1-3 years, but for $CaH_6$, about 10 years passed between the prediction [114] and experimental [115] synthesis! Thus, the field of hydride superconductivity is one of the first branches of superconductivity physics where B. Matthias's last rule, "stay away from theorists", has (partially) lost its significance, and where theoretical calculations received a carte blanche.

We have performed a large number of ab initio calculations of thermodynamically stable phases and their superconducting properties [36] for most of the binary metal-hydrogen systems of the periodic table. By combining these results with experimental data and calculations of other groups, we constructed the distribution diagrams (Figure 58) of the maximum achievable critical temperatures in the pressure region convenient for experiments (≤200 GPa). It was understood quickly that the maximum critical temperatures of superconductivity in metal polyhydrides are observed just for a narrow d-belt (or "superconductivity belt") of metals with 0-2 *d*-electrons ($s^2d^0$ - $s^2d^2$, Figure 58). When we move away from this area in both directions, the superconducting properties disappear because of:

1) the dominance of molecular semiconducting polyhydrides in systems with alkali metals;

2) the absence of higher polyhydrides for the middle and heavier d-elements and the strengthening of the covalent component of metal-hydrogen bonds in them;

3) the paramagnetism of the lanthanide and some d-element (Fe, Co, Ni) atoms.

Regarding the second reason, we should make the following remark, which will be developed below when we talk about electronegativity. It can be shown that superhydrides can exist and be stable in a discharged plasma in the form of charged clusters, in chemical complexes with electron donors (e.g., tungsten polyhydrides with organic ligands $WH_4L$-$WH_6L$), and in electrochemical processes [285-288]. In this case, the hydrogen shell is stabilized by a local electric field. A similar electric field exists in 3D polyhydride crystals. Obviously, the stronger the charge on the hydride-forming atom, the stronger the stabilization of hydrogen shell will be. However, with a small difference in electronegativity between the hydrogen and the metal atom, the charge transfer will be negligible and the formation of a superhydride with a high hydrogen content is unlikely.

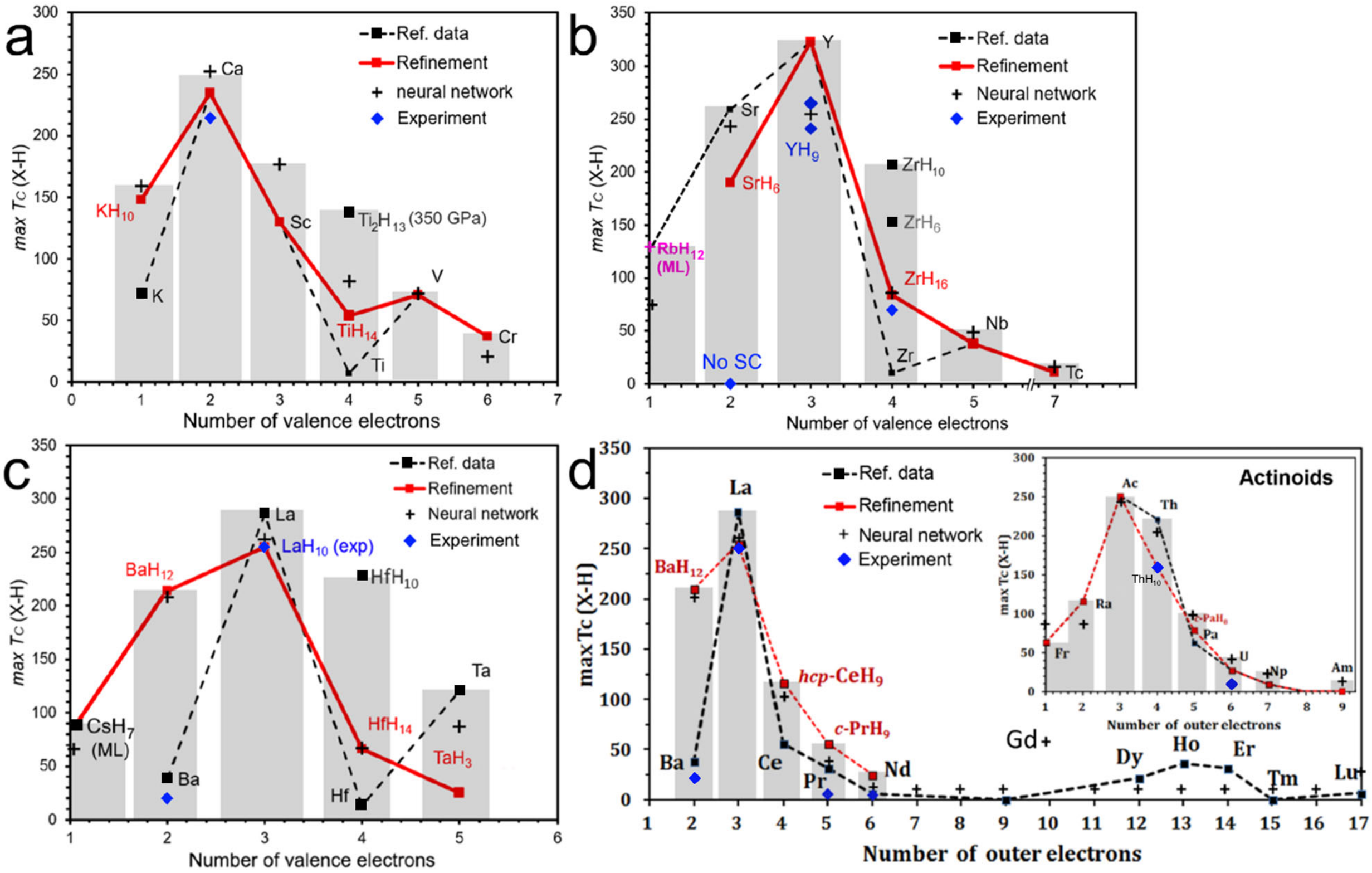

**Figure 58.** Distribution of superconducting properties among metal polyhydrides under pressure. Grey bars show maximum predicted or expected $T_C$ for each system. “Ref. data” means literature data, “Refinement” means that this binary system was reinvestigated using USPEX and newly found superconducting phases were studied. (a) Maximum $T_C$ of hydrides of metals from 4th period. (b) Maximum $T_C$ of hydrides of metals from 5th period. (c) Maximum $T_C$ of hydrides of metals from 6th period. *hcp*-$HfH_{10}$ was taken from Ref. [211]. (d) Maximum critical temperature (max $T_C$) in lanthanoid and actinoid hydrides as a function of the number of outer electrons in the metal atom. One can see predicted by the neural network “secondary wave” of superconductivity in hydrides of heavy lanthanoids. All ab initio calculations shown here were non-magnetic.

In addition to the distribution of superconductivity in hydrides, we also investigated the optimal composition of polyhydrides. As the statistical analysis shows (Figure 59a), the best superconducting properties are found in hydrides with a high hydrogen content of 6 to 12 H-atoms per metal atom. Moreover, the maximum $T_C$ is observed for the composition $XH_n$, where $n = 10$. Indeed, since 2019, when the paper [36] was published on ArXiv, the studies conducted for metal hydrides have shown the validity of the bell-shaped dependence of the maximum achievable critical temperature on the stoichiometry of compounds: $T_C(LaH_{10}) > T_C(YH_9) > T_C(LaH_6) > T_C(YH_6) > T_C(YH_4)$. On the slopes of this bell-shaped dependence, $dT_C/dn$ can reach rather large values of ~70 K per hydrogen atom, whereas near the maximum the efficiency of the next introduced hydrogen atom decreases to 5-15 K per hydrogen atom. Thus, the room temperature of superconductivity can most likely be achieved in metal deca- and nonahydrides $XH_9$ and $XH_{10}$. On the other hand, this pattern limits the possibilities of pressure reduction for superhydrides, because the pressure reduction leads to a loss of some hydrogen and a sharp decrease in superconducting $T_C$. This is well illustrated by the yttrium polyhydride sequence: $YH_3$ ($P = 0$ GPa) - $YH_4$ - $YH_6$ - $YH_9$ ($P = 200$ GPa), in which maximum $T_C$ takes values of 0, 88, 227, and 243 K, respectively.

The bell-shaped dependence of $T_C$ on the hydrogen content allows us to indirectly estimate the critical temperatures of the not yet synthesized or “impossible” polyhydrides. For instance, knowing that $T_C(CaH_6) < T_C(YH_6)$, we can assume that hypothetical $CaH_9$ and $CaH_{10}$ [289] will likely have lower $T_C$ than $YH_9$ and $YH_{10}$. Similarly, as has been shown recently, $ScH_3$ has a higher $T_C$ than $LuH_3$ [247], which allows us to assume that the $T_C(LuH_{9-10})$ will also not reach room temperatures when the hydrogen content is increased. Another argument is the fact that superconductivity is suppressed by impurities of Lu in YBCO [290], La [291], and $ZrB_{12}$ [292]. Moreover, in the latter case, the valence state of lutetium changes, as a result its atoms acquire the magnetic moment (see also [293] and Appendix).

Superconductivity in hydrides is facilitated by the maximum density of electronic states $N_F$ and the contribution of the hydrogen sublattice to $N_F$, which does not cross the critical limit of spin splitting and magnetization. This corresponds to B. Matthias's rule No. 2: "high density of electronic states is good". The study of the electron density of states at the Fermi level $N_F$ for superhydrides indicates the existence of some optimal value for high-temperature superconductivity, close to 1 state per Ry per 1 hydrogen atom (Figure 59b). As the density of states increases above the optimal value, spin splitting in the band structure appears, leading to scattering of the Cooper pairs with spin flipping, their destruction and, as a consequence, suppression of the superconducting state.

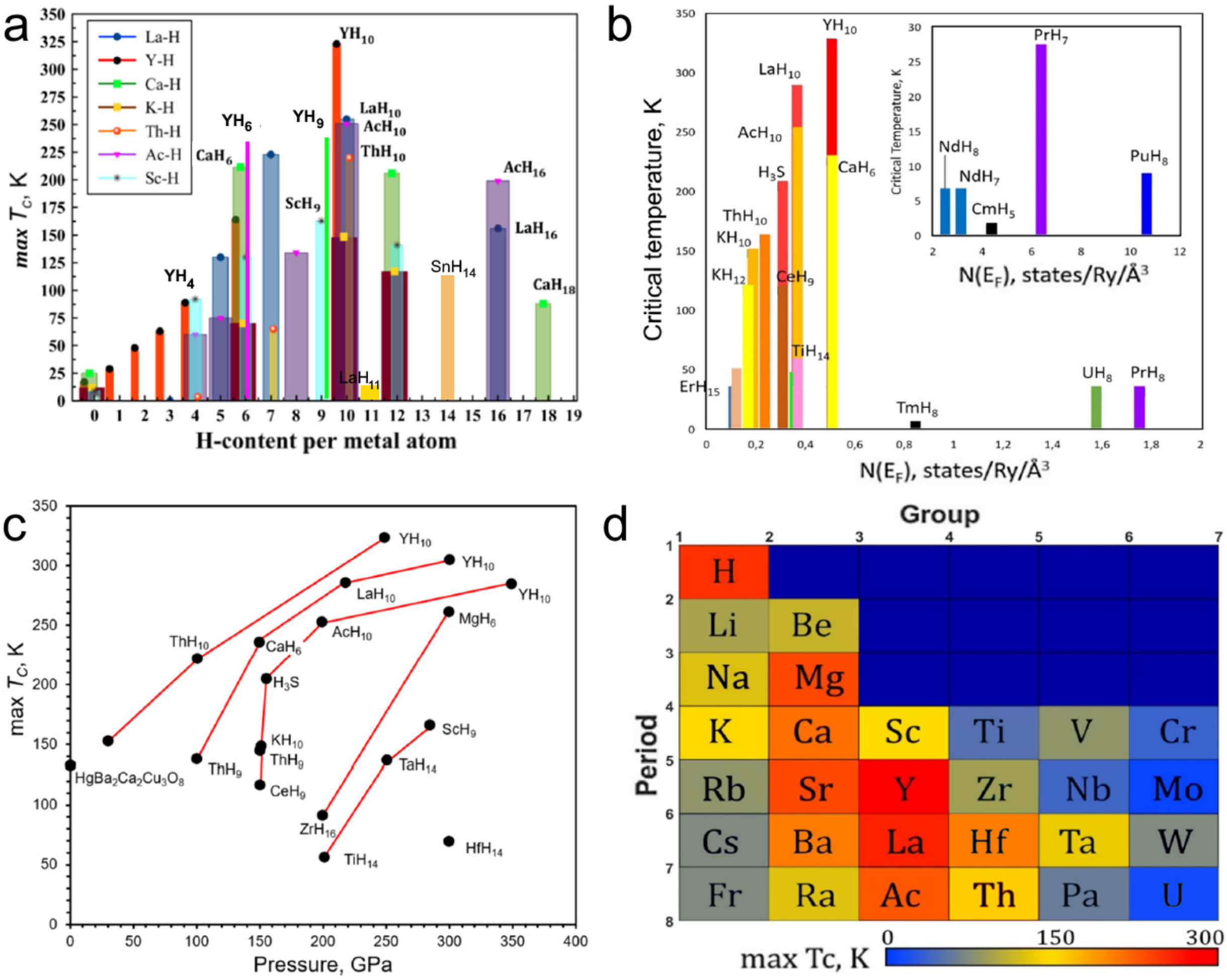

**Figure 59.** Maximum $T_C$ for a given hydrogen content per metal atom in the $XH_n$ hydrides, calculated using the Allen-Dynes formula. (b) Maximum predicted $T_C$ versus the electronic density of states at the Fermi level for the studied hydrides. (c) Ashby plot of the maximum $T_C$ versus pressure for the best-known calculated metal hydrides. (d) The distribution of max$T_C$ of metal hydrides in the left part of Mendeleev's table.

As we know from the BCS theory, the critical temperature of superconductivity is influenced by two main factors: the electron-phonon coupling strength λ and the average phonon frequency (e.g., logarithmically averaged $\omega_{log}$). For each of these factors, we found a distribution over the periodic table using the data for the best superhydrides described in the literature or predicted using the neural network developed in [36] (Figure 60b). As expected, $\omega_{log} \sim (1/m_A)^{0.5}$ (more precisely, $\sim 1/[V_{cell}^{1/3}\times(m_A)^{0.5}]$): an increase in the average phonon frequency of polyhydrides is observed for light elements with a small atomic radius and unit cell volume, and with increased electronegativity (Li, Sc, Be, Mg, Y), whose polyhydride stabilization requires pressures above 200 GPa. This means that the distribution of superconducting properties ($T_C$) in pure metals must correlate with the mass of the atom (see Appendix). Practically, the highest average phonon frequency is expected for metallic hydrogen. Generally, the average phonon frequency increases along with the density and pressure in the matter when the steepness of walls $\partial U/\partial q_i$ of potential wells in which individual atoms are located increases.

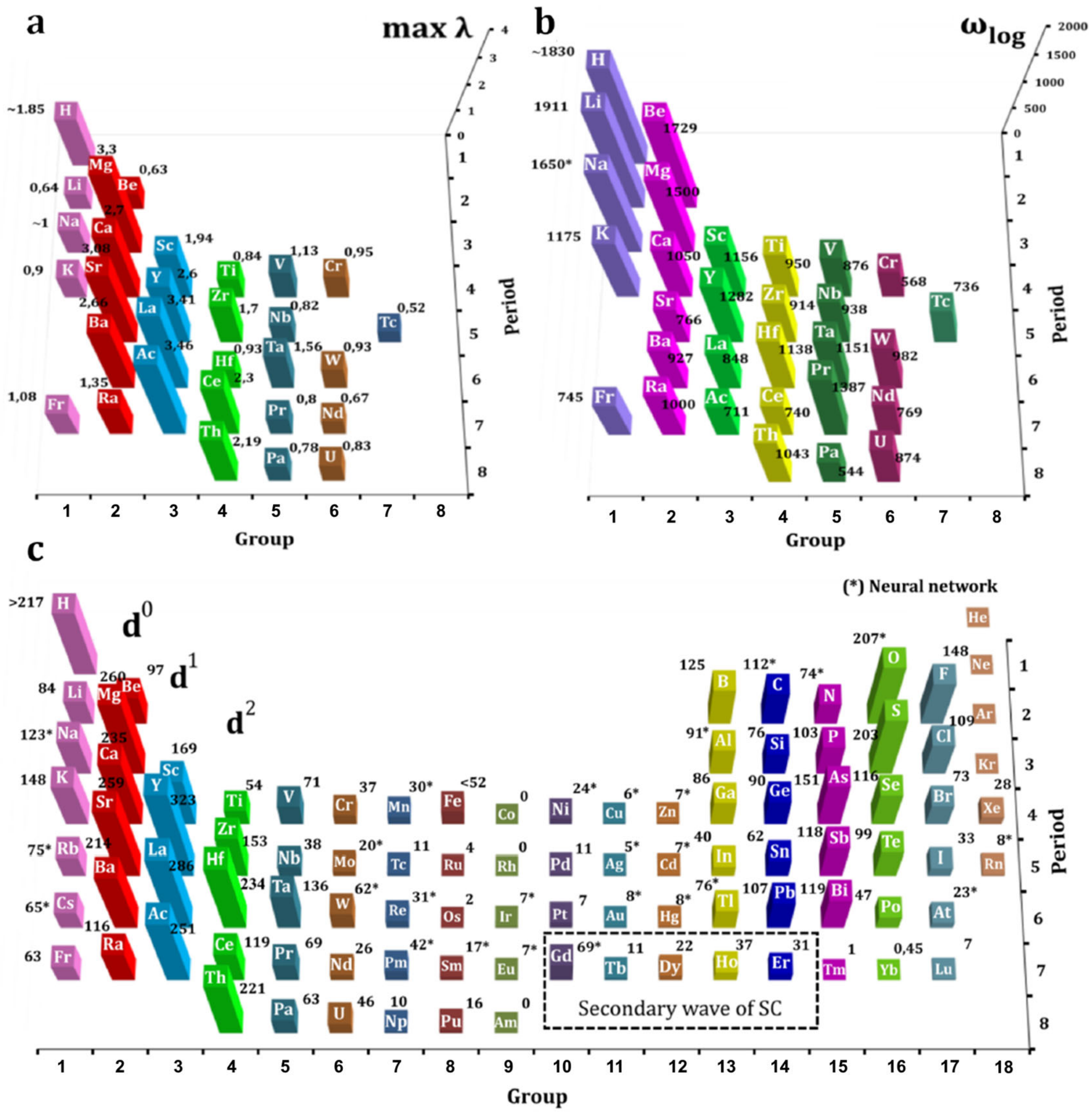

**Figure 60.** Distribution of (a) the EPC coefficient λ and (b) logarithmic average frequency $\omega_{log}$ corresponding to the max$T_C$ points over Mendeleev's Table [36]; the data for Na-H were estimated using the developed neural network [36]. (c) Mendeleev's Table with already studied binary hydrides. The maximum values of $T_C$ were calculated using ab initio methods or were predicted by the developed neural network (marked by *).

The situation with the electron-phonon coupling parameter (λ) is not as straightforward. There is no simple expression that allows us to identify the most promising elements in terms of high values of λ. At the moment, the experimentally measured values of λ are limited to ~ 2.5 - 3 achieved in very soft (at room temperature) alloys like Bi-Pb and in Ga due to low-frequency phonon modes [294]. A further increase in λ is not formally restricted, but does not occur in any of the known compounds. For high-$T_C$ hydrides, the analogy is the melting/destruction of the hydrogen sublattice at the superconducting transition temperature (see also Chapter 5), since the hydrogen sublattice is the most easily meltable component of polyhydrides.

As the pressure increases/decreases, λ(P) changes nonmonotonically, reaching local maxima in the vicinity of phase transitions, where various kinds of structural instabilities arise. In Figure 60a, we can see that the maximum λ is reached in metal polyhydrides belonging to the superconductivity belt ($d^0$-$d^2$ elements). The hydrogen sublattice in such hydrides is an intermediate between the molecular lattice of polyhydrides of alkali and alkaline earth elements and unstable polyhydrides of elements of groups IV-VIII, which cannot be stabilized by the weak charge of the central atom forming covalent bonds with hydrogen. Stabilization of the hydrogen shell for such atoms can be observed only with the introduction of various donor atoms (in complexes with organic phosphines). If there are no other possibilities to stabilize the hydrogen shell, the compound decomposes to form two different phases: molecular hydrogen and metal.

Thus, given the two scenarios described above: 1) formation of a semiconducting molecular polyhydride (e.g., $CsH_{17}$, $SrH_{22}$); 2) formation of a separate phase of molecular hydrogen and metal (decomposition), we can represent the $d^0$-$d^2$ elements as an unstable transition zone between these two variants. Local lattice oscillations in superhydrides of metals from the "superconductivity belt" can cause both local formation of $H_2$ molecules (as observed, for instance, in the higher polyhydrides of $LaH_{11}$-$LaH_{12}$) and decomposition of a compound with loss of $H_2$ with decreasing pressure, as observed in $LaH_{10}$ below 140 GPa.

It is interesting to note that the "superconductivity belt" in compressed metal polyhydrides also coincides with the series of metals having the highest hydrogen content in lower hydrides at ambient pressure ($YH_3$ - $LaH_3$ - $CeH_3$ - $Th_4H_{15}$, Figure 61).

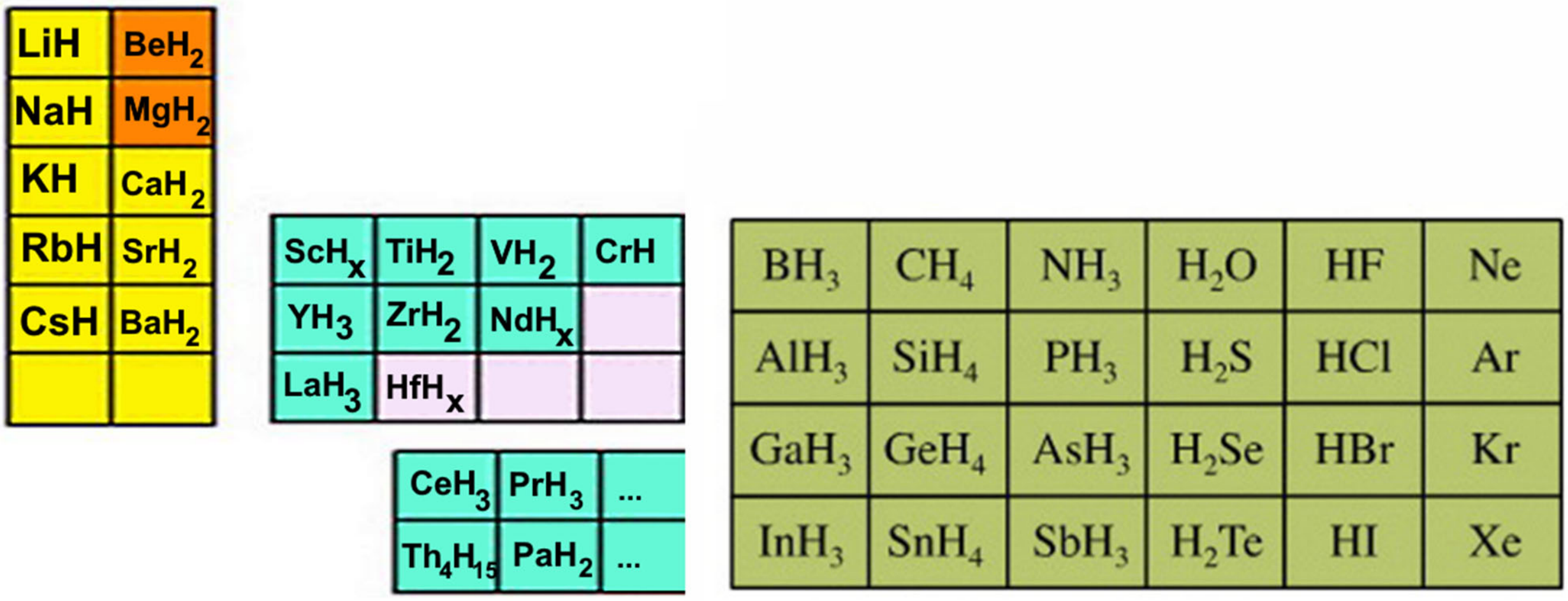

**Figure 61.** Fragment of the Mendeleev Table with known lower hydrides formed by elements and hydrogen at ambient pressure. $AB_x$ - corresponds to non-stoichiometric hydrides.

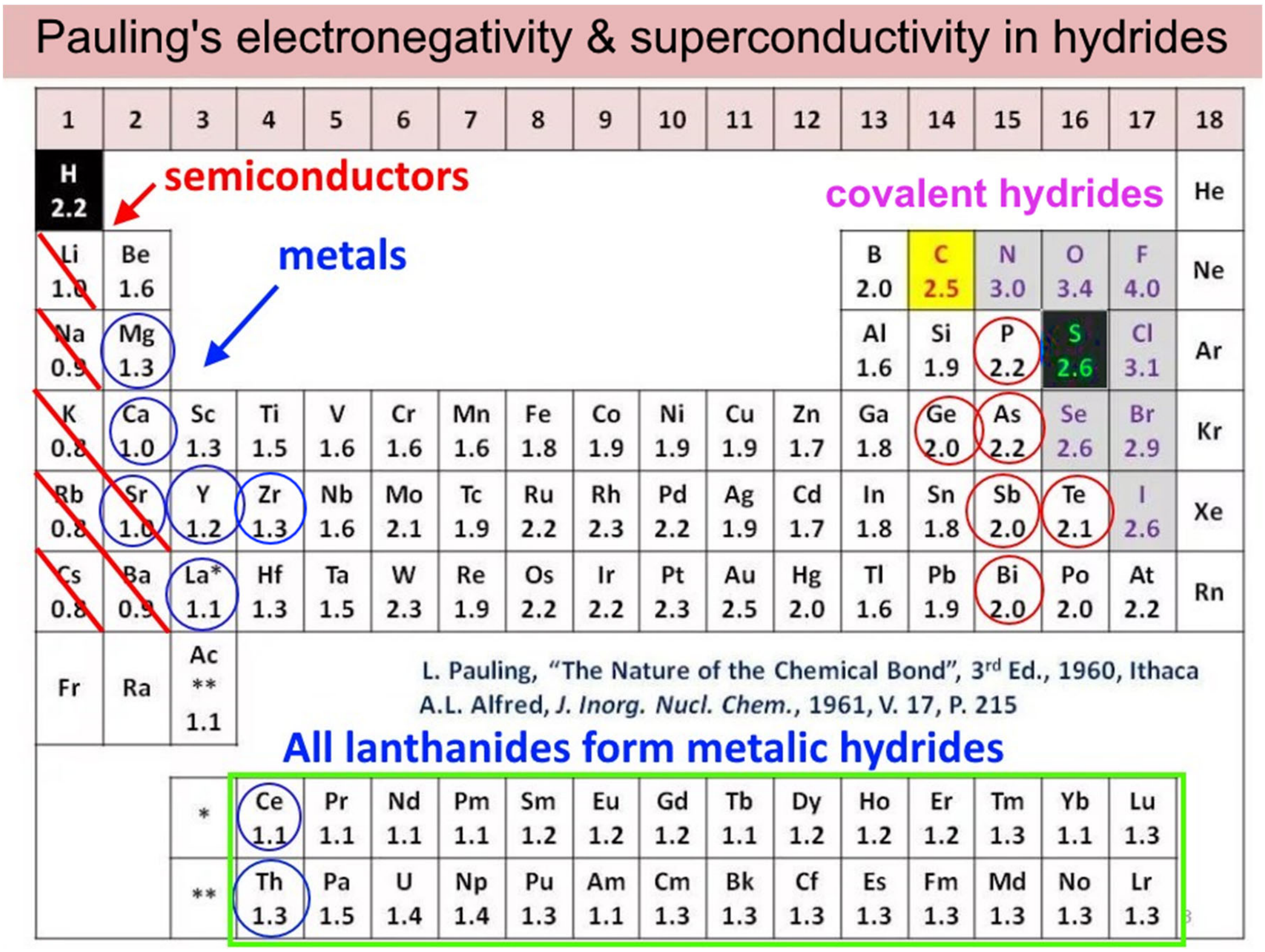

**Figure 62.** Mendeleev Table with values of electronegativity according to Pauling [295, 296]. Those elements that form semiconducting polyhydrides under pressure are crossed out. Blue circles highlight the most promising hydride-forming metal elements. Red circles highlight the nonmetals that may be of interest for the design of ternary superhydrides.

Finally, there is a correlation between the superconducting properties of hydrides and the electronegativity ($\chi$) of the hydride-forming element (Figure 62). For instance, alkali and alkaline earth elements with low electronegativity form molecular polyhydrides with semiconducting properties (Li, Na, K, Rb, Cs, Sr, Ba), in which the hydrogen sublattice often has a disordered (glass-like) structure and can rather be described as a hydrogen shell [45]. Elements with electronegativity $\chi$ = 1-1.3 under pressure form metallic superhydrides with an atomic hydrogen sublattice and a strong electron-phonon interaction. An increase in electronegativity leads to an increase in the degree of covalent bonding in the hydrides. As a result, p-elements and late d-elements do not form higher polyhydrides with some exceptions ($SnH_{14}$ [47]). This approach allows us to discuss the possibility of forming some hydrides for which we do not yet have reliable experimental data. For example, the electronegativity of beryllium is $\chi$ = 1.57-1.6, which speaks in favor of the essentially covalent nature of the Be-H bond, and does not allow us to expect the formation of higher polyhydrides in the Be-H system even at high pressure. To date, only $BeH_3$ has been obtained [297]. Similarly, titanium has a higher electronegativity ($\chi$ = 1.5) than Zr and Hf, so we expect that the formation of polyhydrides, including superconducting compounds, will also be difficult in this case despite theoretical predictions of $TiH_{12}$ and $TiH_{14}$ stability [298]. So far, only the $TiH_{2.5\text{-}3}$ compound has been obtained at 120 GPa (see Figure 63).

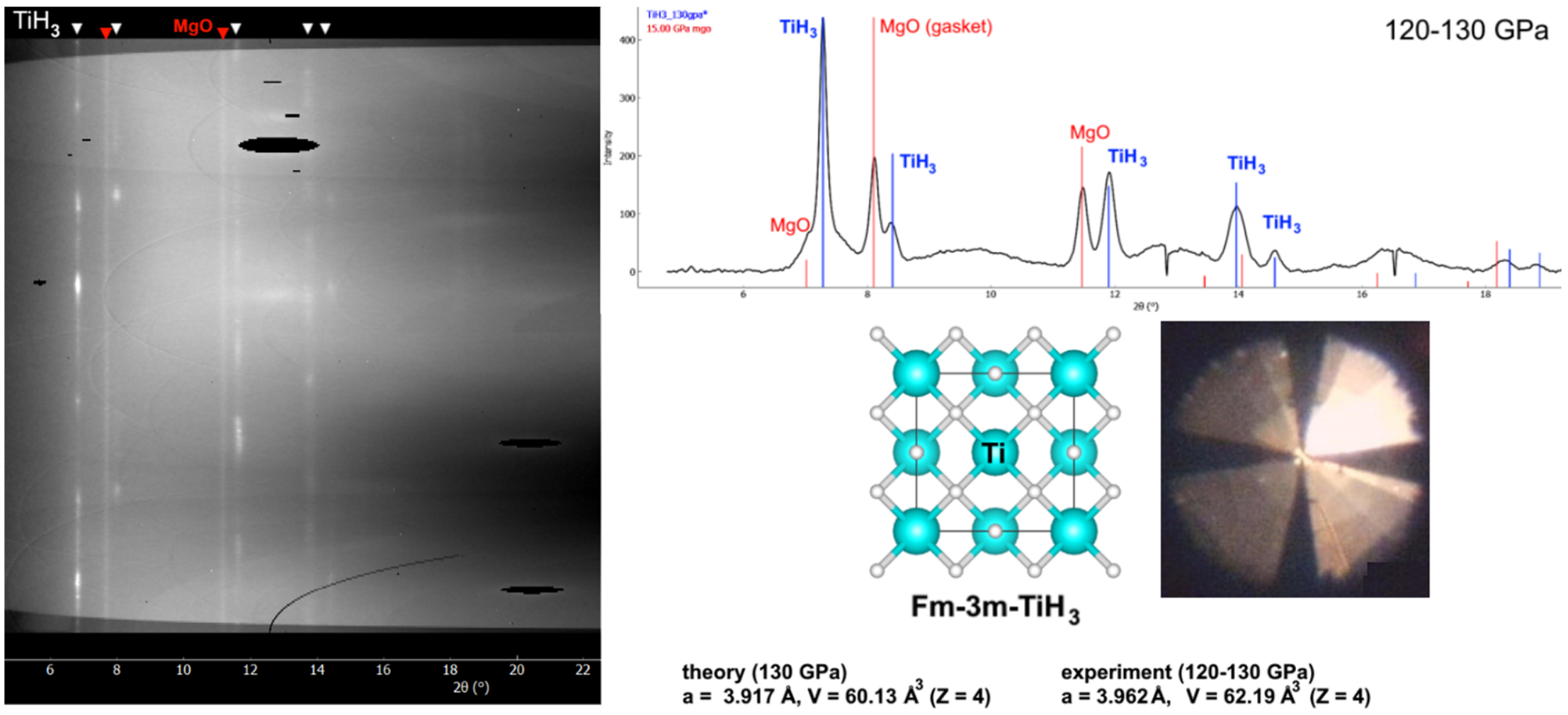

**Figure 63.** X-ray diffraction pattern of a titanium hydride sample obtained by heating the $TiH_2/H_2$ target at 120-130 GPa. MgO was used as a gasket. The obtained data indicate the formation of the previously predicted $Fm\bar{3}m$-$TiH_3$ [298], the sample of which does not exhibit superconducting properties. This experiment was carried out jointly with Dr. M. Kuzovnikov.

Finding out the distribution law of superconductivity in polyhydrides can also play an important role in a search for similar distribution laws for maximum $T_C$ in metal borides, carbides and nitrides (under pressure), allowing us to further extrapolate the laws found to all classes of superconducting compounds with the conventional electron-phonon coupling. As we have seen, B. Matthias's rules are also valid for superhydrides. The need for high Debye temperature and average phonon frequency, following from the BCS theory, excludes obtaining superconducting polyhydrides at low pressures. For most superhydrides, average $T_C \sim P$ (in GPa) is numerically equal to the pressure in GPa, and does not exceed 2P: $T_C < 2P$ (in GPa).

## 6.2 Distribution of superconducting properties in ternary polyhydrides

By now (2022), the majority of the most promising binary metal-hydrogen systems have already been investigated and the hopes of researchers have turned to the synthesis of ternary polyhydrides. Specialists in hydride superconductivity are now pursuing two main goals: 1) reducing the stabilization pressure while maintaining outstanding superconducting properties of hydrides, and 2) achieving room-temperature superconductivity.

One of the simplest methods of constructing ternary superhydrides is replacing one of the metal atoms in the supercell of a known superhydride with another element. This preserves the high symmetry of the crystal, the complexity of calculations increases insignificantly, but we, of course, know nothing about whether such an artificial structure will exist in experiment. In this thesis, we would like to illustrate such "template" calculations using three examples: *Pm*-3*m*-$YXH_{12}$ (derived from *Im*-3*m*-$Y_2H_{12}$), *R*-3*m*-$YXH_{20}$ (derived from $Fm\bar{3}m$-$YH_{10}$), and *R*-3*m*-$LaXH_{20}$ (derived from $Fm\bar{3}m$-$LaH_{10}$) at 180 GPa. In this approach, it is important to keep the calculation parameters constant. We used the tetrahedral method (tetrahedra_opt) with a 16×16×16 grid of k-points and a 2×2×2 grid of q-points. For X = Y, we obtained $T_C(Y_2H_{12})$ = 206 K, lower than the experimental value (Table 12). The accuracy of such calculations is not great, but the computing speed allows us to cover a large number of systems and obtain a map of distribution of the superconducting properties, the gaps of which can then be filled using neural networks (see below).

Figure 62 shows the graphical distribution of critical temperatures in *Pm*-3*m*-$YXH_{12}$ hydrides for the area of the "superconductivity belt" (Figure 62a) and for nonmetals (Figure 62c). Although structures like *Pm*-3*m*-$YXH_{12}$ may be unstable for many of the X atoms listed in Table 12, we still find these formal calculations useful as they show the combination of yttrium with which atoms can lead to the best SC properties. As a side result, we will also get the information on which $X_2H_{12}$-type hydrides would be good superconductors (if stable). In particular, the Li-Y-H, Ca-Y-H, Sr-Y-H, and Sc-Y-H systems would be the most promising in terms of superconductivity. For all these systems, the cubic $XYH_{12}$ hydride has a significantly higher calculated $T_C$ than the prototype *Im*-3*m*-$YH_6$. However, despite considerable efforts, the extrapolation with the neural network to the other elements of the periodic table shows unsatisfactory results (Figure 63). For instance, the neural network erroneously ignores the suppression of superconductivity for the heavy lanthanides and actinides, and gives a low $T_C$ value for $LaYH_{12}$.

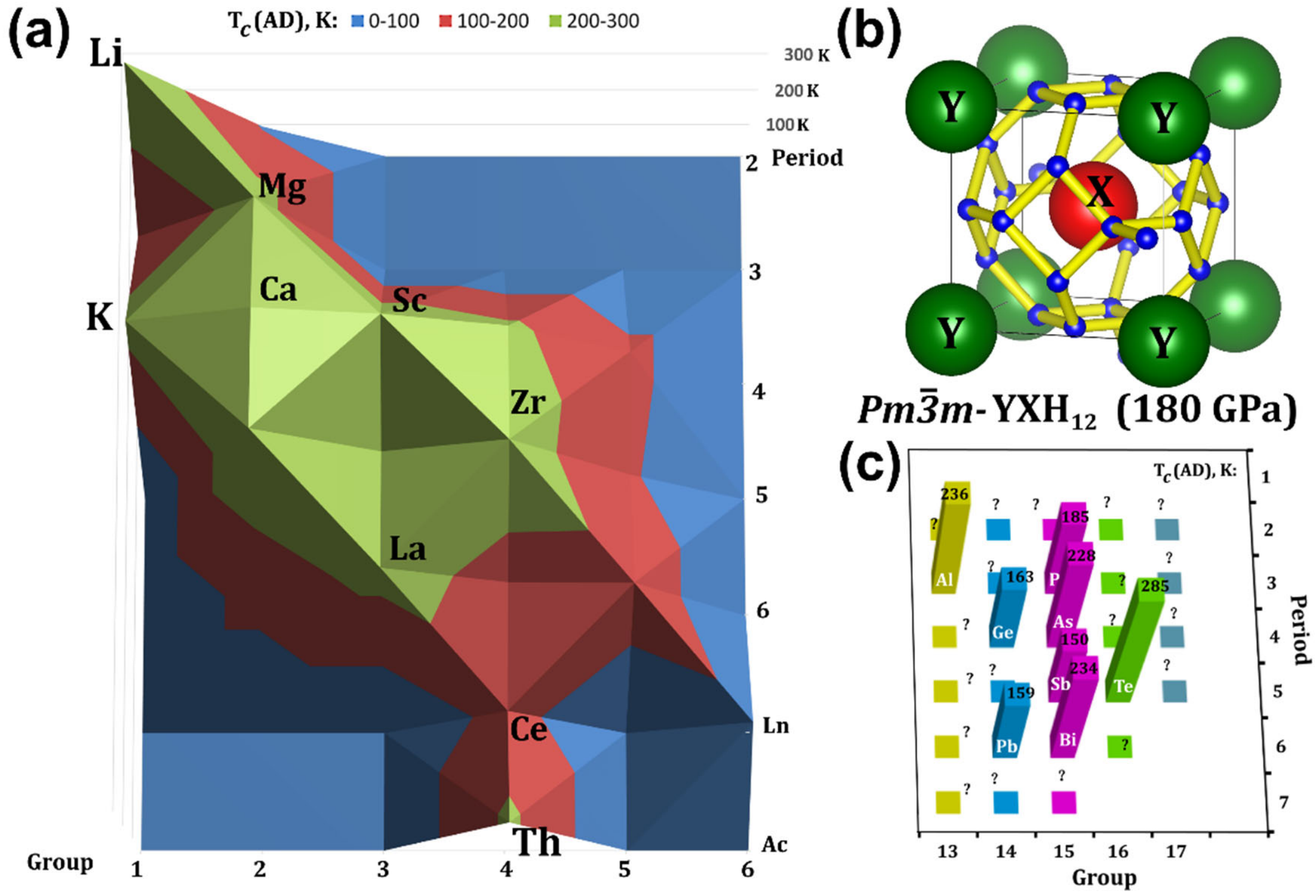

**Figure 62.** Distribution of the superconducting properties in the formally constructed class of $YXH_{12}$ hydrides at 180 GPa for (a) metals with $d^0$-$d^2$ number of valence electrons, and (c) nonmetals. An example of a "template" approach to the search for new superconducting hydrides.

Calculation results for the systems $R$-3$m$-La$XH_{20}$ and $R$-3$m$-Y$XH_{20}$ at 180 GPa are shown in Figure 64. The promising system is La-Y-H, however, the experiment [72] does not confirm a significant increase in $T_C$ in this system. The most promising system is Y-Sr-H ($T_C$ = 297 K), which also gives good results for $YSrH_{12}$.

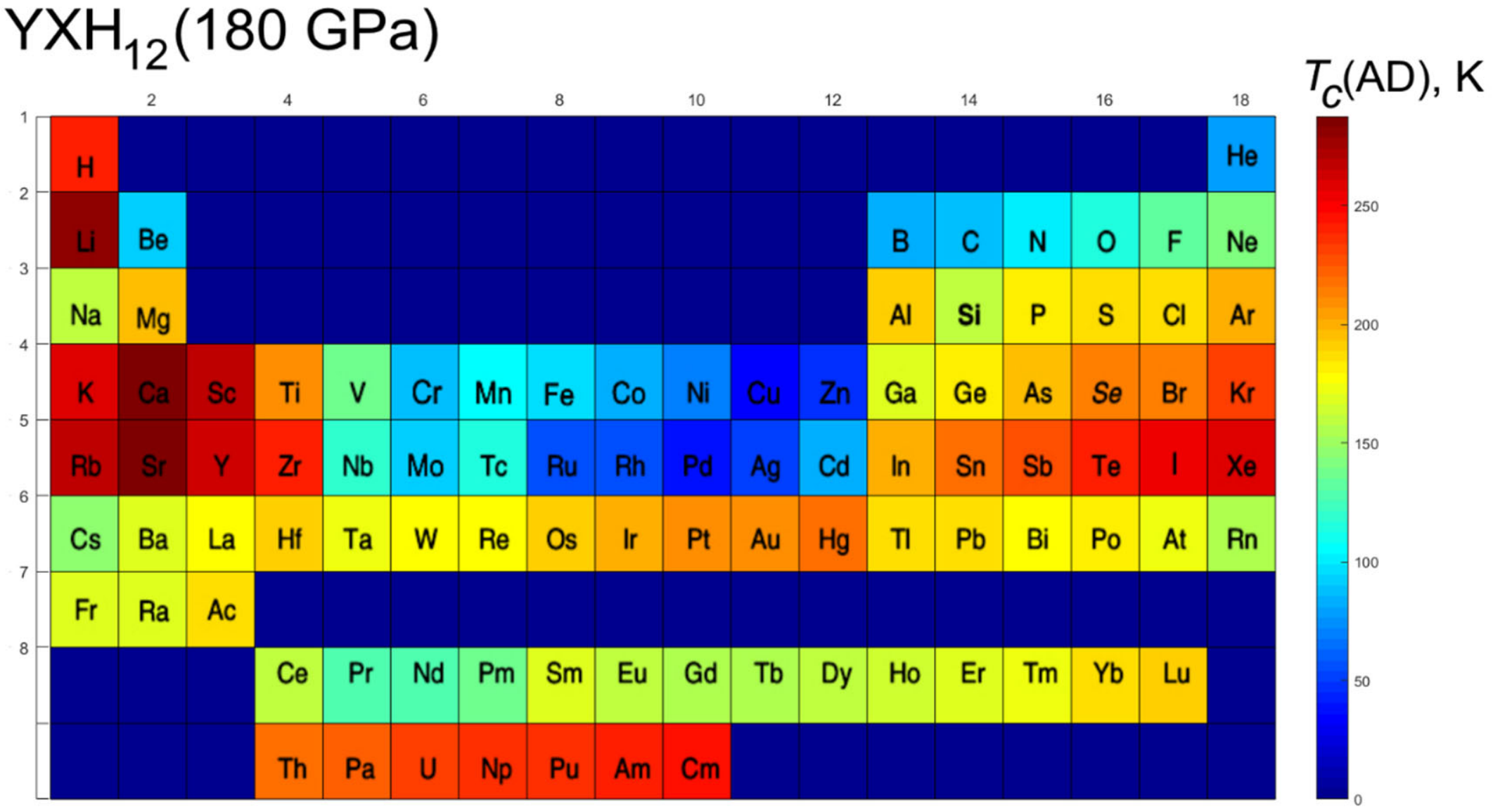

**Figure 63.** Distribution of the critical temperature of superconductivity in cubic ($Im\bar{3}m$) ternary hydrides $YXH_{12}$ obtained by replacing 50 % of Y atoms by X atoms at 180 GPa. This extrapolation is obtained using neural networks.

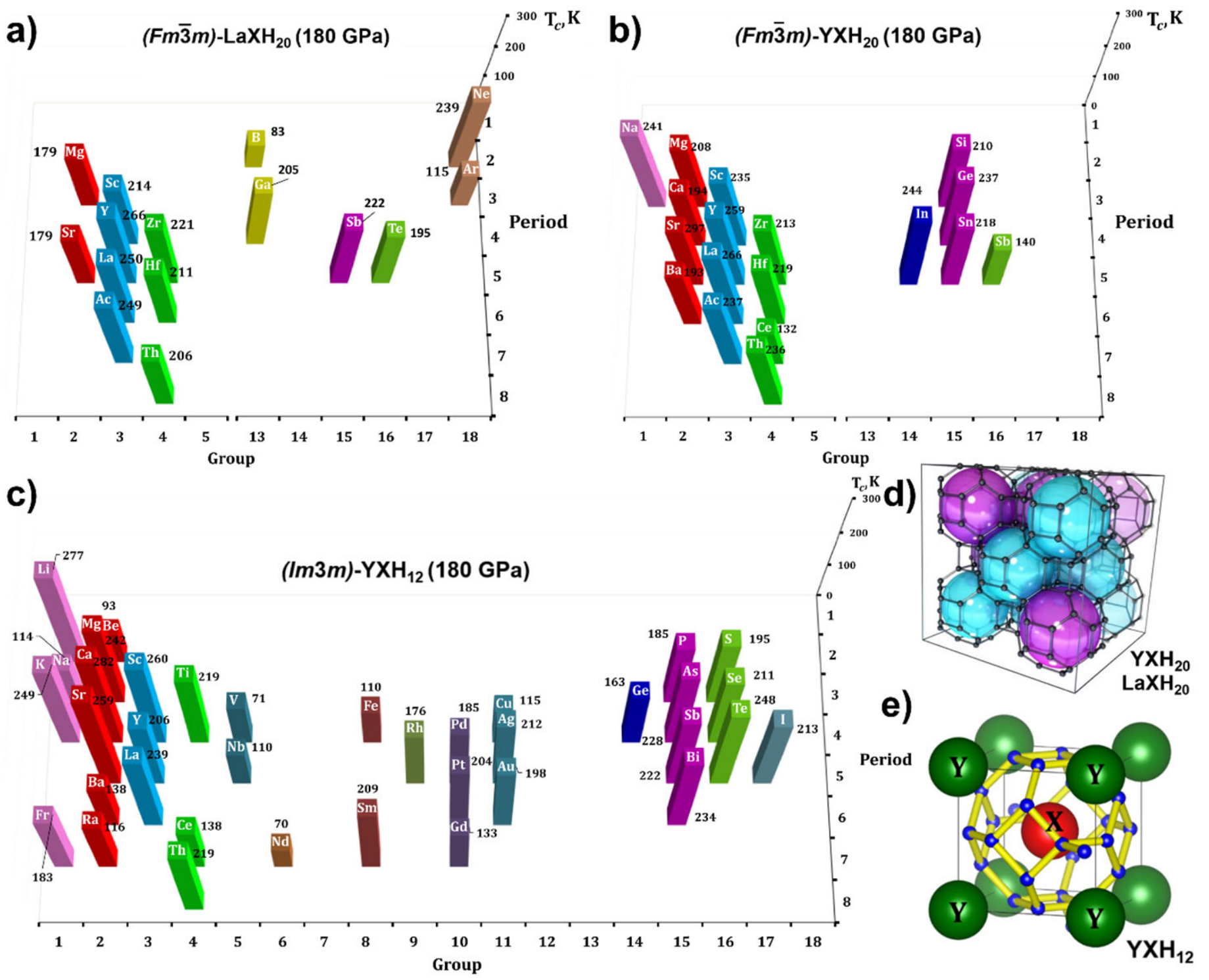

**Figure 64.** Distribution of superconducting properties in formally constructed classes of hydrides: (a) *R*-3*m*-$LaXH_{20}$, (b) *R*-3*m*-$YXH_{20}$ and (c) *Pm*-3*m*-$YXH_{12}$. Different colors in this figure refer to elements of different groups of the Periodic table.

**Table 12.** Parameters of the superconducting state of structures *Pm*-3*m*-$YXH_{12}$ with various X atoms. Calculations were performed using the tetrahedral method in Quantum ESPRESSO with the same grid k = 16×16×16, q = 2×2×2 at 180 GPa. These calculations were performed together with Dr. Di Zhou.

| Atom X | λ | $\omega_{log}$, K | $T_C$(A-D), K | Atom X | λ | $\omega_{log}$, K | $T_C$(A-D), K |
|---|---|---|---|---|---|---|---|
| **Li** | 3,32 | 1012 | 277 | **Sc** | 2,35 | 837 | 260 |
| **Na** | 1,04 | 1447 | 114 | **Y** | 2,24 | 926 | 206 |
| **K** | 3,39 | 829 | 249 | **La** | 2,89 | 712 | 239 |
| **Be** | 1,09 | 1084 | 93 | **Ce** | 1,45 | 919 | 138 |
| **Mg** | 2,47 | 636 | 242 | **Nd** | 1,2 | 720 | 70 |
| **Ca** | 1,97 | 1090 | 282 | **Sm** | 3,48 | 704 | 209 |
| **Sr** | 1,98 | 1100 | 297 | **V** | 1,19 | 795 | 125 |
| **Ba** | 1,5 | 1059 | 138 | **Nb** | 1,54 | 705 | 110 |
| **Ti** | 3,01 | 880 | 219 | **Ta** | 3,15 | 935 | 168 |
| **Zr** | 3,33 | 930 | 255 | **Al** | 2,64 | 1052 | 236 |
| **Hf** | 2,97 | 659 | 170 | **Fe** | 1,92 | 636 | 110 |
| **Th** | 2,69 | 750 | 219 | **Ge** | 2,57 | 724 | 163 |
| **Pb** | 3,3 | 610 | 159 | **P** | 2,74 | 771 | 185 |
| **Te** | 3,68 | 849 | 248 | **As** | 3,57 | 782 | 228 |
| **Bi** | 3,54 | 826 | 234 | **Sb** | 3,42 | 792 | 222 |

The number of possible ternary hydrides is very large (~4608), which excludes the possibility of systematic studies by *ab initio* calculations or with similar methods and by experimental techniques. The main goal of this chapter is to apply artificial intelligence methods to define ternary A-B-H systems with an optimal combination of the following parameters: maximum $T_C$ reached in the pressure range 0-300 GPa (max$T_C$), and minimum stabilization pressure (*minP*). This will allow us to narrow the region of ternary high-$T_C$ superconductors (HTSC) for further detailed theoretical and experimental searches. We also monitored the maximum hydrogen content as a chemically important characteristic.

To predict maximum $T_C$, minimum stabilization pressure $P$ and maximum H-content $N_H$ of ternary A-B-H systems a regression model $F_i$(A,B) was created on the basis of a fully connected neural network (each next layer was connected with all outputs of the previous one). It was found empirically that the optimal topologies consist of 15 layers with 16 neurons in a layer for max$T_C$ (Figure 65), 7 layers with 4 neurons each for max$N$, and 14 layers with 20 neurons per layer for min$P$. The number of *s*-, *p*-, *d*, and *f*-electrons in an atom and atomic number were used as inputs for the network. The neural network was trained on a set of 180 reference max$T_C$ values of binary (A-A-H, A-H-H) and ternary hydrides using the RMSProp (Root Mean Square Propagation) method in 10000 steps. The boundary condition F(A,1) = F(1,A) = F(A,A) and symmetry condition F(A,B) = F(B,A) were included in the training set. The symmetry rule was also implemented by averaging the results of symmetrical inputs. The *tensorflow* and *keras* libraries were applied to create the model. After the training, the network was validated on 20% of randomly selected points, which were not used in the training set. Test data set was prepared using linear congruential generator (variation of pseudorandom number generator algorithms). The standard deviation for max$T_C$, min$P$ and H-content (defined for $A_xB_yH_z$ as z/(x+y)) was 33.5 K, 70 GPa and 1.7 respectively (see Supporting Information).

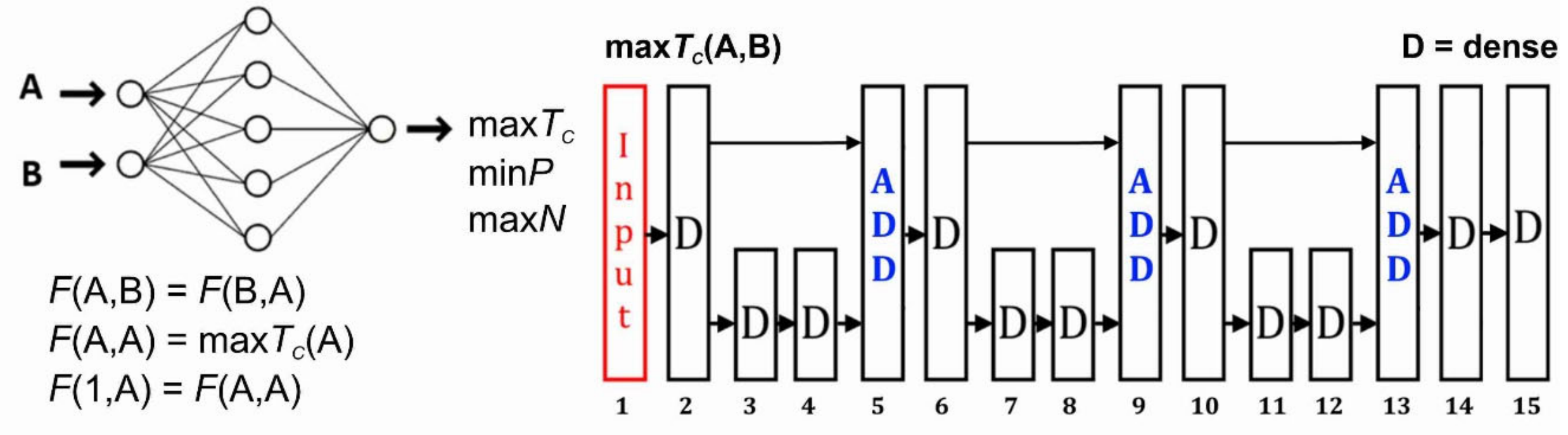

**Figure 65.** Developed neural network topology and boundary conditions used for predicting the critical parameters of ternary hydrides. "ADD" means summation layer, "Dense" (D) means fully-connected layer of neurons. This neural network was developed together with Dr. Igor Savkin (MSU).

One of the prerequisites for high-$T_C$ hydride superconductivity is a high hydrogen content in the compound [36], more than dedicated by atomic valences. Analysis of the results obtained after applying our neural network to an array of parameters of the experimentally known or *ab initio* predicted binary and ternary hydrides in the pressure range of 0-300 GPa is shown in Figures 66A, B. The pronounced maximum of the hydrogen content ($[AB]H_{12-14}$) is observed for ternary compounds containing heavy lanthanoids and *6d*-elements (Hf, Ta, W, Re), see Figure 66B. The statistical analysis of the hydrogen content in ternary compounds shows the majority of elements cannot form superhydrides in this range of pressure, see Figure 66C. A relatively high population of hydrides observed for the [AB]$H_{3-4}$ and [AB]$H_{8-10}$ compositions has many experimental evidences. Since studies of higher (n > 16) polyhydrides are currently quite rare, the neural network erroneously predicts very low or zero statistics of their formation.

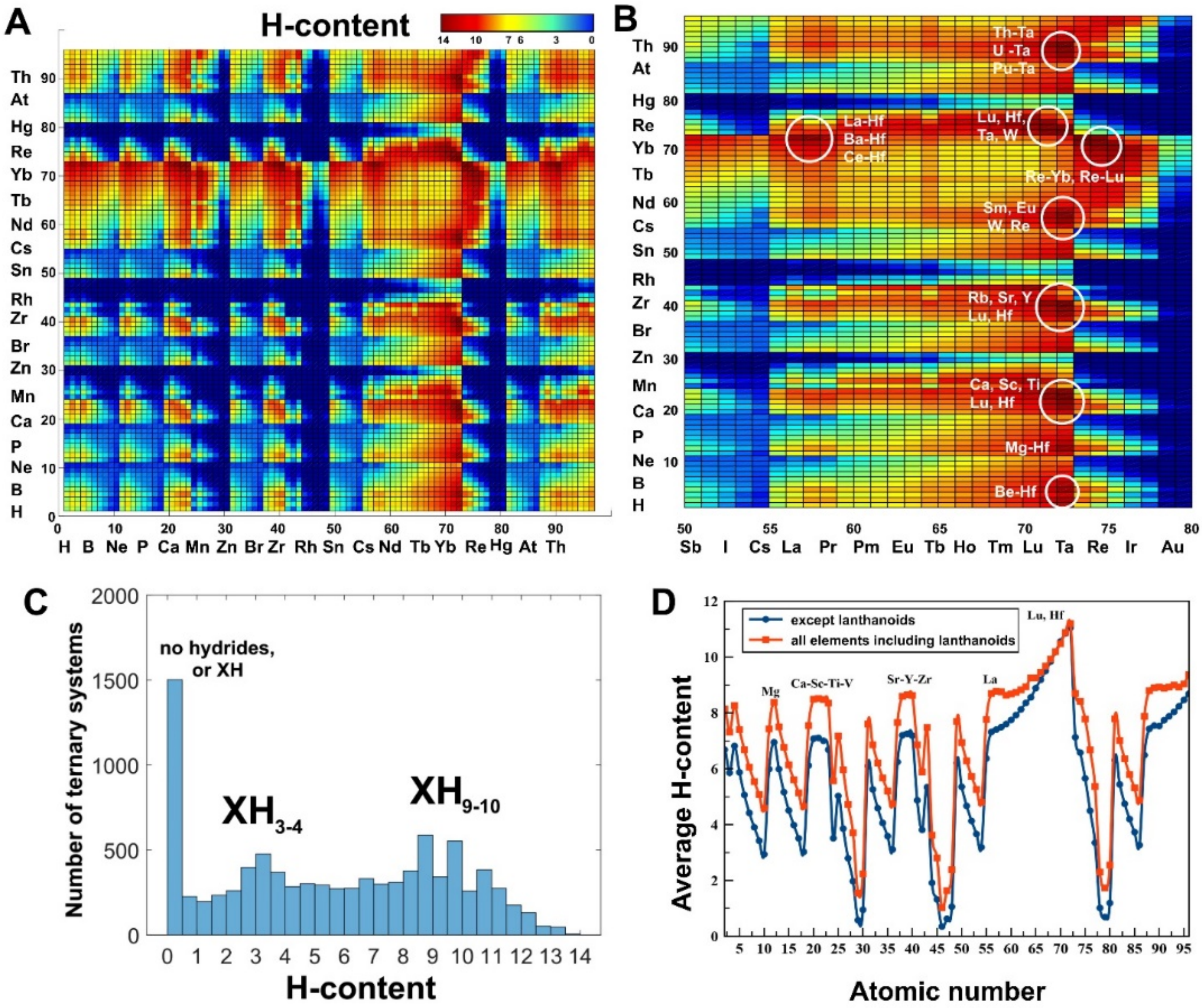

**Figure 66.** A) Distribution of the maximum hydrogen content per (defined as the ratio of hydrogen atoms to the sum of other atoms); B) zoom-in of a group of ternary lanthanoid hydrides. The most promising combinations of elements are marked by circles; C) number of possible ternary systems depending on the hydrogen content; D) maximum hydrogen content in ternary hydrides $A_xB_yH_z$ average over all elements B as a function of element A.

The dependence of the maximum hydrogen content, averaged over all possible hydrides of a chemical element (i.e. $\Sigma_B \max N_H(A,B) / 96$), on the atomic number of the element is shown in Figure 66D. Heavy lanthanoids form higher polyhydrides much easier than all other elements. The elements from the $d^0$, $d^1$ and $d^2$-belts (Mg, Ca, Sr, Sc, Y, La, etc.) can be used in combination with heavy lanthanoids to increase the H-content. We unexpectedly observed many promising ternary polyhydrides of Tm, Yb, Lu, Ta, Zr, W, and other elements located in the vicinity of $d^0$ and $d^1$-belts, which have no promising binary hydrides [36, 299].

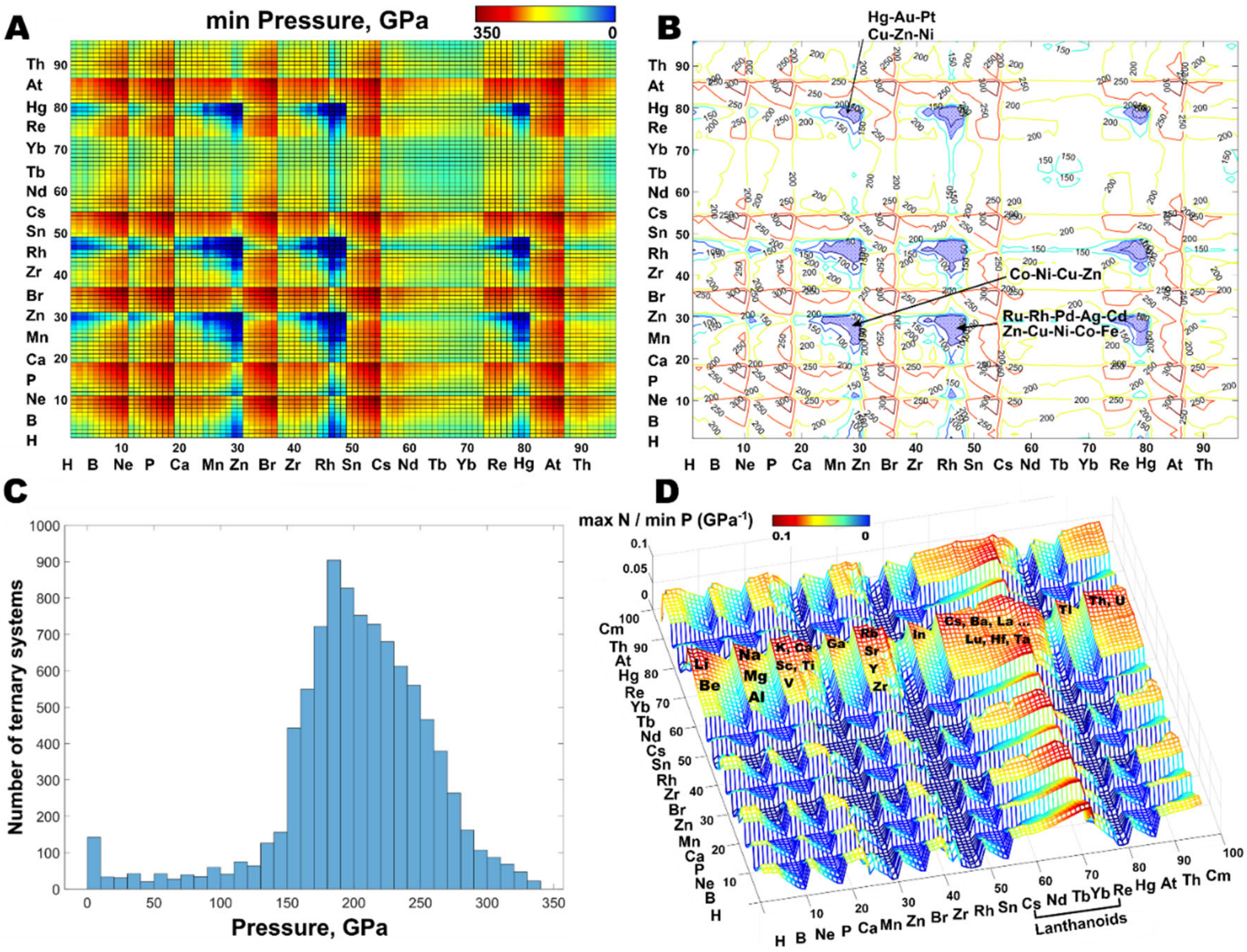

**Figure 67.** A) Distribution of stabilization pressure over ternary superhydrides; B) isobaric map of stability of ternary hydrides. Combinations of elements forming stable hydrides at the lowest pressure are shown explicitly; C) number of ternary systems, which potentially may form superhydrides at various pressures (area histogram of A); D) distribution of the ratio of maximum hydrogen content to the stabilization pressure of ternary hydrides (divergences are excluded).

One of the most important parameters for the synthesis of a particular metal hydride is the pressure needed for its stabilization. The training of our neural network on the published minimum values of the stabilization pressure for highest binary and ternary hydrides leads to the distribution of stabilization pressures for all ternary hydrides (Figure 67). To be thermodynamically stable, almost 90% of ternary hydrides with max$N_H$ content require a pressure above 120 GPa (Figure 67C). The remaining 10%, late 3*d*-5*d* metals corresponding to the "wells" on the stabilization map (Figure 67A), do not form superhydrides (see Figure 66A), and, thus, they have little chance to possess high-$T_C$ superconductivity. The highest stabilization pressures (> 300 GPa) correspond to the hydrides of noble gases.

The most interesting regions of the obtained map (Figure 67A) are related to the superhydrides, having the lowest possible stabilization pressure and maximum H-content. To define these regions, we obtained the distribution of the ratio of maximum hydrogen content to minimum stabilization pressure max*N* / min*P* (Figure 67D, divergences are excluded). Physically this ratio means the average molar content of hydrogen per 1 GPa of applied pressure. The maximum value of max*N*/min*P* reaches ~0.08-0.1 for lanthanoids and *5d* elements (Cs, Ba…Hf, Ta) in combination with alkaline and alkaline earth metals. This means that pressures of ~100 GPa (or more) are needed to stabilize superhydrides with H-content 10, and at least 60 GPa pressure is needed to achieve H-content 6.

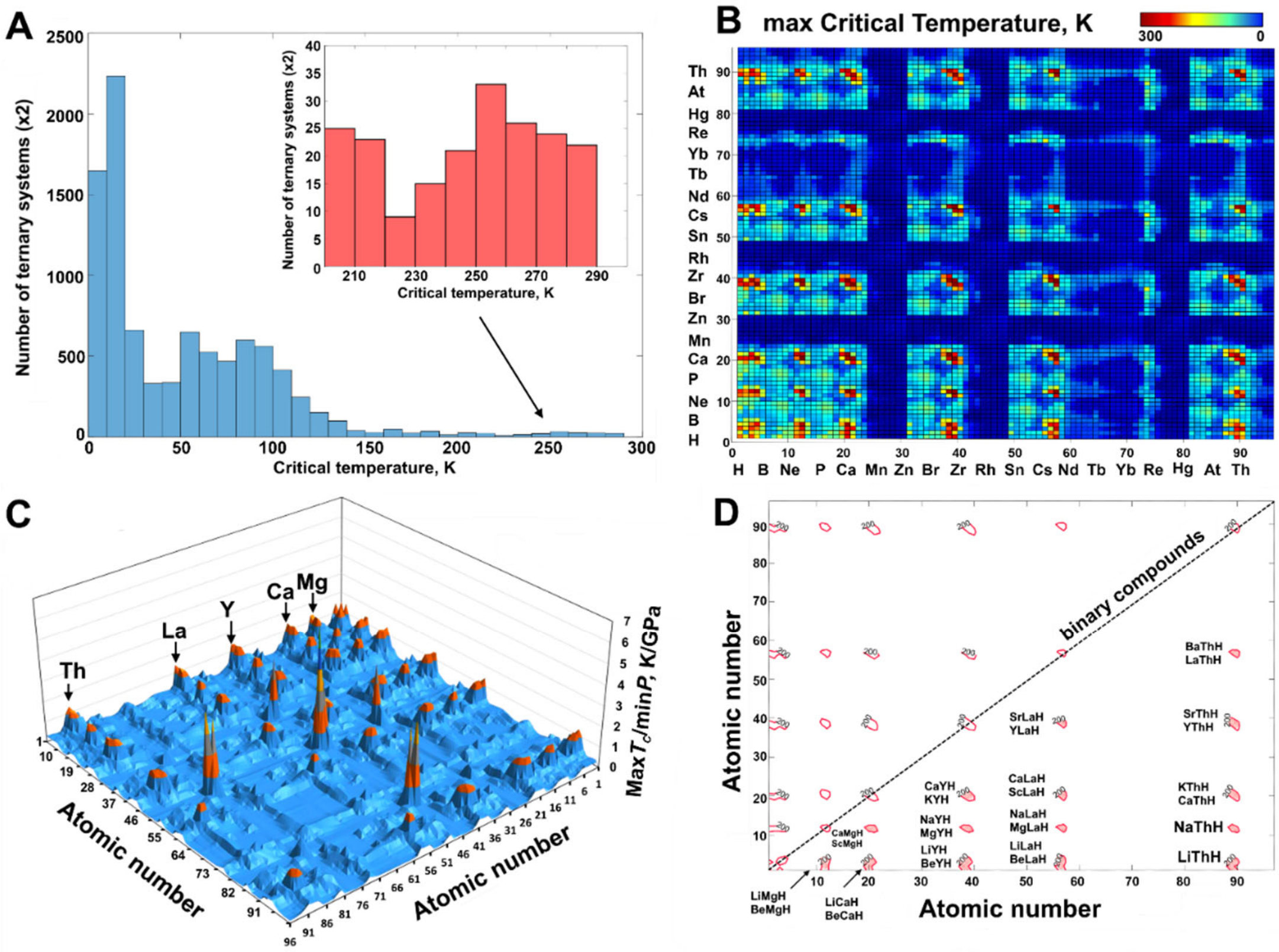

**Figure 68.** A) Number of ternary systems as a function of the critical temperature; B) map of the maximum $T_C$ over the ternary hydrides; C) distribution of the max$T_C$ / min$P$ ratio showing the efficiency of applied pressure to certain systems; D) map of the most promising high-$T_C$ ternary hydrides.

The most interesting parameter predicted using this neural network is the critical temperature of superconductivity ($T_C$). The number of interesting high-$T_C$ ternary hydrides (like $LaH_{10}$ or better) is quite small (Figure 68B-D). More than 95% of ternary hydrides have $T_C$ smaller than 150 K (Figure 68A). The total number of promising systems with $T_C > 250$ K is about 50-60, see Table 13.

Another important parameter is the max$T_C$ / min$P$ ratio showing the efficiency of the pressure applied to a particular system in order to form a high-$T_C$ superhydride (Figure 68C, the points corresponding to min $P \sim 0$ are discarded). All promising regions are related to the combinations of elements from $d^0$, $d^1$ and $d^2$-belts of the Periodic Table. The maximum predicted values of max$T_C$ / min$P$ are 1.3-1.5 K/GPa. Other peaks related to the ternary hydrides of Cu, Zn, Rh, Pd, Au, and other elements mainly come from the existence of lower hydrides stable at almost zero pressure. Such compounds do not possess high-$T_C$ superconductivity.

The most promising directions of further studies of high-$T_C$ hydrides are listed in Table 13. For instance, it was found that ternary hydrides containing Th and an alkali metal (e.g., Li, Nb or K) have better $T_C$ than binary hydrides of $ThH_{10}$ [12,13] or alkali metals, and in fact, we expect for these ternary hydrides $T_C$ comparable with that of $LaH_{10}$ [4,5] at pressures lower than 200 GPa. On the other hand, the most efficient way to increase $T_C$ in $LaH_{10}$ is to have ternary hydrides, where Li, Mg, Ca and Y play the role of the second metal, which could potentially lead to $T_C$ of around 285 K. The results obtained using our neural network show that ternary hydrides of Y, where the role of the second metal is played by Li, Mg or Ca, will have $T_C$ of ~290 K, which is better than $T_C$ of $YH_6$ [30].

**Table 13.** Most promising high-$T_C$ ternary A-B-H systems. The hydrides of short-lived radioactive Ac, Pa and Pm were discarded. The systems suggested for the experimental synthesis are underlined.

| Entry No | First atom (A) | Second atom (B) | Expected $T_C$, K | Expected stabilization pressure, GPa | Expected H-content (per A+B) |
|---|---|---|---|---|---|
| 1 | Th | Li | 257 | 195 | 10 |
| 2 | | Na | 265 | 196 | 9.5 |
| 3 | | K | 258 | 195 | 9.5 |
| 4 | | Ca | 265 | 209 | 10.5 |
| 5 | | Sr | 271 | 208 | 10.5 |
| 6 | | Y | 240 | 214 | 11 |
| 7 | | Ba | 260 | 206 | 10.5 |
| 8 | | La | 276 | 213 | 11.5 |
| 9 | Ra | Sc | 251 | 203 | 10 |
| 10 | | Y | 229 | 201 | 10 |
| 11 | | Zr | 257 | 207 | 10.5 |
| 12 | La | Li | 267 | 190 | 9 |
| 13 | | Be | 250 | 204 | 9.5 |
| 14 | | Na | 259 | 190 | 9 |
| 15 | | Mg | 281 | 205 | 9.5 |
| 16 | | K | 228 | 189 | 9 |
| 17 | | Ca | 280 | 204 | 10 |
| 18 | | Sc | 271 | 210 | 10.5 |
| 19 | | Sr | 268 | 203 | 10 |
| 20 | | Y | 286 | 209 | 10.5 |
| 21 | Ba | Be | 253 | 199 | 9 |
| 22 | | Mg | 241 | 199 | 9 |
| 23 | | Sc | 275 | 204 | 10 |
| 24 | | Y | 268 | 203 | 10 |
| 25 | | Zr | 265 | 208 | 10.5 |
| 26 | Y | Li | 269 | 188 | 9 |
| 27 | | Be | 231 | 203 | 9.5 |
| 28 | | Na | 269 | 189 | 9 |
| 29 | | Mg | 280 | 204 | 9.5 |
| 30 | | K | 253 | 189 | 9 |
| 31 | | Ca | 285 | 205 | 10 |
| 32 | | Sc | 248 | 210 | 10.5 |
| 33 | Sr | Be | 278 | 199 | 9 |
| 34 | | Mg | 274 | 200 | 9 |
| 35 | | Ca | 260 | 200 | 9 |
| 36 | | Sc | 282 | 205 | 10 |
| 37 | Ca | Li | 252 | 182 | 8 |
| 38 | | Be | 280 | 197 | 9 |
| 39 | | Mg | 283 | 199 | 9 |
| 40 | Sc | Li | 247 | 186 | 9 |
| 41 | | Na | 264 | 188 | 9 |
| 42 | | Mg | 264 | 203 | 9.5 |
| 43 | | K | 254 | 188 | 9 |
| 44 | | Ca | 280 | 205 | 9.5 |
| 45 | Mg | Li | 257 | 180 | 8 |
| 46 | | Be | 271 | 195 | 9 |
| 47 | | Na | 237 | 182 | 8 |

It is necessary to say a few words about the limitations of this approach to predicting the properties of ternary hydrides. Every month, 3-4 new papers with ab initio calculations of ternary metal-hydrogen systems appear. This leads to the need for regular retraining of existing neural networks and reanalysis of their predictions. For this reason, it makes sense to highlight only the most promising direction in the study of ternary polyhydrides, without dwelling on details, which change with the emergence of more and more new data. Another point is that the constructed neural network does not take into account the crystal symmetry of compounds, their thermodynamic and dynamic stability, whereas all calculations from the literature are made using different methods, with different pseudopotentials, accuracy, and at different pressures. To train the neural network, a large amount of homogeneous data obtained under the same conditions is required. To date, out of more than 4,500 ternary hydrides, the DFT methods have been applied to only about 100 of them. Under these conditions, we cannot speak about any accuracy of the neural network predictions. Therefore, the results obtained should be treated with caution.

## 6.3 Concluding remarks and ways to increase the critical temperature

Many experts in the field of superconductivity have noticed that the cuprate superconductors and the best hydride superconductors have the same elements in their composition. This opens up the possibility of a statistical analysis based on the known literature (the database [287] was used), assuming that the systems with the best superconducting properties have always attracted the most attention of researchers. As can be seen from Figures 69-70, such elements as Be, Mg, Sc, alkali metals, Ti, Zr, Hf, have not attracted much attention of researchers, whereas Ca, Sr, Ba, La, Y, Ce, Bi and As - are the most important elements for superconductor design. The same situation is observed for polyhydrides - all attention is focused on the "superconductivity belt" of $d^0$-$d^2$ elements.

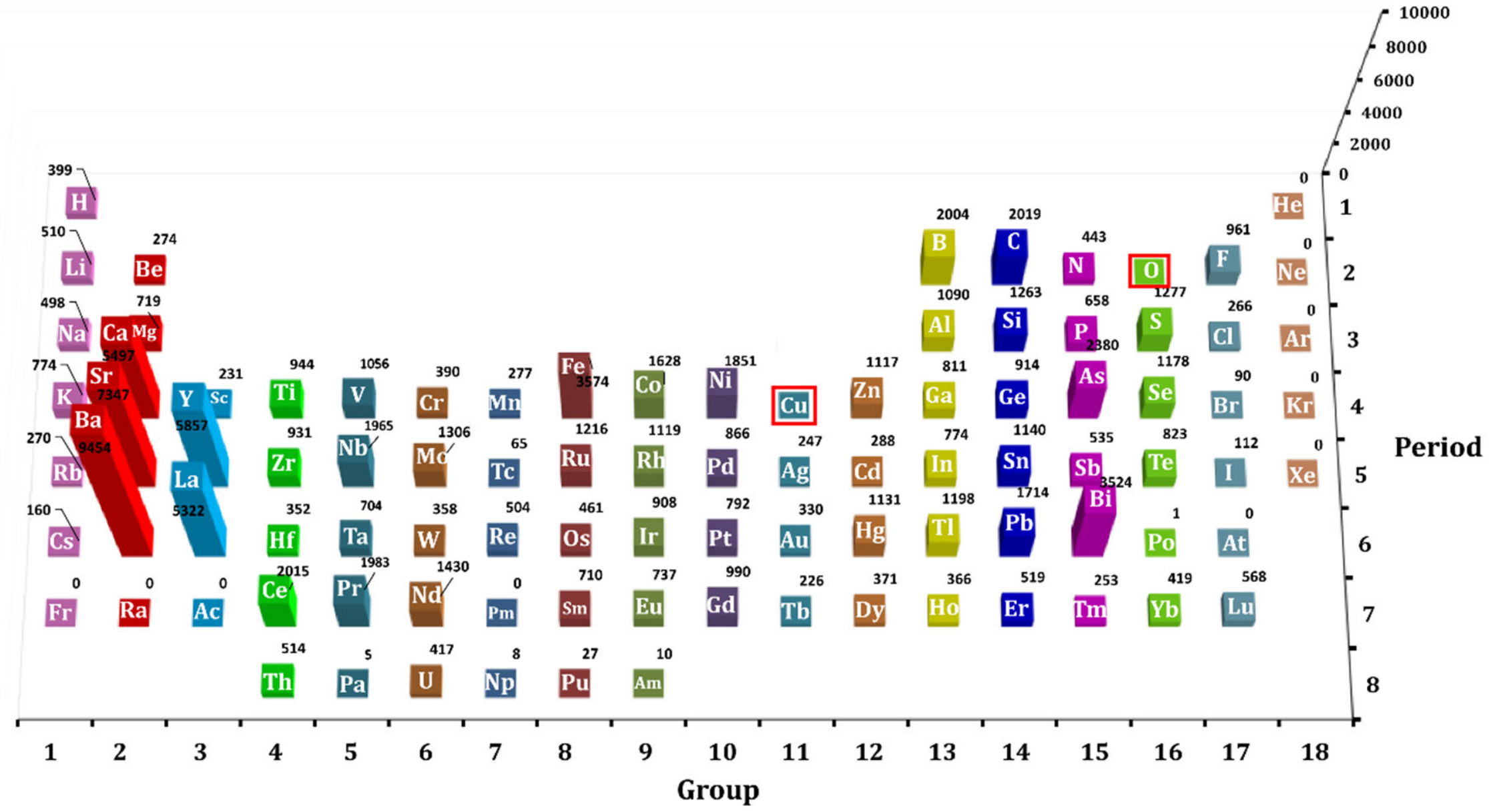

**Figure 69.** Number of entries containing the selected element of the Mendeleev table in the database of superconducting compounds [300]. Data for copper and oxygen are not shown, since these elements are present in all compounds of the cuprate series.

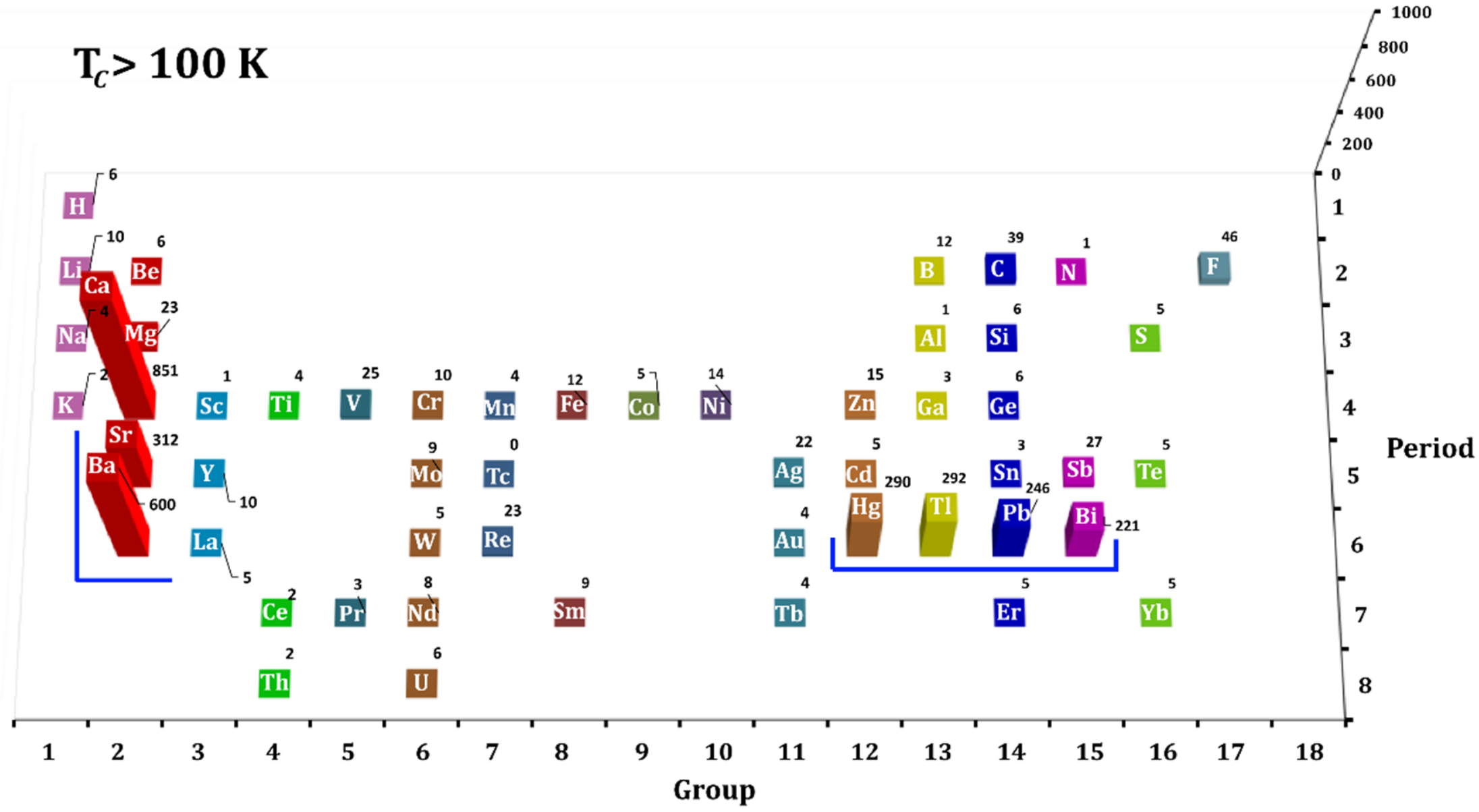

**Figure 70.** Statistics of records (compounds) with the critical temperature $T_C$ > 100 K that contain the selected element of the periodic table in the database [300].

A set of the most used elements for obtaining superconducting materials with different mechanisms of superconductivity has significant similarities with the set of elements from the "superconductivity belt". Analysis of the $T_C$ distribution in ternary hydrides using neural networks (see paragraph 6.2) also shows that the most promising are the ternary hydrides constructed from the elements of the "superconductivity belt". Thus, I believe that the synthesis of ternary hydrides of $d^0$-$d^2$ elements, despite the initial failures with solid solutions, is still one of the most promising directions in hydride superconductivity.

Another interesting direction to increase the critical temperature is the construction of hybrid structures containing covalent and metal bonds simultaneously (e.g., the Y-Te-H and Y-Bi-H systems), atomic and molecular hydrogen (which increases $\omega_{log}$), the introduction of polymer chains of light atoms to stabilize molecular hydrogen (Figure 71), and condensed fullerenes with metal atoms inside, which are full analogues of clathrate polyhydrides [301-303]. In the latter case, the light atoms solve the problem of filling the intermediate frequency region in the Eliashberg function between the low-frequency zone of the metal sublattice and the high-frequency vibrations of the hydrogen sublattice (Figure 71).

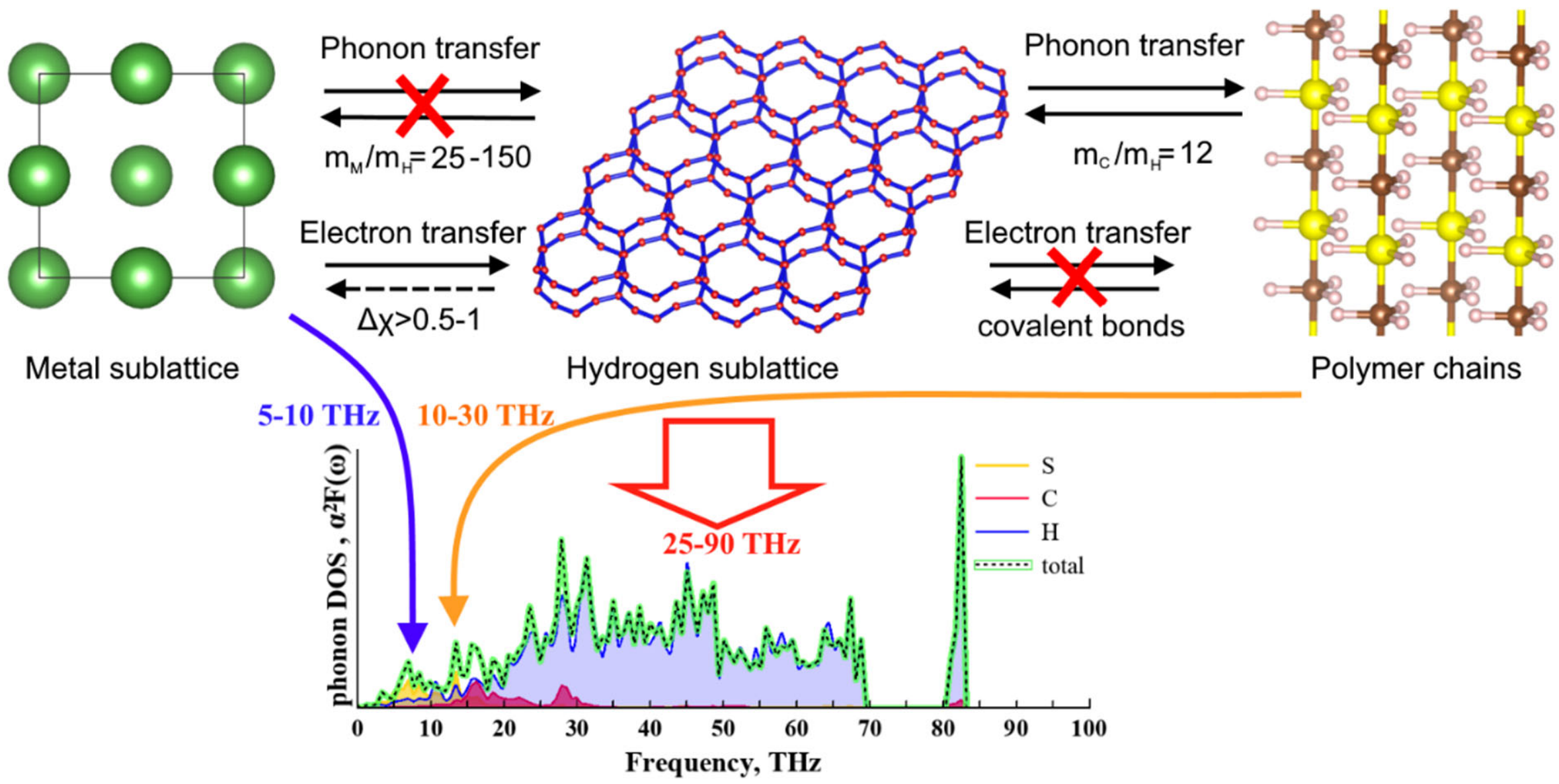

**Figure 71.** Scheme of the possible electron-phonon interaction of the metal sublattice, molecular hydrogen layers, and polymer chains of light atoms (-$CH_3$-$SH_3$-) in the formation of the Eliashberg function.

Tremendous recent progress in achieving near-room temperature superconductivity is primarily due to the use of a new material, metallic hydrogen, stabilized in superhydrides at high pressures. Using the elementary approach $T_C \sim m_A^{0.5}$, it is possible to achieve $T_C$ = 29 K and more in simple metals, for instance, in calcium at 216 GPa [304]. Considering hydrogen as an ordinary metal, $T_C(H) \sim T_C(Ca) \times m_{Ca}^{0.5}$ = 183 K, not so far from $T_C$ in polyhydrides. It means that, by the order of magnitude, superconductivity in polyhydrides is limited just as it is limited in ordinary metals. Considering a sequential pressure rise for simple metals such as La, it is easy to verify by direct calculation that $\lambda(P \rightarrow \infty)$ is 0. For example, for pure *fcc* lanthanum under pressure: $\lambda$(100 GPa) ≈ 1, $\lambda$(130 GPa) ≈ 0.65 [259], whereas at 1000 GPa we have $\lambda$(1 TPa) = 0.17, $\omega_{log}$ = 870 K, and $T_C$ = 0 K (see also a similar work on superconductivity in sulfur [305]). The idea that superconducting polyhydrides do not differ in their properties from ordinary metals is a red thread running through my thesis, so there must be

**Asymptotic hypothesis:** $T_C(P \rightarrow \pm\infty) = 0$ for any material. The superconductivity temperature of any material tends to zero as the pressure increases starting from some pressure value.

Moreover, in 2022, in a remarkable work by E. Yuzbashyan and B. Altshuler [306], the limitation of maximum possible values of the electron-phonon interaction parameter ($\lambda_{max} = 3 - 3.7$) has been proved in superconducting materials that can be described within the Migdal-Eliashberg theory. In principle, this limits the critical temperature $T_C$ of the conventional phonon-assisted superconductivity that can be achieved in nature because the averaged phonon energy, the Debye temperature (as well as $\omega_{log}$) is also limited by the fundamental proton/electron mass ratio $m_p/m_e$ [307]. Considering that the physical density of polyhydrides under pressure does not differ much from that of pure metals (as was already discussed) and that the speed of sound cannot exceed 36100 m/s [307], we obtained an estimate of the maximum value of $\omega_{log}$ = 2000–2500 K, which leads to an estimate for the maximum possible critical temperature of superconductivity max $T_C$ = 600–900 K (see the improved Allen–Dynes formula in Appendix). This estimate is still far from practical use because it does not take into account the negative correlation between $\lambda$ and $\omega_{log}$, which does not let these two parameters take their maximum values simultaneously. Nevertheless, the importance of these works is to confirm the fundamental possibility of achieving room-temperature superconductivity in polyhydrides with reachable electron–phonon interaction parameters: $\lambda \approx 3$ and $\omega_{log} \approx 1000$ K.

The estimates of max $T_C$ can be improved by applying various models for the Eliashberg function that allow us to establish a direct relationship between $\lambda$ and $\omega_{log}$. An example of such a model is the rectangular Eliashberg function of polyhydrides: $\alpha^2F(\omega) = \text{const.} = a$ in the range $\omega \subset (\omega_1, \omega_2)$, and $\alpha^2F(\omega) = 0$ outside this range. The frequency $\omega_1$ corresponds to the vibration frequency of the metal sublattice (5–10 THz), whereas $\omega_2$ corresponds to the vibration frequency of the hydrogen sublattice (≈ 60 THz for many superhydrides). In this case, $\omega_{log} = (\omega_1\omega_2)^{0.5}$, $\lambda = 2a\cdot\ln(\omega_2/\omega_1) < 3–3.7$, or $a < 1$ [308]. Accepting maximum $a = 1$ and expressing $\omega_{log}$ through $\lambda$ (as a variable we consider soft modes, $\omega_1$, responsible for phase transitions in polyhydrides), we get $\omega_{log} = \omega_2\cdot\exp(-\lambda/4a) = \omega_2\cdot\exp(-\lambda/4) \approx 2900\exp(-\lambda/4)$ [308]. This leads to the estimate of the maximum achievable critical temperature of superconductivity max $T_C \sim 400$ K, which calls into question some of recent studies in the field of hydride superconductivity [32, 221].

A few words about possible applications of superconducting polyhydrides. At the moment, it is impossible to create more or less long polyhydride wires (> 100-200 μm) when synthesizing on diamond anvils. Therefore, the only application area open to polyhydrides is their use as sensors. First of all, we are talking about the application of micron samples of polyhydrides for photon counting as single-photon detectors operating in a wide range of incident light wavelengths. Directly, the scheme for implementing such a detector is reduced to adding a focusing lens to the diamond cell design, cooling the system below the critical temperature, and using a near-critical current with detection of the resulting potential difference in a 4-contact circuit (Figure 72a). When a single photon hits the polyhydride sample, it is absorbed by the highly dispersed hydride powder with destruction of the superconducting state and the appearance of a finite potential difference between the electrodes.

Another application of superhydrides is their use for the design of SQUID magnetometers on diamond anvils that do not require deep cooling. However, because of the small loop size (d~20-40 μm, Figure 72c) such a SQUID will not have a record sensitivity ($\Phi_0 \times S \approx 6$ μT). Narrow superconducting transitions ($\Delta T_C$ = 3-4 K) also allow detection of changes in external parameters: temperature, magnetic field and current, by recording the electrical resistance in 10 μΩ steps (typical resistance noise level). With this approach, we can expect to confidently detect changes in the magnetic field $\Delta H = 10^{-5}$ T and in the temperature of $5\times10^{-5}$ K. Temperature detection is of greatest interest, because in this case the superconductor can be used to study the thermal effects of phase transitions in DACs. Such a sensor can be placed directly on a diamond anvil, next to the object under study. Despite the narrowness of the phase transition region, it can be shifted to the desired region by using higher electric current, applying an external magnetic field, changing the pressure or by introducing magnetic impurities into the superconductor.

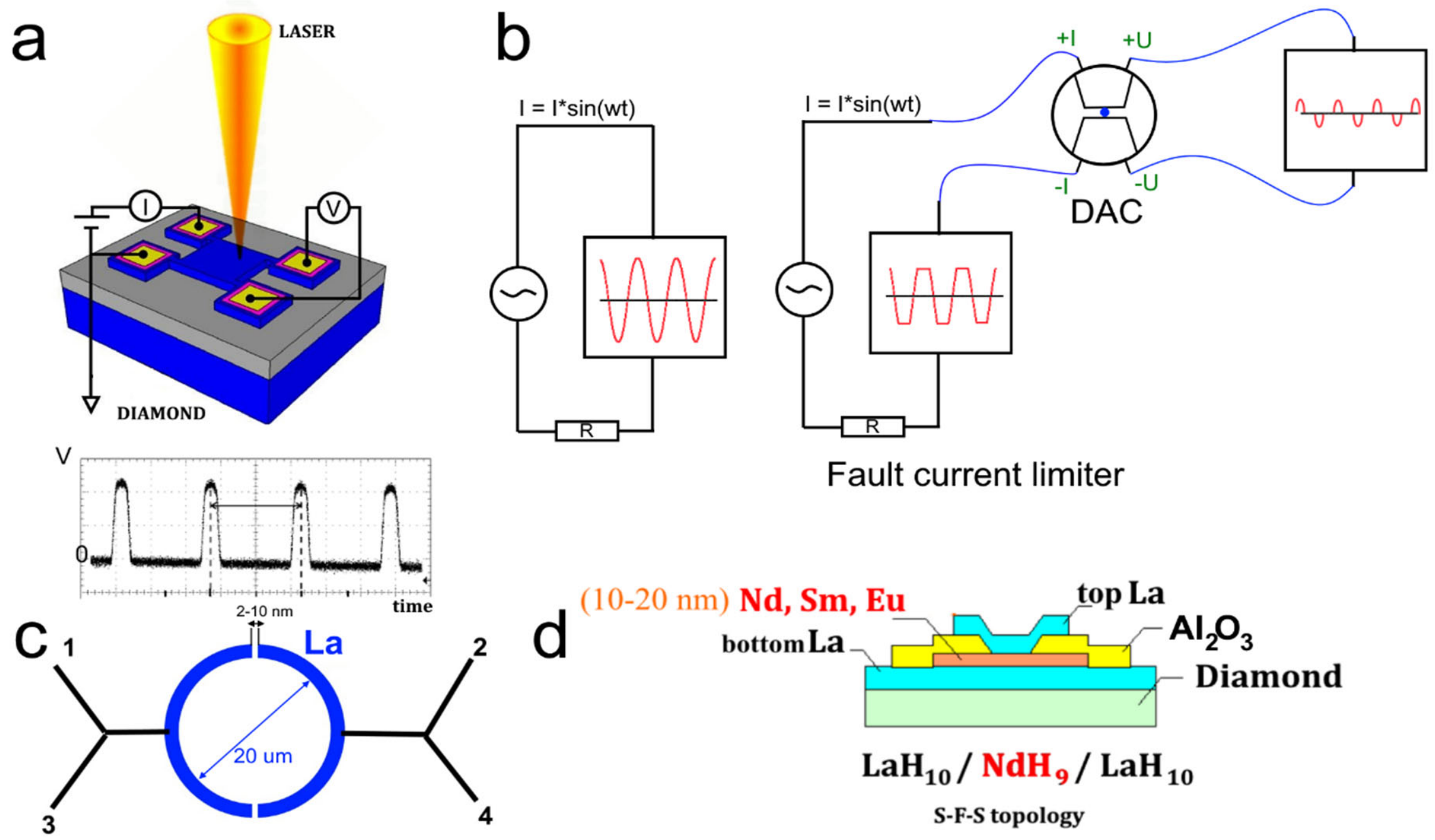

**Figure 72**. Possible applications of superhydrides as (a) single-photon detectors; (b) current limiters and nonlinear electronic elements; (c) SQUID magnetometers and magnetic field sensors; (d) gates controlled by magnetic field in SFS contacts.

Superconductors with high $T_C$, a sharp superconducting transition, and a low upper critical magnetic field are very important for applications. For such materials we can control the SC transition by readily available magnetic fields. Alternatively, such control can be achieved by fabricating S-F-S contacts (Figure 72d), where a magnetic superhydride layer ($GdH_x$ or $EuH_x$) will be placed between thin layers of superconducting hydrides. Such contacts in high-pressure diamond cells can also be used as a nonlinear electronic element, together with the use of superhydrides as current limiters (Figure 72b).

The problem here, however, will be the large resistance of the feeding wires (~ 60-80 Ω) compared to the resistance of the sample itself (0.1-1 Ω). This factor also limits the study of electrochemical processes in high-pressure DACs: in order to apply a potential difference of 0.5-1 V to the sample, a current of at least 0.5-1 A should pass through the wires, which is fraught with destruction of copper contacts.

## 6.4 Conclusions of Chapter 6

1. Most of the high-temperature superconducting metal hydrides are concentrated in the "superconducting belt" of the periodic table. The metals forming the high-$T_C$ hydrides are Sc-Y-La-Ac ($s^2d^1$ elements), Mg-Ca-Sr-Ba-Ra ($s^2d^0$ elements), Ce ($s^2f^1d^1$), and Zr-Hf-Th ($s^2d^2$).

2. The optimal hydrogen content in superconductors corresponds to $XH_n$ ($n = 10 \pm 4$), a composition that can be formed at pressures of about 150–250 GPa. Both $T_C(P)$ and $T_C(H)$ are dome-shaped dependencies. If the composition deviates from the optimal values, the superconductivity parameters rapidly decrease.

3. The superconducting properties of hydrides greatly diminish as the number of *d*- and *f*-electrons increases because of electron scattering on anisotropic centers (metal atoms) with a change in spin and appearance of magnetic ordering.

4. The pressure required for the stabilization of polyhydrides decreases going down the periodic table along with a decrease in electronegativity and an increase in the atomic radius. The importance of the averaged phonon energy makes it difficult to reduce the pressure without losing the SC properties of thermodynamically stable hydrides. Numerically, $T_C$ [K] ~ $P$ [GPa] and does not exceed $2P$ [GPa].

5. The rules of B. Matthias [284, 309], developed for the search of low-temperature BCS superconductors, remain valid for polyhydrides:

- The best superconducting polyhydrides have cubic and hexagonal crystal structures.
- High density of electronic states, which does not go into spin splitting and magnetization, contributes to superconductivity in hydrides. The electronic density of states at the Fermi level $N_F$ of the highest-$T_C$ superconducting hydrides is ~1 state/Ry/H atom.
- Magnetic atoms suppress superconductivity. Lanthanide polyhydrides are mostly non-superconducting.
- Molecular polyhydrides are bad superconductors.

6. The study of superconductivity in ternary hydrides shows that the most interesting properties should be exhibited by compounds composed of elements of the "superconductivity belt". The most promising are cubic and hexagonal ternary hydrides with a hydrogen content expressed as $(X,Y)H_n$ ($n = 10 \pm 4$), obtained by regular substitution of metal atoms in clathrate binary superhydrides at pressures of 200-250 GPa.

# Conclusions

Over the six years of research (2015-2021) following the prediction and subsequent experimental discovery of the unique properties of $H_3S$ [5], polyhydrides have become an established new class of superconducting materials with record critical parameters. There is no doubt that many more exciting discoveries await us in this field. Hydrogen is an ideal element for the realization of high-temperature superconductivity with the electron-phonon mechanism of coupling. We just need to find polyhydrides that require minimal pressure for their stabilization and maintain record-high critical temperatures. Ternary hydrides of metals from the "superconductivity belt" and hybrid metastable materials that combine both covalent bonds and a metal-stabilized hydrogen sublattice have a great potential in this regard.

The progress in the field of hydride superconductivity would not have been so fast and bright without the well-developed Bardeen-Cooper-Schrieffer-Migdal-Eliashberg strong electron-phonon interaction theory, evolutionary search methods for thermodynamically stable crystal structures (such as USPEX), and convenient Quantum ESPRESSO and EPW packages for ab initio calculations of the critical parameters of superconductors. The developers of the SSCHA package are noted for their contribution to understanding the importance of anharmonic effects in polyhydrides at ultrahigh pressures.

In this thesis, we tried to emphasize the importance and productivity of the interpenetration of theoretical calculations and experiments with DACs, which complement each other. We always use experiment to better understand theory, and theory to better describe experiment.

Analyzing the data of the measurements in high-pressure diamond anvil cells in Chapter 1, we found that polyhydrides have all the properties of conventional BCS superconductors: the isotope effect, a sharp drop in the electrical resistance at the transition temperature, and $T_C$ dependence on the external magnetic field, current, and presence of magnetic impurities. However, given the very limited set of experimental techniques available for investigating matter at extreme pressures, the main parameters of the superconducting state of hydrides — the EPC parameter $\lambda$, superconducting gap $\Delta$, Eliashberg function $\alpha^2F(\omega)$, and lower critical magnetic field $\mu_0H_{C1}$ — have yet to be determined.

The investigation of thorium polyhydrides at high pressures (Chapter 2), which played an important role in the formation of the Moscow high-pressure collaboration (Crystallography Institute RAS, Skolkovo Institute, and Lebedev Physical Institute), showed that high-temperature superconductivity in polyhydrides is not a unique property of $H_3S$ [5], discovered in 2015, but is inherent to a wide class of compounds. Our work allowed us to form an idea of individual families of polyhydrides ($XH_{10}$, $XH_9$, $XH_6$, $XH_4$), which were subsequently augmented by other members. Here, for the first time, we also encountered deviations of the experimental superconducting properties ($T_C$) from their predicted values.

The study of the Y-H system at high pressures (Chapter 3) showed that despite the similarity of the chemical properties of lanthanum and yttrium, their interaction with hydrogen at high pressures proceeds differently. Instead of La-type hydrides, such as $XH_{10}$, we observed formation of *Im*-3*m*-$YH_6$ and *P*6$_3$/*mmc*-$YH_9$. In our work, we pointed out the importance of considering anisotropy in calculations of the critical parameters of hexagonal polyhydrides ($YH_9$, $CeH_9$), as well as the anomalous properties of $YH_6$: an extremely high upper critical magnetic field and unexpected suppression of superconductivity, which cannot be explained within the known models.

In Chapter 4, we characterized a group of magnetic lanthanide superhydrides (Pr, Nd, Eu) which, despite the presence of a metallic hydrogen sublattice, do not display superconducting properties (because of the scattering of Cooper pairs on the magnetic moments of atoms) but can exhibit magnetic ordering at low temperatures. Lanthanide hydrides are extremely difficult for theoretical studies because of the need to take into account the strong correlations in the *f*-shell electrons. This group of metals is very convenient for controlling the properties of superconductors. Small additions of lanthanides make it possible to significantly reduce the critical temperature of superconductivity where necessary. In addition, on the basis of lanthanides (e.g., Gd), it is possible to create S-F-S interfaces that are very sensitive to an external magnetic field.

In addition to binary hydrides, in Chapter 5 we considered the properties of ternary polyhydrides of the La-Y-H, La-Nd-H, and La-Ce-H systems. It turned out that the synthesis of ternary hydrides often leads to formation of solid solutions in the metal sublattice, which have "trivial" properties and obey Anderson's theorem for superconductors. Anderson's theorem, which states that small deviations from the ideal stoichiometry (e.g., $LaH_{10}$) do not affect the value of the critical temperature of superconductivity, can also be applied to polyhydrides of nonstoichiometric composition (e.g., $LaH_{10\pm1}$).

There are at least two scenarios in which the introduction of an additional atom into a binary hydride leads to new results: (1) stabilization of "impossible" compounds that cannot be obtained in binary metal-hydrogen systems, and (2) formation of true ternary polyhydrides, in case the properties of the two metals differ greatly from each other.

In the last part of this thesis, we answered the key question of the materials science of superconducting materials: what elements, in what crystal structure, and under what conditions should be combined to obtain a superconductor with the best properties. For binary and probably ternary hydrides the question is solved by introducing the so-called "superconductivity belt" – a series of $d^0$-$d^2$ elements (13 metals: Mg, Ca, Sr, Ba, Sc, Y, La, Th, Ce, Ti, Zr, Hf, and Lu) which, by themselves and in combinations, form high-temperature superconducting polyhydrides at 150-250 GPa. These polyhydrides with a cubic or hexagonal crystal structure should contain from 6 to 14 hydrogens per metal atom and have an electronic state density of about 1 $Ry^{-1}$ per H atom. This reduces the search space for the best ternary hydrides from >4600 systems to about 60-70 most promising ones that are expected to be investigated experimentally.

Within this work, we developed a useful algorithm for analyzing experimental and theoretical data, which makes it possible to extract maximum information from published transport measurements for superhydrides.

# Appendix

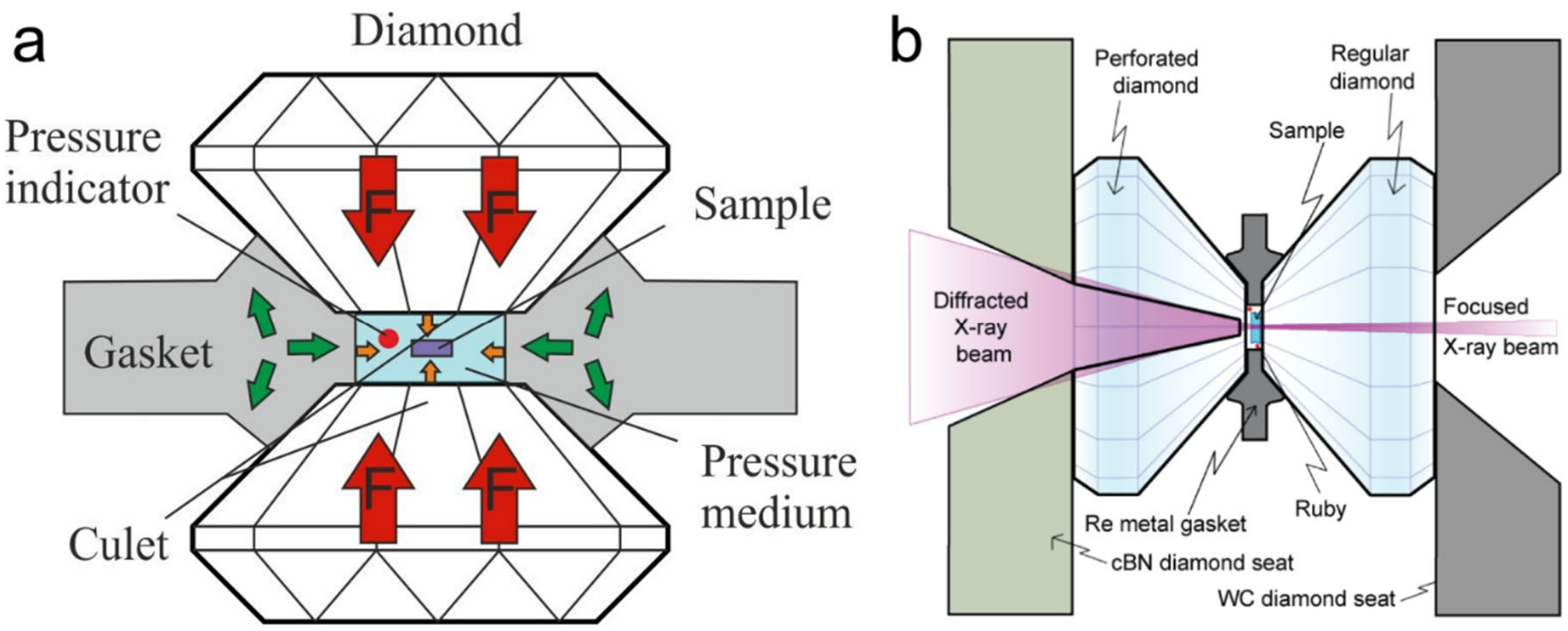

**Figure A1.** High-pressure diamond anvil cell design and X-ray diffraction study scheme.

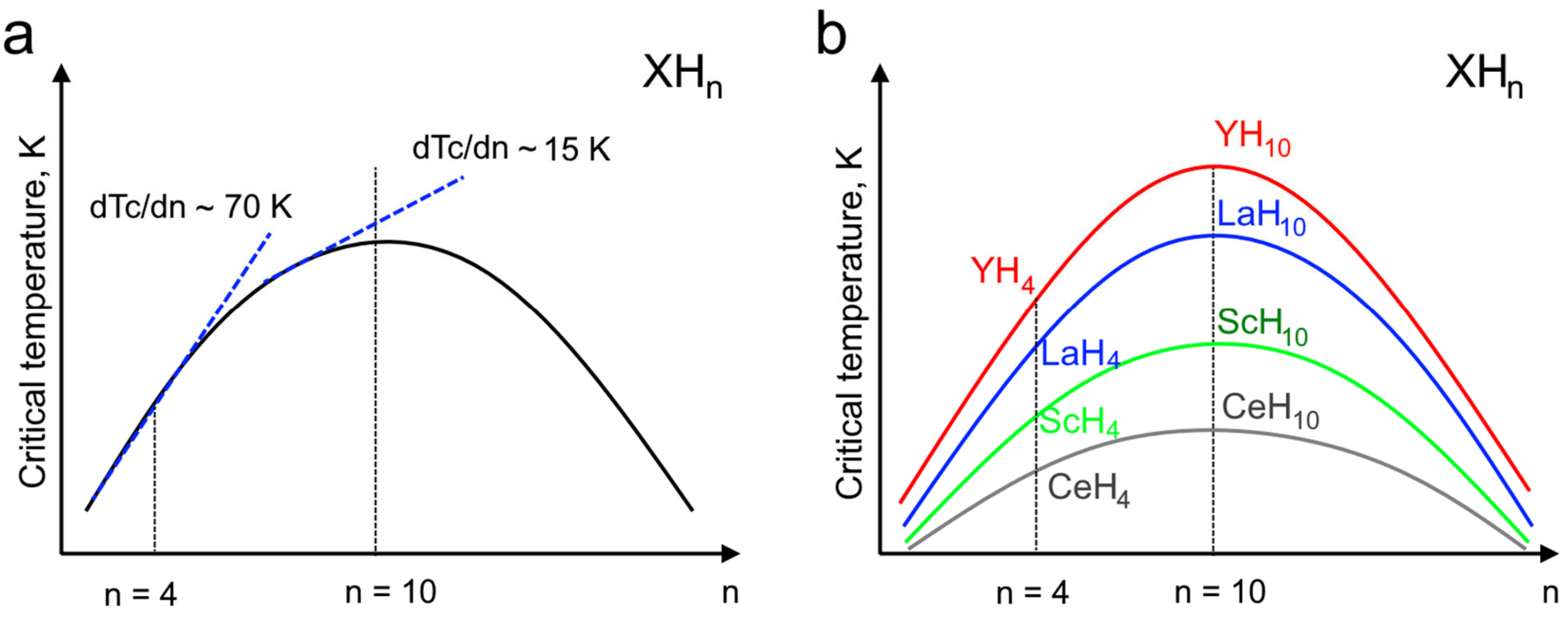

**Figure A2.** Illustrations (qualitative) for the dependence of the critical temperature of binary hydrides on the number of hydrogen atoms in the chemical formula. (a) Bell-shaped $T_C$(n) dependence, which illustrates the impossibility of a significant increase in $T_C$ when going from $XH_9$ to $XH_{10}$ superhydrides. (b) Illustration of the absence of $T_C$(n) intersections for different metals. This idea allows one to search for the most promising superhydrides by comparing their lower hydrides at moderate pressures.

**Calculations of the basic parameters of the superconducting state.** To calculate the isotope coefficient α, the Allen–Dynes interpolation formulas were used:

$$\beta_{McM} = -\frac{d \ln T_C}{d \ln M} = \frac{1}{2}\left[1 - \frac{1.04(1+\lambda)(1+0.62\lambda)}{[\lambda - \mu^*(1+0.62\lambda)]^2}\mu^{*2}\right] \quad \text{(A1)}$$

$$\beta_{AD} = \beta_{McM} - \frac{2.34\mu^{*2}\lambda^{3/2}}{(2.46+9.25\mu^*)\cdot((2.46+9.25\mu^*)^{3/2}+\lambda^{3/2})} - \frac{130.4\cdot\mu^{*2}\lambda^2(1+6.3\mu^*)\left(1-\frac{\omega_{\log}}{\omega_2}\right)\frac{\omega_{\log}}{\omega_2}}{\left(8.28+104\mu^*+329\mu^{*2}+2.5\cdot\lambda^2\frac{\omega_{\log}}{\omega_2}\right)\cdot\left(8.28+104\mu^*+329\mu^{*2}+2.5\cdot\lambda^2\left(\frac{\omega_{\log}}{\omega_2}\right)^2\right)} \quad \text{(A2)}$$

where the last two correction terms are usually small (~0.01).

The superconducting transition temperature $T_C$ was estimated using the Allen–Dynes formula in the following form:

$$T_C = \omega_{\log}\frac{f_1 f_2}{1.2}\exp\left(\frac{-1.04(1+\lambda)}{\lambda - \mu^* - 0.62\lambda\mu^*}\right) \quad \text{(A3)}$$

where

$$f_1 f_2 = \sqrt[3]{1+\left(\frac{\lambda}{2.46(1+3.8\mu^*)}\right)^{\frac{3}{2}}}\cdot\left(1-\frac{\lambda^2(1-\omega_2/\omega_{\log})}{\lambda^2+3.312(1+6.3\mu^*)^2}\right) \quad \text{(A4)}$$

The EPC constant λ, logarithmic average frequency $\omega_{\log}$, and mean square frequency $\omega_2$ were calculated as

$$\lambda = \int_0^{\omega_{\max}} \frac{2\alpha^2 F(\omega)}{\omega} d\omega \quad \text{(A5)}$$

$$\omega_{\log} = \exp\left(\frac{2}{\lambda}\int_0^{\omega_{\max}}\frac{d\omega}{\omega}\alpha^2 F(\omega)\ln(\omega)\right)$$

$$\omega_2 = \sqrt{\frac{1}{\lambda}\int_0^{\omega_{\max}}\left[\frac{2\alpha^2 F(\omega)}{\omega}\right]\omega^2 d\omega} \quad \text{(A6)}$$

where $\mu^*$ is the Coulomb pseudopotential, for which we used widely accepted lower and upper bounds of 0.10 and 0.15.

Recently, new formulas based on machine learning, taking into account new data obtained for high-temperature superconducting hydrides, have been proposed [A1]. For example, instead of the coefficients $f_1$, $f_2$ (formula A4) the following, more accurate, expressions have been proposed

$$f_\omega = 1.92\left(\frac{\lambda - \sqrt[3]{\mu^*} + \frac{\omega_{\log}}{\omega_2}}{\sqrt{\lambda}\cdot\exp(\frac{\omega_{\log}}{\omega_2})}\right) - 0.08, \quad \text{(A4.1)}$$

$$f_\mu = 6.86\left(\frac{exp(-\frac{\lambda}{\mu^*})}{\frac{1}{\lambda} - \mu^* - \frac{\omega_{log}}{\omega_2}}\right) + 1, \quad \text{(A4.2)}$$

as well as a simple formula for $T_C$, containing only one term [A1]

$$T_C^{SISSO} = 0.0953\frac{\omega_{log}\lambda^4}{\lambda^3+\sqrt{\mu^*}}. \qquad \text{(A4.3)}$$

The Sommerfeld parameter was found as

$$\gamma = \frac{2}{3}\pi^2 k_B{}^2 N(0)(1+\lambda) \qquad \text{(A7)}$$

where $N(0)$ is the total density of electronic states at the Fermi level per spin. It was used to estimate the upper critical magnetic field and superconducting gap using the known semiempirical equations of the BCS theory (see Ref. [A2], eq 4.1 and 5.11), working satisfactorily for $T_C/\omega_{log} < 0.25$:

$$\frac{\gamma' T_C{}^2}{(\mu_0 H_{C_2}(0))^2} = 0.168\left[1 - 12.2\left(\left(\frac{T_C}{\omega_{log}}\right)^2 \ln\left(\frac{\omega_{log}}{3T_C}\right)\right)\right] \qquad \text{(A8)}$$

$$\frac{2\Delta(0)}{k_B T_C} = 3.53\left[1+12.5\left(\frac{T_C}{\omega_{log}}\right)^2 \ln\left(\frac{\omega_{log}}{2T_C}\right)\right] \qquad \text{(A9)}$$

where $\gamma' = 2\gamma$.

Regarding eq A8, the upper critical field naturally depends on the amount of impurities and defects and changes significantly from sample to sample. Only the lower limit for $\mu_0 H_{C2}$, which corresponds to an ideal crystal, can be estimated. However, eq A8, originally designed to calculate the thermodynamic critical field $\mu_0 H_C$ of superconductors, also gives a good estimate of the upper critical field of hydrides if the full DOS expressed per mole of a hydride is used in the Sommerfeld parameter ($\gamma'$). In this case, $\mu_0 H_{C2}$ is expressed in teslas in eq A8.

The lower critical magnetic field was calculated according to the Ginzburg–Landau theory [A3]

$$\frac{H_{C1}}{H_{C2}} = \frac{\ln\kappa}{2\sqrt{2}\kappa^2}, \quad \kappa = \frac{\lambda_L}{\xi} \qquad \text{(A10)}$$

where $\lambda_L$ is the London penetration depth, which can be estimated as

$$\lambda_L = 1.0541\cdot 10^{-5}\sqrt{\frac{m_e c^2}{4\pi n_e e^2}} \qquad \text{(A11)}$$

where $c$ is the speed of light, $e$ is the electron charge, $m_e$ is the mass of an electron, and $n_e$ is the effective concentration of charge carriers expressed via the average Fermi velocity $V_F$ in the Fermi gas model:

$$n_e = \frac{1}{3\pi^2}\left(\frac{m_e V_F}{\hbar}\right)^3 \qquad \text{(A12)}$$

The average Fermi velocity can be estimated as

$$V_F = \frac{\pi\cdot\Delta(0)}{\hbar}\xi \qquad \text{(A13)}$$

where $\xi$ is the coherence length calculated from the experimental upper critical magnetic field as $\xi = \sqrt{\hbar / 2e(\mu_0 H_{C2})}$. The average Fermi velocity can also be calculated directly from the band structure:

$$\langle V_{\mathrm{F}}\rangle = \sqrt{\frac{\sum_k \delta(E_k - E_F){V_k}^2}{\sum_k \delta(E_k - E_F)}} = \frac{a}{\pi\hbar}\sqrt{\frac{\sum_k \delta(E_k - E_F)(\mathrm{d}E_k/\mathrm{d}t)^2}{\sum_k \delta(E_k - E_F)}} \quad \text{(A14)}$$

where $\mathrm{d}E_i/\mathrm{d}k$ ($k = -\pi/a \ldots \pi/a$) was replaced with $\mathrm{d}E_i/\mathrm{d}t$ ($t = -1 \ldots 1$).

It is possible to estimate the scattering time τ and scattering length $L$ in discovered superhydrides in the dirty and strong-coupling limits using the interpolation formula proposed by Carbotte [A2] (eq 7.11)

$$\mu_0 H_{\mathrm{C2}}(0) = \frac{\pi k_{\mathrm{B}} T_{\mathrm{C}}(1+\lambda)}{3.561(eD)}\left[1 + 3.3\left(\frac{T_{\mathrm{C}}}{\omega_{\log}}\right) - 4.8\left(\frac{T_{\mathrm{C}}}{\omega_{\log}}\right)^2 \ln\left(\frac{\omega_{\log}}{T_{\mathrm{C}}}\right)\right] \quad \text{(A15)}$$

where $e$ is the electron charge, $D = 1/3eV_{\mathrm{F}}^2\tau$ is the diffusion coefficient, and τ is the scattering time ($\tau \sim L/V_{\mathrm{F}}$).

To process the results of the measurements of the critical temperatures in external magnetic fields, we used the Werthamer–Helfand–Hohenberg (WHH) model simplified by Baumgartner et al. [A4]

$$\mu_0 H_{C2}(T) = \frac{\mu_0 H_{C2}(0)}{0.693}\left(\left(1-\frac{T}{T_C}\right) - 0.153\cdot\left(1-\frac{T}{T_C}\right)^2 - 0.152\cdot\left(1-\frac{T}{T_C}\right)^4\right) \quad \text{(A16)}$$

More accurate fit to the WHH model for $\mu_0 H_{\mathrm{C2}}(T)$ was applied to validate the results of the approximate formula of Baumgartner et al. (eq A16):

$$\ln\frac{T_{\mathrm{C}}}{T} = \sum_{k=-\infty}^{\infty}\left\{\frac{1}{|2k+1|} - \frac{1}{|2k+1| + h\frac{T_{\mathrm{C}}}{T} + \left(h\frac{T_{\mathrm{C}}}{T}\right)^2 \frac{\alpha^2}{|2k+1| + (h+\lambda_{\mathrm{SO}})\frac{T_{\mathrm{C}}}{T}}}\right\} \quad \text{(A17)}$$

$$h = -\frac{4}{\pi^2}\frac{B_{\mathrm{C2}}(T)}{T_{\mathrm{C}}(\mathrm{d}B_{\mathrm{C2}}/\mathrm{d}T)_{T\mathrm{c}}}$$

where α is the Maki parameter which is related to the spin-limited critical field, and $\lambda_{\mathrm{SO}}$ is the spin–orbit interaction parameter.

The critical temperature of the superconducting transition was calculated using the Matsubara-type linearized Eliashberg equations [A5]:

$$\hbar\omega_j = \pi(2j+1)k_{\mathrm{B}}T, \qquad j = 0, \pm1, \pm2, \ldots \quad \text{(A18)}$$

$$\lambda(\omega_i - \omega_j) = 2\int_0^\infty \frac{\omega\alpha^2 F(\omega)}{\omega^2 + (\omega_i - \omega_j)^2}\mathrm{d}\omega \quad \text{(A19)}$$

$$\Delta(\omega = \omega_i, T) = \Delta_i(T) = = \pi k_{\mathrm{B}} T \sum_j \frac{\left[\lambda(\omega_i - \omega_j) - \mu^*\right]}{\rho + \left|\hbar\omega_j + \pi k_{\mathrm{B}} T \sum_k (\mathrm{sign}\,\omega_k)\cdot\lambda(\omega_i - \omega_j)\right|}\cdot\Delta_j(T) \quad \text{(A20)}$$

where $T$ is the temperature in kelvins, $\mu^*$ is the Coulomb pseudopotential, ω is the frequency in hertz, $\rho(T)$ is the pair-breaking parameter, the function $\lambda(\omega_i - \omega_j)$ is related to an effective electron–electron interaction via the exchange of phonons [A6]. The transition temperature can be found as the solution of the equation $\rho(T_{\mathrm{C}}) = 0$, where $\rho(T)$ is defined as max(ρ), provided that Δ(ω) is not a zero function of ω at a fixed temperature.

These equations can be rewritten in a matrix form as [A7]

$$\rho(T)\psi_m = \sum_{n=0}^{N} K_{mn}\psi_n \Leftrightarrow \rho(T)\begin{pmatrix}\psi_1\\ \dots\\ \psi_N\end{pmatrix} = \begin{pmatrix}K_{11} & \dots & K_{1N}\\ \dots & K_{ii} & \dots\\ K_{N1} & \dots & K_{NN}\end{pmatrix} \times \begin{pmatrix}\psi_1\\ \dots\\ \psi_N\end{pmatrix} \quad \text{(A21)}$$

where $\psi_n$ relates to $\Delta(\omega, T)$ and

$$K_{mn} = F(m-n) + F(m+n+1) - 2\mu^* - \delta_{mn}\left[2m + 1 + F(0) + 2\sum_{l=1}^{m} F(l)\right] \quad \text{(A22)}$$

$$F(x) = F(x,T) = 2\int_0^{\omega\max} \frac{\alpha^2 F(\omega)}{(\hbar\omega)^2 + (2\pi k_B T x)^2}\hbar\omega d\omega \quad \text{(A23)}$$

where $\delta_{nn} = 1$ and $\delta_{nm} = 0$ ($n \neq m$), which is a unit matrix. Now we can replace the equation $\rho(T_C) = 0$ with the vanishing of the maximum eigenvalue of the matrix $K_{nm}$: [$\rho$ = max_eigenvalue($K_{nm}$) = $f(T)$, $f(T_C) = 0$].

The calculations within the Migdal–Eliashberg (ME) approach can be done using the full ME equations on the imaginary axis [A5]:

$$\Delta_n = \frac{\pi}{\beta}\sum_{m=-\text{Max}}^{\text{Max}} \frac{\lambda(i\omega_n - i\omega_m) - \mu^*\theta(\omega_c - |\omega_m|)}{\sqrt{\omega_m{}^2 Z_m{}^2 + \Delta_m{}^2}}\Delta_m \quad \text{(A24)}$$

$$Z_n = 1 + \frac{\pi}{\beta\omega_n}\sum_{m=-\text{Max}}^{\text{Max}} \frac{\lambda(i\omega_n - i\omega_m)}{\sqrt{\omega_m{}^2 Z_m{}^2 + \Delta_m{}^2}}\omega_m Z_m \quad \text{(A25)}$$

$$\lambda(z) = \int_0^{\omega_c} \frac{2\alpha^2 F(\omega)}{\omega^2 - z^2}\omega d\omega \quad \text{(A26)}$$

where $\Delta_n$ is the order parameter function, $Z_n$ is the wave function renormalization factor, $\theta(x)$ is the Heaviside function, $\omega_n = \pi k_B T(2n - 1)$ is the $n$th Matsubara frequency, $\beta = k_B T$, $\mu^*$ is the Coulomb pseudopotential, $\lambda(z)$ is the electron–phonon pairing kernel, and $\omega_c$ is the cutoff energy.

In accordance with the Debye model [A8, 9], the Debye temperature $\theta_D$ and both sound velocities were calculated:

$$\theta_D = \frac{h}{k_B}\sqrt[3]{\frac{9n_{f.u.}V_{f.u.}}{4\pi(v_l{}^{-3} + 2v_t{}^{-3})}} = \frac{h\sqrt{N_A}}{k_B}\left\{\sqrt[3]{\frac{9n_{f.u.}}{4\pi}} \cdot \frac{V^{1/6}{}_{f.u.}}{\sqrt{M}} \cdot \frac{\sqrt{GB + (4/3)G^2}}{\sqrt[3]{2(B + (4/3)G)^{3/2} + G^{3/2}}}\right\} \quad \text{(A27)}$$

$$v_l{}^{-3} = \left(\frac{\rho}{B + (4G/3)}\right)^{3/2} \quad \text{(A28)}$$

$$v_t{}^{-3} = \left(\frac{\rho}{G}\right)^{3/2} \quad \text{(A29)}$$

**Experimental details.** To synthesize the La-Y, La-Nd and other alloys, metal ingots and powder were weighed in the stoichiometric proportions in an inert glove box and placed in a $ZrO_2$ crucible filled with toluene to prevent a contact with atmospheric oxygen during heating. The crucible was heated to 1000 °C and kept at this temperature for four hours in an induction furnace in a controlled Ar atmosphere. After cooling, the composition of the melted ingot was analyzed using the XRD, X-ray fluorescence (XRF), and EDS methods. The measurements of the X-ray energy-dispersive spectra (EDS) were performed on FEI Quanta 200 3D scanning electron microscope (SEM) with EDAX Genesys setup.

To load the high-pressure diamond anvil cells (DACs), we took material from the homogeneous regions of alloys with the desired ratio, determined using the EDS and XRF methods. We used the diamond anvils with a 50 μm culet beveled to 300 μm at 8.5°, equipped with four ~200 nm thick Ta electrodes with ~80 nm gold plating that were sputtered onto the piston diamond. Composite gaskets consisting of a rhenium ring and a $CaF_2$/epoxy mixture were used to isolate the electrical leads. Metal pieces with a thickness of ~1–2 μm were sandwiched between the electrodes and ammonia borane $NH_3BH_3$ in the gasket hole with a diameter of 20 μm and a thickness of 10–12 μm. The laser heating of the samples above 1500 K at pressures of 170–180 GPa by several 100 μs pulses led to the formation of binary or ternary metal hydrides whose structure was analyzed using the X-ray diffraction.

The XRD patterns of the polyhydride samples were recorded at various synchrotron beamlines using monochromatic synchrotron radiation and an imaging plate detector at room temperature. The XRD data were analyzed and integrated using Dioptas software package (version 0.5) [A10]. The full profile analysis of the diffraction patterns and the calculation of the unit cell parameters were performed using JANA2006 software [A11] with the Le Bail method [A12]. The pressure in the DACs was determined via the Raman signal of diamond at room temperature [A13, 14].

**Details of the DFT calculations.** The computational predictions of thermodynamic stability of polyhydrides were carried out using the variable-composition evolutionary algorithm USPEX [A15-18]. The first generation typically consisting of 60-100 structures was produced using the random symmetry [A18] and random topology [A19] generators, whereas all subsequent generations (up to 50-80) contained 20% of random structures and 80% of those created using heredity, softmutation, and transmutation operators. The evolutionary searches were combined with structure relaxations using the density functional theory (DFT) [A20, 21] within the Perdew–Burke–Ernzerhof (PBE) generalized gradient approximation (GGA) functional [A22] and the projector augmented wave method [A23, 24] as implemented in the VASP code [A25-27]. The kinetic energy cutoff for plane waves was 600 eV. The Brillouin zone was sampled using Γ-centered *k*-points meshes with a resolution of $2\pi \times 0.05$ Å$^{-1}$. The same parameters were used to calculate the equations of state of the discovered phases. We also calculated the phonon densities of states of the studied materials using the finite displacements method (VASP and PHONOPY [A28, 29]). The calculations of the critical temperature of superconductivity $T_C$ were carried out using Quantum ESPRESSO (QE) package [A30, 31]. The phonon frequencies and electron–phonon coupling (EPC) coefficients were computed using the density functional perturbation theory [A32], employing the plane-wave pseudopotential method and the PBE exchange–correlation functional.

**Details of the SCDFT calculations.** The gap equation in the density functional theory for superconductors (SCDFT) [A33, 34] for evaluating $T_C$ is

$$\Delta_{nk}(T) = -Z_{nk}(T)\Delta_{nk}(T) - \frac{1}{2}\sum_{n'k'} K_{nkn'k'}(T)\frac{\tanh \beta E_{n'k'}}{E_{n'k'}}\Delta_{n'k'}(T) \quad \text{(A30)}$$

with $E_{n'k'} = \sqrt{\xi_{n'k'}^2 + \Delta_{n'k'}^2}$, $\beta = 1/k_BT$, $Z_{nk} = Z_{nk}^{\mathrm{ph}}$, $K_{nkn'k'} = K_{nkn'k'}^{\mathrm{ph}} + K_{nkn'k'}^{\mathrm{el}}$, where the screened Coulomb repulsion is described by $K_{nkn'k'}^{\mathrm{el}}$, the phonon-mediated electron–electron attraction is described by $K_{nkn'k'}^{\mathrm{ph}} = K^{\mathrm{ph}}(\xi_{nk}, \xi_{n'k'})$, and the mass-renormalization by the phonon exchange is described by $Z_{nk}^{\mathrm{ph}} = Z^{\mathrm{ph}}(\xi_{nk})$. The "order parameter" $\Delta_{nk}(T)$, which does not depend on the frequency ω but on the eigenstate of $\widehat{H}_e$ labeled by the band index *n* and crystal wavenumber *k*, is defined in a way different from that in the Eliashberg equation and is proportional to the thermal average $\langle c_{nk\uparrow}c_{n-k\downarrow}\rangle$, with $c_{nk\sigma}$ being the annihilation operator of the spin state *nk*σ [A33]. $\xi_{nk}$ is the energy eigenvalue of state *nk* measured from the Fermi level, as calculated using the standard Kohn–Sham equation for the normal state. $Z_{nk}(T)$ and $K_{nkn'k'}(T)$ represent the electron–phonon and electron–electron Coulomb interaction effects, the formulas for which have been constructed so that the self-energy corrections are almost the same as those in the Eliashberg equations with the Migdal approximation [A5, 35-37]. We calculated the screened electron–electron Coulomb interaction within the random phase approximation [A38], the electronic density of states (DOS) of the normal state was used for solving eq A30. We generated dense $\xi_{nk}$ data points entering eq A30 around $E_F$ by a linear interpolation from the values on a dense k-point mesh. Differently from the Eliashberg equation, frequency ω is not in the SCDFT gap equation. It has been nevertheless shown that the retardation effect [A39] is approximately incorporated [A34]. The absence of ω enables us to treat all electronic states in a wide energy range of about ±30 eV with a feasible computational cost, thanks to which we can estimate $T_C$ without the empirical Coulomb pseudopotential $\mu^*$. We used this gap equation only to estimate $T_C$ as the temperature where its nontrivial solution vanishes.

The *nk*-averaged formulas for $K^{\mathrm{ph}}$ [A34] and $Z^{\mathrm{ph}}$ [A40] were employed to use calculated $\alpha^2F(\omega)$. The *nk*-dependent formula for $K^{\mathrm{ph}}$ [A41] was used, which includes the dynamical and static screening effect on the Coulomb repulsion. The dielectric matrices $\varepsilon(i\omega)$ were represented with the auxiliary plane wave energy cutoff of 12.8 Ry and calculated within the random-phase approximation [A38], and electron–electron Coulomb matrix elements were evaluated as

$$W_{nk,n'k'}(i\omega) = < \psi_{nk}\psi_{n-k}|\varepsilon^{-1}(i\omega)V|\psi_{n'k'}\psi_{n'-k'} > \quad \text{(A31)}$$

where *V* is the bare Coulomb interaction. Low-energy Kohn–Sham states were accurately treated with the weighted random sampling scheme in solving the gap equation. The Kohn–Sham energy eigenvalues and matrix elements for the sampled states were evaluated using the linear interpolation from the data on equal meshes.

**Details of the magnetic calculations.** We explored the magnetic properties of polyhydrides, starting with determining their most stable collinear magnetic ordering — ferromagnetic (FM) or antiferromagnetic (AFM). The AFM configurations were generated using the derivative structure enumeration library enumlib. The converged parameters of our calculations were: 670 eV for the kinetic energy cutoff of the plane wave basis set, $l = 30$ for the automatic generation of Γ-centered Monkhorst–Pack grids as implemented in the VASP code, and the smearing parameter $\sigma = 0.1$ with the Methfessel–Paxton method of order 1. These parameters give us a maximum error of 1 meV/atom with respect to more accurate calculations. The Ising Hamiltonian was used to model the magnetic interactions, considering them only up to third nearest neighbors. We used the Ising Hamiltonian together with the relaxed enthalpies of the FM and AFM configurations to compute the coupling constants $J_{1,2,3}$. The critical temperature of our Heisenberg model was obtained from a Monte Carlo simulation as implemented in VAMPIRE software. The size of the simulation box was 25×25×25 nm after the convergence tests, the transition temperature was taken as the value where the normalized mean magnetization length goes below 0.25.

**Details of the molecular dynamics calculations.** The dynamic stability and phonon density of states of polyhydrides were studied using classical molecular dynamics and the interatomic potential based on machine learning. We used the Moment Tensor Potential (MTP) [A42] whose applicability in calculations of the phonon properties of materials has been demonstrated previously. Moreover, within this approach we can explicitly take into account the anharmonicity of hydrogen vibrations. To train the potential, we first simulated polyhydrides in quantum molecular dynamics in an *NPT*-ensemble at 10, 100, and 300 K, with a duration of 5 ps using the VASP code [A22, 24]. We used the PAW PBE pseudopotentials for H, and $2\pi \times 0.06$ Å$^{-1}$ $k$-mesh with a cutoff energy of 400 eV. For training the MTP, sets of structures were chosen using active learning. We checked the dynamical stability of studied polyhydrides with the obtained MTPs via several runs of molecular dynamics calculations at 300 K. First, the *NPT* dynamics simulations were performed in a supercell with about 1000 atoms for 40 ps. During the last 20 ps, the cell parameters were averaged. In the second step, the coordinates of the atoms were averaged within the *NPT* dynamics with a duration of 20 ps and the final structures were symmetrized as implemented in T-USPEX method. Then, for the structures of polyhydrides relaxed at 10, 100, and 300 K, the phonon density of states (DOS) was calculated within the MTP using the velocity autocorrelator (VACF) separately for each type of atoms:

$$g(\theta) = 4\int_0^{\infty} \cos(2\pi\theta t)\frac{\langle V(0)V(t)\rangle}{\langle V(0)^2\rangle}\mathrm{d}t \qquad \text{(A32)}$$

where $V$ is the velocity of atoms. The calculations were carried out in a 20×20×20 supercell. The velocity autocorrelator was calculated using molecular dynamics, then the phonon DOS was obtained.

The calculation of the anharmonic correction to the critical temperature of superconductivity is based on the constant DOS approximation [A43], which is applicable if the density of electronic states $N(E)$ in the vicinity of the Fermi energy has no pronounced features (e.g., Van Hove singularities) and can be approximately represented as a constant. In this case, the main contribution to the shape of the Eliashberg function is made by the phonon density of states (denoted here for convenience as $F(\omega)$), and an auxiliary function of the electron–phonon matrix elements contribution $\alpha^2(\omega) = \alpha^2F(\omega)/F(\omega)$ is close to a constant. The exceptions are those frequency zones where both functions $\alpha^2F(\omega)$ and $F(\omega) \sim 0$, and where the calculation of their ratio is not sufficiently stable and accurate, as well as in the acoustic frequency region.

The procedure for calculating the Eliashberg "anharmonic" functions of polyhydrides consisted of several simple steps:

**a.** Calculation of the harmonic Eliashberg function and phonon density of states in Quantum ESPRESSO [A30] using the tetrahedral method [A44].

**b.** Calculation of the anharmonic phonon spectrum using the molecular dynamics at a temperature near the superconducting transition ($T_C$).

**c.** Equalization of a unit cell ($N$, the number of atoms in a formula unit) by calibration of integrals

$$\int F_{\text{harm}}(\omega)d\omega = \int F_{\text{anh}}(\omega)d\omega = 3N$$

**d.** Calculation of the contribution of the electron density of states to the Eliashberg function for the harmonic case: $\alpha^2(\omega) = \alpha^2F(\omega)/F(\omega)$.

**e.** Reduction of the function $\alpha^2(\omega) \geq 0$ to the same mesh on the energy scale that was used to calculate the anharmonic phonon spectrum by linear interpolation. All obtained negative values $\alpha^2(\omega) < 0$ were set to zero. Smoothing the obtained $\alpha^2(\omega)$ to eliminate the influence of "empty" frequency zones where both $\alpha^2F(\omega)$ and $F(\omega) \sim 0$.

**f.** Calculation of the Eliashberg anharmonic function as the product of the electronic contribution and the anharmonic density of phonon states $\alpha^2(\omega)F_{\text{anh}}(\omega)$.

**Details of the upper critical magnetic field calculations.** As already mentioned, the upper critical field depends significantly on the quality of the sample. Different research groups obtain different estimates of $\mu_0H_{C2}(0)$ for the same polyhydrides. Therefore, an exact calculation of $\mu_0H_{C2}(0)$ has no practical importance. Nevertheless, some estimating models can be suggested.

In 2018-2021, our group in Moscow actively used the interpolation formula (A8) to estimate $\mu_0H_{C2}(0)$. As Table A3 and Figure A3 show, this formula gives underestimated values and is applicable only for polyhydrides with small $T_C < 100$ K and $\mu_0H_{C2}(0) < 60$ T:

$$\frac{\gamma' T_C^2}{(\mu_0 H_{C_2}(0))^2} = 0.168\left[1 - 12.2\left(\left(\frac{T_C}{\omega_{log}}\right)^2 ln\left(\frac{\omega_{log}}{3T_C}\right)\right)\right]. \quad \text{(A8)}$$

J. Carbotte [A2] proposed formulas to estimate $\mu_0H_{C2}(0)$ within an ideal crystal ("clean" limit, A34) and within a heavily contaminated sample ("dirty" limit, A35). The problem of using these formulas is that they require knowledge of $V_F$ – the averaged Fermi velocity, EPC coefficient λ, and the electron mean free path. Moreover, the dependence on these parameters is very strong, which leads to a large uncertainty in the calculated value of $\mu_0H_{C2}(0)$.

In the "clean" limit, Ref. [A2], eq. 7.17:

$$H_{C2}^{cl}(0) = \left(\frac{\pi}{2}\right)^2 e^{-\gamma+2}\frac{2T_C^2(1+\lambda)^2}{eV_F^2}\left(1 + 1.44\frac{T_C}{\omega_{log}}\right), \quad \text{(A34)}$$

where $\gamma = 0.577$, e – is the absolute electron charge, $V_F$ – the averaged Fermi velocity. For a rough estimate, we always used $V_F = 5\times10^5$ m/s (see also [A45]).

In the dirty limit, Ref. [A2], eq. 7.17

$$H_{C2}^{di}(0) = \frac{\pi}{2}e^{-\gamma}\frac{T_C(1+\lambda)}{eD}\left(1 + 3.3\frac{T_C}{\omega_{log}} - 4.8\left(\frac{T_C}{\omega_{log}}\right)^2 \ln\left(\frac{\omega_{log}}{T_C}\right)\right), \quad \text{(A35)}$$

where the diffusion coefficient $D = 1/3\ V_F^2\times\tau$, and τ – the electron scattering time on impurities and defects.

**Table A1.** Parameters of the superconducting state of pure metals.Fixed the Fermi velocity $V_F = 5\times10^5$ m/s averaged over BZ was used to calculate $\mu_0 H_{C2}(0)$ via eq. A34 and A36.

| Element | λ [Ref. A46] | $\omega_{log}$, K | $T_C$, K | Exp. $\mu_0 H_{C2}(0)$, T | Predicted $\mu_0 H_{C2}(0)$, T |
|---|---|---|---|---|---|
| Nb [A47] | 1.05 | 229.0 | 9.27 | 0.425 | 0.537 |
| V [A47] | 0.83 | 330.0 | 5.385 | 0.302 | 0.140 |
| Sn [A48] | 0.72 | 165.0 | 3.72 | 0.031 | 0.059 |
| Ta [A48] | 0.73 | 213.0 | 4.48 | 0.083 | 0.087 |

Comparison with experimental data for pure metals (Table A1) allows us to refine the proportionality factor in eq. A34 provided that $V_F = 5\times10^5$ m/s:

$$H_{C2}^{cl}(0) = 1.4061 \times 10^{-3} T_C^2 (1+\lambda)^2 \left(1 + 1.44 \frac{T_C}{\omega_{log}}\right), \qquad \text{(A36)}$$

where $T_C$ and $\omega_{log}$ are in Kelvins, $H_{C2}$ – in Tesla. The comparison (Table A1) also shows that a qualitative agreement with the experimental data for pure metals can be achieved. However, the same formula for polyhydrides gives extremely overestimated results, so a different proportionality factor should be chosen for hydrides:

$$H_{C2}^{cl}(0) = 1.855 \times 10^{-4} T_C^2 (1+\lambda)^2 \left(1 + 1.44 \frac{T_C}{\omega_{log}}\right), \qquad \text{(A37)}$$

Comparison with experimental data for polyhydrides (Figure A3) shows satisfactory agreement between model and experiment at high $T_C$ and $\mu_0 H_{C2}(0)$, which is further illustrated in Tables A2, A3, and worse agreement at low critical superconductivity temperatures (below 100 K), where $\mu_0 H_{C2}(0)$ follows a "linear" ($H_{C2}(0) \sim T$) mode. Formula A37 allows further simplification to

$$H_{C2}^{cl}(0) = 2.45 \times 10^{-4} T_C^2 (1+\lambda)^2. \qquad \text{(A38)}$$

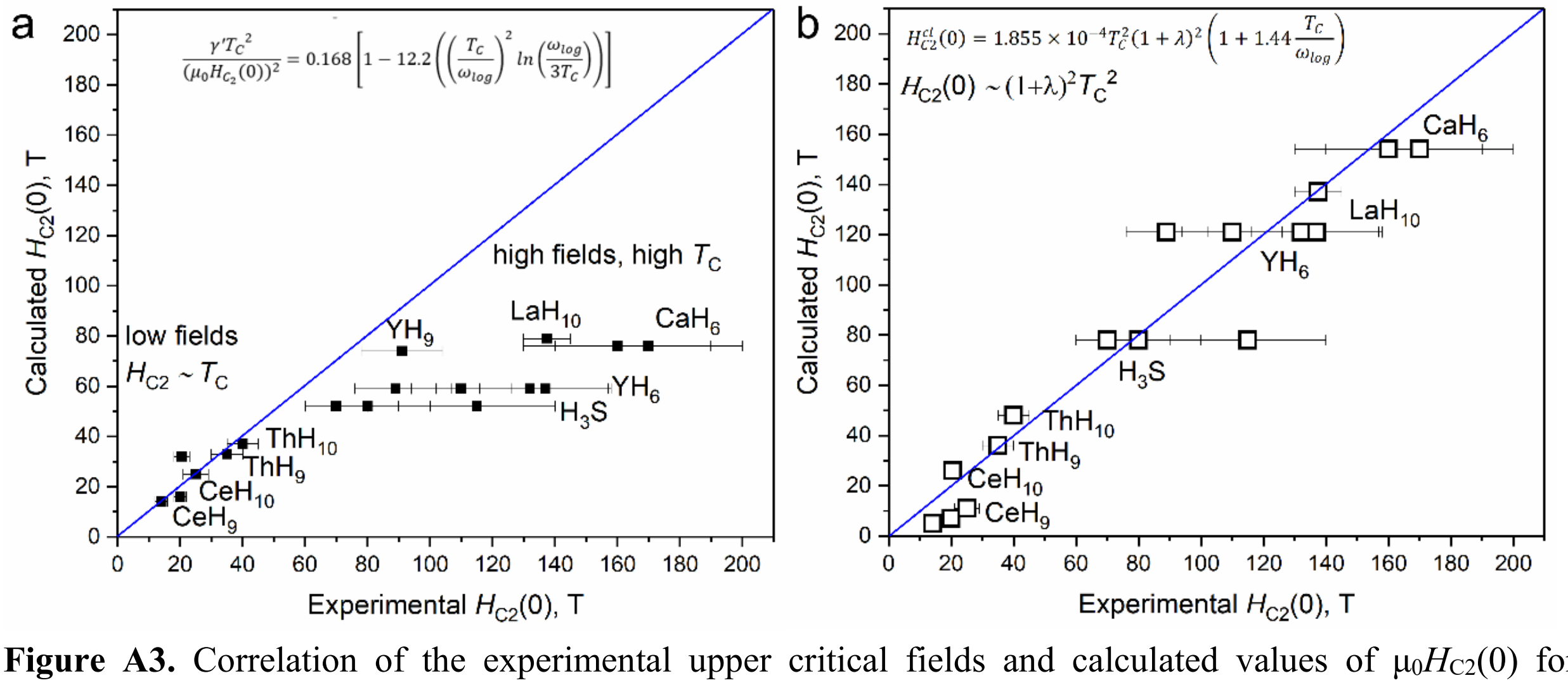

**Figure A3.** Correlation of the experimental upper critical fields and calculated values of $\mu_0 H_{C2}(0)$ for superhydrides obtained using two models: a) linear $H_{C2}(0) \sim N_F^{0.5}(1+\lambda)^{0.5} T_C$ and b) quadratic $H_{C2}(0) \sim [(1+\lambda)T_C]^2$. The data are taken from Table A2.

**Table A2.** Parameters of the superconducting state of polyhydrides. Fixed the Fermi velocity $V_F$ = $5\times10^5$ m/s averaged over BZ was used to calculate $\mu_0 H_{C2}(0)$ via eq. A37 and A38.

| Compound | λ* | $\omega_{log}$, K | $T_C$, K | Exp. $\mu_0 H_{C2}(0)$, T | Predicted $\mu_0 H_{C2}(0)$, T (eq. A37) | Predicted $\mu_0 H_{C2}(0)$, T (eq. A38) |
|---|---|---|---|---|---|---|
| $H_3S$ | 1.84 | 1080.0 | 203.0 | 60-80<br>60-100 | 78 | 81 |
| $LaH_{10}$ | 2.06 | 1340.0 | 250.0 | 90-140<br>130-145 | 137 | 143 |
| $YH_9$ | 2.75 | 885.0 | 243.0 | 78-105 | 215[a] | 203[a] |
| $YH_6$ | 2.24 | 1330.0 | 224.0 | 107-157<br>76-102<br>94-126<br>116-158 | 121 | 129 |
| $ThH_{10}$ | 1.91 | 1210.0 | 161.0 | 35-45 | 48 | 54 |
| $ThH_9$ | 1.73 | 960.0 | 146.0 | 30-40 | 36 | 39 |
| $YH_4$ | 1.10 | 1080.0 | 88.0 | 18-22 | 7.0[b] | 8 |
| $CeH_9$ | 1.46 | 650.0 | 90.0 | 21-29 | 11[b] | 12 |
| $CeH_{10}$ | 2.0 | 1000.0 | 115.0 | 18-23 | 26 | 29 |
| $SnH_x$ | 1.24 | 890.0 | 72.0 | 12-16 | 5.4[b] | 6 |
| $CaH_6$ | 2.69 | 950.0 | 215.0 | 140-200<br>130-190 | 155 | 154 |

[a] discrepancy $H_{exp}$ and $H_{theory}$ means that λ($YH_9$) should be less than 2.75.

[b] this model works not so good for low $T_C$ < 100 K.

Finally, another possibility is to use GLAG theory and sample resistivity values to estimate the average collision frequency. This formula was discussed earlier:

$$H_{C2}^{GLAG}(0) = \text{const} \times T_C \frac{(1+\lambda)\cdot R(T_C)\cdot h\cdot N_F}{V_{cell}}. \quad \text{(A39)}$$

To compare the calculated values of the upper critical field with the experimental $\mu_0 H_{C2}(0)$, we prepared Tables A4, A5. As the estimates of the resistance and thickness (h) of the samples show, the constant in the formula A39 can be chosen ≈ 11. However, uncertainties in sample thickness for $LaH_{10}$ [A56] and $H_3S$ [A50], as well as large variations in the resistivity of hydride samples in different experiments (from 0.01 Ω to 15 Ω), make impossible any accurate determination of the upper critical field within the GLAG model (Table A5).

**Table A3.** Superconducting properties of metal polyhydrides. The upper critical field data were taken from the relevant literature, only the GL and WHH models for $\mu_0 H_{C2}(0)$ were used. The last column of the table is obtained using formula A8. All superconducting state parameters of polyhydrides taken from the literature should be regarded as approximate.

| **Compound** | **Experimental pressure, GPa** | **Estimated $T_C$, K** | **Experimental $T_C$, K** | **EPC parameter (λ)** | **$\omega_{log}$, K** | **N(0), $eV^{-1}$/metal** | **Upper critical magnetic field, T (experiment)** | **Upper critical magnetic field, T (theory, A8)*** |
|---|---|---|---|---|---|---|---|---|
| $Im\bar{3}m$-$H_3S$ | 150 | 200 [A49] | 203 [A50] | 1.84 [A51] | 1080[b] | 0.63 [A52] | 60-80 [a] [A50]<br>60-100 [A53] | 52 |
| $Fm\bar{3}m$-$LaH_{10}$ | 160 | 286 [A54, 55] | 250 [A56] | 2.06 [A57] | 1340[b] | 0.88 [A58] | 90-140 [A56]<br>130-145 [A59] | 79 |
| $P6_3/mmc$-$YH_9$ | 200 | 303 [A55, 60] | 243 [A61] | 2.75 | 885 | 0.73<br>[200 GPa, this work] | 78-105 [this work] | 74 |
| $Im\bar{3}m$-$YH_6$ | 170 | 270 [A62] | 224 [A63] | 1.71 [A63] | 1330[b] | 0.71 [A63] | 107-157 [A61]<br>76-102 [this work]<br>94-126 [this work]<br>116-158 [A63] | 59 |
| $Fm\bar{3}m$-$ThH_{10}$ | 170 | 160–193 [A64] | 161 [A64] | 1.91 | 1210 | 0.53 | 35-45 [A64] | 37 |
| $P6_3/mmc$-$ThH_9$ | 150 | 145-161 [A64] | 146 [A64] | 1.73 | 960 | 0.52 [A64] | 30-40 [A64] | 33 |
| $I4/mmm$-$YH_4$ | 155 | 74-94 [A65] | 88 | 1.10 | 1080 | 0.475 | 18-22 [A65] | 16 |
| $P6_3/mmc$-$CeH_9$ | 110 | 117 [A66, 67] | ~90 [A68] | 1.46 [A68] | 650 | 0.92 | 21-29 [A68] | 25 |
| $Fm\bar{3}m$-$CeH_{10}$ | 100 | 168 [A69] | ~115 [A68] | 2.0 | 1000 | 0.75 [A68] | 18-23 [A68] | 32 |
| $c$-$SnH_x$ | 190 | 81–97 [A70] | 71 [A71] | 1.24 [A72] | 890 | 0.46 | 12-16 [A72] | 14 |
| $Im\bar{3}m$-$CaH_6$ | 170 | 220–235 [A73] | 215 [A74] | 2.69 | 950 | 0.91[A73] | 140-200 [A74]<br>130-190 [A75] | 76 |

a – linear extrapolation gives $H_{C2}(0) > 100$ T for $H_3S$.

b – anharmonic calculations (SSCHA)

**Table A4.** Experimental and calculated parameters of the normal state, resistive transitions, and superconducting state in various polyhydrides. The results of the estimation of the sample thickness (h), corrections to the $T_C$ arising from the anisotropy of the superconducting gap ($T_C^{aniso} - T_C^{iso}$) and the anharmonism of hydrogen sublattice oscillations ($T_C^{harm} - T_C^{anharm}$) are also given.

| Compound | R(300 K)/ R($T_C$) [*dR/dT*], mΩ / K | Debye temp. ($\theta_D$), K | Exp. resistance in normal state R($T_C$) and in SC state [R(0)], Ω | Resistivity in normal ($\rho(T_C)$) and SC state μΩ·cm, [thickness of sample, μm] | $\lambda_{BG}$ [$\omega_{log}$, K] from R(T) | $T_C$ and the width of the SC transition [$\Delta T_C$], K | Broadening of SC transition ($d\Delta T_C/dH$) | Anisotropy ($T_C^{aniso} - T_C^{iso}$), K ($T_C$, SCDFT, K) [$T_C^{harm} - T_C^{anharm}$, K] | ME analysis (QE calcs.) λ, ($\omega_{log}$) [$T_C$(E), μ*=0.1] | N(0), $eV^{-1}$/metal |
|---|---|---|---|---|---|---|---|---|---|---|
| $YH_6$ (166 GPa) | 1.2 [*0.131*] At 200 GPa: 1.5 [*3.7*] | 1200 ± 25 | 0.05 [21×$10^{-6}$] | [< 12] | 2.71 [992] | 224-227 [4] At 200 GPa: 215 [8] | No (166 GPa) 0.31 (213 GPa) | - (181) [23] | 1.71 (1333) [247] | 0.71 |
| $YH_9$ (205 GPa) | 1.33 [*1.17*] | 1275 ± 10 At 213 GPa: 1160 ± 10 | 0.225 [16×$10^{-6}$] | 12 – EPW 1.6×$10^{-3}$ – SC [1] | 2.66 [1093] At 213 GPa: [990] | 235 [5] | no | 21 (246) [-16 ?] | 2.75 (884) [231] | 0.73 |
| $LaYH_{20}$ (180 GPa) | 1.05-1.08 [*0.39*] | 1022 ± 5 | 0.2-0.4 [6-9×$10^{-6}$] | 48-96 7.2-21.6×$10^{-4}$ – SC [1.2-2.4] | 3.63 [845] | 253 [20] | no | - (252) [19] | 3.87 (850-1050) [281-300] | 0.9 |
| $ThH_{10}$ (170 GPa) | 1.35 [0.046] | 1350 ± 5 1466-1500 [A76] | 0.06-0.08 [10-14×$10^{-6}$] | 20-30 4×$10^{-3}$ – SC [3-4] | 1.68-1.71 [1160] | 161 [3] | 0.15 | - (207) [-] | 1.97 (1210) [214] | 0.53 |
| $ThH_9$ (150 GPa) | 1.65 [*0.23*] | 1453 ± 13 [A76] | 0.055 [20×$10^{-6}$] | 16-22 40-45 – EPW 6-8×$10^{-3}$ – SC [3-4] | 1.46 [1200] | 146 [6] | - | - (184) [-] | 1.47 (1186) [161] | 0.52 |

**Table A5.** Polyhydride samples' parameters for use in the GLAG model for the upper critical magnetic field (eq. A39).

| Compound | λ | $R(T_C)$, Ω | $h^a$, μm | N(0), $eV^{-1}$/metal | $V_{cell}$, $Å^3$/metal | $T_C$, K | $H_{C2}(0)$(GLAG), T | Exp. $H_{C2}(0)$, $T^b$ |
|---|---|---|---|---|---|---|---|---|
| $YH_6$ (166 GPa) | 1.71 | 0.05 | < 12* | 0.71 | 22.9 | 224 | 124 | 132 |
| $YH_9$ (205 GPa) | 2.75 | 0.225 | 1 | 0.73 | 26.3 | 243 | 63 | 91 |
| $LaYH_{20}$ (180 GPa) | 3.87 | 0.3 | 1.8 | 0.9 | 32 | 253 | 206 | 117 |
| $ThH_{10}$ (170 GPa) | 1.97 | 0.07 | 3.5 | 0.53 | 32.5 | 161 | 21 | 40 |
| $ThH_9$ (150 GPa) | 1.47 | 0.055 | 3.5 | 0.52 | 32.5 | 146 | 12 | 35 |
| $H_3S$ (150 GPa) | 1.84 | 0.035 [A50] | <10 | 0.63 | 14.8 | 203 | 94 | 80 |
| $LaH_{10}$ (150 GPa) | 2.06 | 0.11 [A56] | < 10 | 0.88 | 33.2 | 250 | 245 | 115 |

[a] thickness of gasket, μm

[b] average value

On the other hand, this GLAG-type model can be simplified by eliminating the product $R(T_C)\times h$, which is not exactly known from the experiment:

$$H_{C2}^{GLAG}(0) = 4.22 \times T_C \frac{(1+\lambda)\cdot N_F}{V_{cell}}, \qquad \text{(A40)}$$

where $T_C$ is in K, $V_{cell}$ – in $Å^3$/metal atom, $N_F$ – is the density of electron states at the Fermi level, in states/eV/metal atom. Table A6 shows that this simplified model may be used to describes the experimental data.

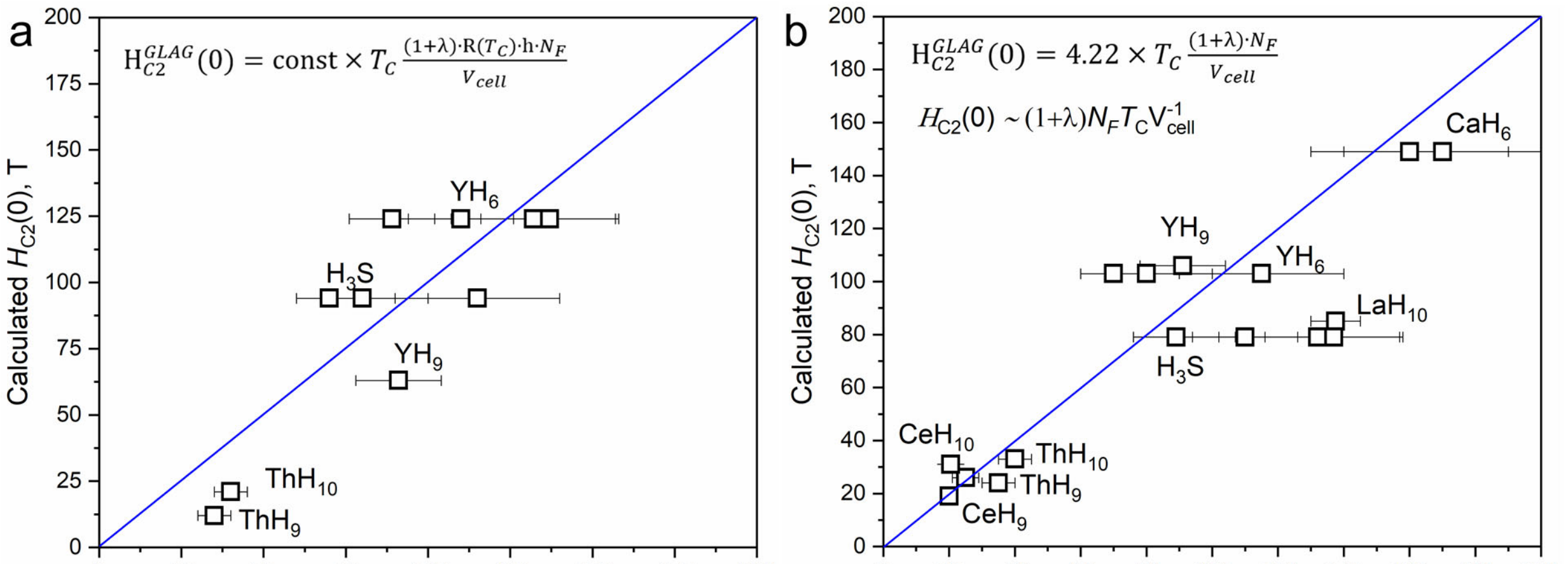

**Figure A4.** Correlation of the experimental upper critical fields and calculated values of $\mu_0 H_{C2}(0)$ for superhydrides obtained using two GLAG-type models: a) full $H_{C2}(0) \sim \rho N_F(1+\lambda)V^{-1}T_C$ and b) simplified $H_{C2}(0) \sim N_F(1+\lambda)V^{-1}T_C$. The data are taken from Tables A5-6.

**Table A6.** Polyhydride samples' parameters for use in the simplified GLAG model for the upper critical magnetic field (eq. A40).

| Compound (Pressure) | λ | Stabilization pressure, GPa[a] | N(0), $eV^{-1}$/metal | $V_{cell}$, $Å^3$/metal | $T_C$, K | Calc. $H_{C2}(0)$, T | Exp. $H_{C2}(0)$, T* |
|---|---|---|---|---|---|---|---|
| $YH_6$ (166 GPa) | 1.71 | 120 [A77] | 0.71 | 22.9 | 224 | 79 | 132 |
| $YH_9$ (205 GPa) | 2.75 | 150 [A55] | 0.73 | 26.3 | 243 | 107 | 91 |
| $LaYH_{20}$ (180 GPa) | 3.87 | ~ 180 [A78] | 0.9 | 32 | 253 | 146 | 117 |
| $ThH_{10}$ (170 GPa) | 1.97 | 100 [A79] | 0.53 | 32.5 | 161 | 33 | 40 |
| $ThH_9$ (150 GPa) | 1.47 | 100 [A64] | 0.52 | 32.5 | 146 | 24 | 35 |
| $H_3S$ (150 GPa) | 1.84 | 103 [A80] | 0.63 | 14.8 | 203 | 103 | 80 |
| $LaH_{10}$ (150 GPa) | 2.06 | 129 [A57] | 0.88 | 33.2 | 250 | 85 | 115 |
| $YH_4$ (155 GPa) | 1.1 | 150 [A81] | 0.475 | 19.5 | 88 | 19 | 20 |
| $CeH_9$ (110 GPa) | 1.46 | 100 [A67] | 0.92 | 32.5 | 90 | 31 | 25 |
| $CeH_{10}$ (100 GPa) | 2.0 | 92 [A69] | 0.75 | 35 | 115 | 31 | 20.5 |
| $CaH_6$ (170 GPa) | 2.69 | 150 [A73] | 0.91 | 20.5 | 215 | 149 | 170 |

[a] Minimum values, mostly found by calculations of phonon spectra

[b] Averaged value

Let us touch upon the influence of the hydrogen sublattice contribution on the total density of electron states at the Fermi level $N_F$. Several typical examples are shown in Figure A5. In all high-temperature superconducting hydrides, the hydrogen contribution to $N_F$ is significant and can sometimes exceed the contribution from metal. In pure metals and lower hydrides, the hydrogen contribution $N_F$(H) is very small, therefore we can call the criterion $N_F$(H) ~ $N_F$(metal) necessary for realization of high-temperature superconductivity in polyhydrides. However, this criterion is not sufficient: as the example of $BaH_{12}$ shows (Figure A5f), where the hydrogen contribution far exceeds that of the metal, $N_F$(H) itself does not lead to high-temperature superconductivity.

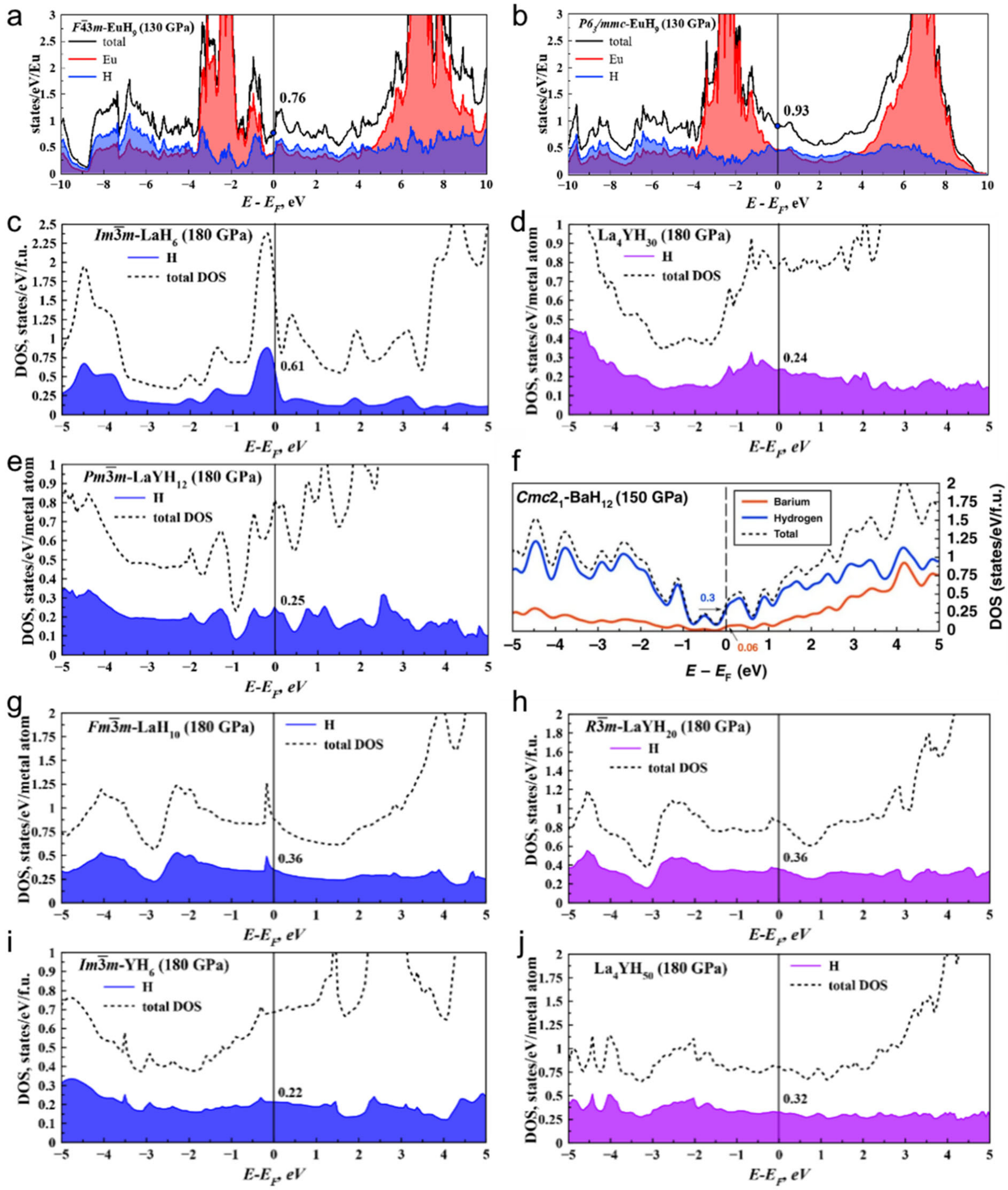

**Figure A5.** Total electron density of various metal hydrides (black line) and the contribution of the hydrogen sublattice (blue and purple). (a, b) Europium hydrides $EuH_9$; (c, g) lanthanum hydrides $LaH_{10}$ and $LaH_6$; (d, e, h, j) lanthanum-yttrium polyhydrides; (f) $BaH_{12}$; (i) $YH_6$.

An important question in the theory of superconductivity is the effect of atomic mass on the superconducting properties of pure elements. According to the Bardeen–Cooper–Schrieffer theory, the critical temperature of elements is proportional to the average phonon energy, which is inversely proportional to the square root of the mass of the atom $M$. However, as seen from Figure A6a (see also [A78]), there is no correlation at ambient pressure between the critical temperature of metals and their atomic mass: the correlation coefficient between $T_C$ and $1/\sqrt{M}$ is –0.25, which means that there is no dependence. However, we know that the mass of the atom is not the only factor that affects $T_C$.

We must also consider the counterbalancing effect of the electron–phonon interaction coefficient λ. A large atomic mass leads to emergence of low-frequency phonon modes and an increase in the electron–phonon interaction constant as $\lambda \sim \omega^{-1}$, then $T_C \propto \omega \cdot \exp(-a\omega)$ or, in other words, $T_C \propto M^{-\frac{1}{2}}\exp(-bM^{-\frac{1}{2}})$, where $a, b$ - some coefficients. This allows us to assume the existence of some optimal value of the atomic mass in terms of achieving maximum $T_C$.

The situation changes when we include pressure, which allows us to "tune" the electronic structure of matter to achieve optimal electron–phonon coupling. For many elements, the maximum critical temperature of superconductivity is reached at high pressures of tens or hundreds of gigapascals (Ca, S, La, Sc, etc.; see Figure 45). However, for many other elements, $dT_C/dP < 0$, and to increase $T_C$, it is necessary to consider the region of formally "negative" pressures (increased unit cell parameters), which can be studied by introducing noninteracting atoms into the lattice of such metals (Figure 6A). When using the available data (e.g., from Figure 45) for the maximum critical temperature of superconductivity of pure elements at elevated pressures, the correlation coefficient $\max_P T_C(P) - 1/\sqrt{M}$ increases to 0.25–0.34, which is quite significant. However, the decisive influence is exerted by $T_C$ of metallic hydrogen. If we take max $T_C(H) = 200$ K, as for $H_3S$, the correlation coefficient increases to 0.89. This means that there is a significant correlation between $\max_P T_C(P)$ and $1/\sqrt{M}$, as we know from the BCS theory.

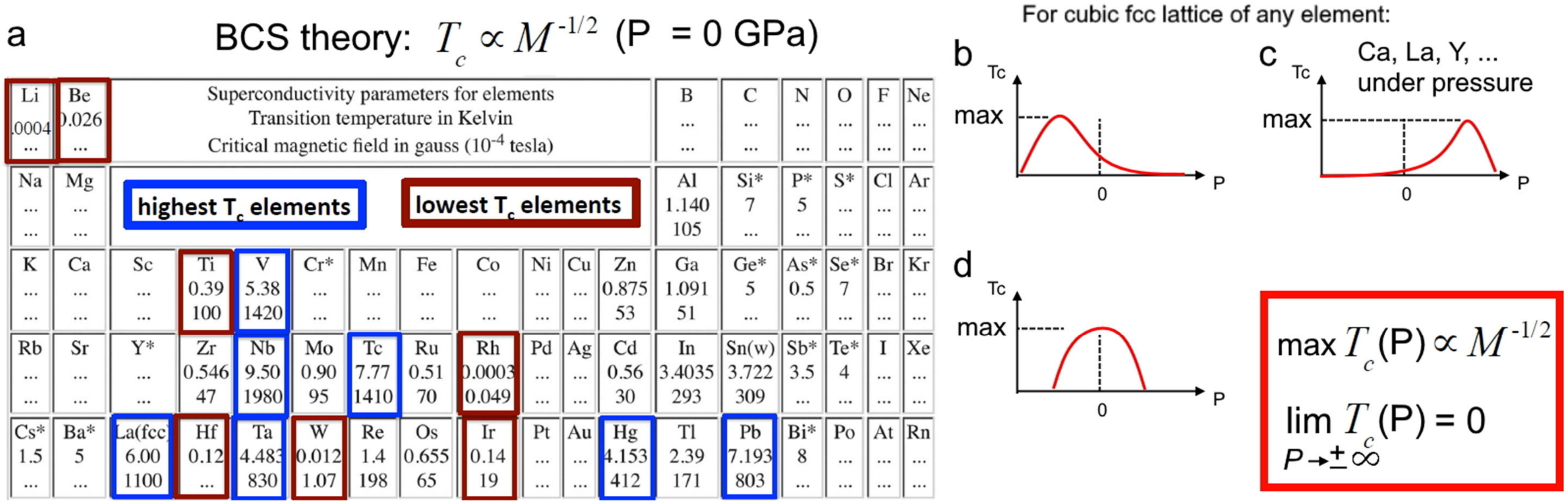

**Figure A6.** (a) Superconducting properties of metals as a function of atomic mass at ambient pressure [A78]. (b, c, d) Schematic representation of possible variants of the dependence of the critical temperature $T_C$(P) on pressure (P).

Another interesting issue is the comparison of dynamic and mechanical stability criteria for polyhydrides. It is known that in harmonic approximation imaginary phonon modes in the structure of *fcc* $LaH_{10}$ appear already at 210-230 GPa [18, 127], while experimental data indicate that cubic $LaH_{10}$ is stable above 140 GPa [175]. We used the mechanical stability criteria (Born's criteria) [1], to investigate the question of the $LaH_{10}$ stabilization pressure. We have found that the mechanical stability criteria give a much better approximation ($P_{min} \approx 152$ GPa) to the experimental data ($P_{min} = 140$ GPa) than the criterion for the appearance of imaginary modes in the harmonic approximation. This work was done together with Daniil Poletaev (Skoltech).

In Table A9, the elastic constants of $LaH_{10}$ at 0 K and six pressures (180, 130, and 100 GPa) calculated from the stress-strain relationships in the cubic unit cell $La_4H_{40}$ as implemented in the thermo-pw package [A83-A85] are shown. The convergence of elastic constants with respect to the k-point density was tested at 130 GPa on $12^3$, $16^3$, $24^3$ , $30^3$, $36^3$, and $40^3$ gamma-centered k-meshes. It was found that calculated moduli become converged with maximum uncertainty less than 1% with $36^3$ k-mesh. In the future it will be necessary to find out whether this is a special case or the mechanical stability criteria always gives more correct estimates of the stability region of polyhydrides.

**Table A7.** Dependence of the screened Coulomb potential ("pseudopotential") on the Debye temperature at $\mu = 0.4$, $E_F = 5$ eV.

| $\omega_D$, K | $\mu^*$ |
|---|---|
| 300 | 0.13 |
| 600 | 0.14 |
| 900 | 0.15 |
| 1200 | 0.16 |

**Table A8.** Isotope effect in various polyhydrides under pressure.

| Compound | Isotope coefficient (α) |
|---|---|
| $H_3S$ | 0.22-0.31 [A82]<br>(theory: 0.35-0.4) [A80] |
| $LaH_{10}$ | 0.46 [A56] |
| $YH_6$ | 0.39-0.4 [A61, 63] |
| $YH_9$ | 0.5 [A61] |
| $CeH_9$ | 0.49 [A68] |

**Table A9.** Zero-kelvin elastic constants calculated with $36^3$ k-mesh in the cubic unit cell of $La_4H_{40}$ at 180, 170, 160, 150, 140 and 130 GPa within the stress-strain approach.

| Pressure, GPa | $C_{11}$, GPa | $C_{12}$, GPa | $C_{44}$, GPa |
|---|---|---|---|
| 180 | 1205.8 | 310.7 | 59.5 |
| 170 | 1157.4 | 295.6 | 34.8 |
| 160 | 1117.1 | 276.8 | 15.3 |
| 150 | 1069.8 | 261.8 | -2.0 |
| 140 | 1024.1 | 246.8 | -24.4 |
| 130 | 979.7 | 228.4 | -11.0 |

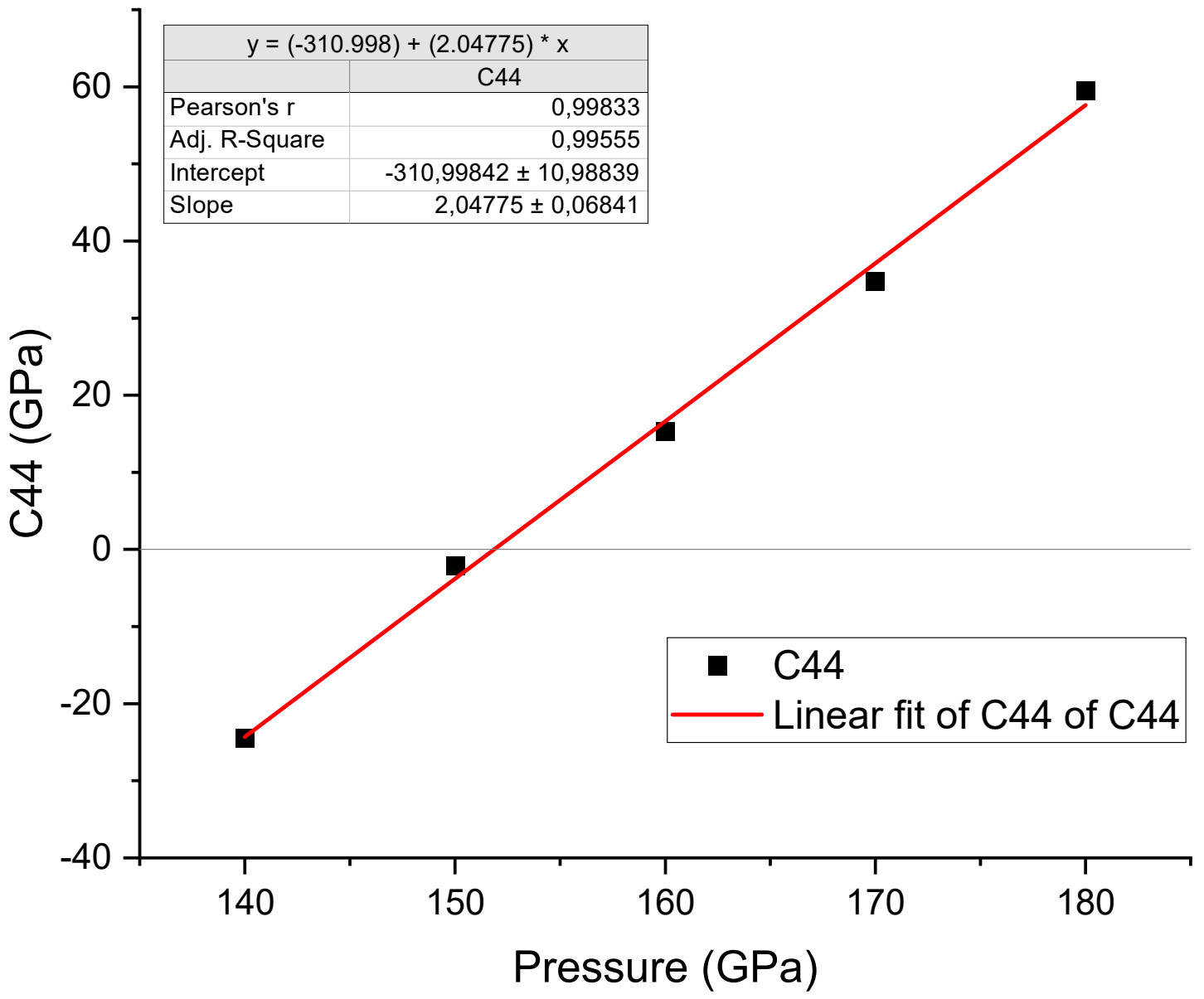

**Figure A7.** The linear fit of $C_{44}$ module of $LaH_{10}$ dependence on pressure. The predicted onset of mechanical instability is at 152 GPa.